\documentclass[useAMS,usenatbib]{mn2e}
\usepackage{times}
\usepackage[T1]{fontenc}
\usepackage{aecompl}
\usepackage{graphicx}
\usepackage{subfigure}
\usepackage{ amssymb }
\usepackage{ gensymb }
\usepackage{rotating}
\usepackage{pdflscape}

\graphicspath{{figures/}}

\title[Temperaments of young stars] {Temperaments of young stars: Rapid mass-accretion rate changes in T Tauri and Herbig Ae stars}
\author[Costigan, Vink, Scholz, Ray, Testi]{G. Costigan$^{1,2,3}$\thanks{E-mail:costigag@gmail.com}, Jorick S. Vink$^{2}$, A. Scholz$^{1,4}$, T. Ray$^{1}$, L. Testi$^{3,5,6}$ \\
$^{1}$ School of Cosmic Physics, Dublin Institute for Advanced Studies, 31 Fitzwilliam Place, Dublin 2, Ireland \\
$^{2}$ Armagh Observatory, College Hill, Armagh, BT61 9DG, Northern Ireland \\
$^{3}$ European Southern Observatory, Karl-Schwarzschild-Str. 2, 85748 Garching bei M\"unchen, Germany \\
$^{4}$ School of Physics \& Astronomy, University of St. Andrews, North Haugh, St Andrews, KY16 9SS, Scotland\\
$^{5}$ INAF-Osservatorio Astrofisico di Arcetri, Largo E. Fermi, I-50125 Firenze, Italy \\
$^{6}$ Excellence Cluster Universe, Boltzmannstr. 2, D-85748, Garching, Germany \\
}

\begin{document}

\date{ }

\pagerange{\pageref{firstpage}--\pageref{lastpage}} \pubyear{2012}

\maketitle

\label{firstpage}

\begin{abstract}

Variability in emission lines is a characteristic feature in young stars and can be used as a tool to study the physics of the accretion process. Here we present a study of H$\alpha$ variability in 15 T Tauri and Herbig Ae stars (K7\,-\,B2) over a wide range of time windows, from minutes, to hours, to days, and years. 
We assess the variability using linewidth measurements and the time series of line profiles. 
All objects show gradual, slow profile changes on time-scales of days. In addition, in three cases there is evidence for rapid variations in H$\alpha$ with typical time-scales of 10\,min, which occurs in 10\% of the total covered observing time. The mean accretion-rate changes, inferred from the line fluxes,are 0.01\,-\,0.07\,dex for time-scales of $<1$\,hour, 0.04\,-\,0.4\,dex for time-scales of days, and 0.13\,-\,0.52\,dex for time-scales of years. 

In \cite{2012MNRAS.427.1344C} we derived an upper limit finding that the intermediate (days) variability dominated over longer (years) variability. 
Here our new results, based on much higher cadence observations, also provide a {\it lower} limit to accretion-rate variability on similar time-scales (days), thereby constraining the accretion rate variability physics in a much more definitive way. 
A plausible explanation for the gradual variations over days is an asymmetric accretion flow resulting in a rotational modulation of the accretion-related emission, although other interpretations are possible as well. 
In conjunction with our previous work, we find that the time-scales and the extent of the variability is similar for objects ranging in mass from $\sim 0.1$ to $\sim$5$\,M_{\odot}$. This confirms
that a single mode of accretion is at work from T Tauri to Herbig Ae stars -- across a wide range of stellar masses.

\end{abstract}

\begin{keywords}
accretion, accretion discs — stars: variables: T Tauri - star: formation - stars: pre-main-sequence
\end{keywords}

\section{Introduction}

Accretion is a vital and central process in the formation of low mass stars, controlling the flow of angular momentum 
and mass from the interstellar medium (ISM) onto the star. 
The accretion history can have a large effect on the long-term properties of the stellar system, such as its luminosity, radius, mass, and rotation 
rate \citep{2009ApJ...702L..27B}. It is also thought to be strongly connected to the mechanisms of wind and jet 
launching \citep{2007prpl.conf..231R}, and of disc clearing and evolution \citep{2005ApJ...625..906M}. 
In order to understand these processes we need to gain a full understanding of accretion. 

The current model for this process is magnetospheric accretion where the stellar magnetic field 
threads the disc, and the material in the disc falls along the field lines onto the surface of the 
star \citep{1991ApJ...370L..39K}. However, this simple model does not explain one of the defining features: its variability. 
The accretion flows and shocks emit continuum emission from the ultraviolet (UV) to the infrared (IR) as well as a number 
of emission lines (e.g., H$\alpha$, Ca\,II, He\,I), which, when observed are all found to be variable on time-scales from hours, to weeks, and 
to years \citep{2005ApJ...626..498M,Nguyen09,2006ApJ...638.1056S}. 

This variability can and has been used to probe the inner regions of these accreting systems, to find 
the source of the variations and to provide stringent constraints on the nature of the accretion process 
itself \citep{2012ApJ...750...73D}. The time-scale of accretion variations and the magnitude of the variations will depend on their source, so by monitoring these accreting systems over different time-scales, we can identify where 
these variations originate. This has been done very successfully for individual objects such as  
AATau \citep{2003A&A...409..169B} and V2129 \citep{2012A&A...541A.116A}, but it has yet to be done 
systematically for a larger sample.

Our previous observing program LAMP (Long-term Accreting Monitoring Program) was designed to systematically explore the possible origin of accretion variability by monitoring H$\alpha$~and the Ca\,II triplet. Two successful runs took place over two months, with a total baseline of 15 months. Over this period 25 young stellar objects (YSOs) in Chameleon I were observed, each object 12 times. Among these targets we found 10 accretors, which covered a spectral range of G2\,-\,M5.75. All the accretors showed variations in their accretion signatures. The average accretion spread as calculated from the H$\alpha$ equivalent width (EW), was found to be 0.37\,dex \citep{2012MNRAS.427.1344C}. Although there are long term accretion rate variations in the sample,  the amplitude of variations reached a maximum or are within 70\% of the maximum after 25 days. This was the shortest time-scale in our sample, which indicates the dominant cause of this variability is occurring on times scales of 2\,-\,3 weeks, or less. 

This result is backed up by other studies, such as a short-term monitoring covering time-scales 
up to $\sim$ 48 hours of $\sim$ 29 sources in Chameleon II, which defined the spread 
in accretion rates as derived from H$\alpha$ to be 0.2\,-\,0.6 dex \citep{2012A&A...547A.104B}. 
Also, \cite{Nguyen09} found a very similar accretion-rate spread over time-scales 
of days and months. 
These and our previous results rule out origins of variability such as those due to a wind form the 
disc or large-scale instabilities in the accretion disc, as these would occur on much longer time-scales 
of months and years. 

These time-scales of $\lesssim$~ 2\,-\,3 weeks are close to the rotation period of these kinds of 
objects (1\,-\,10 days \cite{2007prpl.conf..297H}), which suggests that what we are observing could be 
a rotational modulation of the accretion flow. If there is even a slight offset between the rotation 
and the magnetic field axis (2\,-\,5\degree) it has been shown that an asymmetric accretion flow 
will form \citep{2003ApJ...595.1009R}. As this rotates with the star, different parts of the accretion 
flow will become visible, which will change the accretion signatures that we observe. 

A second possible explanation for variations on these time-scales concerns the existence of instabilities in these systems, either in the magnetic field, or in the inner disc \citep{2013MNRAS.431.2673K, 1998ApJ...492..323G}. These will cause short term, stochastic variations occurring on the time-scales as short as hours. 

If we wish to distinguish between these two possible causes of accretion variations on short time-scales, 
we need to monitor these systems on time-scales close to the rotation period (multiple days) as well as on time-scales 
of hours. Just such a test can be performed using the high signal-to-noise, high-cadence  
linear spectropolarimetry data-set of Vink et al. (2005) on a sample of T Tauri and Herbig Ae stars. 
This data-set will allow us to constrain whether the mass-accretion rate variations indeed involve a slow and gradual variation in 
accretion emission across the rotation period, or if they concern stochastic and rapid variations.

Our combined sample of T Tauri and Herbig Ae stars probing the time-scales of minutes and hours, with LAMP probing the time-scales of days, months, and years, 
we can constrain the dominant accretion variations. Using the H$\alpha$ emission as an accretion indicator, and to measure 
accretion rates, we compare the variations found on {\it all time-scales}.

Furthermore, we can extend our work to higher stellar masses (up to 2-3 $M_{\sun}$ for Herbig Ae stars). 
This way, we can test whether the accretion process in Herbig Ae stars might be similar to that in lower mass T Tauri stars, as 
suggested by \citep{2001A&A...371..186N,2002MNRAS.337..356V,2003A&A...406..703V,2006A&A...459..837G,2004ApJ...613.1049E}. Whilst magnetic fields have indeed been claimed in Herbig Ae stars \citep{2004A&A...428L...1H,2005A&A...442L..31W}, the fact that the field incidence in Herbig Ae stars is similarly low to that in post-main sequence objects (e.g. \cite{2012sf2a.conf..401A}) may cast doubt on an extension of the magnetospheric accretion scenario towards higher masses. Therefore the issue of the fundamental accretion process in Herbig Ae stars remains open, and can be tested in this paper. 

The paper is laid out as follows: In Sect.\ref{sec:sample_observations} we discuss the chosen sample, the observations and reduction, Sect. \ref{sec:behaviour} presents the data and the variations present in the H$\alpha$ emission, Sect. \ref{sec:origin} addresses the nature of the H$\alpha$ emission, Sect. \ref{sec:accretion_rates} discuses the derivation of accretion rates and Sect. \ref{sec:discusion} deals with possible causes of the variations observed. Individual sources and their variations are presented in the Appendix. 

\section{Sample and Observations}\label{sec:sample_observations}
Our targets were originally observed for a linear spectropolarimetry study \citep{2005MNRAS.359.1049V} to probe the circumstellar structures around Herbig Ae/Be and T Tauri stars. 
The sample was chosen from \cite{1988cels.book.....H}, based on their relative brightness (V $\lesssim$ 11), and their position on the sky, but not on any known circumstellar geometries, or T Tauri/Herbig Ae type. 
The target list is provided in Table \ref{tab:target_list} and the stellar parameters are listed in Table \ref{tab:stellar_parameters}.

The observations were obtained during the nights of December 10\,-\,13$^{th}$ 2003 and December 26\,-27$^{th}$ 2001 with the ISIS spectrograph on the 4.2\,m William Herschel Telescope, La Palma. Each target was monitored over $\lesssim$ 1 hour blocks, in a few cases, targets were observed in multiple blocks in a single night. The number of exposures in each block of observations, the exposure times, and dates of observations are all given in Table \ref{tab:stellar_parameters}. A slit width of 1.0 arcsec was used for all observations, along with the MARCONI2 CCD detector, and the R1200R grating. 
This grating has a spectral range of 1000~\AA~centred on 6500~\AA, with a spectral resolution of $\simeq$ 35\,kms$^{-1}$ around H$\alpha$. The ISIS set-up also included the polarization optics of a rotating half-wave plate and a calcite block. 

The data reduction was carried out using IRAF and included bias-subtraction, spectrum extraction and wavelength calibrations. 
Since these spectra were originally used for polarimetry studies, each observation is split into two spectra, of two separate polarisations. The sum of the two extracted spectra were taken as the full H$\alpha$ emission line and continuum was then normalised to 1 before any measurements were taken.

\begin{table*}
\caption{Target list. Magnitudes are given in the V-band. Variations in V band are given where found in the literature. The T Tauri type is taken from \protect\citet{1988cels.book.....H}, where a SU Aur type is given as `A star like SU Aur: type late F to K, weak emission at H-alpha and Ca II, very broad absorption lines (v sin i \textgreater 50 km/s), and relatively high luminosity'. Exposures are given as number of exposures times the exposure time in seconds. References: 1: \protect\citet{2001A&A...378..116M}, 2: \protect\citet{2000A&A...355L..27H},  3: \protect\citet{1979ApJS...41..743C}, 4: \protect\citet{1976ApJS...30..307R}, 5: \protect\citet{1988A&AS...72..505J}, 6: \protect\citet{1988cels.book.....H}, 7: \protect\citet{1989ApJ...341..340B}, 8: \protect\citet{2008A&A...479..827G}, 9: \protect\citet{2013AJ....145...79C}, 10:\protect\citet{2001A&A...380..609D}, 11:\protect\citet{1983ApJ...267..191R}, 12: \protect\citet{1990MNRAS.247..517D}. 
}
\begin{tabular}{@{}llcccllll@{}}
\hline
Name	    &Mag. &$\Delta$Mag.    & SpT&Ref.& Type 	&Date       & Exposures	       & Total \\
        &     &                &    &    &       &           &                  & [hrs] \\
\hline
RY Tau	&10.1 & 	\, \,9.55\,-\,11.56 	& K1 &  3,8 & CTTS  &26-12-01   & 4x120,12x180      & 4.0  \\
	    &     & 		&    &    &       &10-12-03 	& 8x120, 20x240	   &      \\
	    &	  & 		&    &	  &	    &12-12-03	& 8x90, 12x180     &      \\
	    &	  &		&    &  	  &	    &13-12-03	& 8x120, 12x180    &      \\
AB Aur  &7.1  &$\pm$ 0.7	  & A0 & 3,9  &	    &27-12-01	&20x30,16x30,16x30 & 1.83 \\
        &	  & 		&    &	  &       &10-12-03	&12x90		       &      \\
        &	  & 		&    &    &       &11-12-03	&20x60		       &      \\
        &	  & 		&    &    &	    &12-12-03	&20x60		       &      \\
        &	  & 		&    &  	  &	    &13-12-03	&4x120,12x90	       &      \\
T Tau 	&10.3 & \, \,9.75\,-\,10.18& K1 & 3,8  & CTTS	&12-12-01   & 8x45,16x120      & 1.37     \\
	    &     & 		&    &    &       &12-12-03	& 4x60, 18x150     &     \\
SU Aur	&9.0  &\, \,8.92\,-\,10.02 & G2 & 3,8  & SU AUR&27-12-01   &12x180             & 2.30     \\
	    &     & 		&    &    &       &10-12-03	& 8x90,16x150      &  \\
	    &	  &		&    &	  &       &11-12-03	& 4x90,5x240	       &      \\
	    &	  &		&    &	  &       &13-12-03	&12x120		       &      \\
DR Tau	&11.43& 10.76\,-\,12.78	& K5 & 2,1,8& CTTS	&27-12-01   &16x180             &  1.87    \\
	    &     & 		&    &    &       &12-12-03	& 4x60,20x180	   & \\
RW Aur A&10.36& \, \,9.32\,-\,11.75& K1 &2,7,8 & CTTS	&26-12-01   &11x180            &  1.40    \\
	    &     & 		&    &    &       &10-12-03	& 12x150	           &  \\
	    &     &		&    & 	  &       &13-12-03	&4x60,20x150	       &      \\	
GW Ori	&11.1 & \, \,9.74\,-\,10.53 & G5 & 3,8  &CTTS	&11-12-03	&4x60,20x240	   & 1.40 \\
V773 Tau&10.35& 10.82\,-\,11.03& K2 & 4,11 &CTTS   &13-12-03	&4x60,16x240	   & 1.13 \\
UX Tau A&11.3 & 10.64\,-\,12.77 		& K2 & 3,8  &WTTS	&13-12-03	&1x120,24x240	   & 1.63 \\
BP Tau	&12.1 & 	11.67\,-\,12.99	& K7 & 3  &CTTS	&12-12-03	&4x90		       & 2.13 \\
	    &	  &		&    &  	  &	    &13-12-03	&4x30,24x300	   &      \\
BF Ori  &12.2 &\, \,9.82\,-\,11.67 & A0 & 3,10  & HAe   &27-12-01   &16x180            & 0.80 \\
LkH$\alpha$ 215&10.8 & 10.36\,-\,10.47& B1 & 3,12  & HAe   &12-12-01   &16x120            & 0.53 \\
MWC 480 &7.6  &	-	& A3 & 5  & HAe   &26-12-01   &16x75,16x30       & 1.03 \\
        &     & 		&    &    &       &27-12-01   &16x45,8x45,16x60  &      \\
CO Ori  &9.83 &\, \,9.81\,-\,12.73 & F8 & 6,8  & SU AUR &27-12-01   &8x120,8x180       &      \\
        &     & 		&    &    &       &24-12-01   &8x12,8x180        & 1.09 \\

\hline

\end{tabular}
\label{tab:target_list}
\end{table*}

\section{Behaviour of the H$\alpha$ Emission}\label{sec:behaviour}
In the following section, AB Aur is used to represent the typical behaviour of the line emission in the sample. This section will concentrate on the 2003 observations of this object as an example, all of the observations for the remaining 13 targets and the 2001 observations of AB Aur are discussed in the Appendix.
\subsection{H$\alpha$ Profiles}
\begin{figure*}
\begin{tabular}{cccc}
\includegraphics[scale=0.22]{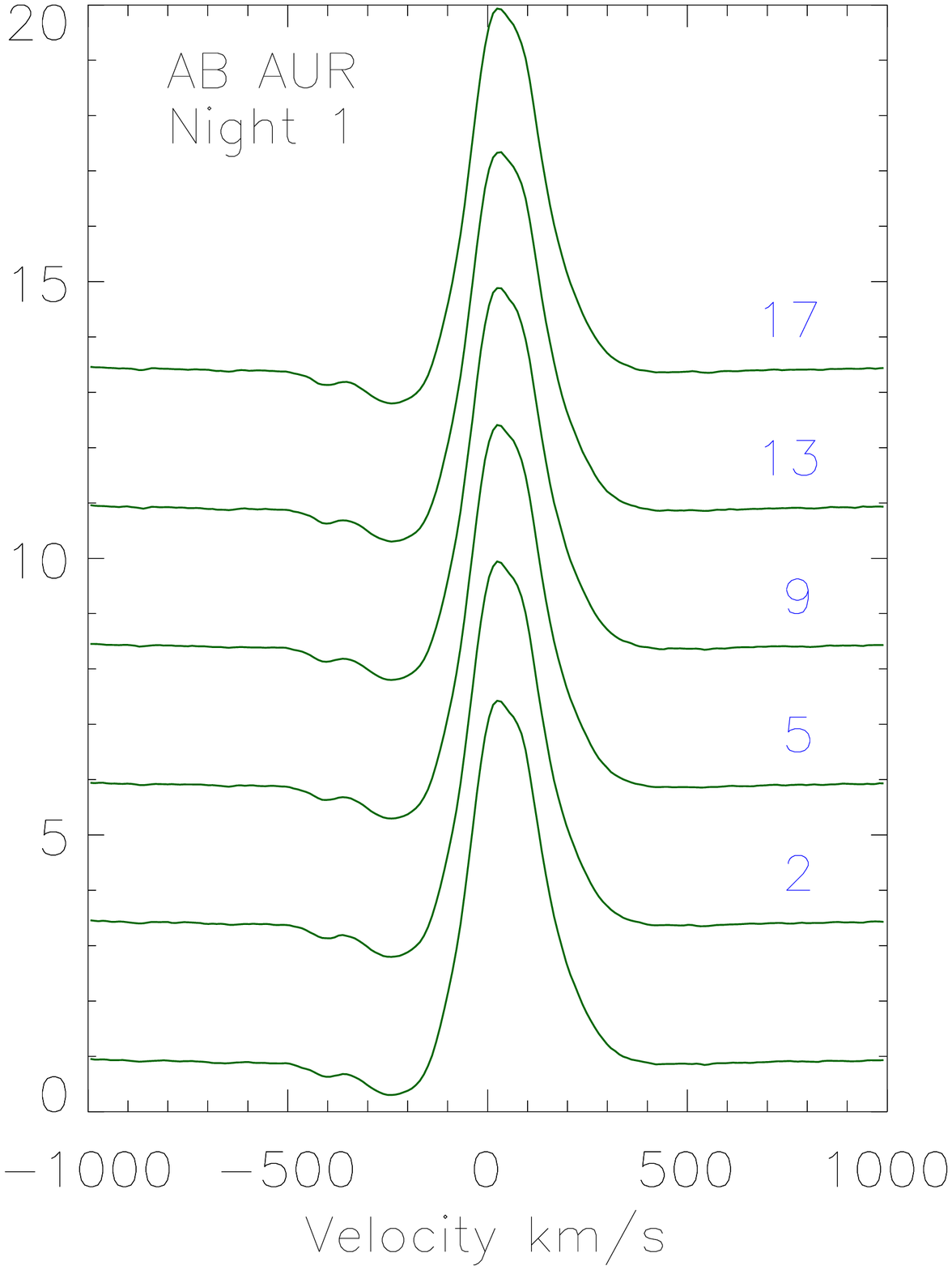}& \includegraphics[scale=0.22]{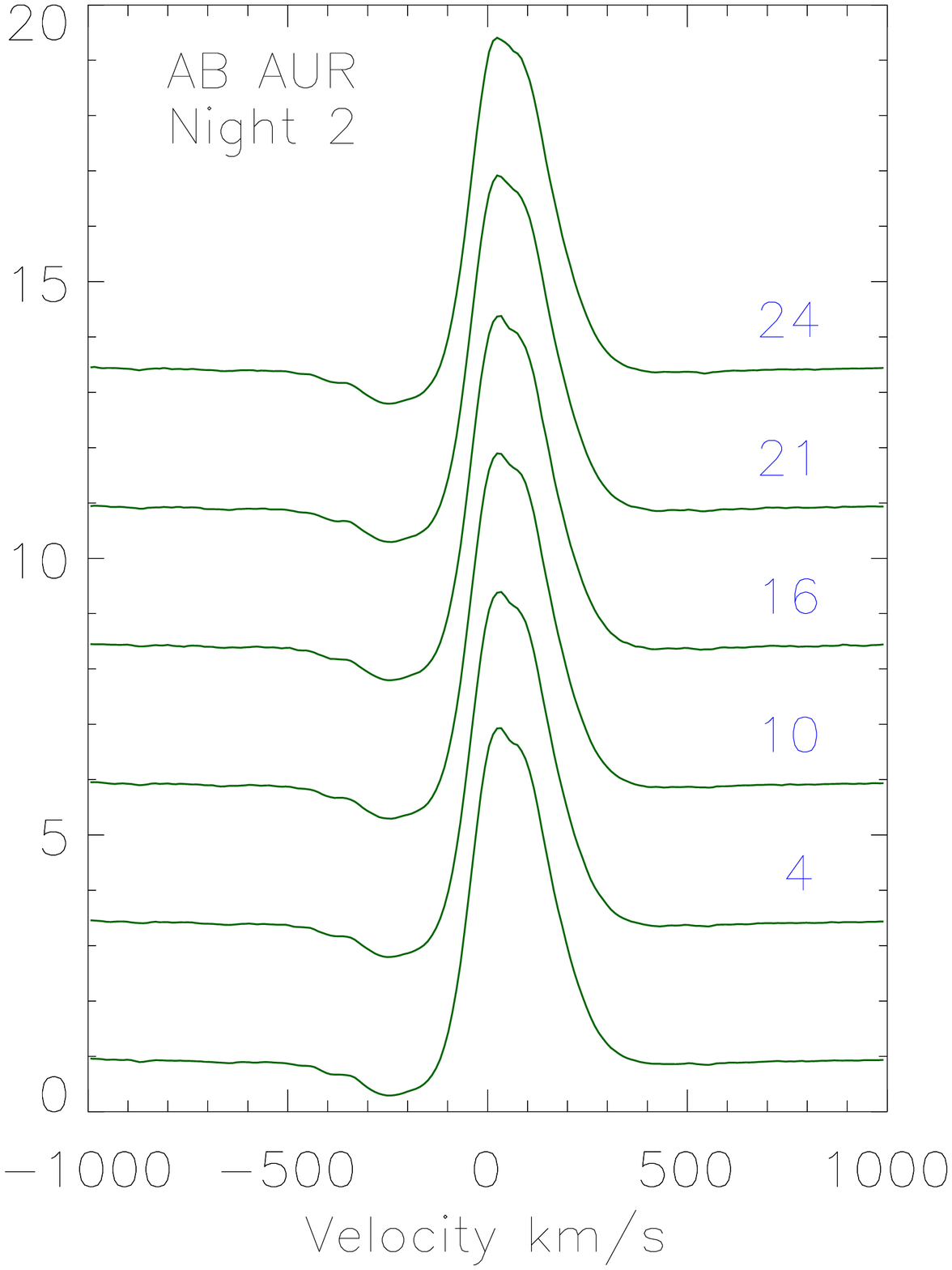}&\includegraphics[scale=0.22]{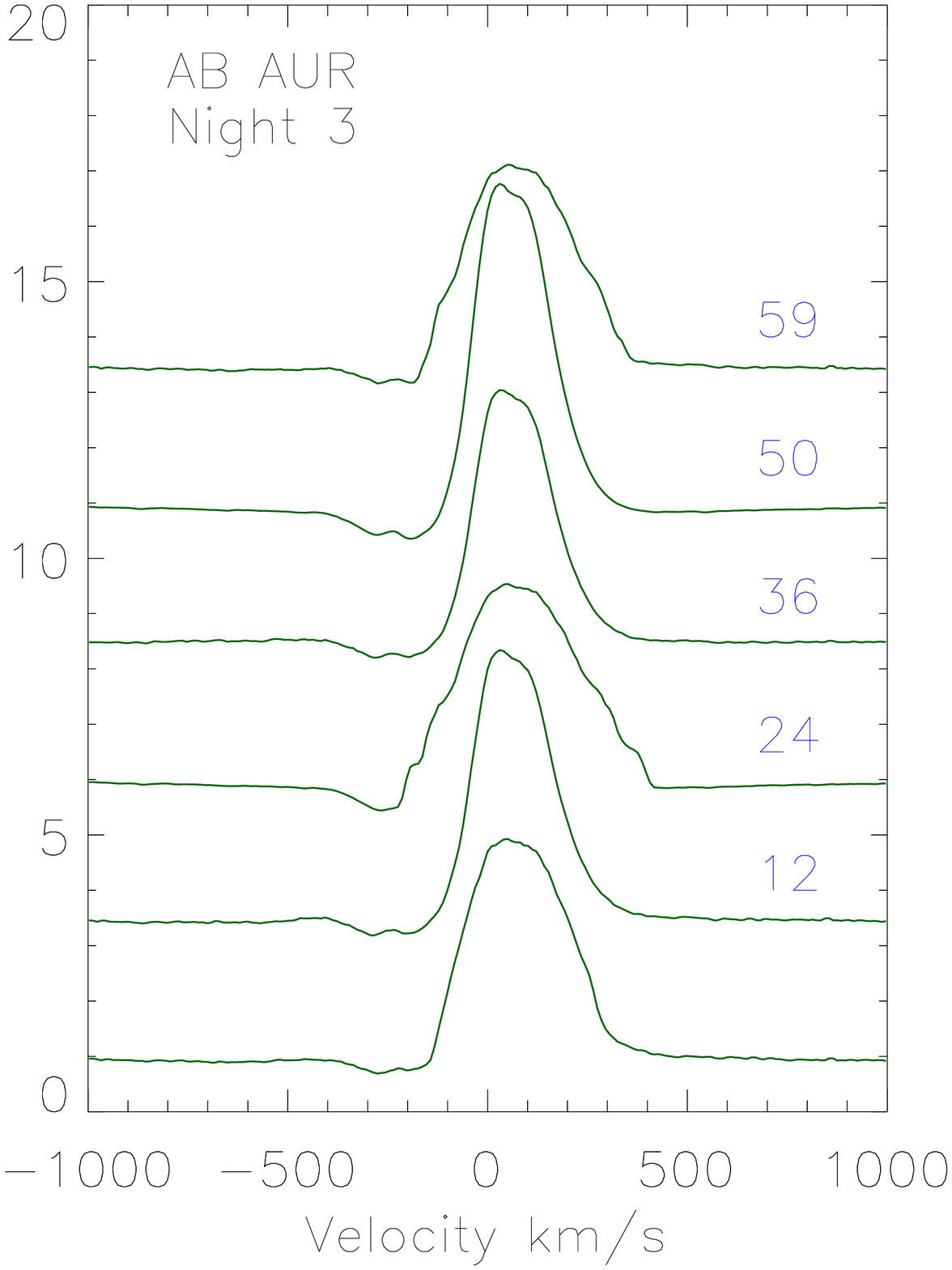} &\includegraphics[scale=0.22]{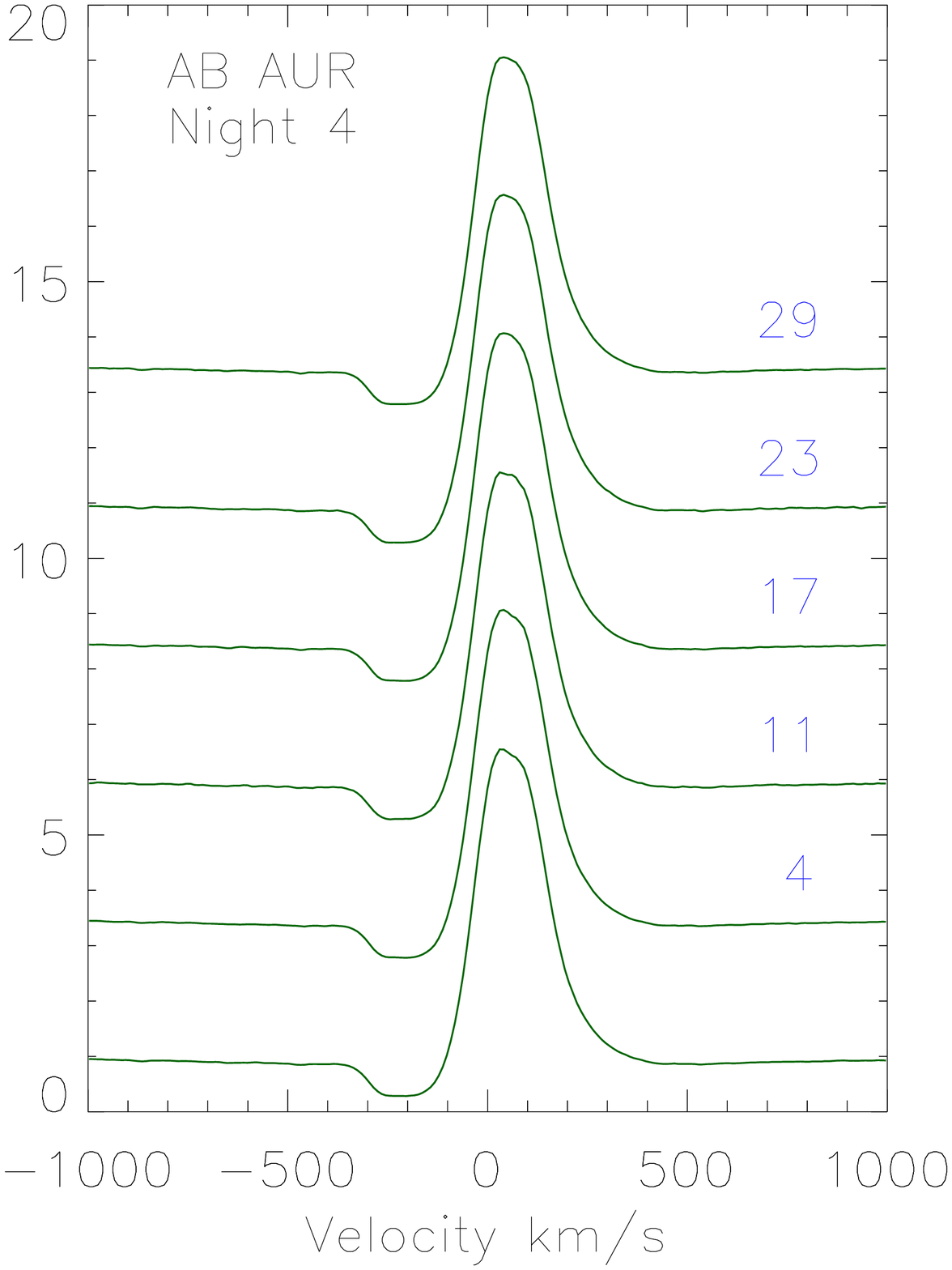} \\
\includegraphics[scale=0.22]{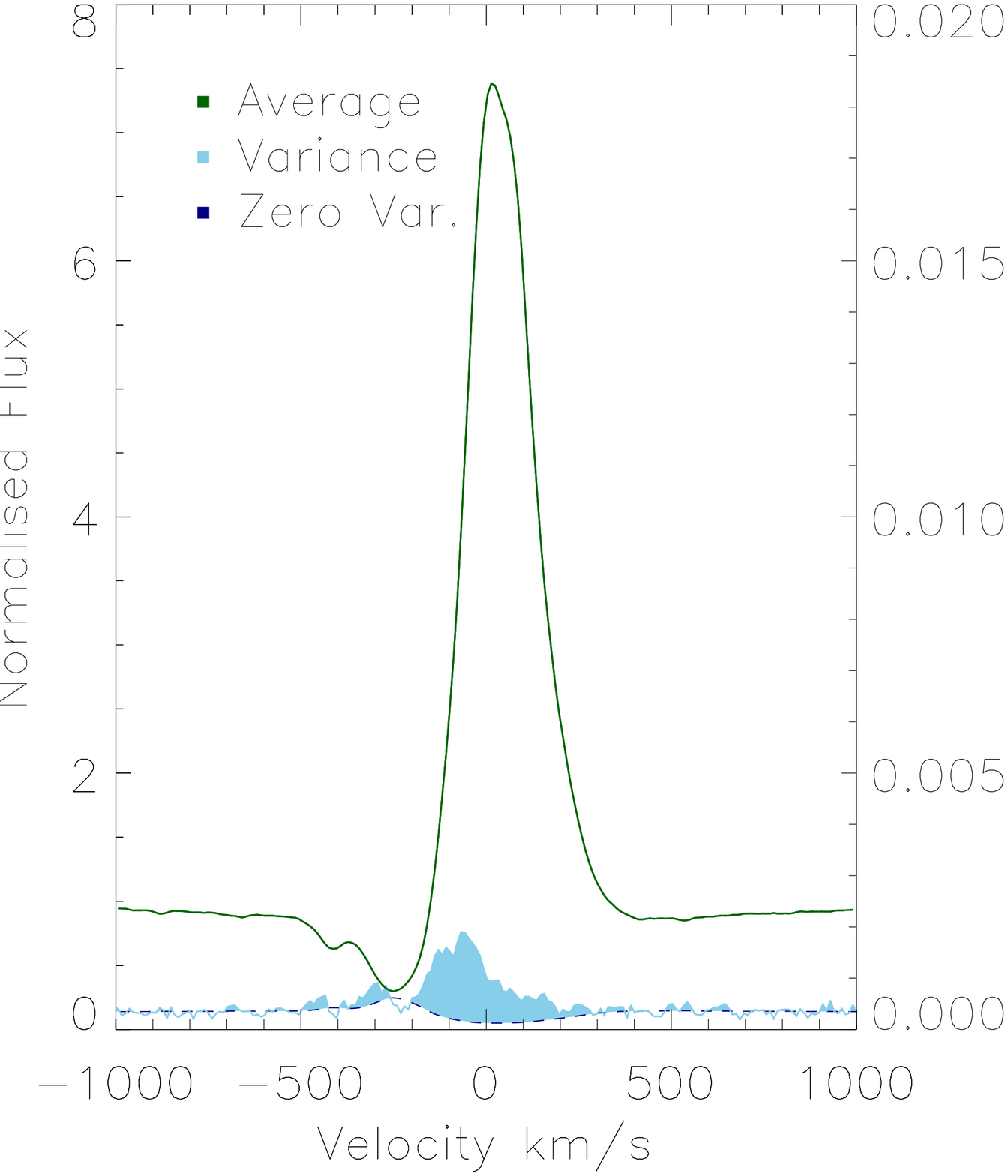} & \includegraphics[scale=0.22]{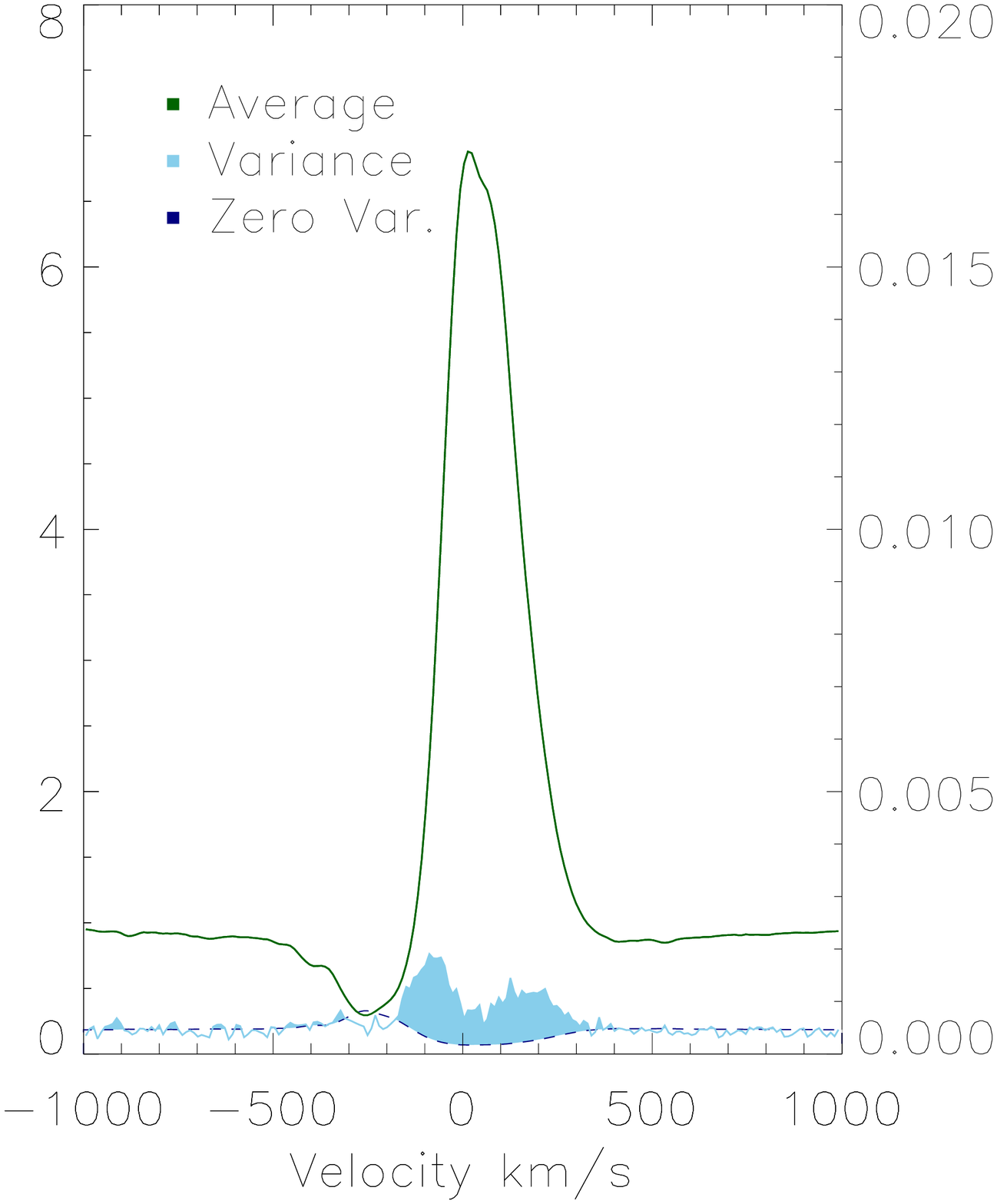}  &\includegraphics[scale=0.22]{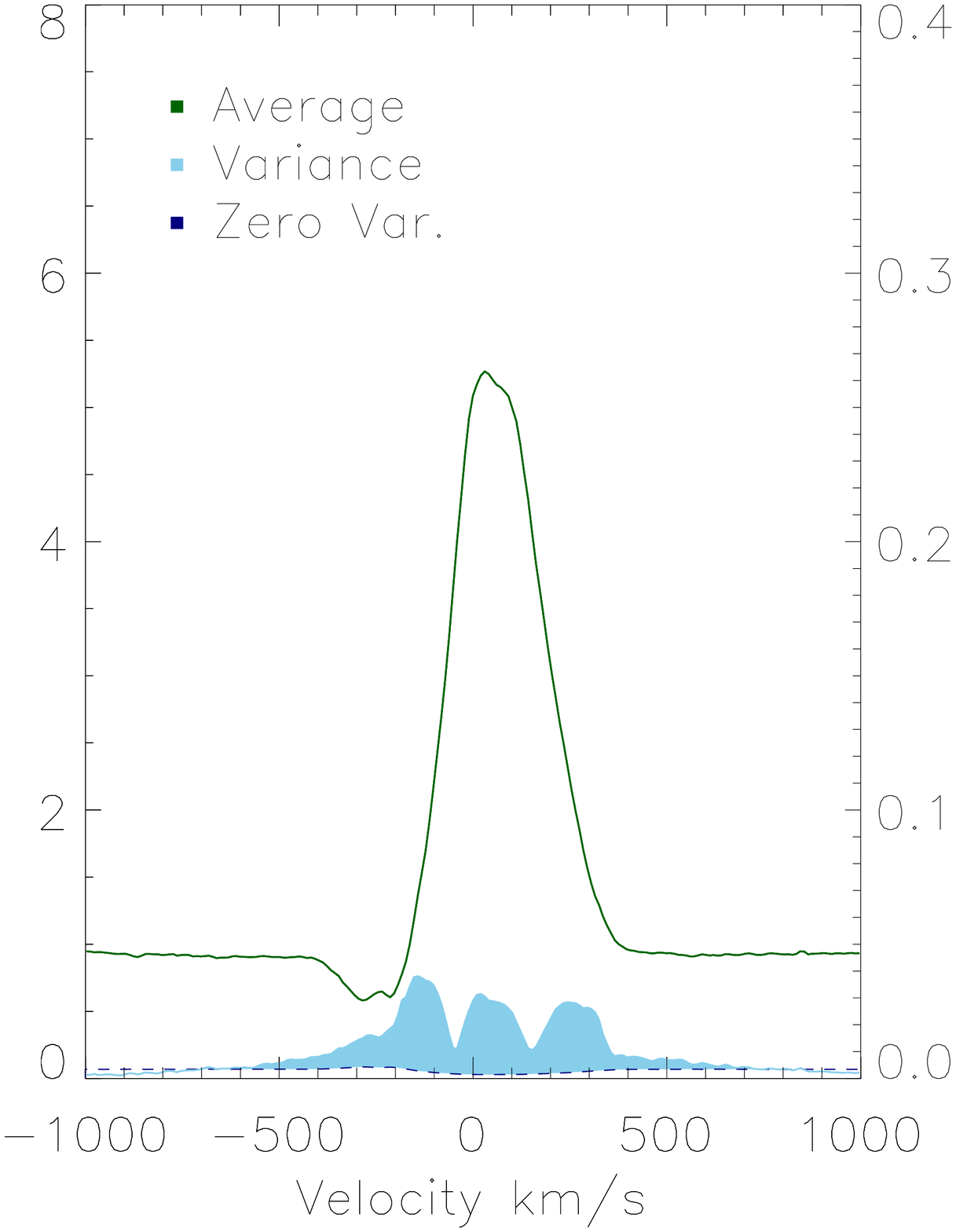}  &\includegraphics[scale=0.22]{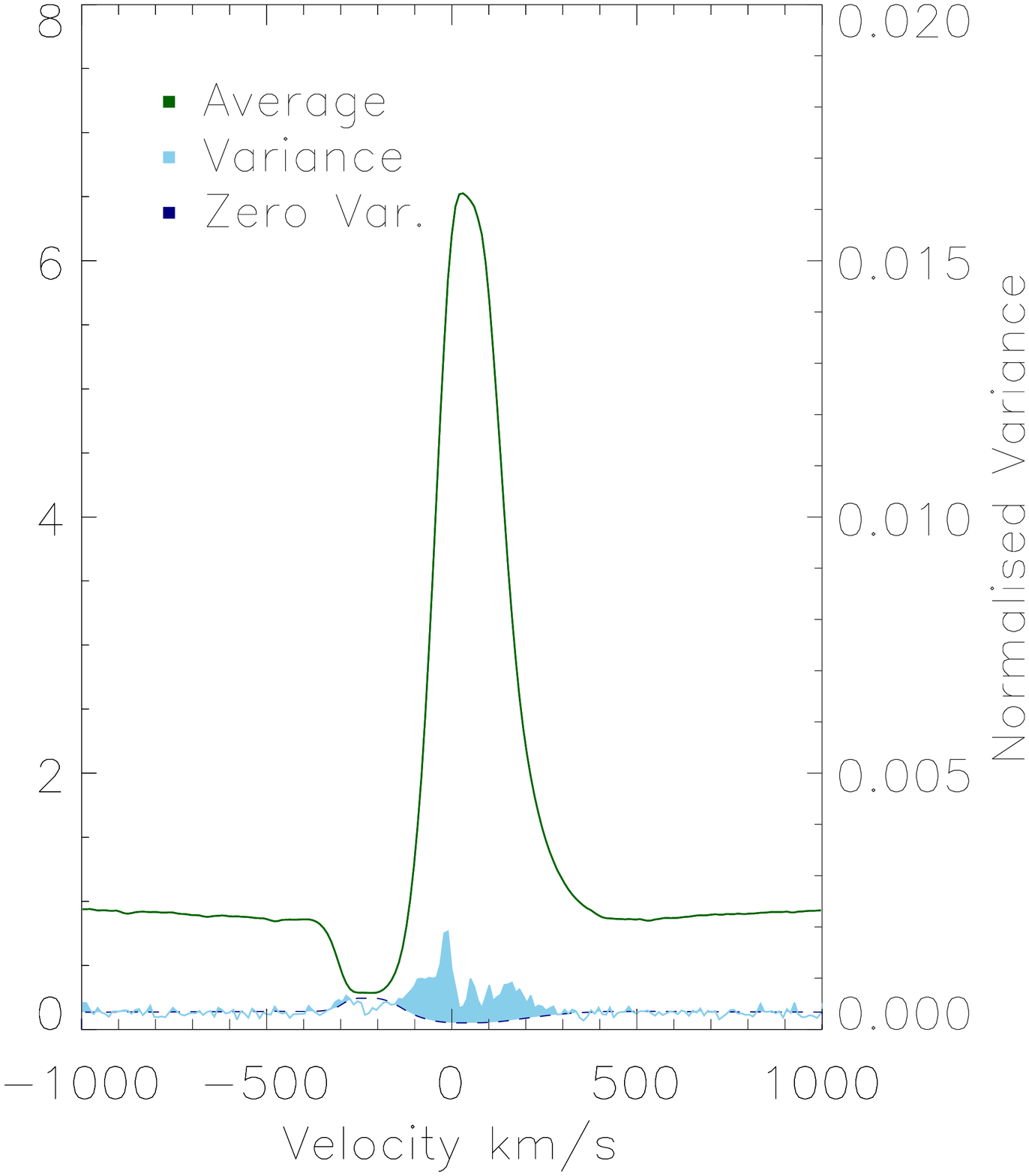} \\
\includegraphics[scale=0.22]{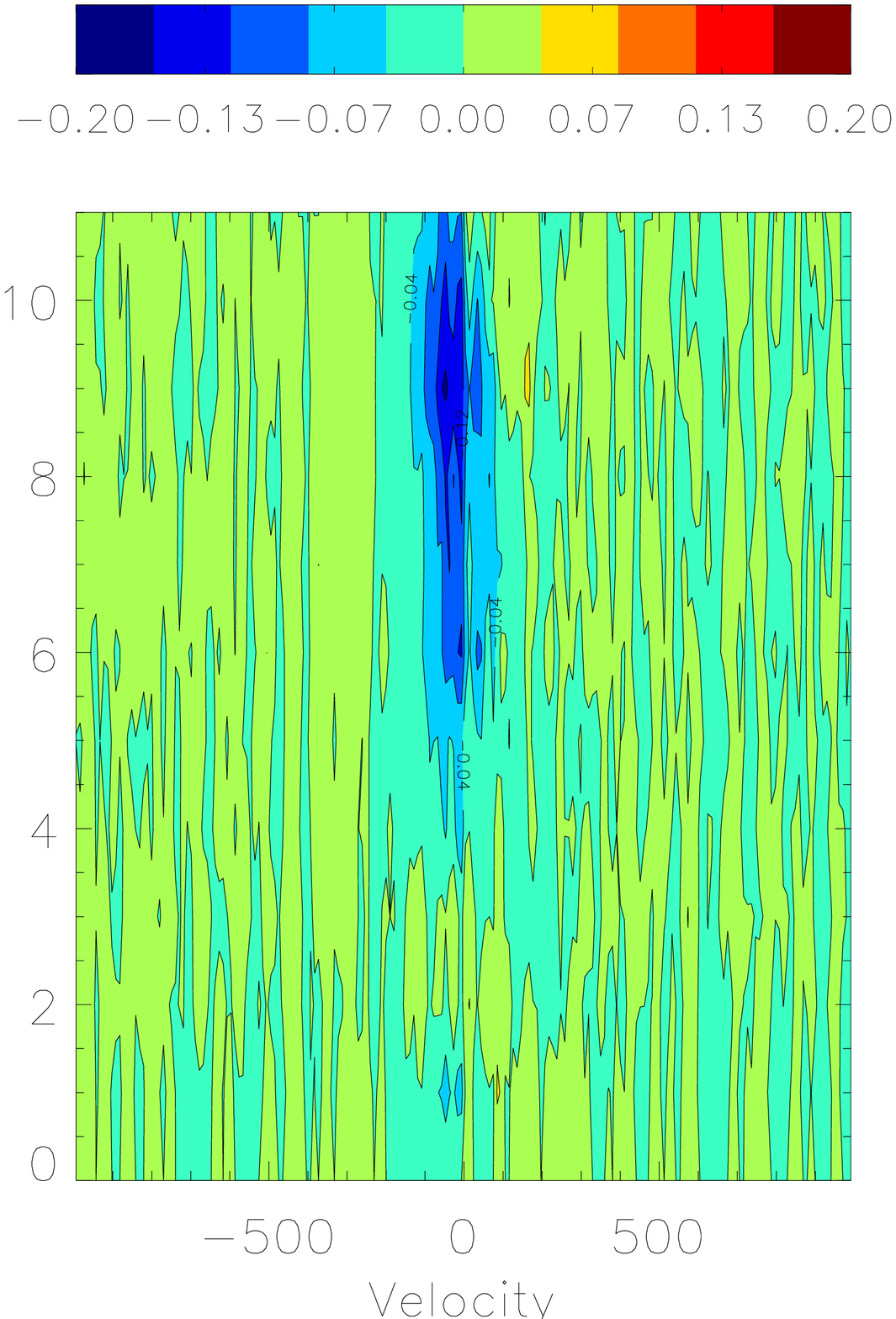}& \includegraphics[scale=0.22]{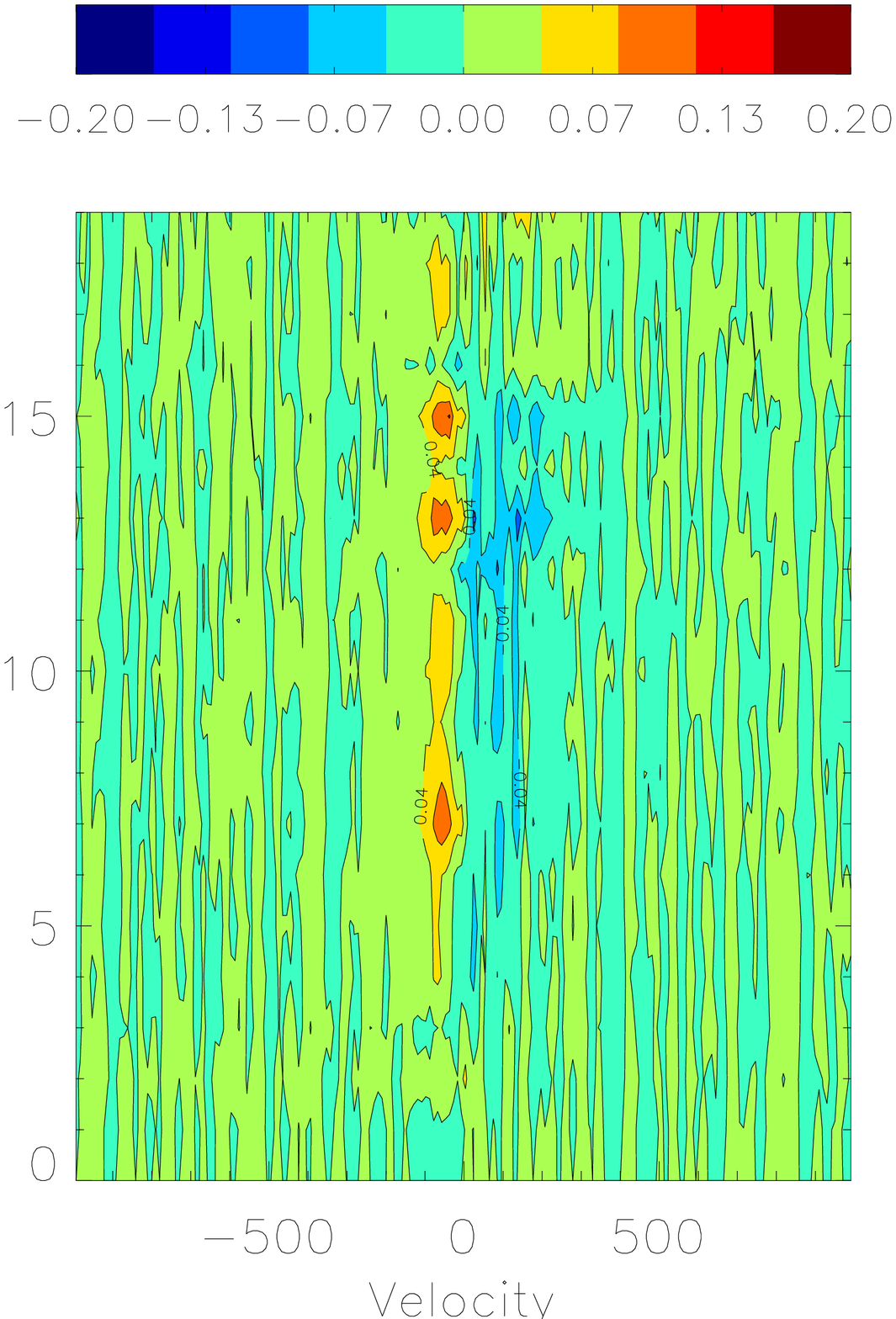} & \includegraphics[scale=0.22]{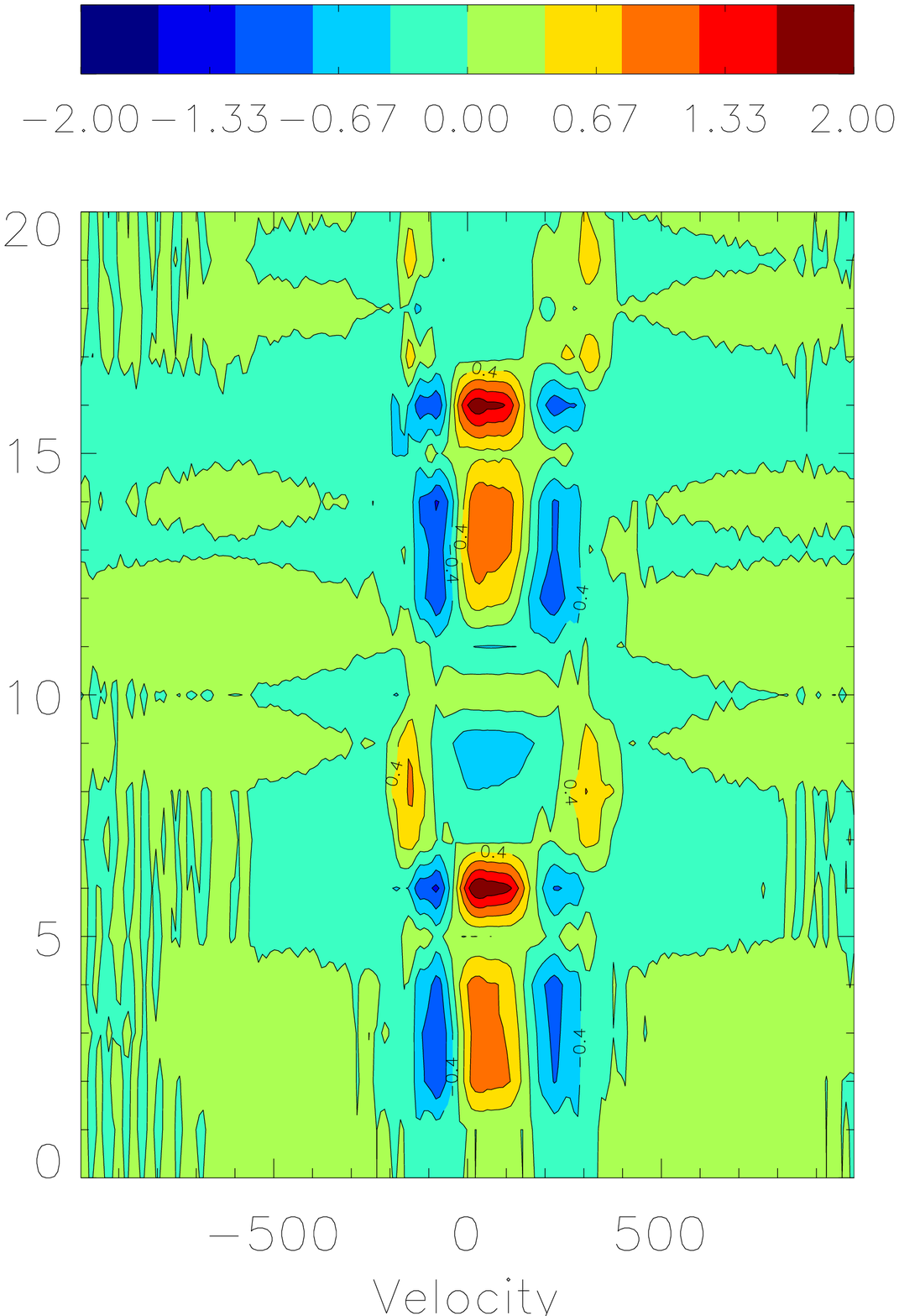} & \includegraphics[scale=0.22]{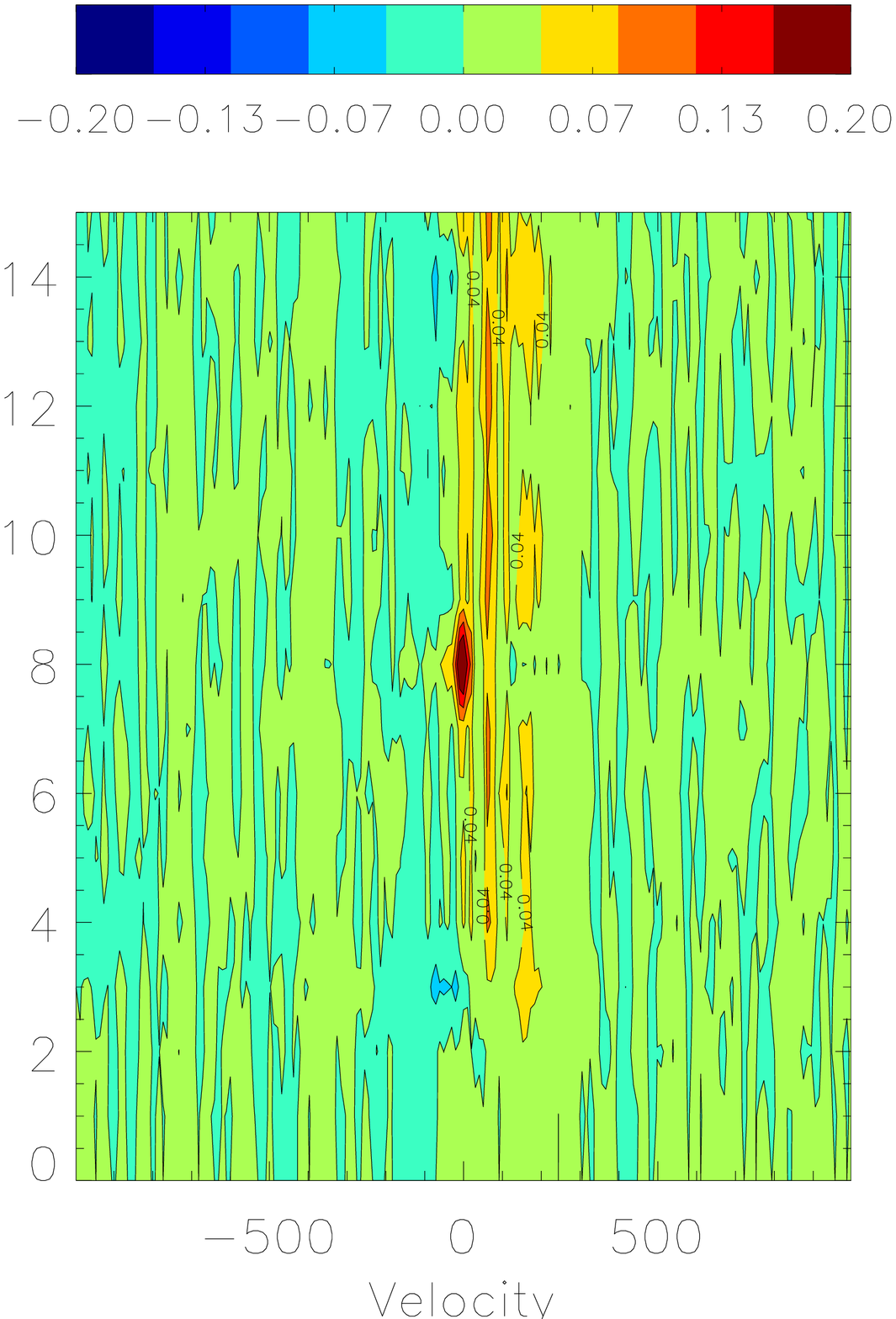}\\
\includegraphics[scale=0.22]{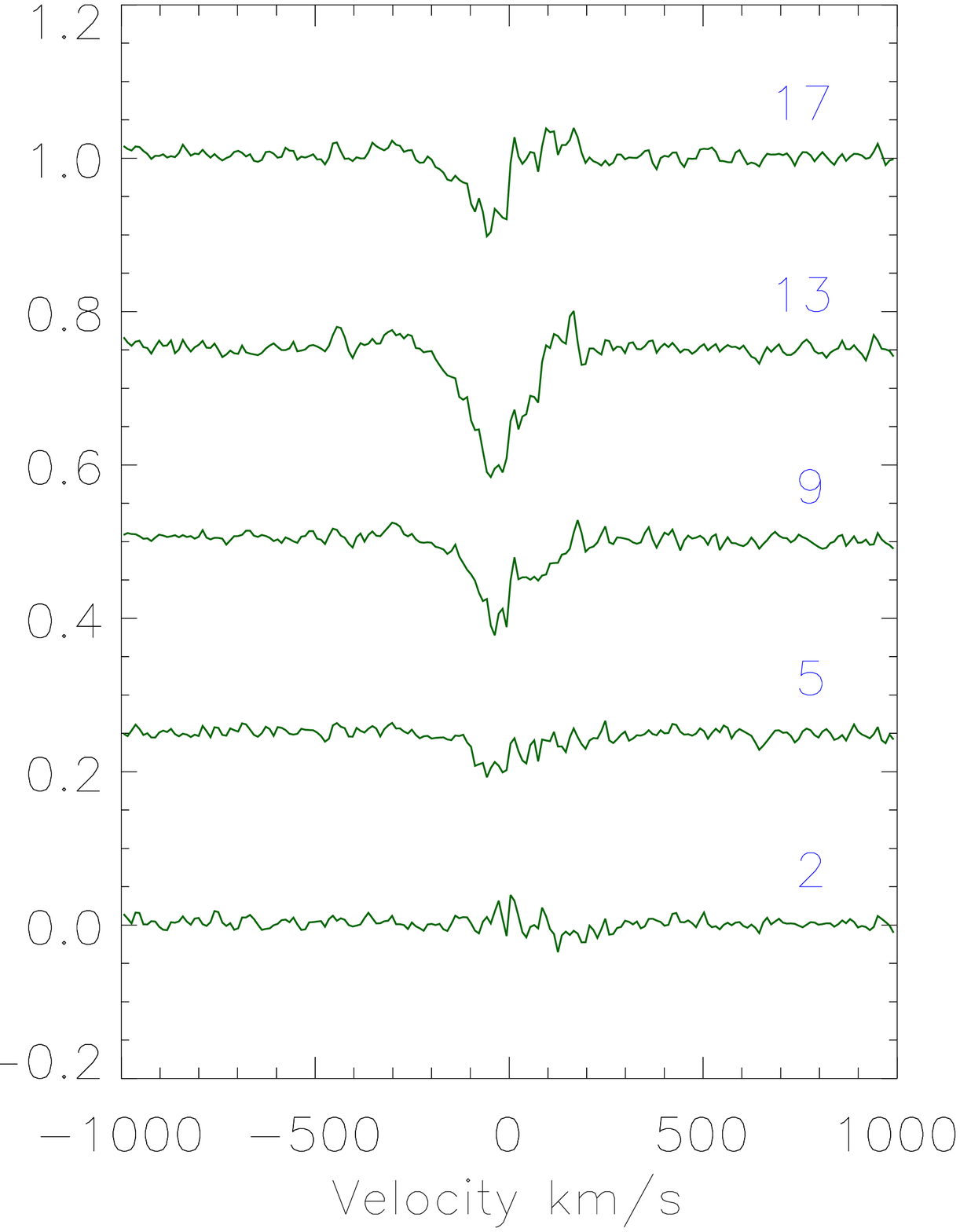} & \includegraphics[scale=0.22]{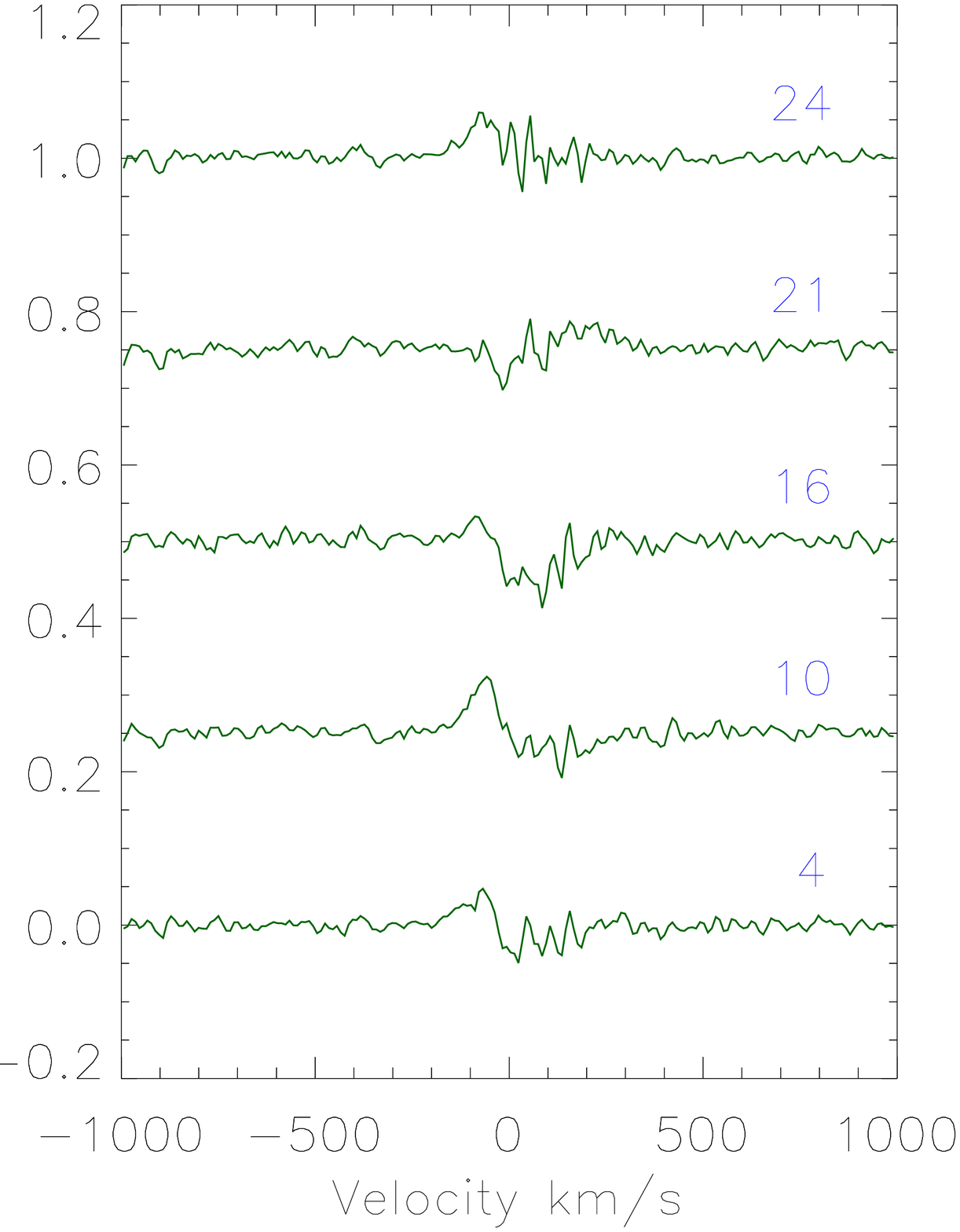} & \includegraphics[scale=0.22]{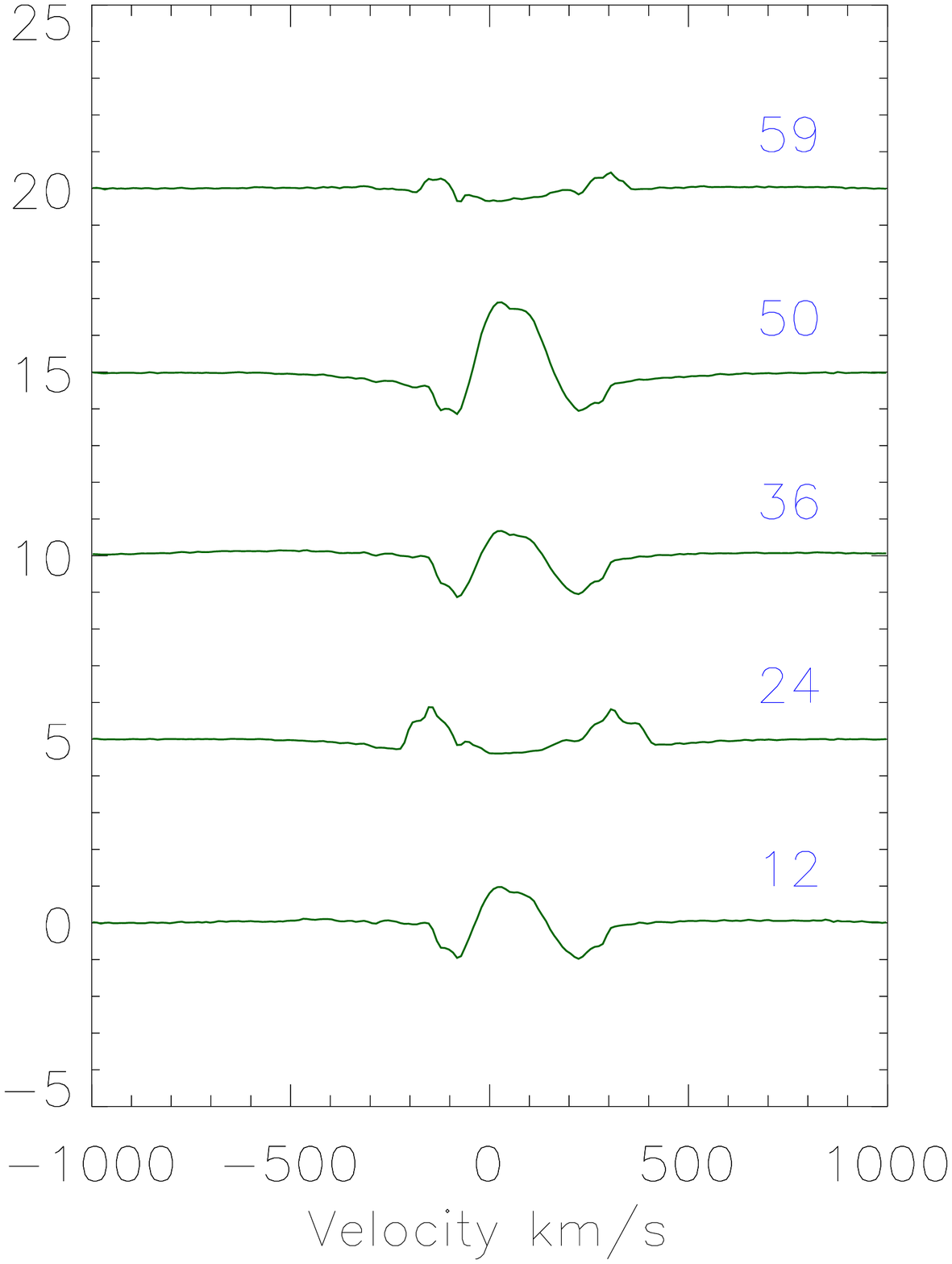} & \includegraphics[scale=0.22]{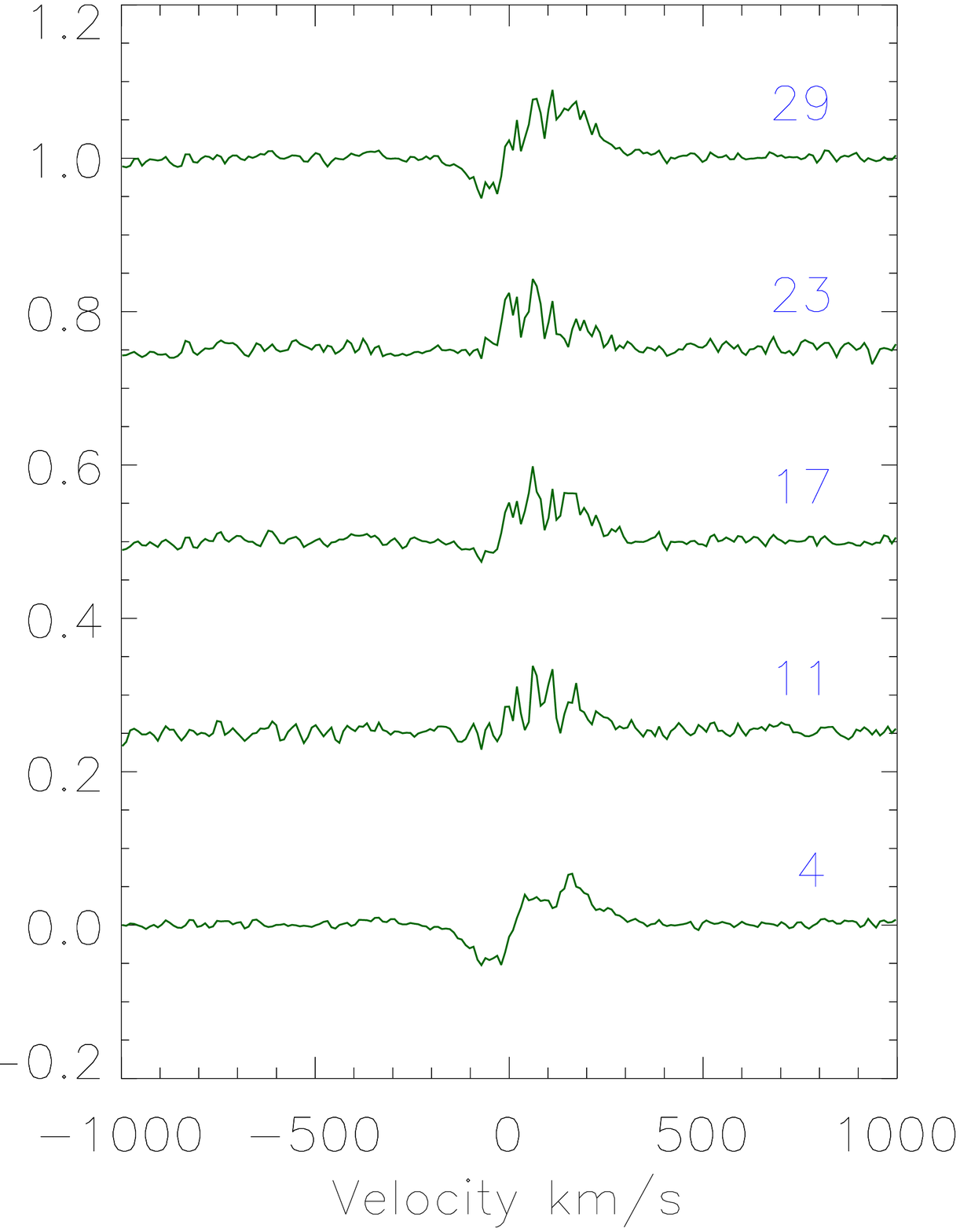}\\

\end{tabular}
\caption{First Row: Sample of H$\alpha$ profiles across the fur nights of of observations of AB Aur in 2003. Each profile is off-set from the previous one for clarity. A time stamp is given to
the right of each plotted spectrum, and it takes the form of the time difference
between the first observation in that block and that spectrum. Second Row: Average and variance profiles. Third Row: A differential surface plot. This plot shows the color coded difference between the first spectra of that night and the preceding spectra. Note scale change on third night. Fourth Row: Time-series of cuts in differential flux plots. This was done in the same way as the surface plots, where the first spectrum of that night was removed from all the rest of the spectra.  These are the same profiles as chosen for the profile time series.  }
\label{fig:ABAUR_profiles_2003_1}
\end{figure*}

For each block of observations of each object there are about 20 different exposures over the course of $\sim$ 1 hour (see Table \ref{tab:target_list}). This provides us with very close temporal coverage of the H$\alpha$ emission. 
The top row in Fig.\,\ref{fig:ABAUR_profiles_2003_1} shows a time series of profiles across the four nights of observations for AB Aur in 2003. These four nights were the 10$^{th}$, 11$^{th}$, 12$^{th}$ and 13$^{th}$ of December, and for simplicity these will be referred to Night 1, 2, 3 and 4 respectively. Each profile is off-set from the previous one for clarity. A time stamp is given to the right of each plotted spectrum, and it takes the form of the time difference between the first observation in that block and that spectrum.

The changes in the profiles during Nights 1, 2 and 4, are typical of what is seen in the majority of objects of this sample. These changes will be referred to as \textit{slow variations}. 
During each of these nights, AB Aur shows a strong central emission with a large blue-shifted absorption in the wing. Within this blue-shifted absorption there is also a small emission peak. 
Within a single night's observation there is very little change seen within the profile.

Larger changes in the profile can be seen from night to night. For example in the case of AB Aur, the small emission peak within the blue absorption noticeably changes in wavelength. 
The strength of the emission, the depth of the absorption, and the red wing also change between the nights observations.

It is only on Night 3 of the observations that we see large changes in the profile from one exposure to the next. These changes occur on the time-scales of {\it minutes}, and they take the form of a drop in intensity across the line, and a broadening of the emission. These changes are referred to as \emph{rapid events}. 
They are not common in the sample. These \emph{rapid events} also occur in the spectra of RY Tau and RW Aur, however they are much weaker than those seen in AB Aur, and show different behaviour.

\begin{table*}
\centering
\caption{Published Stellar parameters. References: 1: \protect\citet{1988ApJ...330..350B}, 2: \protect\citet{1986ApJ...303..311P}, 3: \protect\citet{2003ApJ...590..357D}, 4: \protect\citet{1993A&AS..101..629B}, 5: \protect \citet{2001A&A...378..116M}, 6: \protect \citet{1987AJ.....94..150H},  7: \protect\citet{1995AJ....109.2800J}, 8: \protect\citet{1986ApJ...309..275H}, 9: \protect\citet{2006ApJ...653..657M}, 10: \protect\citet{2009ApJ...704..531K}, 11: \protect\citet{1995AJ....110..776W}, 12: \protect\citet{1983ApJ...267..191R}, 13: \protect\citet{2007ApJ...670L.135E}, 14: \protect\citet{2001A&A...378..116M}, 15: \protect\citet{1997ApJ...490..792M}, 16: \protect\citet{1999A&A...345..884C}, 17: \protect\citet{2001A&A...369..993P}, 18: \protect\citet{2004AJ....127.1682H}, 19: \protect\citet{2001ApJ...556..265W}, 20: \protect\citet{1998ApJ...492..323G}, 21: \protect\citet{1991AJ....101.2184M}, 22: \protect\citet{1986A&A...165..110B}, 23: \protect\citet{2007ApJ...670.1214B}, 24: \protect\citet{1995ApJS..101..117K}, 25: \protect\citet{1989A&A...211...99B}, 26: \protect\citet{1984BAAS...16..998V}, 27: \protect\citet{1992ApJ...397..613H}, 28: \protect\citet{2004AJ....128.1294C}, 29: \protect\citet{2011A&A...536A..45H}, 30: \protect\citet{2003SPIE.4838.1037A}, 31: \protect\citet{2004ApJ...605L..53F}, 32: \protect\citet{2002ApJ...566.1124A}, 33: \protect\citet{2003A&A...405L...1L}, 34: \protect\citet{2000ApJ...545.1034S}
}
\begin{tabular}{@{}llllllllll@{}}
\hline
 Object &  M$_{*}$    &  Teff  &R$_{*}$     &    P$_{rot}$ & \emph{v} sin \emph{i} & Inclin. & Determination of  & Ref. \\ 
        &  $[M_{\odot}]$ & [K] & [$R_{\odot}$] &   [days]     &  [km\,$s^{-1}$]        &  & Inclination                  &    \\ 
\hline
RY Tau   & 2.0 & 6300  & 2.7  &5.6         & 55$\pm$3           & 30$^{\circ}$  &Inteferometric& 1,33,7,6,5 \\
AB Aur   & 2.5 & 9332  & 2.5  &1.4         & -                   & 30$^{\circ}$  &H-Band Imaging& 2,4,5,31,16 \\     
T  Tau   & 2.0 & 5250  & 3.3  &2.8         & 19.5$\pm$2.5       & 29$^{\circ}$  & Inteferometric& 1,19,6,32  \\    
SU Aur   & 2.0 & 5550  & 3.1  &1.7\,-\,2.7 & 59 $\pm$ 1         & 63$^{\circ}$  & Inteferometric& 1,3,7,6  \\
DR Tau   & 1.0 & 4060  & 1.2  &5.1         & \textless 10       & -             &  -           & 1,20,7   \\ 
RW Aur   & 1.0 & 4700  & 2.7  &5.6         & 16\,-\,40          & 45$^{\circ}$  &Jet Inclination& 1,17,33  \\  
GW Ori   & 3.7 & 5700  & 2.5  &3.2         & 40\,-\,43          & 15$^{\circ}$  & Inteferometric& 9,21,22,10,25 \\     
V773 Tau & 1.2 & 4900  & 2.4  &3.43        &  41.4              &   -            &   - & 10,27,23,11,12    \\
UX Tau   & 1.3 & 4900  & 2.0  &2.7         & 25                 & 50$^{\circ}$  & Photometry & 10,24,13,25 \\
BP Tau   & 0.8 & 4000  & 3.0  &7.6         & \textless 10       & 30$^{\circ}$  & Inteferometric& 1,19,1,26,8,34 \\ 
BF Ori   & 2.5 & 8912  & 1.3  &  -         & 37 $\pm$ 2         & -             &  -          & 9,27,14  \\ 
LkH$\alpha$ 215 & 4.8 &14125 & 5.4  &  -         &\textless 200       & -             &   -           & 18,9,27  \\       
MWC 480  & 2.3 & 8890  & 2.1  & 0.5        & 90                 & 30$^{\circ}$  &Interferometric& 15,29,15 \\              
CO Ori   & 2.5 & 6030  & 4.3  &  -         & 40\,-\,50          & -             &   -           & 28       \\ 
\hline

\end{tabular}

\label{tab:stellar_parameters}
\end{table*}

\subsection{Average and Variance Profiles}
In order to better discern the where the changes in the emission profiles occur, average and variance profiles are utilised. The average profile in this case is simply the average flux at each wavelength and is calculated for each night of observation separately. The variance ($\sigma$) at a wavelength $\lambda$ is calculated using the following:
\begin{equation}
  \sigma^{2}({\lambda}) =\frac{1}{n-1} \displaystyle\sum_{i=1}^{n}(I_{\lambda,i} - \bar{I}_{\lambda})^{2}
\end{equation}
\noindent where $I_{\lambda,i} $ is the flux at wavelength $\lambda$ for spectrum number $i$, $\bar{I}_{\lambda}$ is the average flux at that wavelength and $n$ is the total number of spectra for a given object \citep{1995AJ....109.2800J}. The normalised variance is then given by $ \sigma_{N}^{2}(\lambda) =  \sigma^{2}(\lambda)/\bar{I}_{\lambda} $. The variance profile allows is to distinguish which parts of the emission profile are changing, and which are remaining the same. 

The average and normalised variance profiles for AB Aur are given in the second row of Fig.\,\ref{fig:ABAUR_profiles_2003_1}. The horizontal dashed line in these plots represents the zero variability level which is given by
\begin{equation}
 \sigma_{N}^{2}(\lambda) = \sigma_{N,0}^{2}\left(\frac{\sqrt{\bar{I_{\lambda}}}}{\bar{I_{\lambda}}}\right)^{2}
\end{equation}
where $\sigma_{N,0}^{2}$ is the normalised variance in the continuum \citep{2007ApJ...671..842S,2006ApJ...638.1056S}. The zero variability level can be considered the level below which variations are not significant. The area in the variance profile above the zero variability level i.e. significant peaks of variations are shown as a filled (blue) colour.

Across the four nights there different parts of the H$\alpha$ profile change.  A constant change in intensity across the entire line would result in a variance profile that had the same shape as the emission profile. However for AB Aur there are distinct structures within the variance profile for each night. 

For Nights 1 and 2, the changes are concentrated in the wings of the profile. This behaviour is representative of the slow variations seen in the majority of the sample, where most of the 
changes occur in the line wings rather than the line centre. 

On the third night of observations, the rapid variations noted in the time series of 
the H$\alpha$ profile manifest themselves as three distinct regions of variations: one peak in the blue wing, 
one peak in the line centre, and another one in the red wing. (Note the scale change for the variance profile on the third night). 
This is unusual behaviour within this sample, these large changes are only seen in three observation blocks for three separate objects over the course of these observations. However, no other object shows such strong, distinct peaks in the variance profile as AB Aur.  

\subsection{Distinguishing Variations}
The majority of the variations in the profiles are small. So in order to perceive 
these changes more clearly, differential surface plots and differential time series of the profiles are given for each observation block. (See third and fourth row of Fig.\,\ref{fig:ABAUR_profiles_2003_1}). The surface plots were created by finding the difference between the first spectrum of that observation block and all the preceding spectra. These are then plotted as a surface where the difference in flux is color coded. As you move from the bottom of the plot to the top, you are moving through the observation block, and the changes in the surface plot correspond to the differential changes in the profiles across the block.

The last row in Fig.\,\ref{fig:ABAUR_profiles_2003_1}, represents cuts across these surface 
plots. As with the profiles given in the top left panels,  the time stamp shows the difference in minutes between the first observation in that block and that spectrum.

\subsection{Line Measurements}
In order to quantify the changes in the H$\alpha$ emission, two measurements of 
the emission line were taken, the H$\alpha$ equivalent width (EW), and the H$\alpha$ 10\% width (10\%w). The H$\alpha$ EW gives a measure of the strength of the emission line, is measured in \AA~and 
is given by the following

\begin{equation}
 EW = \sum \frac{ F_{c} - F_{\lambda}}{F_{c}} \Delta\lambda
\end{equation}
where $F_{c}$ is the continuum flux, and $F_{\lambda}$ is the flux at wavelength $\lambda$. The H$\alpha$ 10\%w is simply the full width of the line at 10\% of the peak height, where the units are km s$^{-1}$. 

The integration windows for each source differ and are chosen on the basis of the breadth of the H$\alpha$ emission. The window size for each object is given in Table \ref{tab:EW10w}, along with the average, max, min and standard deviation of both the H$\alpha$ EW and 10\%w measurements.

From these measurements it can be noted that the \emph{slow variations} observed across the entire 
sample do result in small changes in the EW and the 10\%w. Much more significant variations occur between each nights observations.

Error analysis was performed by varying the measurement parameters for the EW. The integration window was varied by $\pm$0.5, 1, 1.5, 2, 2.5, 3\,\AA~and continuum measurements were varied by $\pm$1, 2, 3\,$\sigma$. The mean differences between each of these variations and the `real' measurement were added in quadrature. The square root of this sum was taken as the error estimate for that spectrum. This was performed on a sub-sample of spectra for each object, and then averaged across the sub-sample of spectra to give an estimate of the errors in the EW measurements. The 10\%w is not as sensitive to these parameter variations as the EW is, and is more difficult to get an proper estimate of the measurement errors using this method. In this case twice the spectral resolution (10\,km\,s$^{-1}$) can be taken as a rough estimate of the 10\%w errors. These EW errors are given in Table. \ref{tab:EW10w2001} and Table \ref{tab:EW10w}.

\begin{table*}
\centering
\caption{Average H$\alpha$ EW and 10\%w measurements for 2001 observations. Also given are the max, min measurements and the standard deviations ($\sigma$) across the range of measurements. Window is the integration window over which EW and 10\%w were calculated. No 10\%w measurements are given for MWC 480 as the entire blue wing is in absorption. }
\begin{tabular}{@{}lcccccccccccc@{}}
\hline 
Object  & Night &$\overline{EW}$& Max  &  Min.  & $\sigma$& Er.  &$\overline{10\%w}$ & Max  &  Min.  & $\sigma$& Win \\
        &     &    [\AA]	    &      &       &   &      & [km\,s$^{-1}$]           &      &      &       & [\AA]   \\ 
\hline

RY Tau  &ALL  & 10.38           & 10.64& 10.21  & 0.12 & (0.08) & 650    & 660  & 640    & 6 & 25 \\
AB Aur  &ALL  & 24.80			& 26.81& 24.80  & 0.56& (0.28) & 360	   & 360  & 350    & 4   & 25  \\
        & 1   & 26.55           & 26.81& 26.35  & 0.14 & & 360    & 360  & 355    & 2     &     \\
        & 2.1 & 25.89           & 26.36& 25.60  & 0.16 & & 360    & 360  & 350    & 4     &     \\
        & 2.2 & 25.81           & 26.03& 25.51  & 0.17 & & 360    & 360  & 360    & 0     &     \\
        & 2.3 & 24.97           & 25.15& 24.80  & 0.11 & & 360    & 360  & 350    & 4     &     \\
T Tau   &ALL  & 71.42           & 86.51& 55.96  & 12.22& (0.55) & 520    & 525  & 510    & 6  & 25  \\
        & 1   & 86.10           & 86.51& 85.79  & 0.18 & & 520    & 525  & 510    & 8     &     \\
        & 2   & 61.64           & 64.28& 55.96  &1.93  & & 515    & 525  & 510    & 4     &     \\
SU Aur  & ALL & 4.65            & 5.43 & 3.91   & 0.69 & (0.08)& 550    & 560  & 535    & 6   & 20  \\
		& 1   & 3.98            & 4.05 & 3.93   &0.04  & & 555    & 560  & 545    & 5     &     \\
		& 2   & 5.33            & 5.43 & 5.24   &0.06  & & 545    & 555  & 535    & 7     &     \\ 
DR Tau  &ALL  & 65.32           & 78.69& 58.50  & 7.53 & (0.68) & 659    & 689  & 625    & 20 &25  \\
        & 1   & 78.16           & 78.69& 77.49  & 0.43 & & 685    & 690  & 670    & 8     &     \\
        & 2   & 61.31           & 63.45& 58.50  & 1.86 & & 650    & 680  & 625    & 15    &     \\
RW Aur  &ALL  & 74.04           & 81.11& 67.88  & 3.95 & (3.6) & 720    & 725  & 715    & 5   & 40  \\
GW Ori  & ALL & 19.33           & 19.52& 19.06  & 0.14 & (0.16) & 505    & 505   & 490    & 6 &25   \\
BF Ori  & ALL & 3.88            & 4.21 & 3.44   & 0.23 & (0.86)& 495    & 505  & 490    & 5   & 25  \\
LkH$\alpha$ 215 & ALL & 30.45           & 30.72& 29.87  & 0.22 & (0.22)& 690    & 700  & 680    & 5   & 30   \\
MWC480 & ALL & 11.85           & 15.11& 9.65   & 1.87 & (0.36)& -      & -    &  -     & -   &   30  \\
        & 1   & 11.21           & 11.74& 10.48  & 0.37 & & -      & -    &  -     & -     & \\
        & 2   & 10.53           & 11.54& 9.65   & 0.74 & & -      & -    &  -     & -     & \\
        & 3   & 14.86           & 15.11& 14.65  & 0.13 & &-       & -    &  -     & -     & \\
        & 4   & 10.71           & 11.75& 9.85   & 0.78 & &-       & -    &  -     & -     & \\
CO Ori  & ALL & 7.57            & 7.91 & 7.22   & 0.19 & (0.1)& 585    & 600  & 560    & 10    & 20  \\
 
\hline
 \end{tabular}
 
\label{tab:EW10w2001}
 \end{table*}

\begin{table*}
\centering
 \caption{Average H$\alpha$ EW and 10\%w measurements for 2003 observations. Also given are the max, min measurements and the standard deviations ($\sigma$) across the range of measurements. Window is the integration window over which EW and 10\%w were calculated.}
\begin{tabular}{@{}lccccccccccc@{}}
\hline 
Object  &Night&$\overline{EW}$& Max &  Min.  & $\sigma$ & Er. &$\overline{10\%w}$  & Max     &  Min.  & $\sigma$   & Win \\
        &   &    [\AA]	      &       &    &    &      & [km\,s$^{-1}$]                &         &        &       & [\AA]   \\ 
\hline
RY Tau  & ALL & 17.62         & 24.52 & 14.85  & 3.59 & (0.11) & 670         & 710  & 640 & 18 & 25 \\
        & 1   & 23.08         & 24.52 & 20.37  & 1.43 & & 690         & 710  & 680 &  8 &  \\
        & 2   & 14.98         & 15.16 & 14.85  & 0.11 & & 675         & 680  & 665 &  4 &  \\
        & 3   & 15.90         & 16.02 & 15.73  & 0.08 & & 655         & 660  & 640 &  5 &  \\	
AB Aur  & ALL & 22.87	      & 25.92 & 19.99  & 1.84& (0.21) & 385         & 550  & 340 & 54 & 25 \\
        & 1   & 25.56         & 25.92 & 25.30  & 0.21 & & 370 	    & 370  & 370 & 0  &   \\					
        & 2   & 23.65         & 23.78 & 23.47  & 0.09 & & 360 	    & 360  & 360 & 0  &   \\
        & 3   & 22.72	      & 23.49 & 21.17  & 0.60 & & 445 	    & 550  & 345 & 62 &   \\		
        & 4   & 20.10	      & 20.26 & 19.99  & 0.08 & & 345 	    & 345  & 340 &   2&   \\
T Tau   & ALL & 47.03	      & 48.02 & 46.51  & 0.47 &(0.39) & 445	        & 450  & 445 & 2  & 25 \\
SU Aur  & ALL & 5.71	      & 8.09  & 3.60   & 1.91     & (0.09) & 575         & 610  & 540 & 23 & 20 \\
        & 1   & 7.96	      & 8.09  & 7.56   & 0.13     & & 590         & 600  & 580 & 4 &  \\
        & 2   & 4.34	      & 4.49  & 3.93   & 0.16     & & 550         & 565  & 540 & 6 &  \\
        & 3   & 3.68	      & 3.95  & 3.60   & 0.09     & & 560         & 610  & 585 & 8 &  \\
DR Tau  & ALL & 80.48         & 81.47 & 78.74  & 0.70 &(0.29) & 640	        & 645  & 630 & 5  & 50 \\
RW Aur  & ALL & 64.66         &77.13  &  34.57 & 13.54&(3.18)  & 720         & 740  & 705 & 9  & 40 \\
        &  1  & 48.54         &52.32  &  36.57 & 3.67 &  & 730         & 740  & 720 & 5  & \\
        & 2   & 75.41         &77.13  &  73.34 & 0.93 &  & 715         & 720  & 705 & 5  &  \\
GW Ori  & ALL & 33.16         & 33.32 & 32.99  & 0.07 &(0.12)  & 405	        & 410  & 400 & 3  & 25 \\
V773    & ALL & 1.85          &1.98   &1.73    & 0.06 &(0.04)  & 455         & 475  & 440 & 7  & 25 \\
UX Tau  & ALL & 8.72	      & 9.23  &8.28    & 0.29     &(0.05)  & 440         & 455  & 425 & 9  & 25 \\
BP Tau  & ALL & 92.57         & 103.49& 88.63  & 4.50  &(0.44)  & 480         & 500  & 450 & 13 & 30 \\
        & 1   &103.25         &103.49 &102.97  & 0.22  &  & 455         & 460  & 450 & 5  &   \\
	    & 2   & 91.04         & 94.42 &88.63   & 1.97  &  & 485         & 500  & 470 & 9  &    \\

\hline
 \end{tabular}

\label{tab:EW10w}
 \end{table*}

\subsection{Plausibility Checks}
\begin{figure}
\begin{tabular}{lcr}
\includegraphics[scale=0.22]{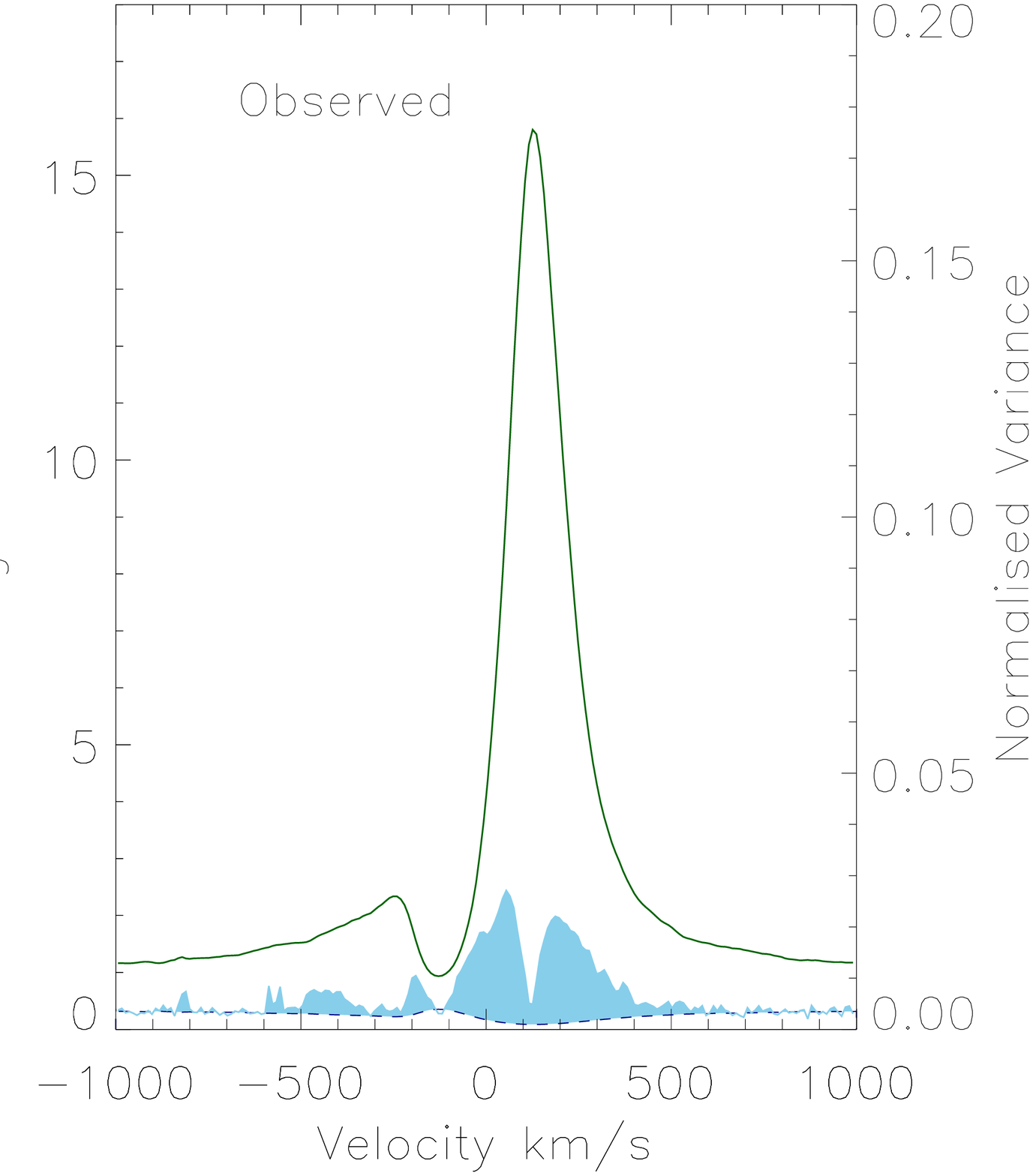}& &\includegraphics[scale=0.22]{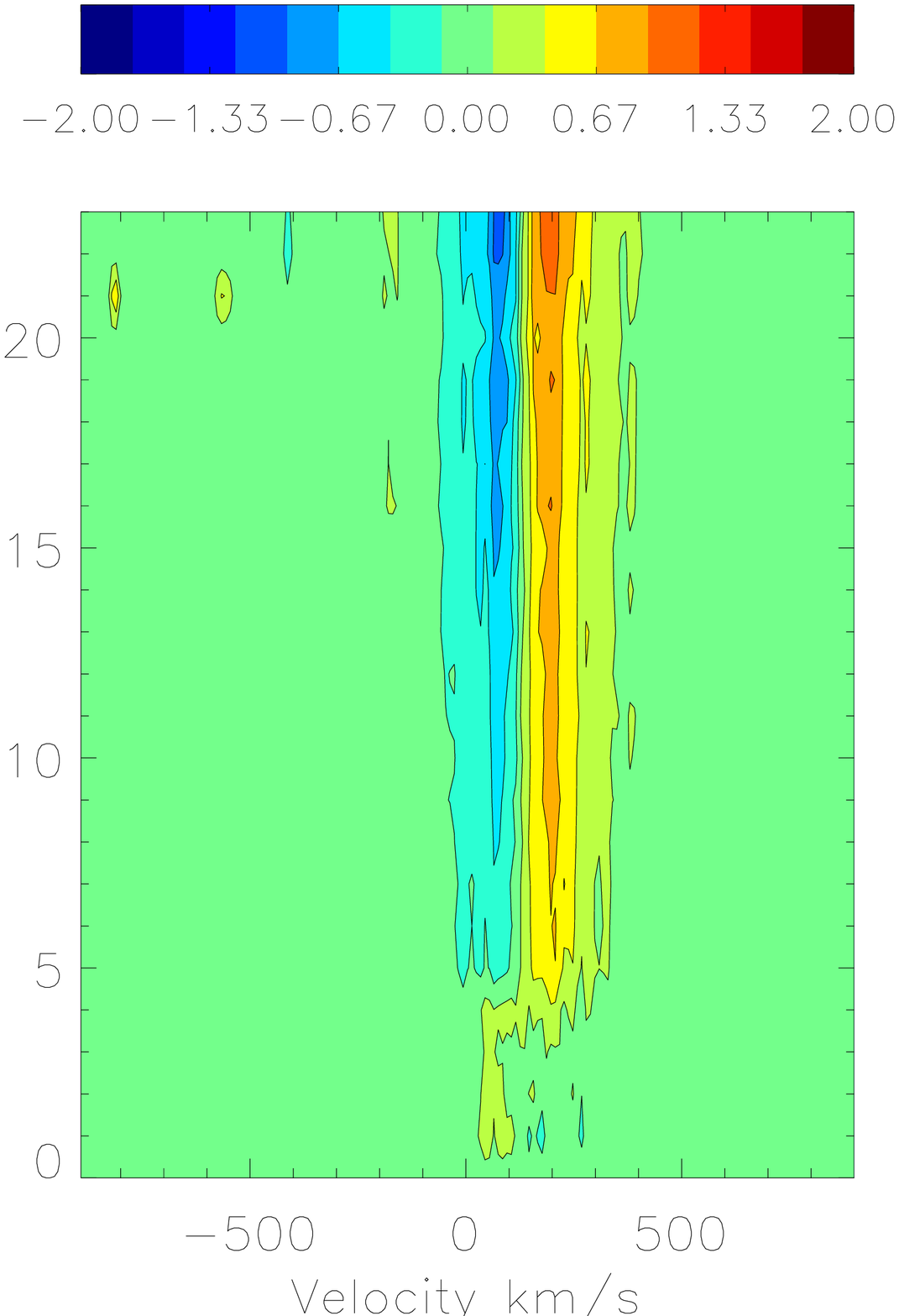} \\
\includegraphics[scale=0.22]{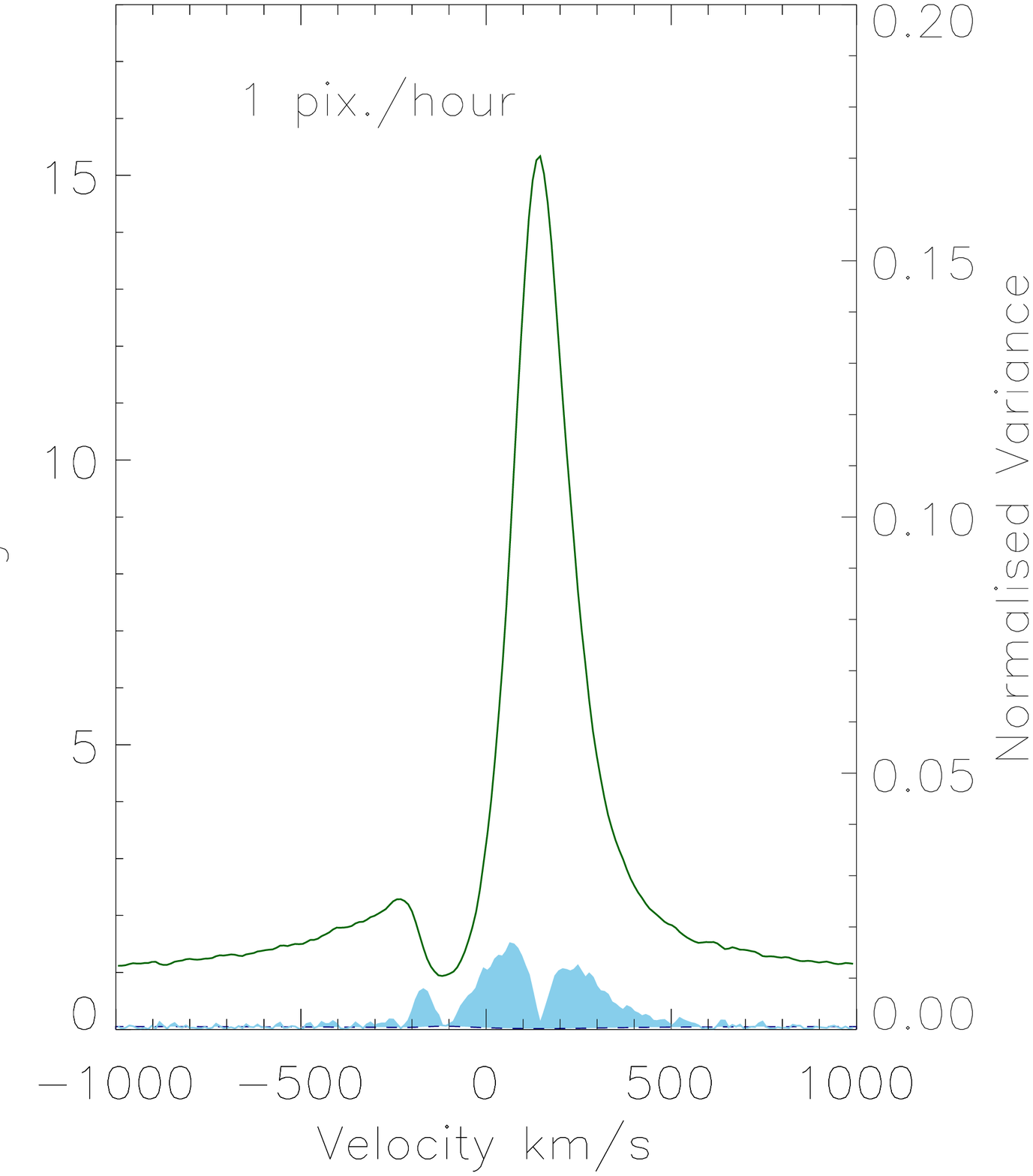}& & \includegraphics[scale=0.22]{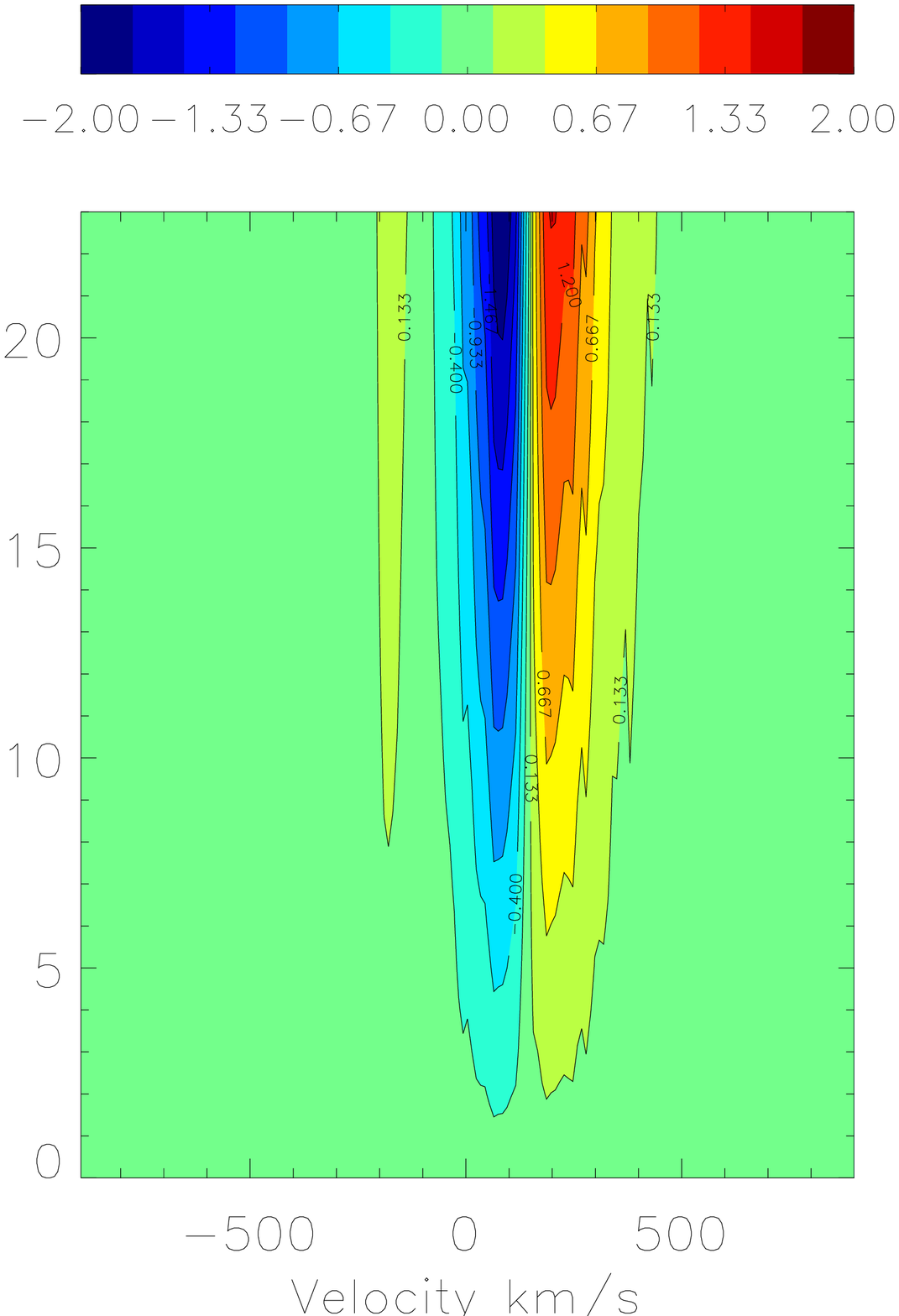} \\
\end{tabular}
\caption{Top: Observed Variations in DR Tau in 2003. Bottom: Taking the first spectrum of the 2003 DR Tau observations, a drift rate of 1 pixel shift per hour was applied to simulate telescope flexure. This is equivalent to the instrument flexure, and results in a similar morphology in the variance profile and the surface plot as is observed. However the EW variations remain an order of magnitude lower than what is observed.} 
\label{fig:DRTAU_profile_shift}
\end{figure}

A number of tests were performed to determine whether the \emph{slow variations} observed in the sample are 
real changes within the line profile, and not because of some changes in the instrumental set up or 
weather. These are listed below. 
\begin{itemize}
 \item The changes in the line measurements were compared to the changes in the signal-to-noise (S/N) 
 in each observations. In the majority of cases, the changes in the EW or 10\%w do not
  follow the changes in the S/N.  There is one object, RY Tau for which changes in the intensity across the line correspond to changes in the S/N  (See Appendix, Fig.\,\ref{fig:RYTAU_plots_2001}). This occurs in the first nights observations in 2003, but there is a similar change in S/N during the second night without any corresponding change to the EW measurement, so the two events are not considered to be connected. 
   \item With regards to the \emph{rapid events} in the profile of AB Aur, tests were also run on the sky 
subtraction around H$\alpha$. This was found not to be the cause of these changes. 
\item Instrumental set-up was checked for any changes between observations, no differences were found.
 The dispersion axis changed between the 2001 and 2003 observations, and as a result the wavelength 
 range is smaller ($\sim$ 200\AA) in the 2001 observations. However this does not effect our 
 observations of the H$\alpha$ emission line.  
\item There were changes in the weather conditions across the four nights, but these would result in 
a change in the intensity across the entire profile rather than a change in the profile structure. 

\item The stability of the instrument was tested with a number of objects and using the changes due to telescope flexure (15 $\mu$m/hour) compared to the pixel size \footnote{See http://www.ing.iac.es/Astronomy/instruments/isis/flextest.html}. In the 2001 observations the pixel size is 24 $\mu$m and in the 2003 observations, the pixel size is 13.5 $\mu$m. The difference arises from a change in detector.  
Based on this, a drift rate of 1 pixel per hour was used to estimate how this would affect our line profiles. For some objects the morphology of the variance profile and the appearance of the surface plot in the shifted series of spectra are very similar to what is observed. In the cases tested, though the difference in flux in the lines matched that observed, the variance profiles remained lower than the observed variance profiles. Also any changes in the EW were well below those observed for these sequences, and remained within the experimental error. 
The top panels of Fig.\,\ref{fig:DRTAU_profile_shift}, show the observed variations in DR Tau in 2003. By shifting the spectra by a total of 1 pixel per hour (as expected from flexure), a very similar morphology in the variance and surface difference plot are seen (lower panels  Fig.\,\ref{fig:DRTAU_profile_shift}). However the variance profile remains lower than is observed. Also the changes in EW due to these shifts are 0.04\,\AA, which is far below our estimated experimental error and the measured variations. To produce the same level of variations in the EW measurements, a drift rate of over six times the instrument flexure is needed. This suggests that though some of these variations could be due to instrument flexure, it is on the low level, and measured changes in the EW are unlikely to come from these instabilities.

\end{itemize}
As a result of these checks, the variations observed in the profiles are believed to be true variations 
in the emission from these objects.

\section{Origin of H$\alpha$ Emission}\label{sec:origin}
The most dominant known sources of H$\alpha$ emission in young stellar objects are chromospheric 
activity and accretion. All the objects are likely candidates for accretion, as they are all thought 
to have circumstellar discs at sufficiently young ages. There is a number of ways to determine whether
 the emission is solely from chromospheric activity or not. 

\subsection{H$\alpha$ Measurements}\label{sec:halpha_measurements}
The most common method to determine whether a stellar object is accreting or not is to use the 
H$\alpha$ EW and 10\%w measurements. The standard limits between an accreting and a non-accreting 
T Tauri star are EW of 10\,\AA~and a 10\%w of 270\,km$s^{-1}$ \citep{2003ApJ...582.1109W}. 
The limits in the EW come from the fact that these young stars are active and show some level of 
H$\alpha$ emission from chromospheric activity alone. 
This activity is much weaker than the emission we expect from accretion especially in the case of T Tauri and Herbig Ae/Be stars \citep[e.g.][]{2013A&A...551A.107M}. As chromospheric H$\alpha$ emission depends strongly on spectral type, for the T Tauri stars in the sample we use a a spectral type dependent cut-off as given by \citet{2003AJ....126.2997B}.

The 10\%w provides a measure of the broadening of the emission line. As the material in the accretion flows is free-falling to the stellar surface, it can reach velocities of 100s of km\,s$^{-1}$. These high velocities are then reflected in the broadening of the H$\alpha$ line, and can be estimated by the 10\%w of the emission line. 

Using the EW cut-off based on spectral types by \citet{2003AJ....126.2997B}, and from H$\alpha$ 10\%w \textgreater 270\,km\,s$^{-1}$ we can classify all of our sample objects as accretors. The mean H$\alpha$ EW and 10\%w measurements for each object are given in Table \ref{tab:EW10w}. 

For Herbig Ae objects, the chromospheric contribution to the H$\alpha$ emission is likely to be very insignificant, as the emission decreases in earlier spectral types. In fact, it is more important to take consideration of the photospheric absorption in these objects. For example, in comparison to main sequence objects of the same spectral type, \citet{1998PASP..110..863P} give an absorption strength of 98\,\AA~for A0, 93\,\AA~for B1, 32\,\AA~for F8, 25\,\AA~for G2 and 0\,\AA~for K5. This is quite significant compared to the emission measured in some of the early targets, and hence the EW measurements are underestimating the strength of the H$\alpha$ emission in the sample. 
To account for this photospheric absorption, it is common to fit the observed spectrum with a main sequence spectrum to estimate the photospheric contribution. \citet{2011A&A...529A..34M} removed the photospheric absorption in this way for a number of objects in this sample. For BF Ori, \citet{2011A&A...529A..34M} measured a mean EW of 9.9\,\AA~compared to 3.8\,\AA~here, for CO Ori they found 21.3\,\AA~compared to 7.6\,\AA~here and for RY Tau 15.3\,\AA~compared to 10.4\,\AA~in 2001 and 17.6\,\AA~in 2003 here. 

This shows the removal of the absorption results in a significant difference for the weaker lines but not for the stronger emission lines. The EWs given in Table \ref{tab:EW10w} are not corrected for photospheric absorption, however before accretion rates are calculated a correction is applied. See Sect.\ref{sec:accretion_rates} for further details.

\subsection{Average and Variance Profiles}
A \emph{typical} characteristic of an accreting system is an asymmetric H$\alpha$ emission line, often with more than one emission peak and overlying absorption features. Under the magnetospheric accretion model, the H$\alpha$ emission originates in a structured flow of material. This results in emission across a large velocity range. Stellar winds and jets are associated with accreting systems \citep{2007prpl.conf..231R}, and can contribute both in the form of excess emission to the H$\alpha$ line and overlying absorption features (see next Sect. \ref{sec:wind}). 
This can been seen in many of H$\alpha$ profiles in our sample. The average profiles for the sample show that the majority of the profiles are asymmetric, with many of them showing large absorption features. 

Over-plotted in blue on each average profile in the second row of Fig.\,\ref{fig:ABAUR_profiles_2003_1} are the variance profiles for each object and each observation block. These profiles indicate clearly which parts of the profiles show the most changes. In most cases where large changes occur they are confined to discrete  wavelength ranges. This is also further evidence that there are multiple components within the emission lines and that it is not all from a single point i.e., the central star.

\subsection{Chromospheric Activity versus Accretion}\label{sec:chromo_accretion}
There have been limited studies into the activity in accreting stars, mainly because both accretion and chromospheric activity cause similar emission in certain emission lines (H$\alpha$, Ca, He etc.) as well as UV and X-ray excess emission \citep{2008A&A...487..293F,2011A&A...526A.104C}. However there has been some recent work on chromospheric activity by \citet{2013A&A...551A.107M}, focusing on lower mass objects than in this sample.

Flares have distinct rise and decay profiles by comparison to accretion events, making this one of the most commonly used attributes to distinguish between the two events. So, if a rapid rise in EW occurs, followed by a slow decay, it is usually attributed to a flare. \citet{1998ASPC..154.1701G} found accretion and flare events using spectroscopic monitoring of WTTS and CTTS. From $\sim$ 9500 spectra of CTTS, and 7200 of WTTS in Chamaeleon and Taurus they found 24 flares in WTTS and 15 in the CTTS. The largest flare was found in a WTTS. Whilst they claim that flare activity on T Tauri stars is more common than solar flares, these numbers show that these events do not occur very often in either CTTS or WTTS. \citet{1998ASPC..154.1701G} also observed many slow variations in EW across a night's observation which they attribute to rotation, or variations in accretion rate.

One definitive attribute of flares is that the H$\alpha$ emission associated with it is nearly entirely limited to the line centre with some extension in the wings up to \textless 100\,km\,s$^{-1}$ \citep{1981ApJS...46..159W,1990ApJS...74..891R}. This can be seen in the observations of the CTTS by \citet{1998ASPC..154.1701G}, where the flaring produces relatively small symmetric emission peaks (20\% above the continuum). In some cases small asymmetries in the emission lines were seen, which are thought to be connected to coronal mass ejections (CMEs). For H$\alpha$ this took the form of excess emission in the blue wing which moved through the line as the material is ejected from the atmosphere of the star. However, in the case of the WTTS IM Lupi, the authors argued that the asymmetries observed 
in the H$\alpha$ profile are more likely to be associated with accretion activity rather than a flare \citep{2010A&A...519A..97G}.

In T Tauri stars, the range of time-scales for flare events is 10$^{2}$\,-\,10$^{4}$\,s \citep{1981ApJ...244..520W}. Although the sample does not have exhaustive time coverage, the \emph{rapid events} do not show the typical fast rise and slow exponential decay expected of flares. Stellar flares can rise to a peak within a few minutes, while the decay can also happen within half an hour or so \citep{2010A&A...514A..94L}, but it is much more usual to have flares lasting hours to days \citep{1999Natur.401...44S,2000A&A...353..987F,1996A&A...311..211K}.

Finally, the strength of the H$\alpha$ emission from accretion is much larger than emission from the majority of flares. This, it is expected that any change in accretion rate will have a much stronger effect on the emission line profile than any flare activity. 

Thus the variations and morphology of the emission seen in this sample, make it unlikely that the majority of the events are from chromospheric activity.

\subsection{Emission from Outflows}\label{sec:wind}
The traditional tracers of outflows and jets in young stellar objects are forbidden emission lines such as [OI]\,$\lambda$6300, 6363\,\AA, [NII]\,$\lambda$6583\,\AA, and [SII]\,$\lambda$6716, 6731\,\AA~\citep{1995ApJ...452..736H}. However, it has also been shown using spectro-astrometry that these outflows can contribute to the H$\alpha$ emission. This mostly takes the form of additional emission on the blue side, but cases have been found in which there is excess emission from the outflows found in both the blue and the red wings \citep{2003A&A...397..675T,2009ApJ...691L.106W}. 

Within the sample, signatures of [OI]\,$\lambda$6300 and 6363\,\AA~were found in many of the targets (see Table \ref{tab:winds}), indicating that there is an outflow from these objects. In the majority of cases the emission line was very narrow and weak ($\sim$ 1\,\AA). This suggests that though there is probably some wind emission contaminating the H$\alpha$ line, it is going to be a weak contribution.

Since the density of the out-flowing material is much lower than the densities in the accretion flows, a small change in an accretion flow will produce the same change in emission as a large change in the outflow rate. This, along with the fact that the accretion emission will always be stronger in the H$\alpha$ line than the outflow emission in an accreting object, makes it more likely that it is an accretion rate change. However, we cannot rule out that changes in the outflow have some contribution to the variations, so any accretion rate variations we derive are upper limits. 

\begin{table}
\centering
\caption{Detections of other emission lines.  [OI]\,$\lambda$6300 and  $\lambda$6363 \AA~are associated with wind emission. The wavelength coverage is slightly smaller in the 2001 observations, so the [OI]$\lambda$6300 is not covered by these observations.  No [SII] ($\lambda$6715, 6729\AA) or [NII] ($\lambda$6548.4, 6583.4\AA)  emission were detected in any of the observations, are usually associated with stellar jets. The emission line HeI $\lambda$ 6678.2\AA~ which was detected in a number of targets, which is associated both with outflows and accretion. }
\begin{tabular}{@{}lcccccccc@{}}
\hline
Name	& Year & [OI] $\lambda$6300\AA & [OI] $\lambda$6363\AA & HeI $\lambda$6678.2\AA 		 \\
\hline
RY Tau	& 2003 &           Yes      &    No             &  No                           \\
      	& 2001 &            -       &    No             &  No                           \\	 
AB Aur  & 2003 &           Yes      &    Yes            &  Yes                           \\
        & 2001 &            -       &    Yes            &  Yes                           \\  
T Tau 	& 2003 &           Yes      &    No             &  No                           \\
		& 2001 &            -       &    No             &  No                           \\
SU Aur	& 2003 &           No       &    No             &  No                           \\  
		& 2001 &            -       &    No             &  No                           \\	 
DR Tau	& 2003 &           Yes      &    Yes            &  Yes                           \\
        & 2001 &            -       &    Yes            &  Yes                           \\    	 
RW Aur  & 2003 &           Yes      &    Yes            &  Yes                           \\
        & 2001 &            -       &    Yes            &  Yes                           \\  
GW Ori	& 2003 &           Yes      &    Yes            &  No                            \\ 
        & 2001 &           -        &    No             &  No                            \\ 	 
V773 Tau& 2003 &           Yes      &    No             &  No                            \\   
UX Tau A& 2003 &           Yes      &    No             &  No                            \\   
BP Tau	& 2003 &           Yes      &    No             &  Yes                            \\     
BF Ori  & 2001 &           -        &    Yes            &  No                            \\
LkH$\alpha$ 215& 2001 &           -        &    No             &  No                            \\   
MWC 480 & 2001 &           -        &    No             &  Yes                            \\   
HD141569& 2001 &           -        &    No             &  Yes                            \\   
CO Ori  & 2001 &           -        &    No             &  Yes                            \\ 

\hline
\end{tabular}

\label{tab:winds}
\end{table}

As a potential caveat, interferometric observations of the Herbig Be star, MWC 297 have found over 90\% of the Br$\gamma$ emission to originate in a disk wind \citep{2007A&A...464...43M} . Further high-resolution spectra were taken, and modelling of the Br$\gamma$ observed emission lead to the conclusion that the Br$\gamma$, H$\alpha$ and H$\beta$ all formed in a disc wind, with the majority of the H$\alpha$ and H$\beta$ emission originating in the polar regions.  
Optical spectra were taken of MWC 297 from which an EW of 649\,\AA~was found for H$\alpha$ \citep{1997MNRAS.286..538D}. Note this is an order of magnitude higher than any H$\alpha$ measured in this sample. \citet{2007A&A...464...43M} compared their wind model to these spectra and found the intensity of the H$\alpha$ and H$\beta$ lines to agree to within 10\% of observed intensities, and also found a good agreement between the model line widths and those observed. \citet{2008A&A...489.1157K} also found connections between B$\gamma$ emission and stellar wind in four out of five Herbig Ae/Be stars using interferometric data. In particular they find the emitting area of the Br$\gamma$ emission line to be correlated with the H$\alpha$ profile. If a star shows an inverse P-Cygni profile, and a high accretion rate, the Br$\gamma$ line is more likely to come from the accretion column, whereas if the H$\alpha$ emission line is single or double peaked, Br$\gamma$ is emitted in a greater area consistent with a disc or stellar wind.
These interferometry results \citep{2008A&A...489.1157K} are in line with the earlier H$\alpha$ spectro-polarimetry work of \citet{2002MNRAS.337..356V}, where it was found that the later Herbig Ae stars have a compact line-forming region, similar to that seen in T Tauri stars \citep{2003A&A...406..703V, 2005MNRAS.359.1049V}, whilst the earlier type Herbig Be stars have a larger line-forming region. 

As this sample does not contain Herbig Be stars, it can be argued that it is not necessary to be overly concerned about disk winds providing the line emission in the sample. 
Moreover, recent studies of accretion diagnostics in Herbig Ae/Be stars (F0\,-\,O9) using the large wavelength coverage of X-Shooter (300\,-\,2500\,nm) have also taken place. \citet{2012AN....333..594P} found that accretion rates derived from H$\alpha$ to be in good agreement with accretion rates derived from indicators which can only originate in the high temperature region close to the stellar surface. 
The variations in the H$\alpha$ accretion rate were also found to follow the accretion rates from the other indicators, and showed a similar magnitude of variations. This is a strong suggestion that the changes in the H$\alpha$ emission of Herbig Ae/Be stars follow the changes in accretion rate, or if H$\alpha$ is indeed primarily a wind indicator for all these objects, then the wind emission is very closely connected to the accretion emission as suggested by magnetospheric accretion models. 

Other studies have shown that, though there is a strong correlation between U band excess and H$\alpha$ luminosity, the variations in both do not follow each other in every case \citep{2011A&A...535A..99M}. \citet{2013ApJ...776...44M} found a close agreement between the accretion rates derived from the Balmer excess and H$\alpha$ emission in two Herbig Ae stars (one is also in this sample, MWC 480). 
Despite this agreement, they suggest the variations in the strength of the H$\alpha$ emission to be uncorrelated to the variations in the Balmer excess. This is an indication that there is either a time delay in the variations in the emission lines that is not seen in the low mass accretors, or the variations in the lines originate from some other process than accretion. Nonetheless, the level of the variations in these cases are low, so the correlation could be washed out by any noise in the observations. More sensitive observations are needed to confirm this. 

Multi-wavelength studies with X-shooter have also been undertaken with low mass T Tauri stars \citep{2012A&A...548A..56R}, which again find the H$\alpha$ emission to be correlated with the UV excess and other indicators. There is substantial evidence confirming the effectiveness of using the H$\alpha$ emission as an accretion rate estimator across the whole mass range of this sample, even if it is partially contaminated by wind emission.

\subsection{Variations from Continuum Changes}
Changes in the accretion rate will lead to changes in the veiling in the stellar continuum. This can lead to measured changes in the EW while the line emission remains steady. As the spectra in this work are not flux calibrated, changes in the continuum cannot be accounted for. The majority of the targets in this sample are optically variable (i.e.\,\,see \citet{2007A&A...461..183G} for light curves of 8 targets), and so it can be expected that some changes in the continuum level occur over the course of these observations. 

However a change in the continuum will lead to a change in the emission across the whole line. This is not what is seen in this sample (i.e.\,see Fig.\,\ref{fig:ABAUR_profiles_2003_1}). The variance profiles attest to the fact that the variations measured originate in a narrow wavelength ranges within the profiles. This strongly suggests that even if there are continuum changes during the observations they are not significant compared to those changes that occur within the profile emission. 

It has been reported that continuum changes in these objects are also likely to come from occultations of the star by orbiting circumstellar clouds \citep{2011A&A...535A..99M,1994A&A...292..165G}. However, apart from some known exceptions (see \citet{1994A&A...292..165G}), these occultations are expected to cause both a fall in the continuum and line emission.

\subsection{Previous Observations}
The objects within the sample are mostly well studied objects. Many of them have previous observations of other accretion related emission lines, as well as confirmed accretion with UV excess measurements, which is considered to be the most direct indicator of ongoing accretion. These detections are discussed on an individual object basis in the Appendix. 

\smallskip
The above arguments point towards accretion being the primary source of the emission observed and the variations in the emission lines. Through the following sections, it is assumed that the contribution of chromospheric and wind emission to the H$\alpha$ line is negligible, and that any variations observed are the result of changes in the accretion rate.

 \section{Accretion Rates}\label{sec:accretion_rates}
Making the assumption that all of the H$\alpha$ emission originates in accretion flows, we can calculate accretion rates from the H$\alpha$ EW measurements for all objects in the sample. 

\subsection{Derivation of Accretion Luminosity Relation}
A number of relations have been derived between the H$\alpha$ line luminosity and accretion luminosity (e.g. \cite{2008ApJ...681..594H}). Since the mass range of our sample lies in in between those of the majority of these relations, we derived a new relation between H$\alpha$ luminosity ans accretion luminosity using the readily available data sets of \cite{2012A&A...543A..59M} and \cite{2014A&A...561A...2A}.  The relation of \cite{2012A&A...543A..59M} was derived using photospheric corrected H$\alpha$ emission (See Sect. \ref{sec:halpha_measurements}) and quasi simultaneous U band photometry in a sample range of $\sim$ 1.5\,-\,5 M$_{\odot}$. \cite{2014A&A...561A...2A} derived accretion luminosities by fitting X-shooter spectra with a slab model, and used the H$\alpha$ emission in the same spectra to derive H$\alpha$ luminosities across a sample range of $\sim$0.03 to $\sim$ 1.2 M$_{\odot}$.

\begin{figure}
\includegraphics[scale=0.44]{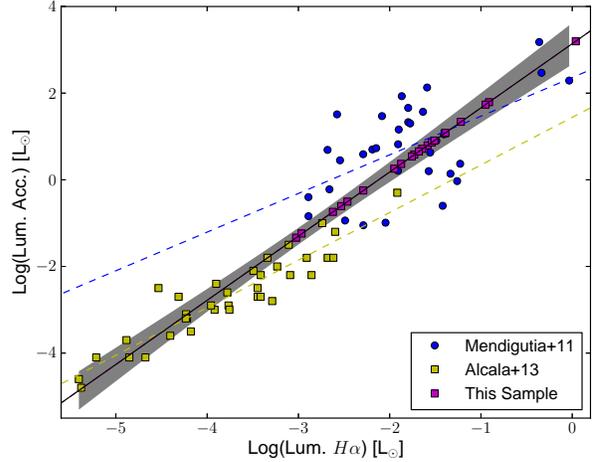}
\caption{Accretion luminosity versus H$\alpha$ line luminosity. Yellow points indicate those taken from \protect\citet{2014A&A...561A...2A} and blue points from \protect\citet{2011A&A...535A..99M}. Yellow dashed line and blue dashed line represent the linear fits to these separate data sets. The solid black line indicates the fit to both data sets. The shaded grey region indicates the 2-$\sigma$ confidence interval around this fit. The purple data points in this plot represent the targets within these works, where 2003 and 2001 data are represented separately. }
\label{fig:accretion_relation}
\end{figure}

The accretion luminosities and H$\alpha$ luminosities from both these samples are plotted in Fig. \ref{fig:accretion_relation}. A least-squares regression was applied to the combined sample, using the \emph{python} procedure \emph{stats.linregress} (\emph{scipy} library), and the derived relations took the form of 
\begin{equation}\label{eq:derived_relation}
 \log(L_{\mathrm{acc}}) = \mathrm{A} + \mathrm{B}\cdot \log(L_{\mathrm{H\alpha}})
\end{equation}
\noindent where $\mathrm{A} = 3.14\pm0.19$ and $\mathrm{B} = 1.48\pm0.06$. This fit is shown as a solid black line in Fig.\,\ref{fig:accretion_relation}. The grey shaded region around this fit represents the 2-$\sigma$ confidence interval. 
The relation found here is similar to that found by \citet{2014A&A...561A...2A} where $\mathrm{A} = 1.50\pm0.26$ and $\mathrm{B} = 1.12\pm0.07$ (represented by a yellow dashed line in Fig.\,\ref{fig:accretion_relation}). Also, the relation found here is also very similar to that found by \cite{2008ApJ...681..594H} where $\mathrm{A} = 2.0\pm0.4$ and $\mathrm{B} = 1.20\pm0.11$. However, the higher mass relation found by \cite{2012A&A...543A..59M} is shallower, with $\mathrm{A} = 2.28\pm0.25$ and $\mathrm{B} = 1.09\pm0.16$, as indicated by the blue dashed line.

\subsection{Derivation of Accretion Rates}
Before deriving the accretion rates from the EW, they were corrected for photospheric absorption. This was done by adding the strength of the absorption in the nearest spectral type as given by \citet{1998PASP..110..863P}. The photospheric corrections used in each case are given in Table \ref{tab:assumed_parameters}.

In order to convert an H$\alpha$ EW to luminosity, it is first necessary to convert the EW to a flux. 
This was done by selecting the closest Kurucz model \citep{1979ApJS...40....1K} in spectral type and surface gravity to each of the sample members. (For some of the sample surface gravities were not available so they were assumed see Table \ref{tab:assumed_parameters}). A line was fitted to the model continuum over the wavelength ranges 6450\,-\,6500  and 6620\,-\,6720\,\AA, and was used to find a continuum flux estimate at the line centre (6562.81\AA). Using these continuum estimates (given in Table \ref{tab:assumed_parameters}) the EW can be transformed to a flux per unit area on the stellar surface ([ergs sec$^{-1}$cm$^{-2}$]), which is then converted to a luminosity ($L_{H\alpha}$) using the published stellar radii given in Table \ref{tab:assumed_parameters}. This step assumes that the source of the line luminosity is distributed uniformly across surface of the star. As the actual distribution of the accretion flow is unknown, this is commonly taken as a reasonable approximation.

Eq. \ref{eq:derived_relation} is then used to determine accretion luminosities, however the spread coefficients is not taken into account as they originate from the empirical derivation of this relation. Assuming that all the gravitational energy from the accretion is converted into luminosity, the accretion luminosity is related to the accretion rate ($\dot{M}$) as follows \citep{2008ApJ...681..594H}:
\begin{equation}
 \dot{M} = \left(1 - \frac{R_{*}}{R_{\mathrm{in}}}\right)^{-1} \cdot \frac{L_{\mathrm{acc}}R_{*}}{G M_{*}}
\end{equation}

\noindent Here $R_{*}$ is the stellar radius, $R_{\mathrm{in}}$ is the in-fall radius and $M_{*}$ is the stellar mass. \citet{1998ApJ...492..323G} approximates the factor $(1 - \frac{R_{*}}{R_{\mathrm{in}}})^{-1} \sim 1.25$, assuming $R_{\mathrm{in}} \sim 5R_{*}$, which is appropriate for T Tauri stars. An infall radius of $R_{\mathrm{in}} \sim 2.5R_{*}$ has been calculated to be more suitable for the higher rotation velocities in Herbig Ae stars \citep{2004ApJ...617..406M}.  Observations of inner hole sizes were provided in the linear spectro-polarimetry study of \citet{2005MNRAS.359.1049V}, these were used where possible, along with other published values. For the rest of the sample, an infall radius of $5R_{*}$ was assumed for the lower mass stars, and in keeping with observations of Herbig Ae stars, an infall radius of $3R_{*}$ was assumed for the higher mass objects. The assumed infall radii, continuum estimates from the models, and the published stellar mass and stellar radii used in the accretion calculations are all given in Table \ref{tab:assumed_parameters}. 

The mean accretion measurements for each object and for each night are give in Table \ref{tab:accretion_rates2001and2003}. Also given, are error estimates, calculated by carrying the H$\alpha$ EW estimated errors through the accretion relations. 

\begin{table}
\centering
\caption{Parameters used in the derivation of accretion rates. The first column contains photospheric absorption correction taken from \protect\citet{1998PASP..110..863P}. ${^*}$:In the case of RY Tau, T Tau and RW Aur no K1 spectral type was available in catalogue so K0 was used in it's place. References for stellar masses and radii are given in Table. \ref{tab:stellar_parameters}. For some objects in the sample, estimates of inner radii from observations were available. For the majority of the sample, an inner radii was assumed, similarly for surface gravity estimates. The continuum under the H$\alpha$ line was estimated from the \citet{1993yCat.6039....0K} stellar atmosphere models. Fluxes are in units of [ergs sec$^{-1}$cm$^{-2}$\AA$^{-1}$].   References: A: Assumed (See Sect. \ref{sec:accretion_rates} for more.), 1: \protect\citet{2002ApJ...581..357K}. 2: \protect\citet{2005MNRAS.359.1049V} 3: \protect\citet{2002ApJ...566.1124A}. 4: \protect\citet{2009A&A...495..901M} 5: \protect\citet{1993A&AS..101..629B},6: \protect\citet{2009PASJ...61..251T} 7: \protect\citet{2003ApJ...590..357D}, 8: \protect\citet{2011A&A...536A..45H}, 9: \protect\citet{1999ApJ...516..900J}  }
\begin{tabular}{@{}lccccccccc@{}}
\hline
Object  &Phot. H$\alpha$ &M$_{*}$ &  R$_{*}$ & R$_{in}$ & log(g) &  F$_{\mathrm{cont}}$           \\
        & EW [\AA]      & $[M_{\odot}]$ & [$R_{\odot}$]&       [R$_{*}$]& [cm.s$^{-2}$] & [10$^{6}$]    \\
\hline
RY Tau  & 1.3*   & 2.0 & 2.7 & 10$^{1}$ & 4.2$^{4}$     & 8.439            \\
AB Aur  & 9.8   & 2.5 & 2.5 &  5$^{2}$ & 4.1$^{5}$     & 3.513 \\
T Tau   & 1.3*   & 2.0 & 3.3 &  5$^{A}$ & 4.1$^{6}$     & 4.858   \\
SU Aur  & 2.5   & 2.0 & 1.0 &  3$^{2,3}$ & 3.9$^{7}$     & 8.355 \\
DR Tau  & 0.0    & 1.0 & 1.2 &  10$^{1}$ & 2.8$^{6}$    & 1.621  \\
RW Aur  & 1.3*   & 1.0 & 2.7 &  5$^{A}$  & 3.0$^{A}$       & 3.026  \\
GW Ori  & 1.9   & 3.7 & 2.5 &  3$^{2}$ & 4.0$^{A}$       & 8.439 \\
BF Ori  & 9.8   & 2.5 & 1.3 &  3$^{A}$ & 3.8$^{4}$     & 2.796  \\          
LkH$\alpha$ 215& 9.8 &4.8 & 5.4 &  3$^{A}$ & 4.0$^{A}$ & 59.47 \\
MWC 480 & 9.3   & 2.3 & 2.1 &  3$^{A}$ & 3.5$^{8}$     & 2.789  \\
CO Ori  & 3.2   & 2.5 & 4.3 &  5$^{A}$ & 3.1$^{4}$     & 9.936   \\
V773 Tau& 0.6    & 1.2 & 2.4 &  5$^{A}$ & 4.3$^{6}$     & 4.853  \\
UX Tau  & 0.6    & 1.3 & 2.7 &  5$^{A}$ & 4.1$^{6}$     & 4.858 \\
BP Tau  & 0.0    & 0.8 & 7.6 &  5$^{A}$ & 3.7$^{9}$     & 1.615  \\
\hline
\end{tabular}

\label{tab:assumed_parameters}
\end{table}

\begin{table*}
\caption{Accretion rates derived from H$\alpha$ EW. The average for all observations in 2001 and 2003 campaign is given, as well as the average for each observation block in the cases of multiple observations. Following each average is the [max\,-\,min] spread in accretion rates over the observation period. Units are $M_{\odot}$.yr$^{-1}$. Errors are calculated by carrying the H$\alpha$ EW errors through the accretion rate calculation. }
\centering
\begin{tabular}{@{}lccccccc@{}}
\hline
\multicolumn{7}{c}{2001} \\
\textbf{2001} & Log($\dot{M}$)&    Night 1	&    Night 2 &   2.2  &   2.3   & Error	         \\
       &  Av.\, \, \,[Sprd]   &  Av.\, \, \,[Sprd]    &   Av.\, \, \,[Sprd]    &  Av.\, \, \,[Sprd]    &        Av.\, \, \,[Sprd]    &  \\
\hline
RY Tau & -7.059\,  \,[0.024] &                     &                     &   &  & (0.004) \\
AB Aur & -5.604\,  \,[0.036] & -5.591\,  \,[0.008] & -5.603\,  \,[0.008] & -5.604\,  \,[0.009] & -5.620\,  \,[0.005]  & (0.005) \\
T Tau  & -5.851\,  \,[0.275] & -5.724\,  \,[0.005] & -5.936\,  \,[0.087]  & & & (0.005) \\
SU Aur & -8.955\,  \,[0.135] & -9.016\,  \,[0.012] &  -8.895\,  \,[0.016]&  &  &  (0.007) \\
DR Tau & -8.111\,  \,[0.191] & -7.991\,  \,[0.010] & -8.148\,  \,[0.052]  &  &  &  (0.007) \\
RW Aur & -6.169\,  \,[0.112] &                     &                       &  &  &  (0.031) \\
GW Ori & -6.945\,  \,[0.014] &                  &                        &  &  &  (0.005) \\
BF Ori & -7.232\,  \,[0.035] &                 &                         &  &  &  (0.005) \\
LkH$\alpha$ 215& -4.067\,  \,[0.014]&         &                          &  &  &  (0.004)  \\
MWC 480& -6.274\,  \,[0.163] & -6.292\,  \,[0.040] & -6.314\,  \,[0.061] & -6.186\,  \,[0.012] & -6.308\,  \,[0.061] & (0.011) \\
CO Ori & -6.252\,  \,[0.041] &                 &                         &  &  &  (0.006) \\

\hline 
\multicolumn{7}{c}{2003} \\
\textbf{2003} &Log($\dot{M}$) &   Night 1 &    Night 2 &   Night 3 &   Night 4  &  Error \\
       &  Av.\, \, \,[Sprd]    &  Av.\, \, \,[Sprd]     &    Av.\, \, \,[Sprd]     &  Av.\, \, \,[Sprd]    &         Av.\, \, \,[Sprd]    &  \\
\hline
RY Tau &   -6.745\,  \,[0.302] & -6.573\,  \,[0.113] & -6.832\,  \,[0.012] & -6.796\,  \,[0.011]  & &(0.004) \\
AB Aur & -5.660\,  \,[0.111] & -5.610\,  \,[0.010] & -5.645\,  \,[0.006] & -5.661\,  \,[0.049] & -5.713\,  \,[0.004] & (0.004) \\
T  Tau &  -6.105\,  \,[0.020] &                     &                     &     & & (0.005) \\
SU Aur & -8.880\,  \,[0.354] & -8.708\,  \,[0.033] & -8.981\,  \,[0.054] & -9.047\,  \,[0.035]& & (0.007)\\
DR Tau & -7.973\,  \,[0.021]   &                     &                     &      & & (0.002) \\
RW Aur &-6.269\,  \,[0.410] & -6.439\,  \,[0.160] & -6.157\,  \,[0.031] &      & & (0.032) \\
GW Ori &-6.622\,  \,[0.006]  &                     &                     &      & & (0.002) \\
V773 Tau  & -7.663\,  \,[0.066]  &               &                      &      & & (0.011)\\
UX Tau & -7.325\,  \,[0.065]  &               &                      &      & & (0.003) \\
BP Tau   & -6.166\,  \,[0.104] & -6.091\,  \,[0.003] & -6.177\,  \,[0.040] & & &(0.003) \\

\hline
\end{tabular}

\label{tab:accretion_rates2001and2003}
\end{table*}

\begin{table}
\centering
\caption{Comparison of Accretion rates from 2003 and 2001 observations. Here the average for each years observations are given, along with the spread [max - min]. Also given is the average over both periods and the spread in accretion rate estimates.  Units are $M_{\odot}$.yr$^{-1}$. }
\begin{tabular}{@{}lcccccc@{}}
\hline 
Object & Log($\dot{M}$) 2001    &  Log($\dot{M}$) 2003               &  Both  \\
       &  Av.\, \, \,[Sprd]    &  Av.\, \, \,[Sprd]    &  Av.\, \, \,[Sprd]   \\
\hline
RY Tau & -7.059\,  \,[0.024] & -6.745\,  \,[0.302] & -6.826\,  \,[0.519] \\
AB Aur & -5.604\,  \,[0.036] & -5.660\,  \,[0.111] & -5.632\,  \,[0.129] \\
SU Aur & -8.955\,  \,[0.135] & -8.880\,  \,[0.354] & -8.900\,  \,[0.354]  \\
RW Aur & -6.169\,  \,[0.112] & -6.269\,  \,[0.410] & -6.248\,  \,[0.443]  \\
DR Tau & -8.111\,  \,[0.191] & -7.973\,  \,[0.021] & -8.037\,  \,[0.212] \\
T  Tau & -5.851\,  \,[0.275] & -6.105\,  \,[0.020] & -5.941\,  \,[0.391] \\
GW Ori & -6.945\,  \,[0.014] & -6.622\,  \,[0.006] & -6.710\,  \,[0.334] \\
\hline
\end{tabular}

\label{tab:accretion_rates_comparison}
\end{table}

Note, this relation (Eq. \ref{eq:derived_relation}) was derived using the entire line emission. Therefore it includes excess emission from wind and chromosphere, the effects of which will have been averaged out over the calibrating sample. The absolute value is then close to what would be measured from the UV excess, but the variations can still be affected by these extra sources of emission.

The mean accretion rate spread for a single night for H$\alpha$ EW is  0.01\,-\,0.07\,dex. Half of the sample have multiple nights observations so a measure of the spread in accretion rates from one night's observation to the next could be obtained. This inter night spread, 0.04\,-\,0.4 dex, is an order of magnitude larger than the spread on a single night. There are 7 targets that have observations in both 2001 and 2003 (See Table \ref{tab:accretion_rates_comparison}), when comparing the accretion rates for these 7 objects over the 2 years the spread is 0.13\,-\,0.52 dex.

\subsection{Comparison with Other Accretion Rate Relations}\label{sec:other_acc_relations}

\begin{figure}
\centering
\includegraphics[scale=0.44]{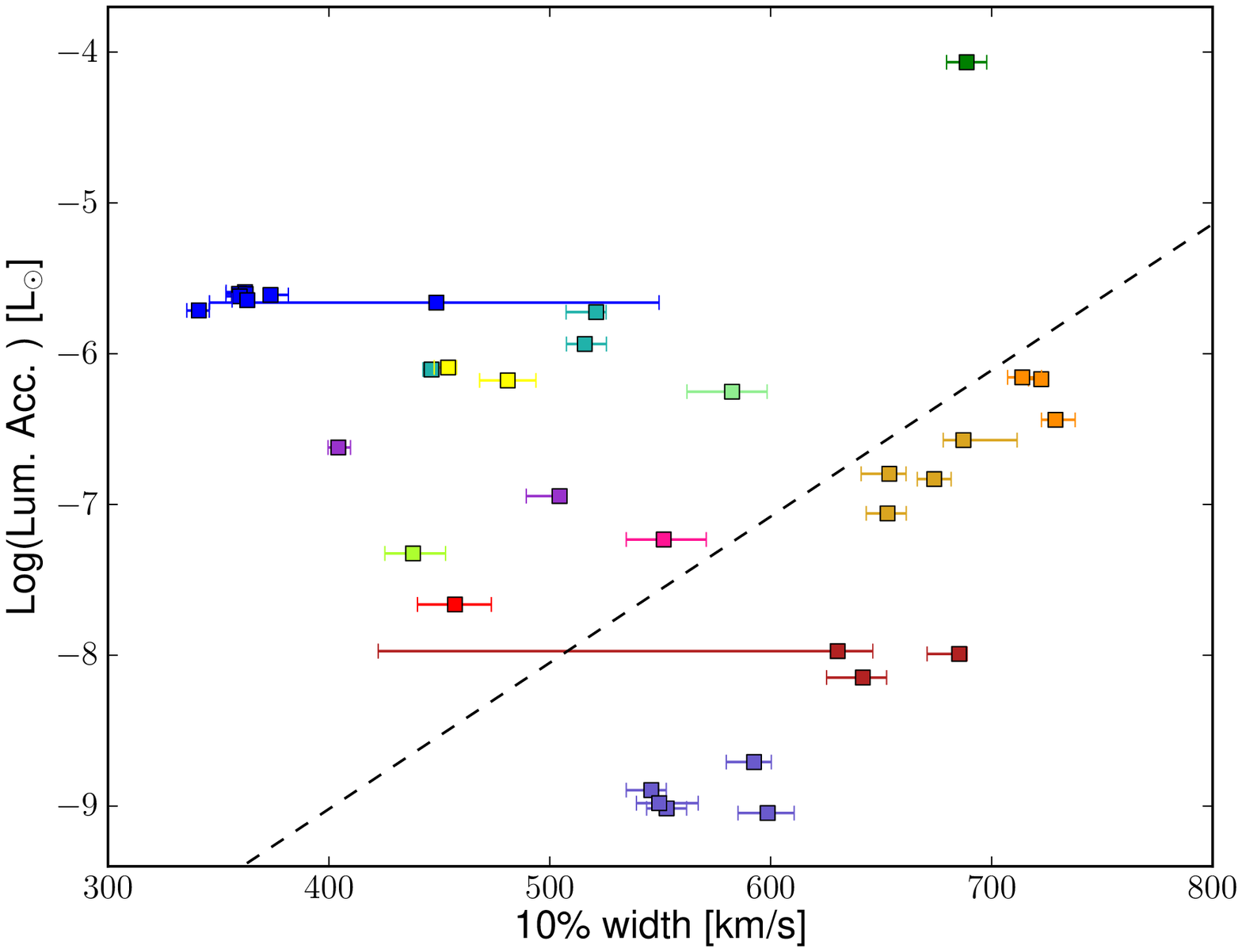}
\caption{Average H$\alpha$ EW derived accretion rate versus the average H$\alpha$ 10\%w. Error bars represent the max - min spread in each. Measurements for each observation block of each object are compared, and the colours represent different objects. The over-plotted dashed line is the accretion rate to 10\%w relation $Log(M_{acc}) \, \sim -12.9 + 0.0097 \cdot (10\%w)$ empirically derived by \citet{2004A&A...424..603N} for brown dwarf and T Tauri objects. As found previously by \citet{2011A&A...529A..34M}, this plot shows that the two measurements are not correlated in the higher mass regime. }
\label{fig:10w_acc}
\end{figure}

The H$\alpha$ 10\%w can also be used to estimate the accretion rate. A relation for low mass T Tauri and brown dwarfs was derived by \citet{2004A&A...424..603N}, and is useful in the cases where little or no stellar continuum can be detected. 
It has been proved to be a very good indicator of accretion \citep{2003ApJ...582.1109W}, however,  as was shown in \citet{2012MNRAS.427.1344C}, its accuracy as a measurement of accretion has been called into question. As we showed in the sample of \citet{2012MNRAS.427.1344C}, the H$\alpha$ 10\%w had much larger variations than the two other accretion indicators, H$\alpha$ EW and Ca\,II EW, demonstrating it is more likely to be affect by wind emission and line broadening processes than the free-fall velocities within the accretion flow.  \citet{2011A&A...529A..34M} show that in higher mass Herbig Ae/Be stars, the 10\%w is strongly correlated with rotational velocities, and no correlation is found between the accretion rate and 10\%w measurements. To test for this within our sample, the derived H$\alpha$ EW derived accretion rate is plotted against the corresponding 10\%w in Fig.\,\ref{fig:10w_acc}. This plot again confirms that there is very little correlation between the two accretion indicators, and for this reason no accretion rates derived from the 10\%w are discussed in this work.

\section{Discussion}\label{sec:discusion}
This paper sets out to constrain the variations in the H$\alpha$ emission of accreting stars. The aim here was to isolate the relative variability of the mass accretion rate as derived from the emission in each star. Thus, we do not take into account any systematic uncertainties from stellar parameters or line luminosity conversions (which will effect the comparison of the accretion values measured for different stars). Only the accuracy in our line measurements are considered in order to ascertain whether the observed variations we measure are above the measurement errors. Though some of the variations may come from sources other than accretion such as winds or from secondary accretion effects such as continuum changes (see Sect.\,\ref{sec:origin}), these measurements allow us to put an upper-limit on the accretion rate changes as observed in the H$\alpha$ emission line.

As previously mentioned, there are two distinct types of behaviour observed in this sample, \emph{slow variations} and \emph{rapid events}. In the following section possible origins of each are discussed.

\subsection{Slow Variations: Rotational Modulation of the Accretion Rate}
In the majority of the cases in this sample, small changes are seen in the profiles across the time-scale of our observation blocks. These variations are referred to as \emph{slow variations}. These occur in discrete wavelength ranges, and take the form of a gradual change in the profile emission. These variations do not translate to large accretion rate changes and on average they result in changes of the accretion rate derived from H$\alpha$ EW of 0.01\, - \,0.07\,dex. When the changes between different nights of observations are examined in this sample, the spread in accretion rate variations increases slightly to 0.04\,-\,0.4\,dex (see Table \ref{tab:accretion_rates2001and2003}). Comparing the difference in derived accretion rates for the objects with observations in 2001 and 2003 this spread does not increase from the individual observation periods remaining at 0.13\,-\,0.52\,dex. This is a strong indication that the time-scales of days are the dominant time-scales for these variations. 

\begin{figure*}
\centering
\centering
\begin{tabular}{ccc}

\includegraphics[scale=0.4]{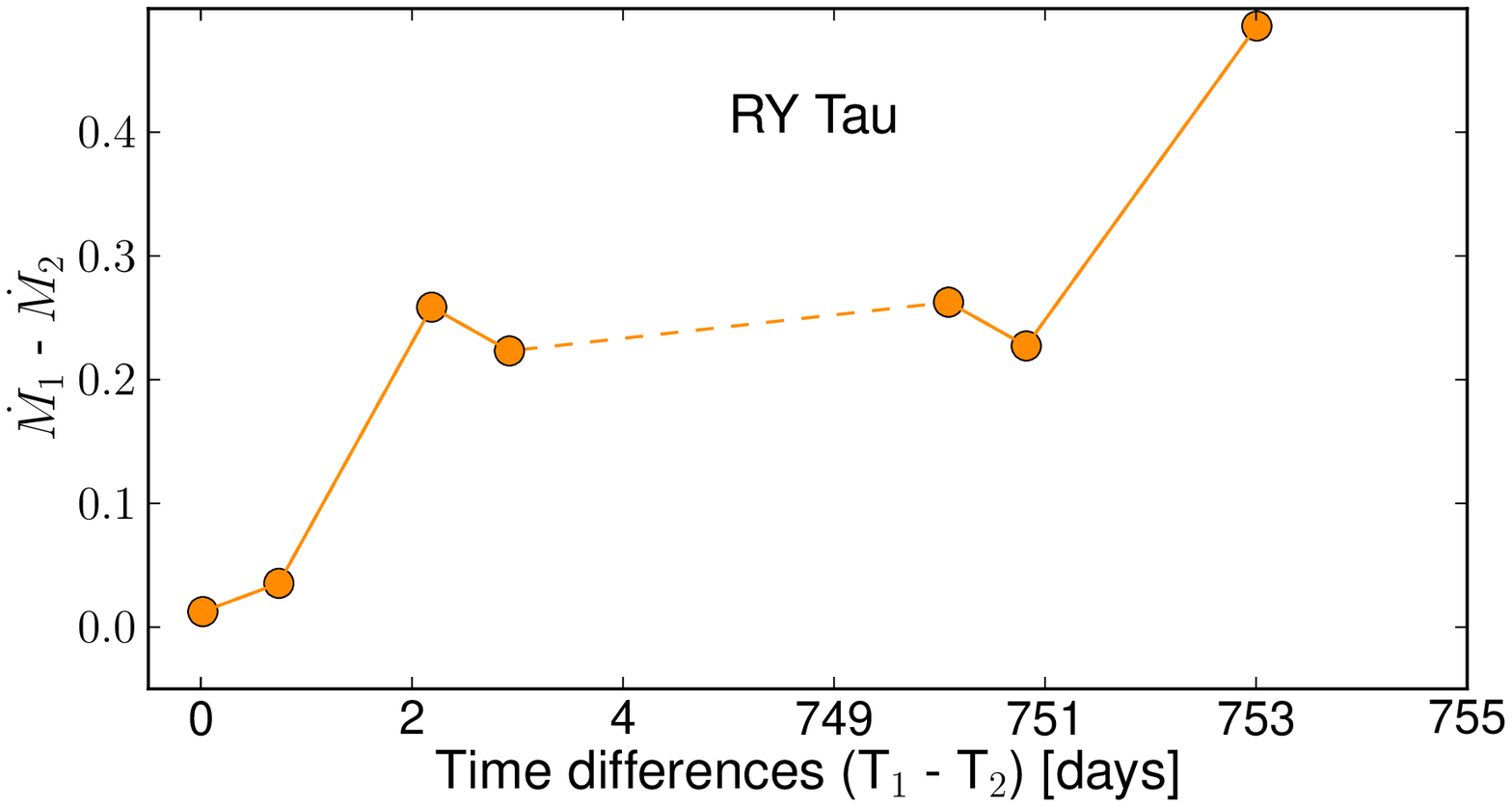} & \includegraphics[scale=0.4]{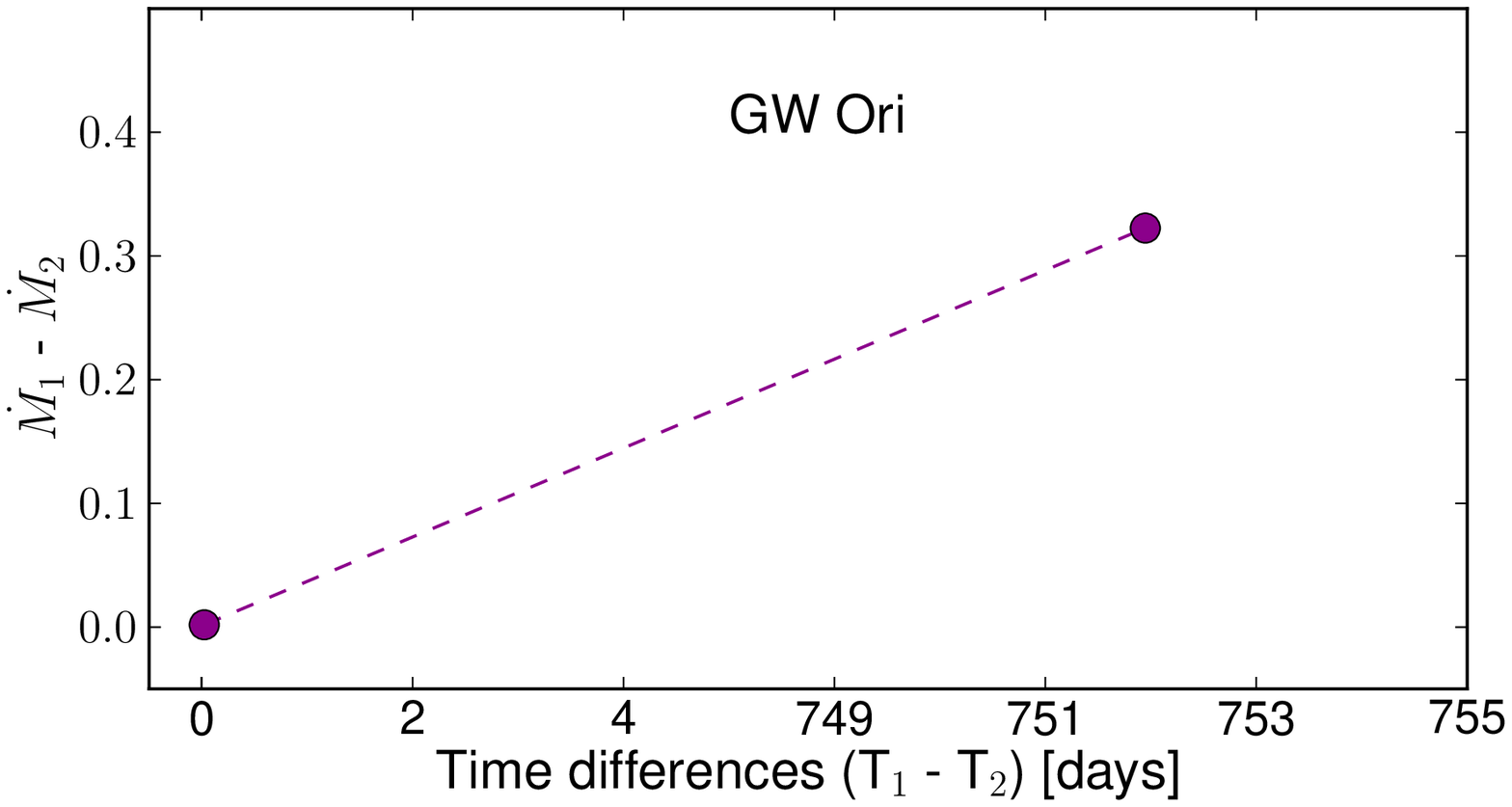} \\
\includegraphics[scale=0.4]{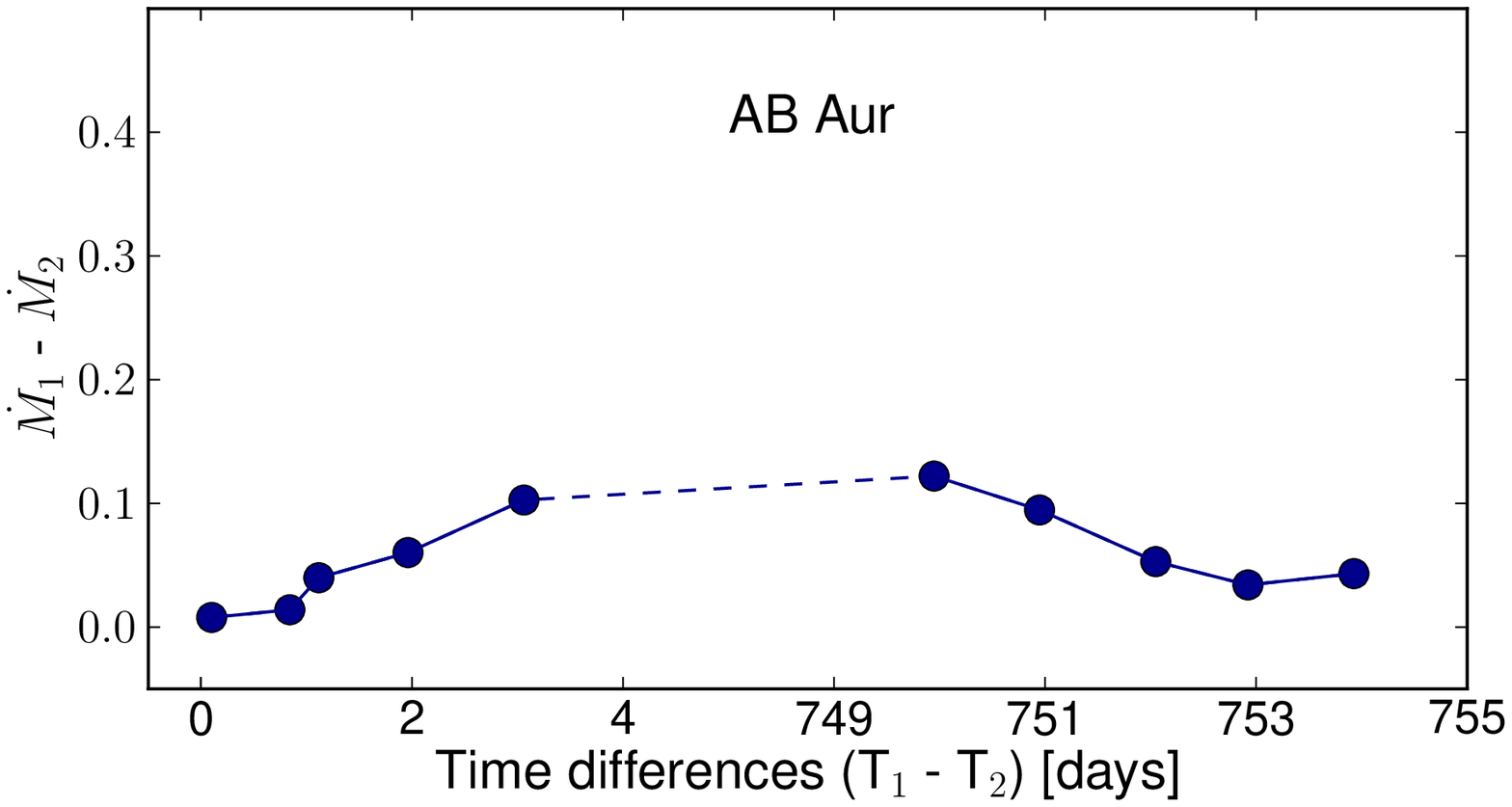} &\includegraphics[scale=0.4]{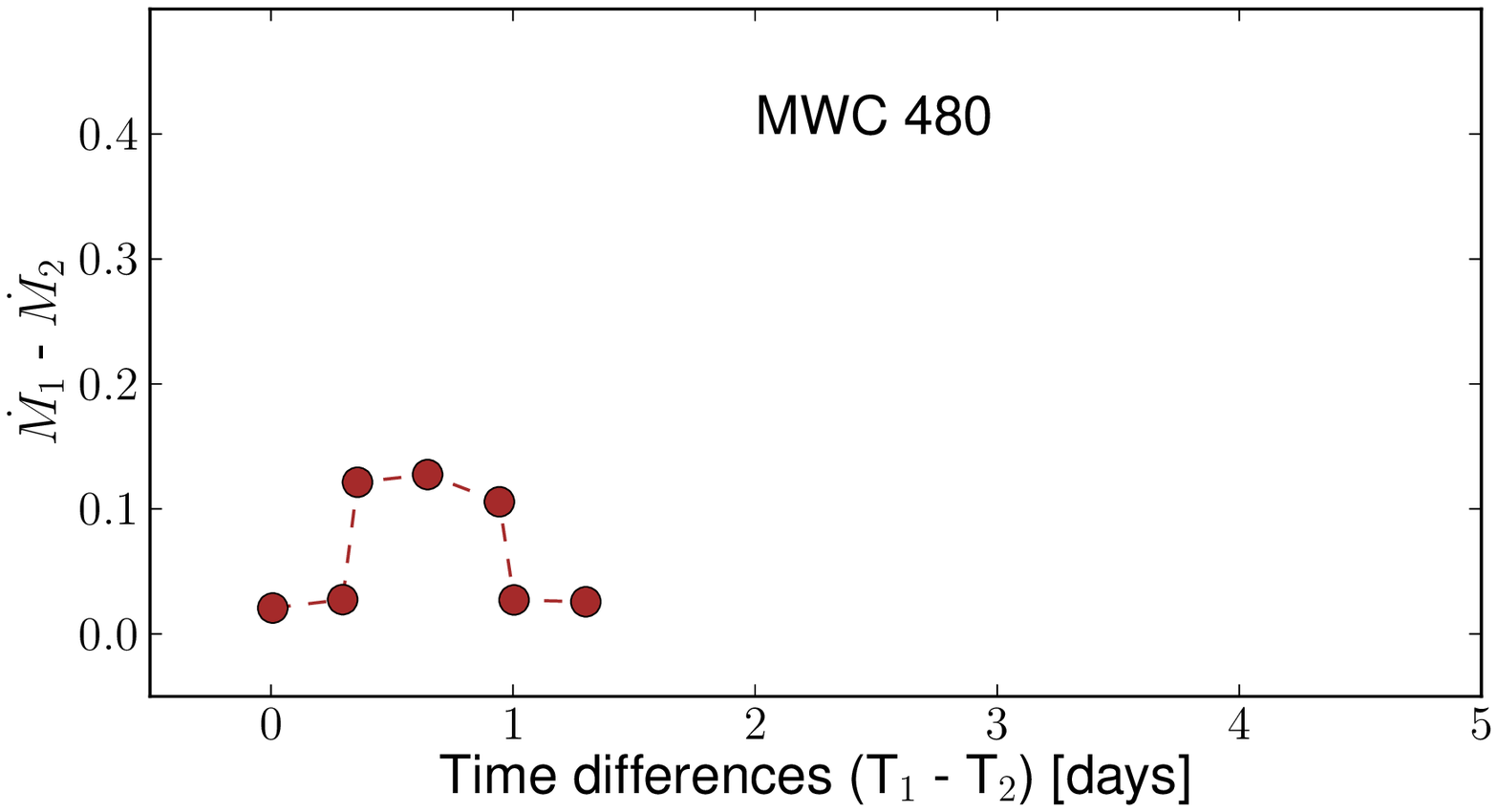}  \\
\includegraphics[scale=0.4]{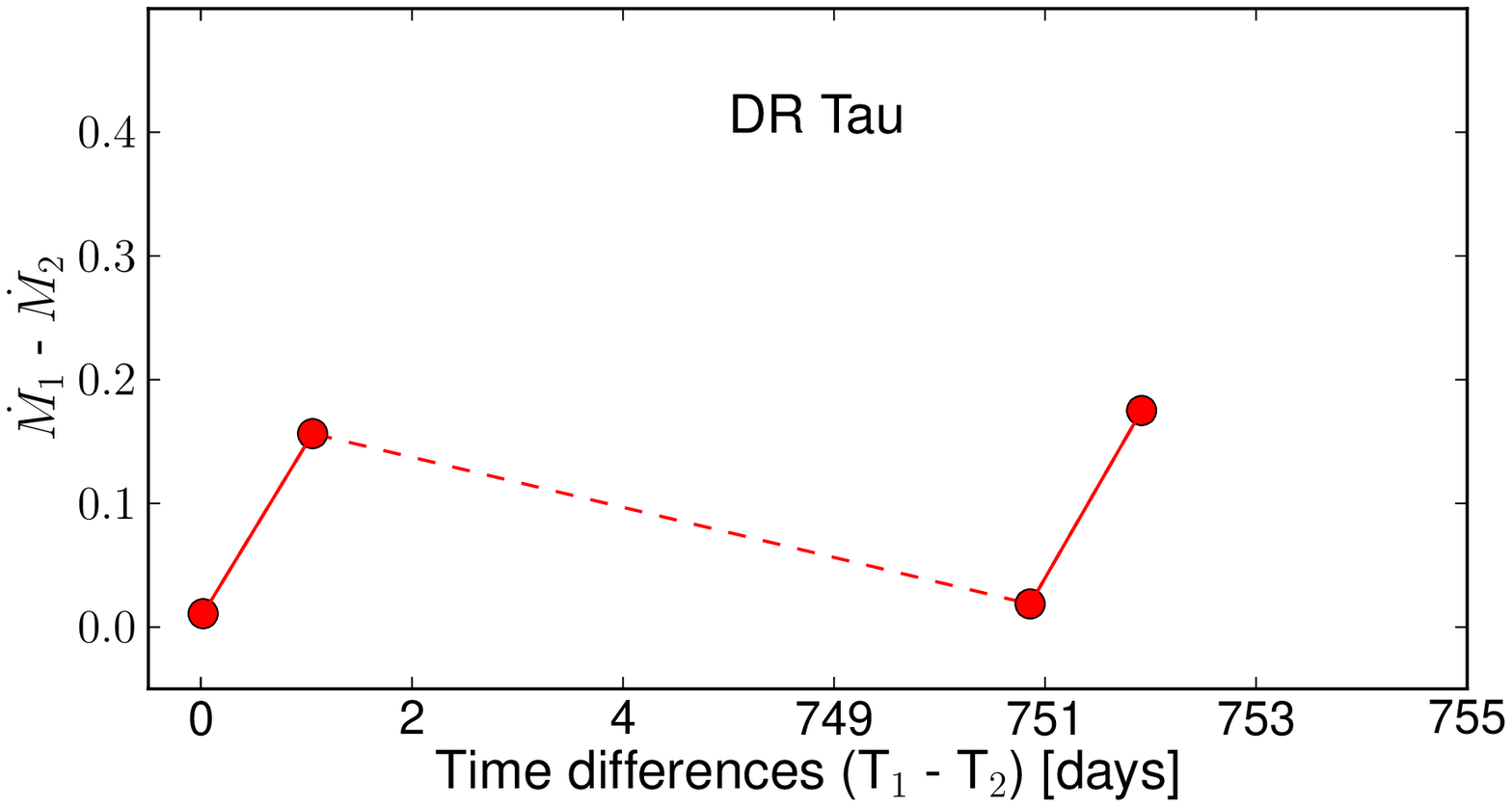} &\includegraphics[scale=0.4]{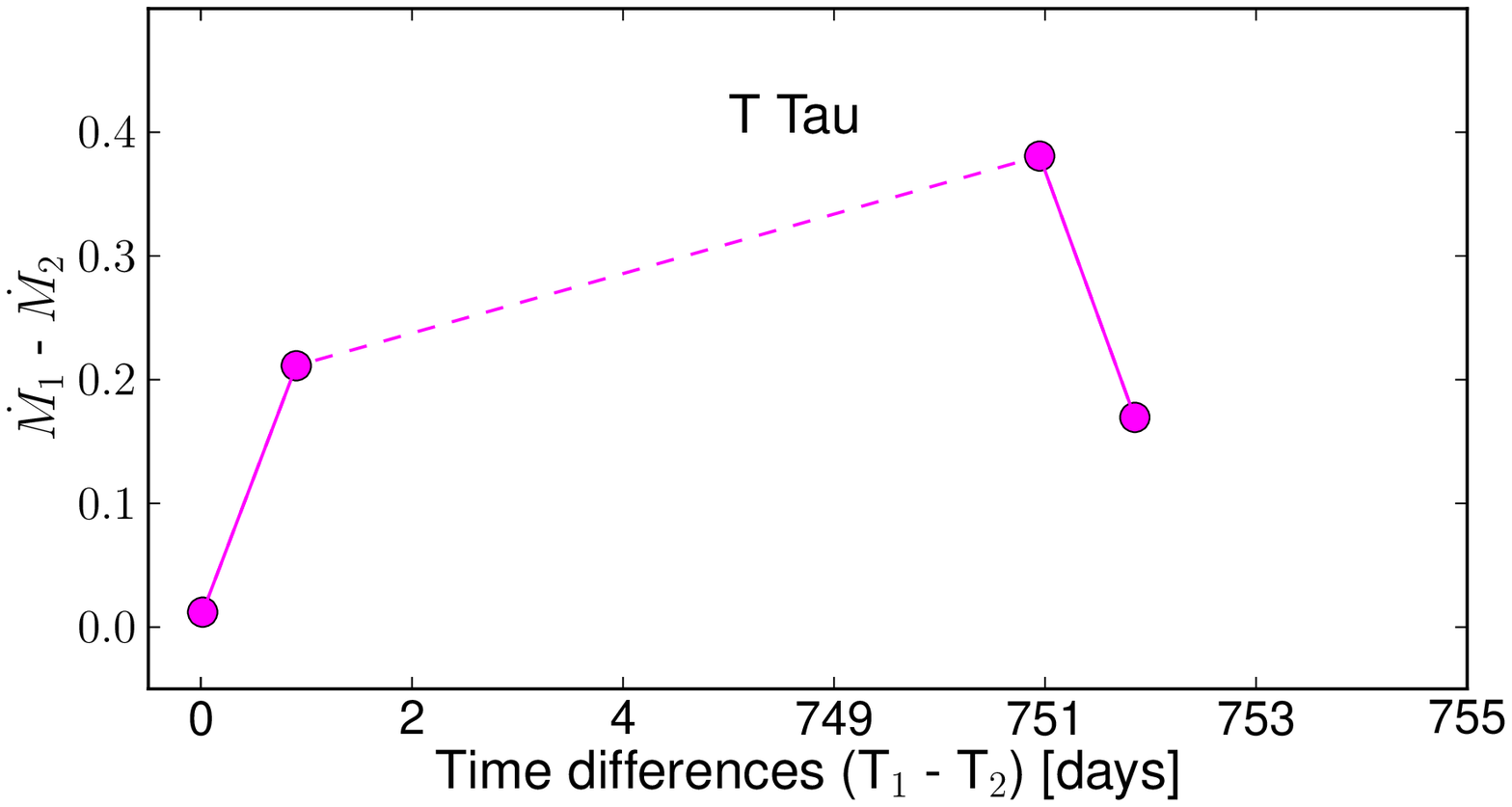} \\
\includegraphics[scale=0.4]{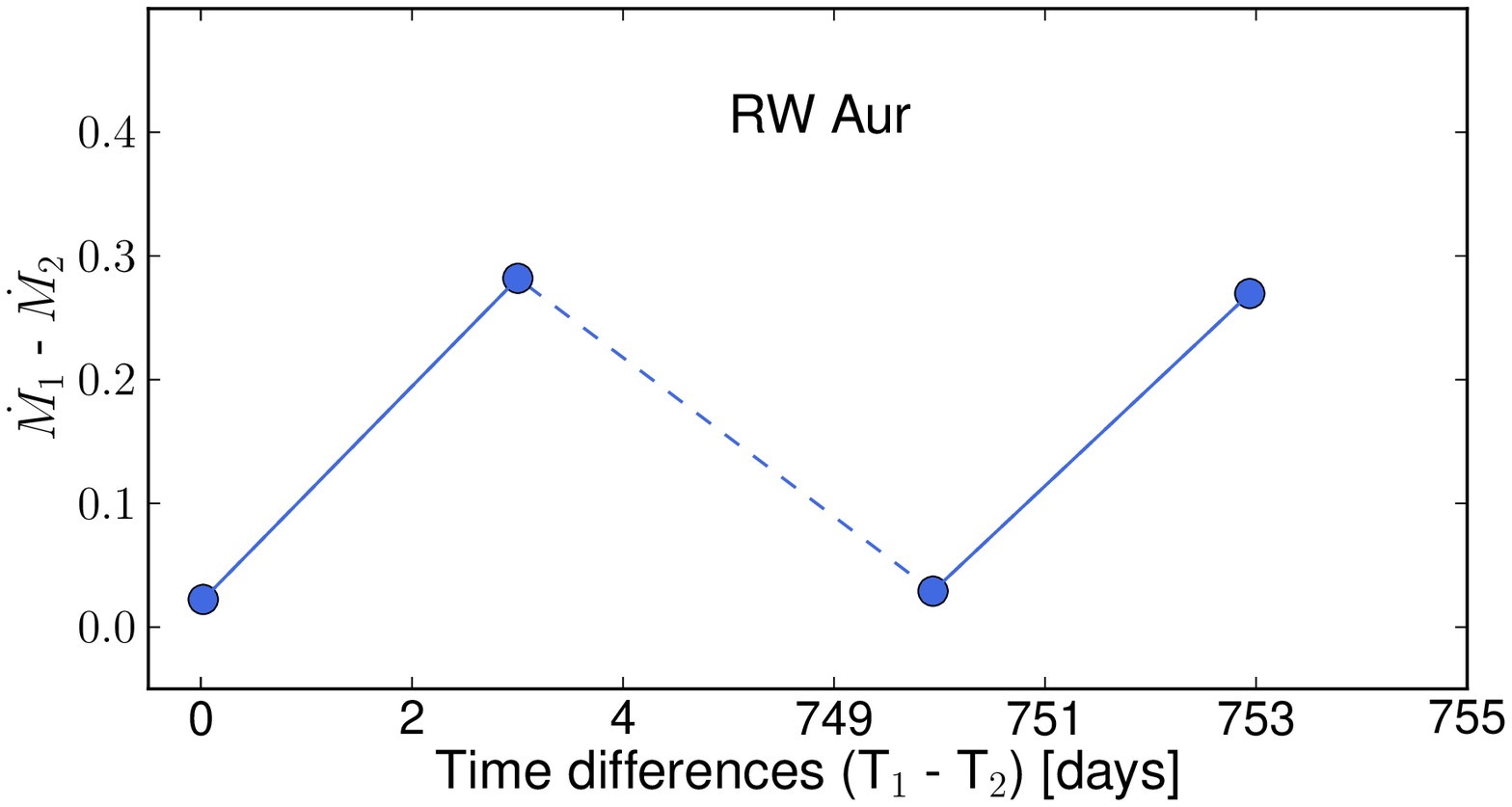} & \includegraphics[scale=0.4]{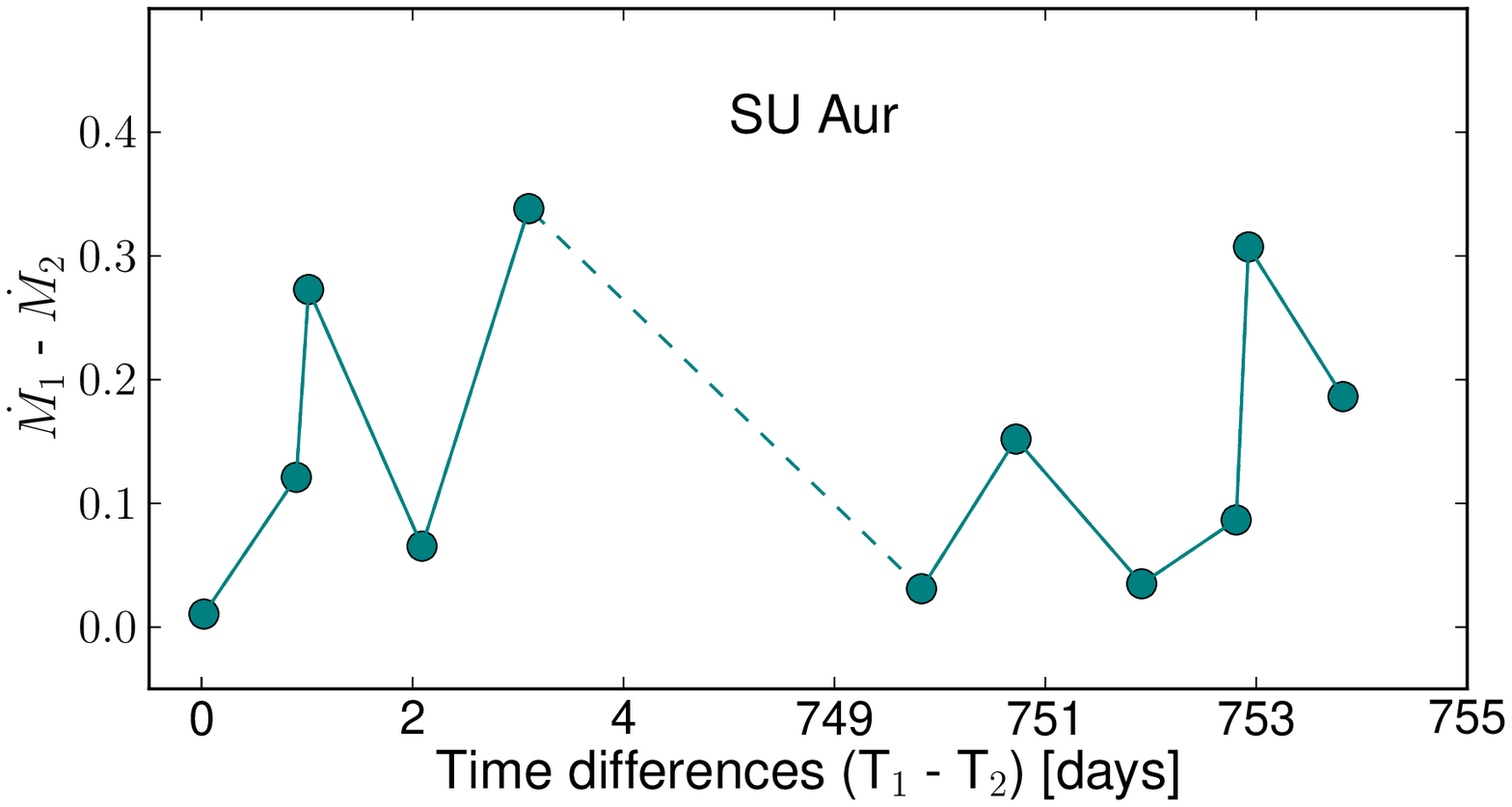} \\
\end{tabular}

\caption{ Mean differences in accretion rates~\lbrack Log(M$_{\odot}$ yr$^{-1}$)\rbrack~versus the time difference \lbrack days\rbrack. In this case we compare all accretion rates for each object, in order to cover all time-scales within the sample. The difference between all accretion rates was calculated, and the mean is plotted for each time bin. The time bins vary from object to object depending on the number of observations blocks. The same plot for all objects where the accretion rate differences are not binned can be seen in Fig.\,\ref{fig:accretion_timescales_all_pointsA} and Fig.\,\ref{fig:accretion_timescales_all_pointsB}.}
\label{fig:accretion_timescales_mean}
\end{figure*}

Fig.\,\ref{fig:accretion_timescales_mean} shows a comparison of accretion rate variations on all the time-scales in the sample. For each object every accretion rate measurement is compared with every other accretion rate measurement for that object. In this way all the time-scales available in the sample can be exploited.  The mean accretion rate is then plotted for each time bin for that object. In all cases but two, the variations reach a maximum after the first few days of observations. It shows that the dominant time-scale of variations in this sample is on the order of days. Two objects do show a rise in variations on the year time-scales, GW Ori and RY Tau. In the case of GW Ori, the sharp rise could be due to the lack of observations  over consecutive days, and is probably not a real increase in accretion variations on the longer time-scales (see Fig.\,\ref{fig:accretion_timescales_mean}). For RY Tau, it could be a real rise in accretion variations, but it is more likely to be a single event observed at the end of the time-series. (See Fig.\,\ref{fig:accretion_timescales_all_pointsA} and Fig.\,\ref{fig:accretion_timescales_all_pointsB} for un-binned comparison of accretion variations and time-scales for all objects.)

The time-scales over which the accretion rate variations reach a maximum are comparable to the rotation periods within this sample, which all lie between 1\,$\sim$\,5 days (see Table \ref{tab:stellar_parameters}). These \emph{slow variations} in the profiles are consistent with what one would expect from a rotational modulation of the accretion rate. In the case of an asymmetric accretion flow, as the star rotates, different parts of the accretion flow will come into view, changing how much of the accretion flow is visible and what velocities within the flow are visible. Apart from three cases, no major changes are seen in the profiles within the single blocks of observations. These blocks represent time-scales of about an hour or less. In this sample it appears that the small, gradual changes we see within these observation blocks accumulate to larger variations between the different nights of observations. 

Our previous work on a sample of T Tauri stars in Chameleon supports this result. We monitored the H$\alpha$ variations over time periods of weeks\,-\,15 months, and also found that it was the short time-scales, of a few weeks or less, that were the dominant time-scales within the sample \citep{2012MNRAS.427.1344C}. In Chameleon, we found the average spread in accretion rates to be 0.37\,dex, which is very close to the variations we find in this sample of both T Tauri and Herbig Ae stars.

Simulations of magnetospheric accretion have found that even a slight offset (2\,-\,5\degree) will result in an asymmetry in the accretion flow \citep{2003ApJ...595.1009R}. This is thought to be in the case for V2129 Oph \citep{2012A&A...541A.116A}. The authors used MHD simulations of the observed magnetic octupole and dipolar fields of V2129 Oph, and radiative transfer codes to reproduce the observed spectral line profiles. Earlier observations of the magnetic field found an offset between the octupole and dipole fields (15 and 25\degree) and the rotation axis \citep{2011MNRAS.412.2454D}. The modelling of these fields result in two ordered flows of material onto the star very close to the poles. The derived profile variations are similar in magnitude to the observed profiles, however the changes in the profile shape are not. At an inclination angle of 60\degree~to the viewer, the models result in a change in 8\,\AA~in H$\alpha$ EW across the rotation period. 

The changes that are observed in this sample over multiple days, are on the same order of magnitude as those modelled for V2129 Oph. For example AB Aur, RY Tau, SU Aur, BP Tau all have EW ranges of  5\,-\,15\,\AA~between multiple nights observations. RW Aur has a much larger spread of 35\,\AA, which may mean that this model of two rotating flows is probably not sufficient to explain all of the variations seen in this object. 

We can expect that rotationally-induced apparent accretion rate change will depend on the inclination of the systems to our line of sight. \citet{2006MNRAS.370..580K} showed through MHD simulations that as the inclination of an accreting system increases, the EW of the H$\alpha$ emission decreases. This is due to the fact that we see less of the accretion flow at higher angles. In these models, as the inclination increases from 10\degree~to 80\degree~the EW changes from 32\,\AA~to 21\,\AA. 
One can also expect to see more changes in the H$\alpha$ EW if the system is inclined to our line of sight, making it more likely that the accretion flow will move in and out of view. In the 2003 observations the three stars with the largest range in accretion rates in the sample are three of the most inclined systems. RY Tau, SU Aur, and RW Aur all have inclinations of 45$^{\circ}$ or over. DR Tau is also highly inclined, but does not show very large variations. However since the system is close to edge on, we may not have a full view of the accretion flow, and the disc may obscure a lot of the light coming from the accretion flows. The inclination angles are given in Table \ref{tab:stellar_parameters}.

This assertion no longer holds true when the 2001 observations are considered. The three objects (RY Tau, SU Aur, and RW Aur) show much less variability than the other objects, whereas DR Tau shows some of the largest accretion-rate variations, and T Tau and MWC 480 with disc inclinations of $\sim$ 30\degree~also show large variations. This suggests that a picture of a stable rotating asymmetric flow is probably too simplistic to describe the full variations in these objects. 

Long term photometric monitoring of accreting objects also support the scenario of a more complicated accretion flow as irregular light curves have been observed in many accreting T Tauri stars \citep{1994AJ....108.1906H,2009MNRAS.398..873S,2004A&A...419..249S,2007A&A...461..183G}. Over the time-scales of years multiple different types of variations can been seen in one object. In many of the cases were variations occur on the time-scales of days, the simple explanation of a rotational modulation cannot explain the full behaviour. 

In their spectro-polarimetric observations of BP Tau, \citet{2008MNRAS.386.1234D} found strong signatures of rotational modulation in the accretion related emission lines. The narrow emission lines associated with accretion (He\,I, Fe\,II and narrow component of Ca\,II IR triplet) varied strongly with rotation period. However, the broad emission lines H$\alpha$, H$\beta$ (and also the broad component of the Ca\,II IR triplet) were found to vary on the level of 10\%\,-\,20\% with rotation, with the remainder of the variations coming from seemingly other sources. This could be explained by the narrow emission component originating close the base of the accretion flow, with the H$\alpha$ emission originating in the bulk of the accretion flow, which may be more sensitive to instabilities. Excess contributions in the H$\alpha$ emission from outflows could also play a role. The authors also suggest, that changes in the rotational modulation of the longitudinal magnetic field between the two observation periods in February and December 2006 implies that the large-scale field topology was distorted by variability in the system between the two periods. This suggests though rotational modulation accounts for the majority of the variations, there are other ongoing processes.

There are many kinds of instabilities that can occur in the disc and accretion flow that may account for the changes in the variations observed over time, e.g.: the build up of material in the circumstellar disc \citep{2012MNRAS.420..416D}. \citet{2013MNRAS.431.2673K} show that in the case where Rayleigh-Taylor instabilities exist in the inner disc, unstable accretion flows can form. These flows change in size, shape, and numbers, meaning there is always a accretion column visible. This results in a constantly observed red-shifted absorption in the Balmer lines, more particularly in the higher Balmer lines such as H$\gamma$ and H$\delta$. This has been observed in RW Aur \citep{1994AJ....108.1056E}, where spectroscopic monitoring also confirmed the presence of an asymmetric accretion flow \citep{2001A&A...369..993P}, as well as for SU Aur \citep{1995ApJ...449..341J,1996A&A...314..821P}. 

It is possible that these instabilities exist in some if not all of the objects in this sample. Comparing the 2001 and the 2003 sample, some changes are seen in the H$\alpha$ profiles, but also differences in the derived accretion rates and their variations. These instabilities could account for changes in the H$\alpha$ emission if they change the form of the accretion flows.

\subsection{Rapid Events}\label{sec:rapid_events}

\begin{figure}
\centering
\begin{tabular}{cc}
\includegraphics[scale=0.22]{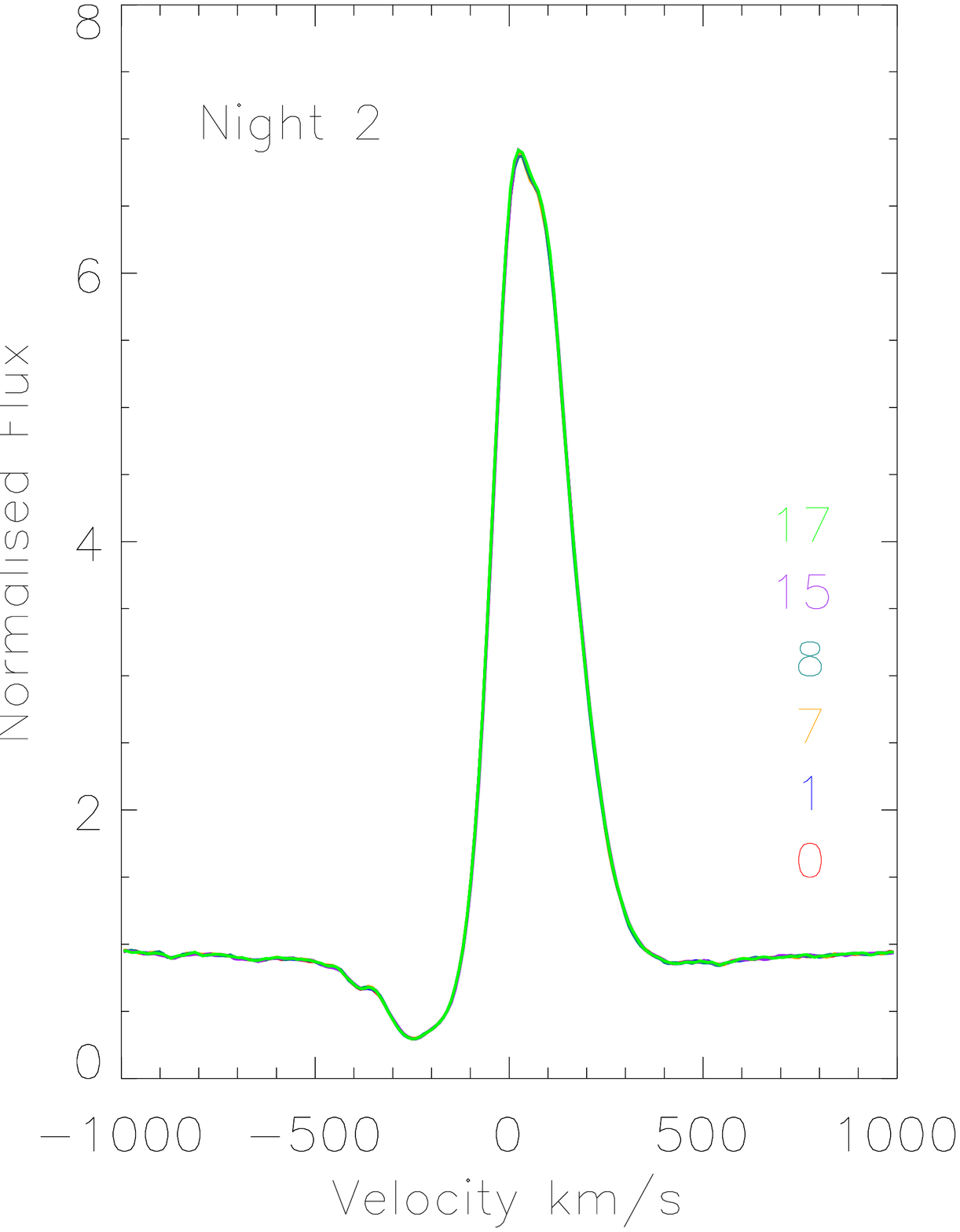}&
\includegraphics[scale=0.22]{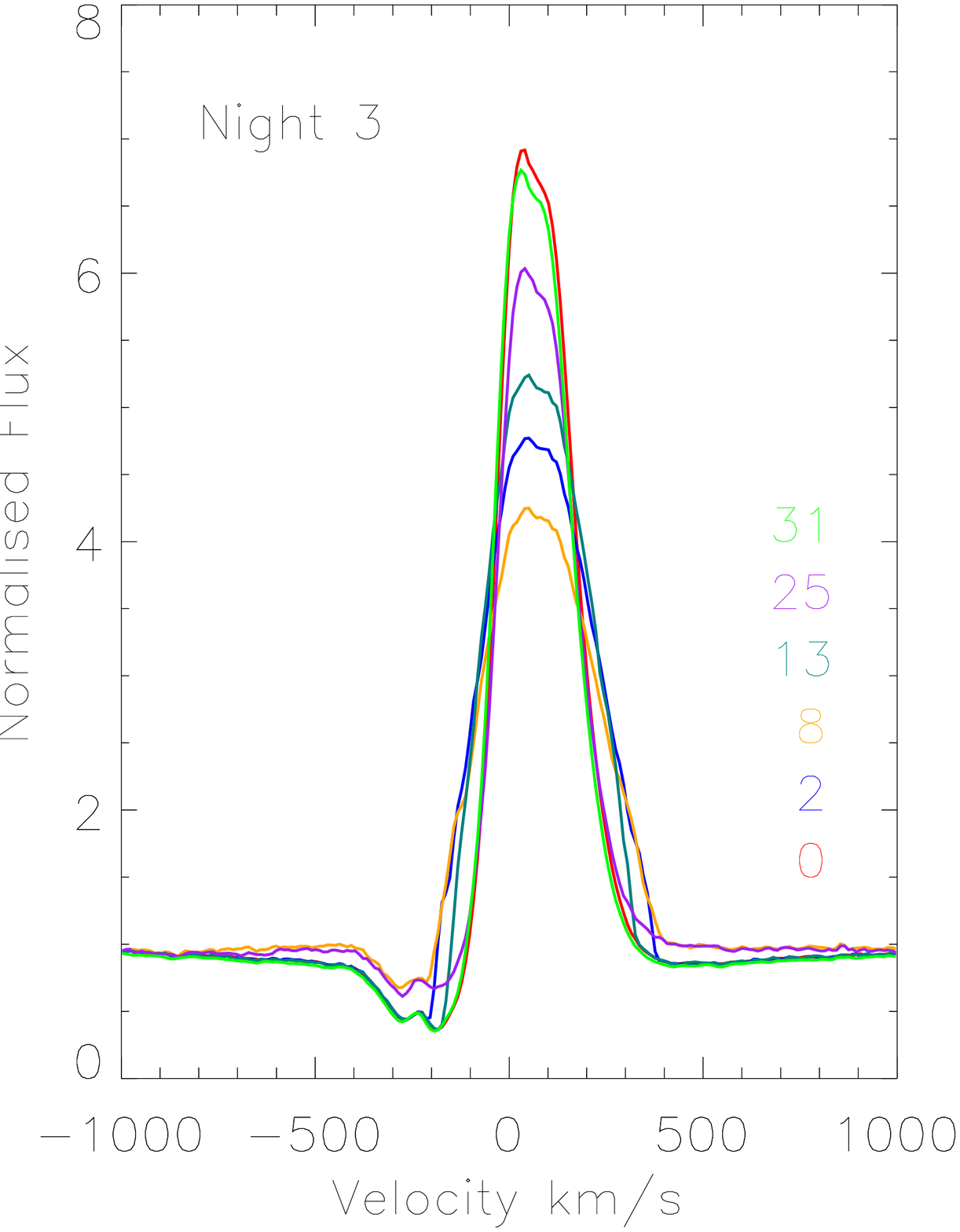}\\
\end{tabular}
\caption{A comparison between a sub-set of profiles on two separate nights of the 2003 observations for AB AUR. On the second night of observations in 2003, no significant change within the profile was observed. The minute of the hour in which the spectrum was observed is given to the right of the profiles in the corresponding colour. During the third night of observations there were large changes. Shown here is a sequence of spectra where the emission line falls in strength, and broadens, before returning to the initial strength.}
\label{fig:rapid_changes}
\end{figure}

The \emph{rapid events} observed in this sample do not fit into the frame of rotational modulation.  AB Aur, and to a lesser extent RY Tau and RW Aur, show significant variations in the profile over the time-scale of 1 hour. In the case of each star, these changes only occur during a single night of observations (For AB Aur see Fig.\,\ref{fig:rapid_changes}). A number of short term rapid variations have previously been discovered in objects in our sample and these are presented in the following paragraphs.

Short term striking variations have been observed in the H$\alpha$ profile of AB Aur previously, where changes occurred in both the intensity and profile shape.  \citet{1995A&A...298..585B} described the changes as occurring during each observing night across the emission line but especially in the absorption feature of the P-Cygni profile and the emission peak. They attribute these variations to the motion of circumstellar inhomogeneities. This is similar to the changes observed in the H$\alpha$ profile of AB Aur on night 3, where it oscillates between broad wings and low emission, and strong emission and with narrow wings. However in the case of \citet{1995A&A...298..585B}, their observations take place with hour separations, so they do not have the short term cadence that we have in this sample. 

Rapid variations have been seen in the H$\gamma$ profile of RW Aur, occurring on time scales as short as 10 mins. Over the course of the observations the central absorption component of the H$\gamma$ line increased in EW by a factor to 2 within an hour \citep{1982ApJ...256..156M}. These variations were not reflected in the H$\alpha$ emission, which is thought to be a result of the high optical thickness in the surrounding envelope. \citet{1982ApJ...256..156M} found these variations to be consistent with both flaring and accretion events. 

Variations in H$\alpha$ profile of RY Tau were found on the time-scales of 10\,-\,20 mins without variations in the star's brightness \citep{1975PZ.....20..153K}. Three separate nights observations took place covering time scales of a few hours each night. The profiles changed between nights, but in two out of three occasions they were stable throughout the night. The brightness of the star was lower on the night of the variations than on the other two nights, which is an indication that these variations came from magnetic activity \citep{1997A&A...324..155G}.

With the short wavelength coverage in this sample and no simultaneous alternative observations, the behaviour seen in AB Aur, RY Tau or RW Aur cannot be properly defined, and the origin of these rapid variations is not clear.  \citet{1994A&A...287..131G} argue that in the case of magnetospheric accretion short term variability is to be expected. The in-fall time-scale of gas towards the pole is less than one hour. Any instabilities that occur in the disc or the magnetic field at the point of their interaction will lead to a clumpy flow of material onto the surface.

The changes in RY Tau and RW Aur take the form of a drop in emission across both lines, but not within the line centre, which would be more indicative of a flare (see Sect. \ref{sec:chromo_accretion}). However it is probable that the rapid events that are observed in RW Aur and RY Tau are due to a flare event. This is not the case with AB Aur. The \emph{rapid variations} that occur in the profile of AB AUR are unique in the sample. In no other object do we see these changes in emission line strength, width and surrounding absorption.

It is possible that all the targets have these periods of \emph{rapid variations}. Out of the total 22.6 hours of observations, these \emph{rapid events} only take place with 3 observation blocks, which constitutes 2.4 hours or $\sim$ 10\% of our total observing time. This suggests they are not very common events and it rules out stochastic processes as the primary source of variations within the sample. However these stochastic events could be the cause of the \emph{rapid variations} we see in a small number and probably take the form of instabilities in the magnetic field \citep{1998AIPC..431..533G}, or inner disc \citep{2013MNRAS.431.2673K}.

\subsection{Comparison between Herbig Ae and T Tauri sample}

Herbig Ae stars are the intermediate mass equivalent of T Tauri stars and are thought to go through a similar process of accretion as T Tauri stars. As they are higher mass, they are shorter lived, but retain their circumstellar disc for long enough to accrete material from them onto their surfaces. 

Similar scalings of accretion rate to disc mass exist between T Tauri and Herbig Ae   \citep{2012A&A...543A..59M}. They also show that the NIR colour excess trend is the same across the T Tauri to Herbig Ae mass range, which can be explained by the reprocessing of both the stellar and accretion luminosities by the inner disc. 

However there is inconclusive evidence whether Herbig Ae stars are host to strong magnetic fields. Under our current understanding of magnetospheric accretion these strong, stable fields are essential for maintaining a quasi-stable accretion flow. 

Within this work a large mass range is covered (up to $\sim$ 5 M$_{\odot}$) and similar variations are seen in all objects. One of the larger mass targets, MWC 480 shows an accretion rate spread of 0.012\,-\,0.061 over single observation blocks. Comparing to one of the smallest mass targets in the sample, DR Tau with 0.010\,-\,0.052, shows there is little difference between the two. Indeed, others have found similar accretion rate variations for Herbig Ae stars as is found in this work, and the LAMP sample. For example \citet{2012AN....333..594P} observed accretion variations of amplitude 0.4\,dex over the time-scales of 10 days for one target, while multiple targets showed variations of 0.3\,dex between consecutive days observations.  From 38 Herbig Ae/Be stars, \citep{2011A&A...535A..99M} found a typical upper limit of accretion variations of 0.5\,dex over time-scales from days to months.

Compiling the samples from both studies (this and the LAMP sample), of 10 low-mass T Tauri and 14 intermediate-mass T Tauri/Herbig Ae stars, suggests that it is the same process that produces the H$\alpha$ variations in T Tauri and Herbig Ae stars, across the entire mass range up to $\sim$ 5 M$_{\odot}$. This variability result is entirely consistent with earlier spectro-polarimetry surveys \citep{2005MNRAS.359.1049V,2003A&A...406..703V,2002MNRAS.337..356V}

\section{Summary}

This study was undertaken to put a lower limit on the time-scales of accretion variability in T Tauri and Herbig Ae stars. Our previous work, \citet{2012MNRAS.427.1344C}, used the H$\alpha$ emission in low mass T Tauri stars as a proxy for accretion, through which an upper limit of 8\,-\,25 days for the time-scales of accretion variations was found. These were the shortest time-scales in that sample. 

This data set gave us the opportunity to approach the problem from the other end of the scale, by studying the variations on the time-scales of minutes, hours, days and in a few cases years. The main findings of this work can be summarised as follows:
\begin{itemize}

\item The majority of variations observed in this sample take the form of \emph{slow variations}, where gradual changes in the H$\alpha$ emission occur across the $\sim$ 1 hour observation blocks. These \emph{slow variations} are consistent with what we would expect from accretion rate changes rather than solely chromospheric activity or wind emission. 

\item Calculating the mass accretion rate from the H$\alpha$ emission, the average spread the accretion rate on time-scales of less $\lesssim$ 1 hour is found to be 0.01\,-\,0.07\,dex. The spread increases by an order of magnitude when different nights observations are considered, 0.04\,-\,0.4\,dex. However, when the variations are considered over 2 years, they have not increased and remain the same, 0.13\,-\,0.52\,dex. 

\item Therefore it is the period of days that is the dominant time-scales of these \emph{slow variations}. These results are found to be consistent with a rotational modulation of the accretion rate and are supported by previous results.  

\item \emph{Rapid events} occur in 3 observation blocks, which constitute 2.4 hours or $\sim$ 10\% of our total observing time. These events consist of fast changes within the H$\alpha $ emission line on time-scales of less than an hour. They could be connected to instabilities in the magnetic field or inner disc, which would create more stochastic accretion events than we observe in the majority of the sample.  

\item This data set covers a large range in masses, in which very similar variations and accretion rate changes are found. This strongly suggests that the same process is taking place across the full mass range, from low mass T Tauri stars up to Herbig Ae stars, and that the same model of accretion holds over large stellar mass range.

\end{itemize}

\begin{figure}
\centering
\includegraphics[scale=0.3]{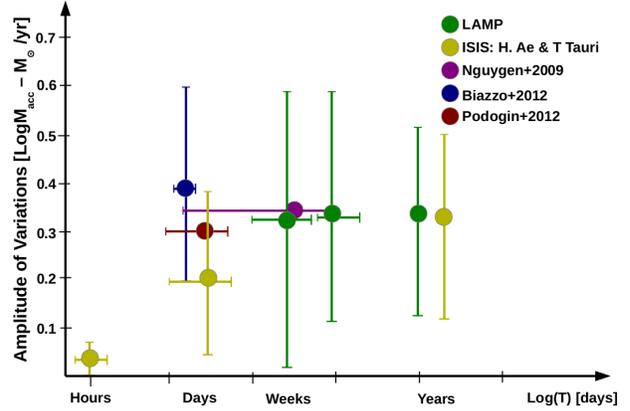}
\caption{Amplitudes of accretion rate variations versus the time-scales over
which they occur. Error bars here indicate the spread in accretion rates, and time coverage in each bin. Data from: \citet{Nguyen09,2012A&A...547A.104B,2012AN....333..594P,2012MNRAS.427.1344C}. }
\label{fig:timescales_versus_Mdot}
\end{figure}

The dominant time-scales of the variations found in this data agree with those found in other studies. Fig.\,\ref{fig:timescales_versus_Mdot} is a graphical comparison of accretion variations found in different samples. The error bars within the plot, represent the parameter space covered by each data set, not the errors. This is not a complete sample of variability studies, but as these studies used similar similar accretion indicators, they are more comparable. For example, \citet{2012AN....333..594P} used a number of emission lines associated with accretion to estimate accretion rates in 8 Herbig Ae stars. For half the sample they have multiple observations covering time-scales of tens of days. By averaging across the accretion indicators, including H$\alpha$ they found variations of 0.1\,-\,0.4\,dex for these four objects. \citet{2012A&A...547A.104B}, with a sample of 12 objects, found variations of 0.2\,-\,0.6\,dex over the two days separating observations.  The last data set in this comparison was covered by \citet{Nguyen09}, who found variations from Ca\,II emission of 0.35\,dex on average over time-scales of days to months. 

Through these studies, it has become clear that typical accretors do not go through large accretion bursts like EX Ors or even FU Ors \citep{1977ApJ...217..693H,2007AJ....133.2679H}. Over the time-scales of days, months and years, no large variations in accretion rate were found.

This work cannot rule out that large changes in the accretion could occur on time-scales of $\sim$ decades or longer (as is the case with FU Ori). However, it can be said that over the time-scales of years, the dominant variations occur on time-scales close to the rotation period. This suggests that observations over the time-scale of \textless \,1 week in typical accreting objects are sufficient to put limits on the expected magnitudes of variations that occur over the time-scales of years. 

These low levels of variations are also significant for the $\dot{M} -M_{*}^{2}$ relation (see \citet{2005ApJ...626..498M}). The mean amplitude of variations in the samples studied here is $\sim$ 0.5\,dex. This demonstrates that the $\sim$2 orders of magnitude scatter around the $\dot{M} -M_{*}^{2}$ relation at a given mass cannot be solely due to accretion variability. Studies of individual star forming regions with low age spreads have shown that this spread is unlikely to be due to evolutionary effects \citep[i.e.][]{2006A&A...452..245N}. This strongly suggests that the $\dot{M} -M_{*}$ relation is due to either limitations in our detections of low accretion rates in high mass objects \citep{2006MNRAS.370L..10C,2011MNRAS.415..103B}, initial conditions \citep{2006ApJ...639L..83A,2006ApJ...645L..69D} or is an indirect correlation where both the accretion rate and stellar mass are linked through a different process.

\section*{Acknowledgments}
JSV would like to thank the UK Science and Technologies Facility Council (STFC) and the the Northern Ireland Department of Culture Arts and Leisure (DCAL) for financial support. AS and TR would like to thank the Science Foundation of Ireland (SFI) for their support under grant numbers 11/RFP/AST/3331 and 10/RFP/AST2780.
 
\appendix

\section{Individual Objects}\label{sec:individual_objects}

In the following, the properties of the individual targets used in the analysis in this paper are presented along with time series of the H$\alpha$ profiles, average and variance profiles and differential surface and spectra plots.

The plots for each target take the same form. The first plot is a time series of H$\alpha$ profiles across the observation block. Each profile is off-set from the previous one for clarity. A time stamp is give to the right of each spectrum, this is the time difference in minutes between the first observation in that block and that particular spectrum. The second plot contains the average and variance profiles. The third plot is a differential surface plot. This plot shows the difference between the first spectra of that night and the preceding spectra. The fourth plot shows a time series of cuts in differential flux plots. This was done in the same way as the surface plots, where the first spectrum of that night was removed from all the rest of the spectra.  These are the same profiles as chosen for the profile time series, and again the time difference between each spectrum and the first observation in that block is is given to the right of each profile.

Plots for each object showing the difference between two accretion rate measurements~\lbrack Log(M$_{\odot}$ yr$^{-1}$)\rbrack~versus their time difference \lbrack days\rbrack~are given at the end of the appendix. These are similar to Fig.\,\ref{fig:accretion_timescales_mean}, however in this case the accretion rate differences in each time bin have not been averaged.

\subsection{RW Aur}

\begin{figure*}
\begin{tabular}{ccc}
 \includegraphics[scale=0.22]{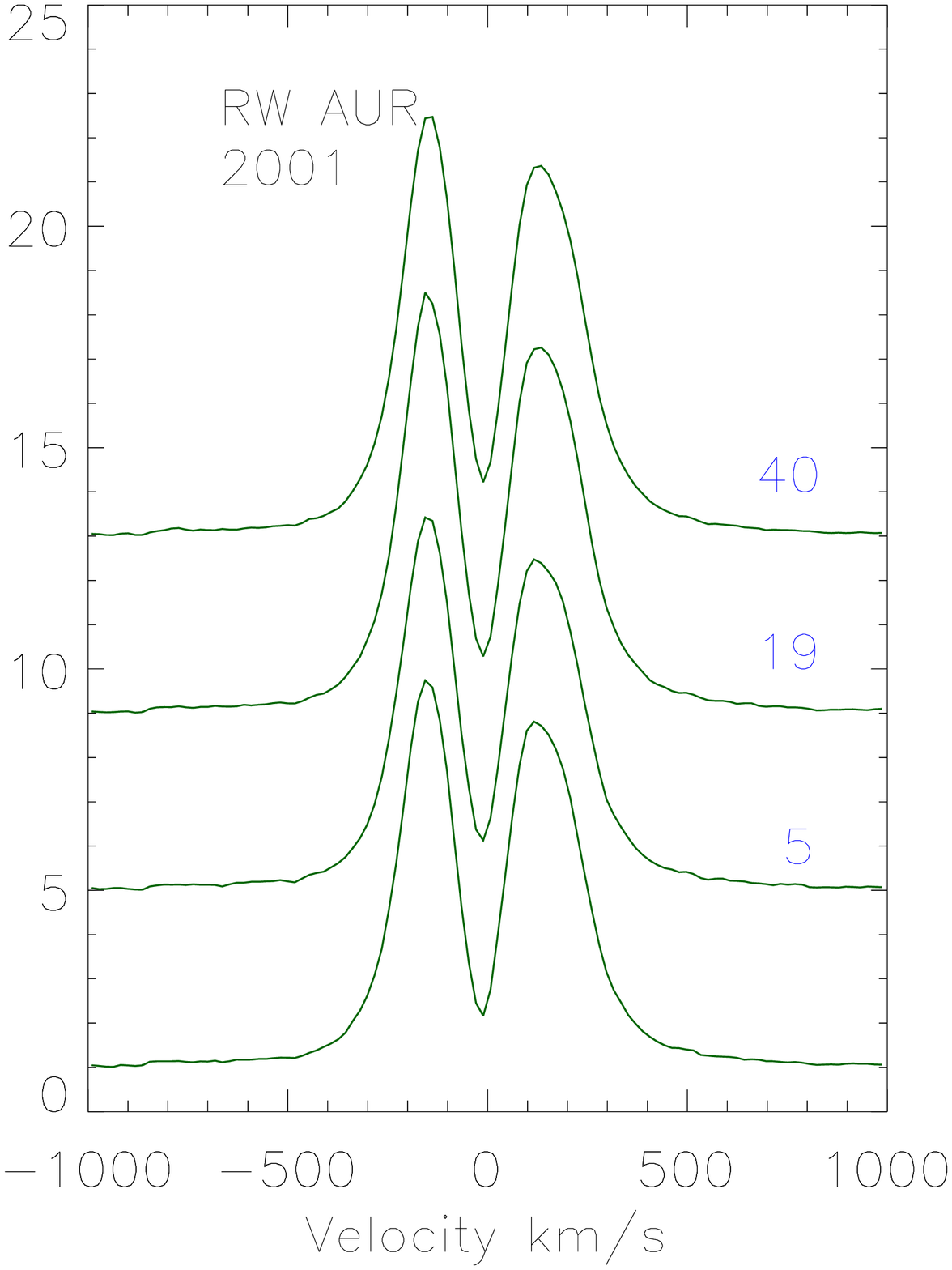}  & \includegraphics[scale=0.22]{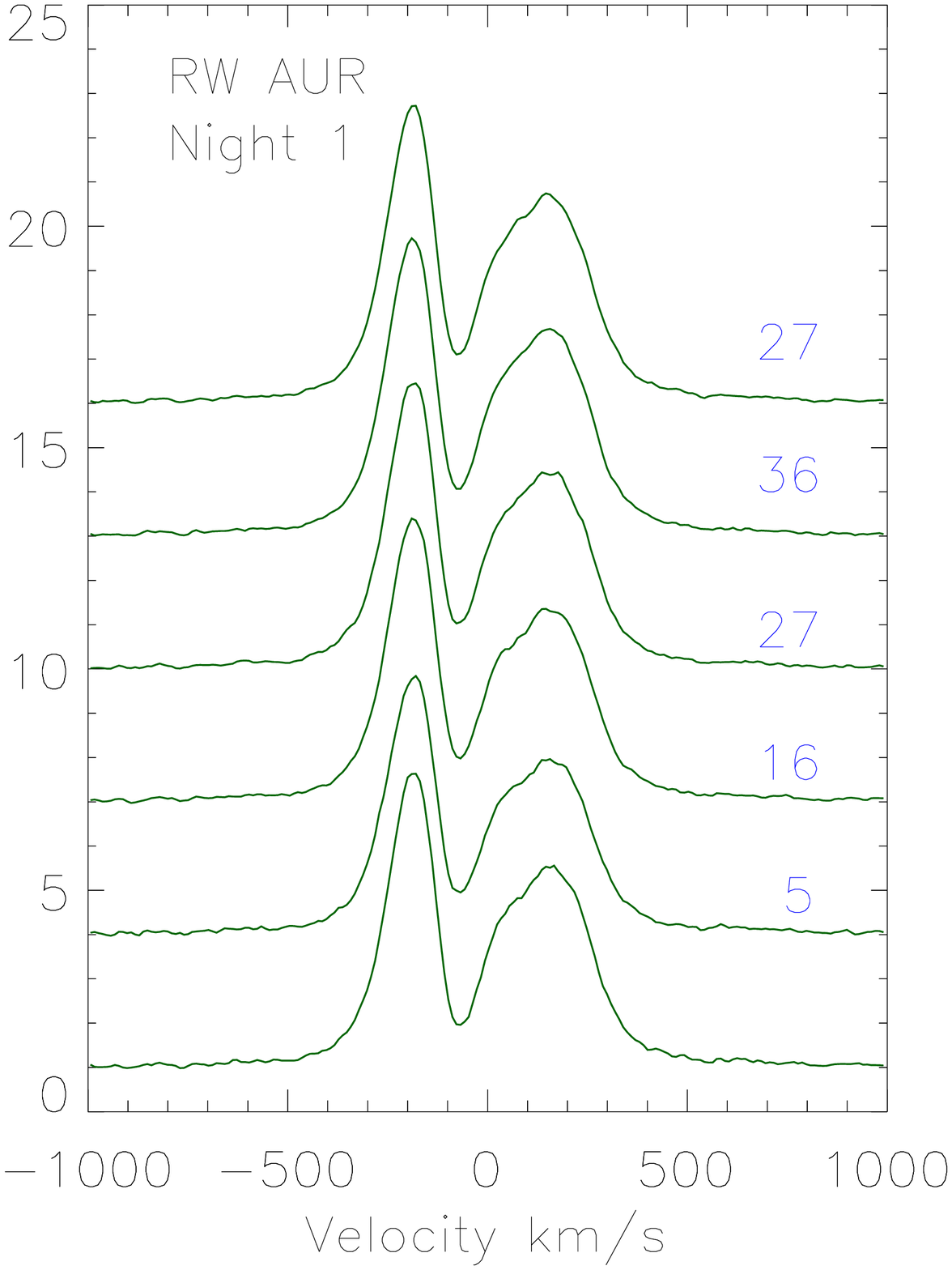}& \includegraphics[scale=0.22]{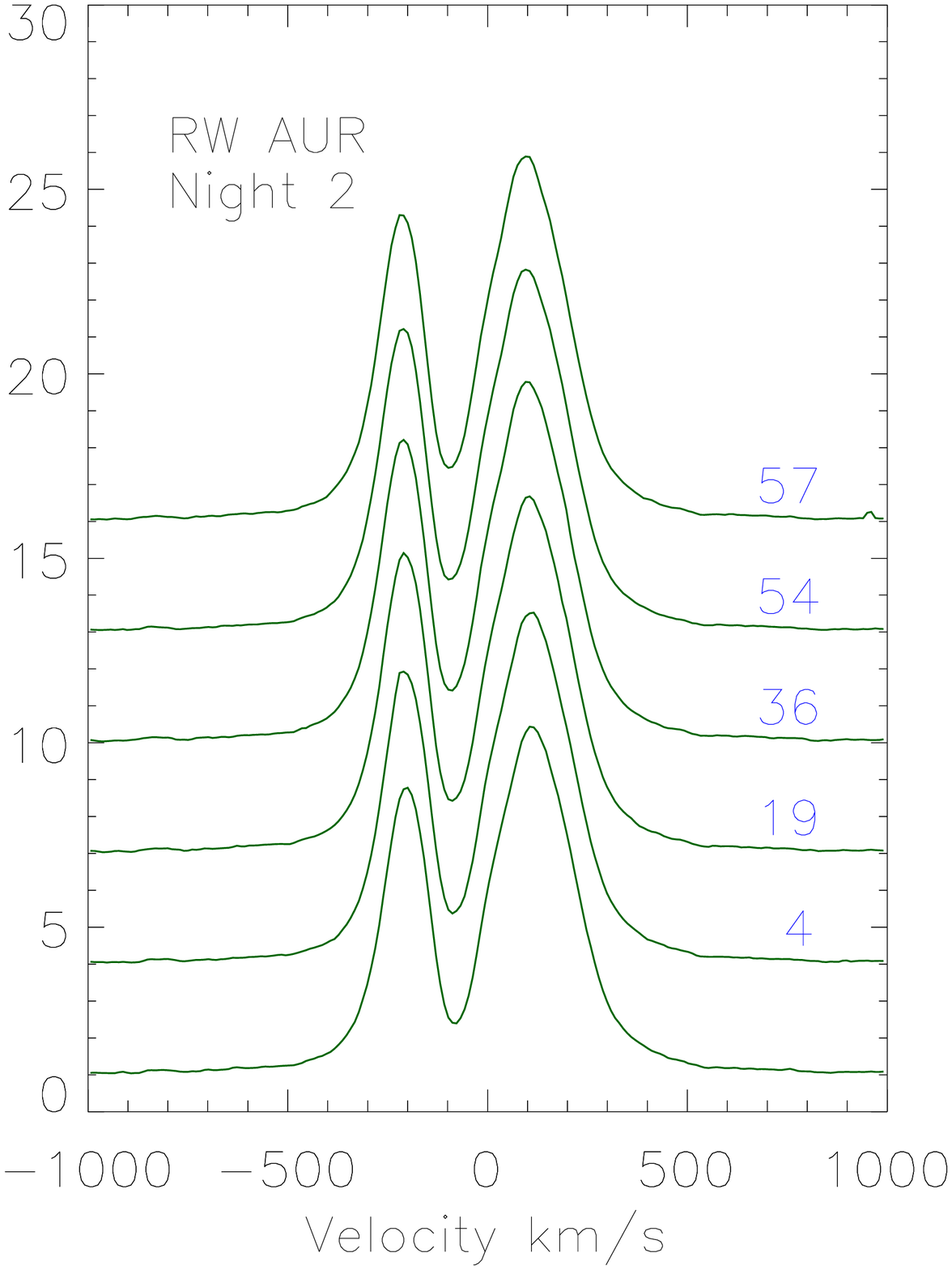} \\
 \includegraphics[scale=0.22]{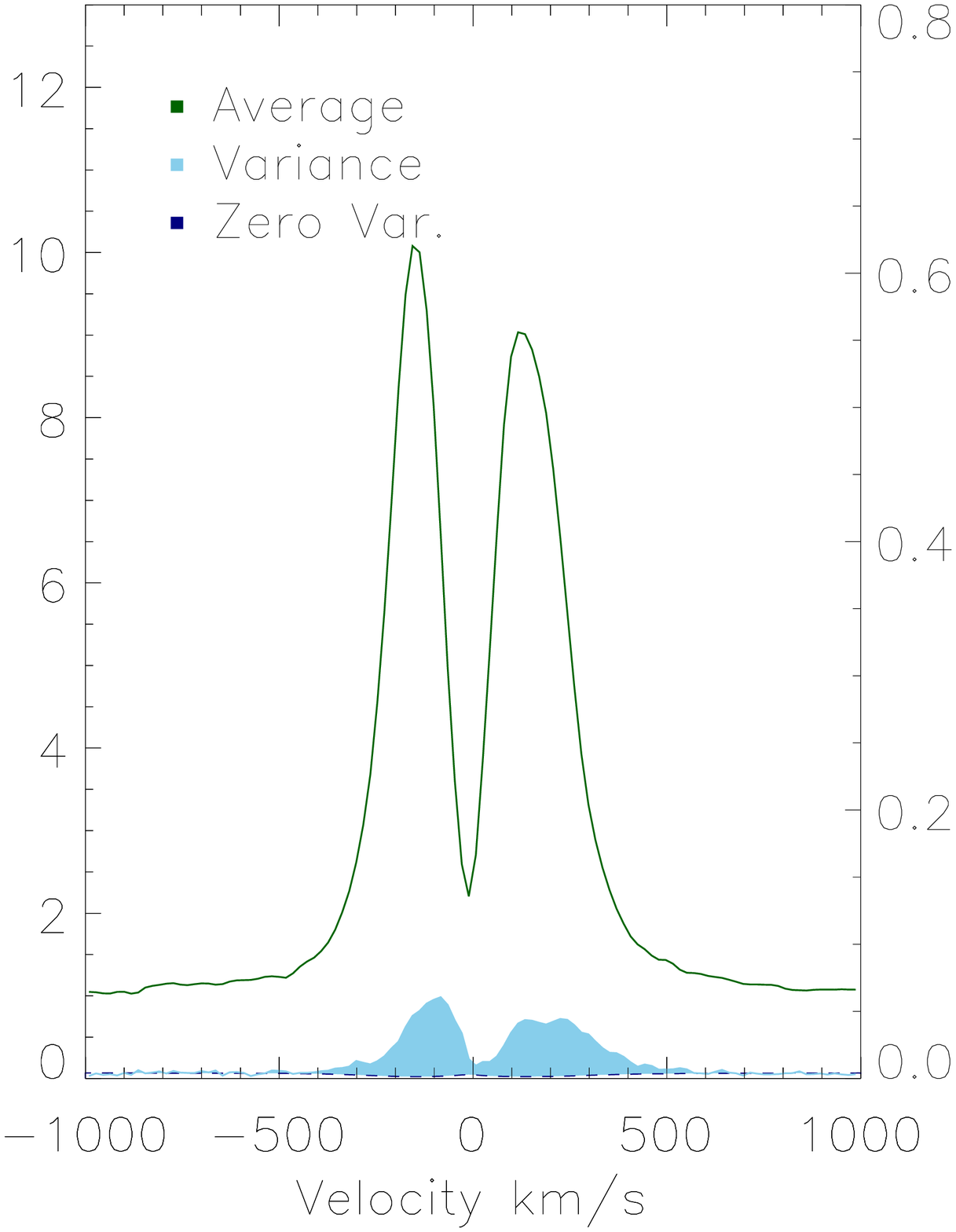} & \includegraphics[scale=0.22]{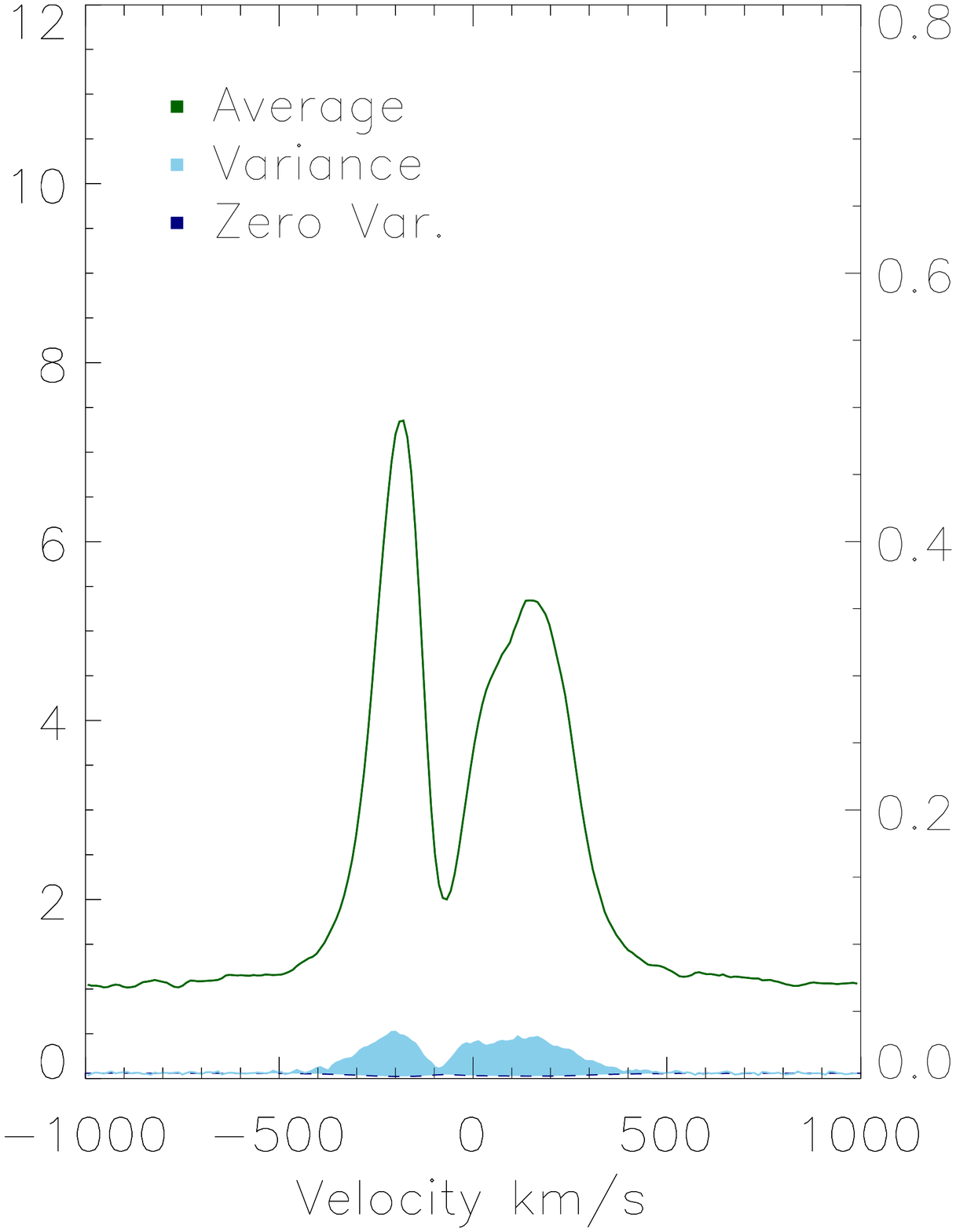}& \includegraphics[scale=0.22]{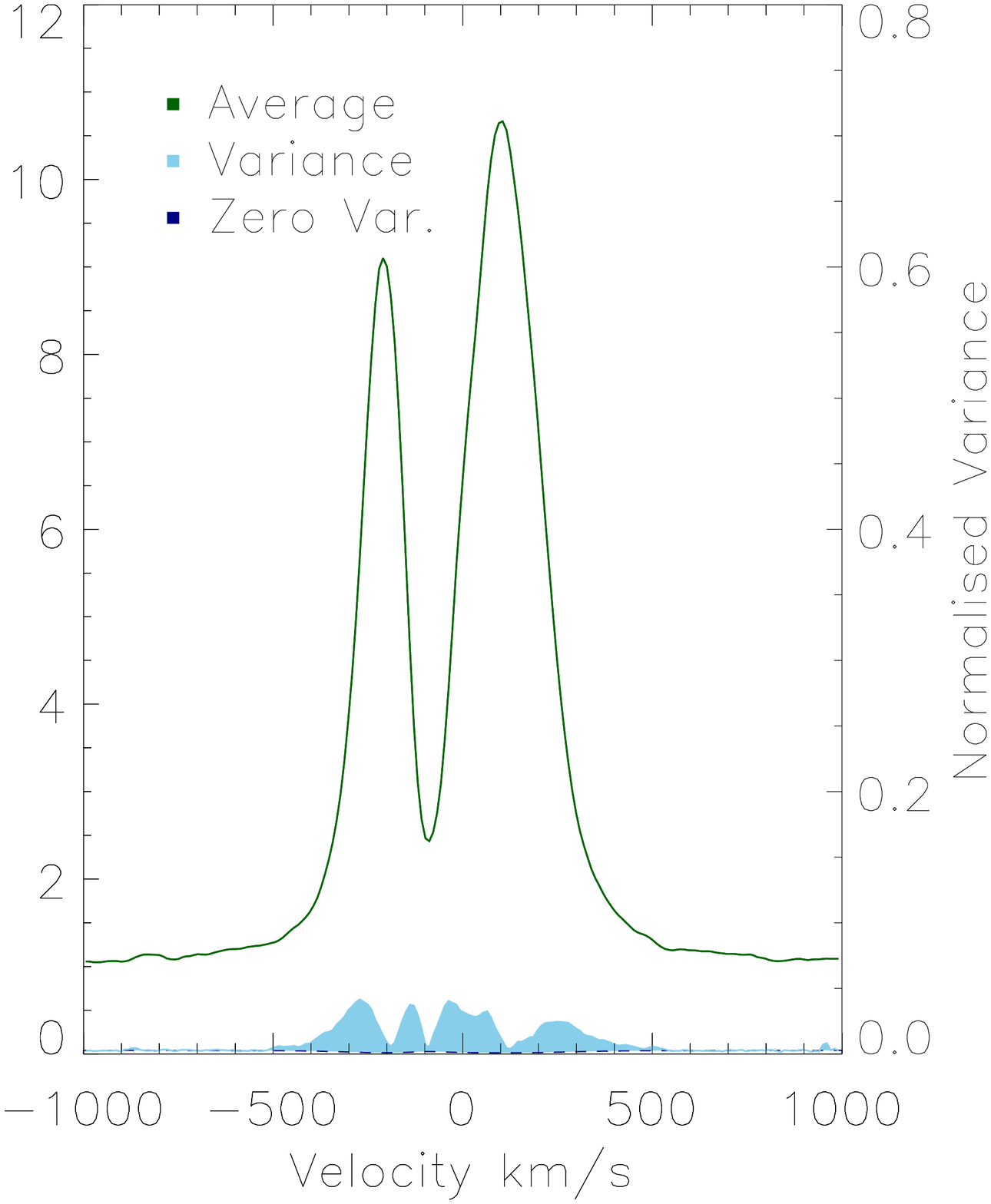} \\
 \includegraphics[scale=0.22]{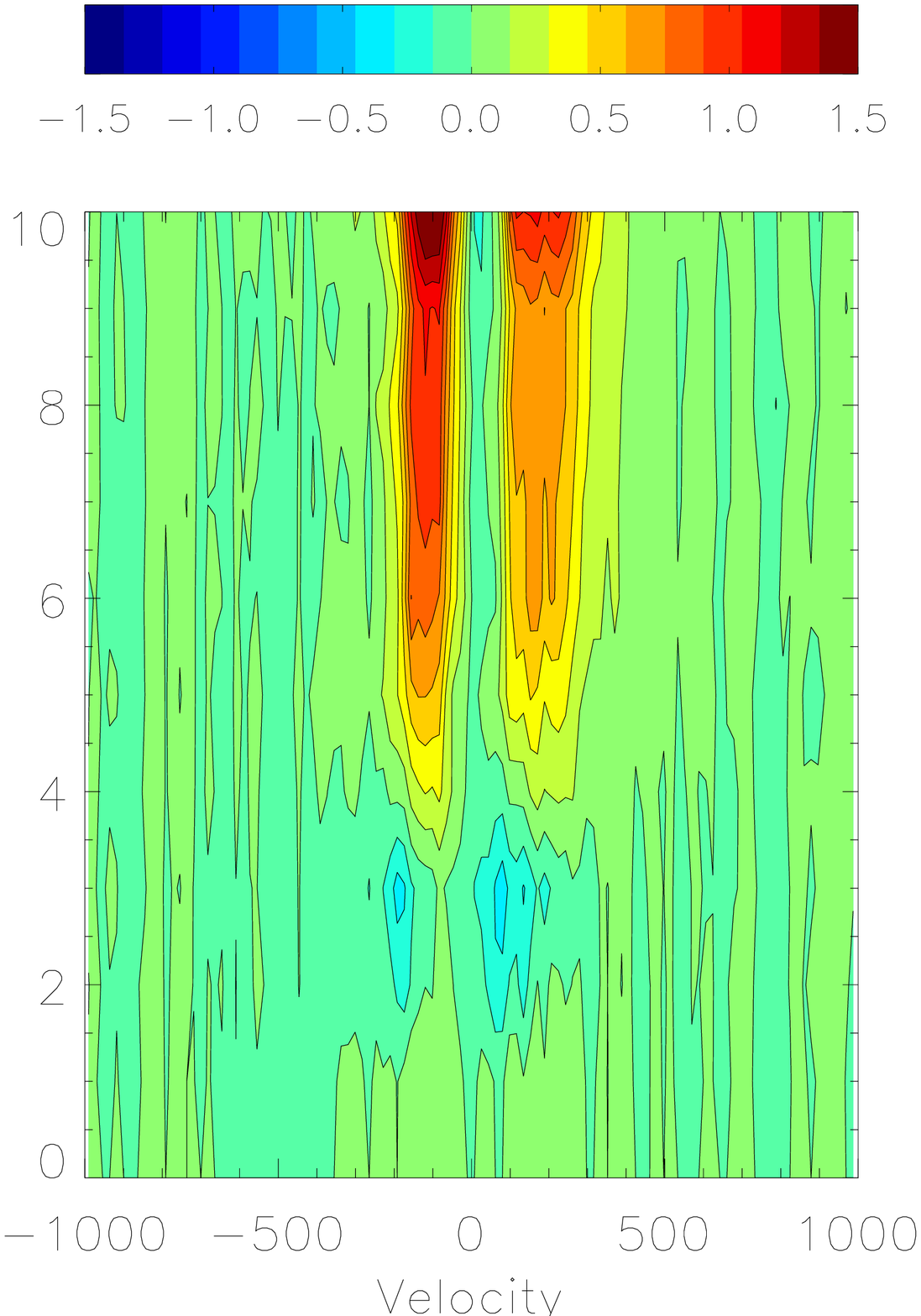} & \includegraphics[scale=0.22]{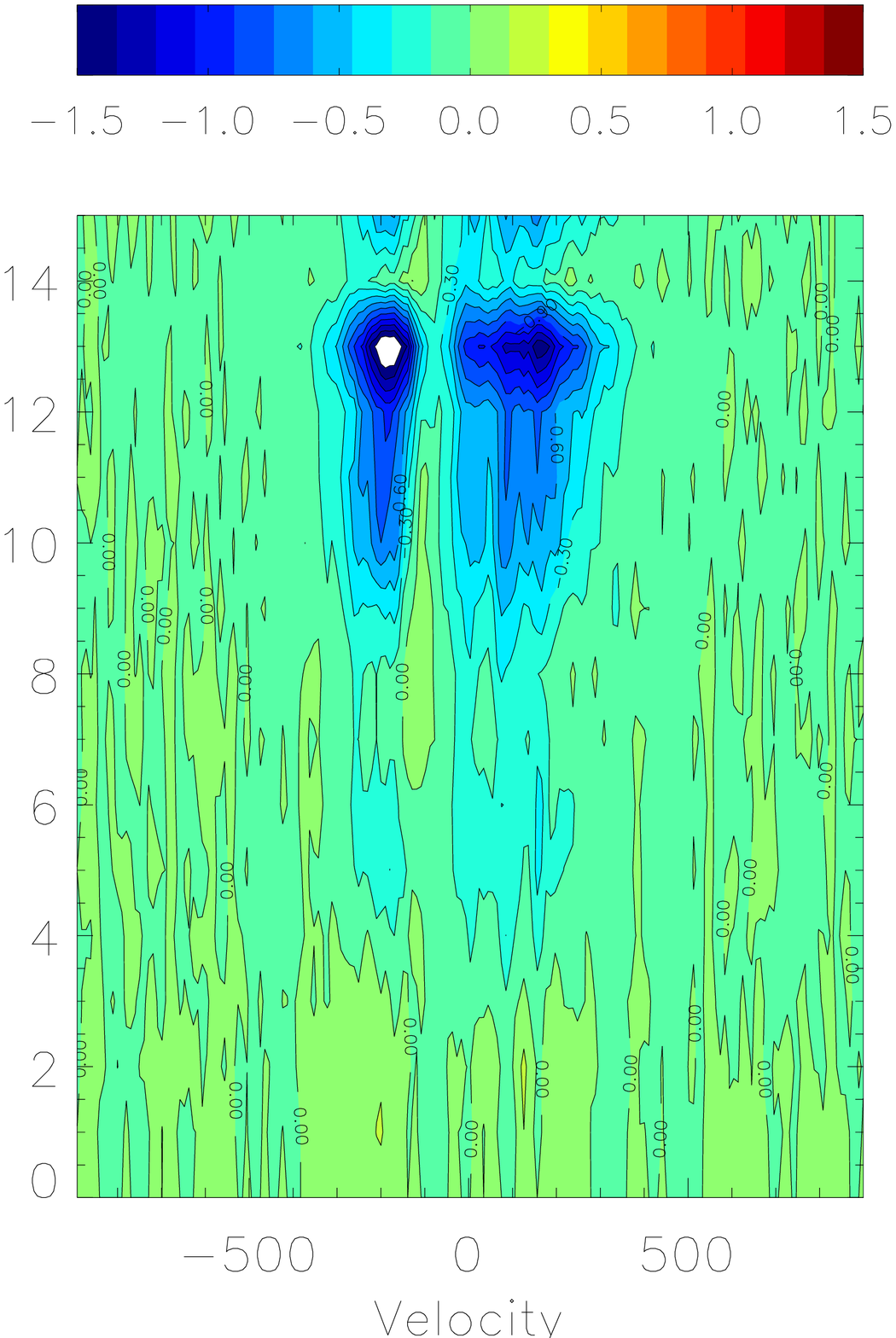}&  \includegraphics[scale=0.22]{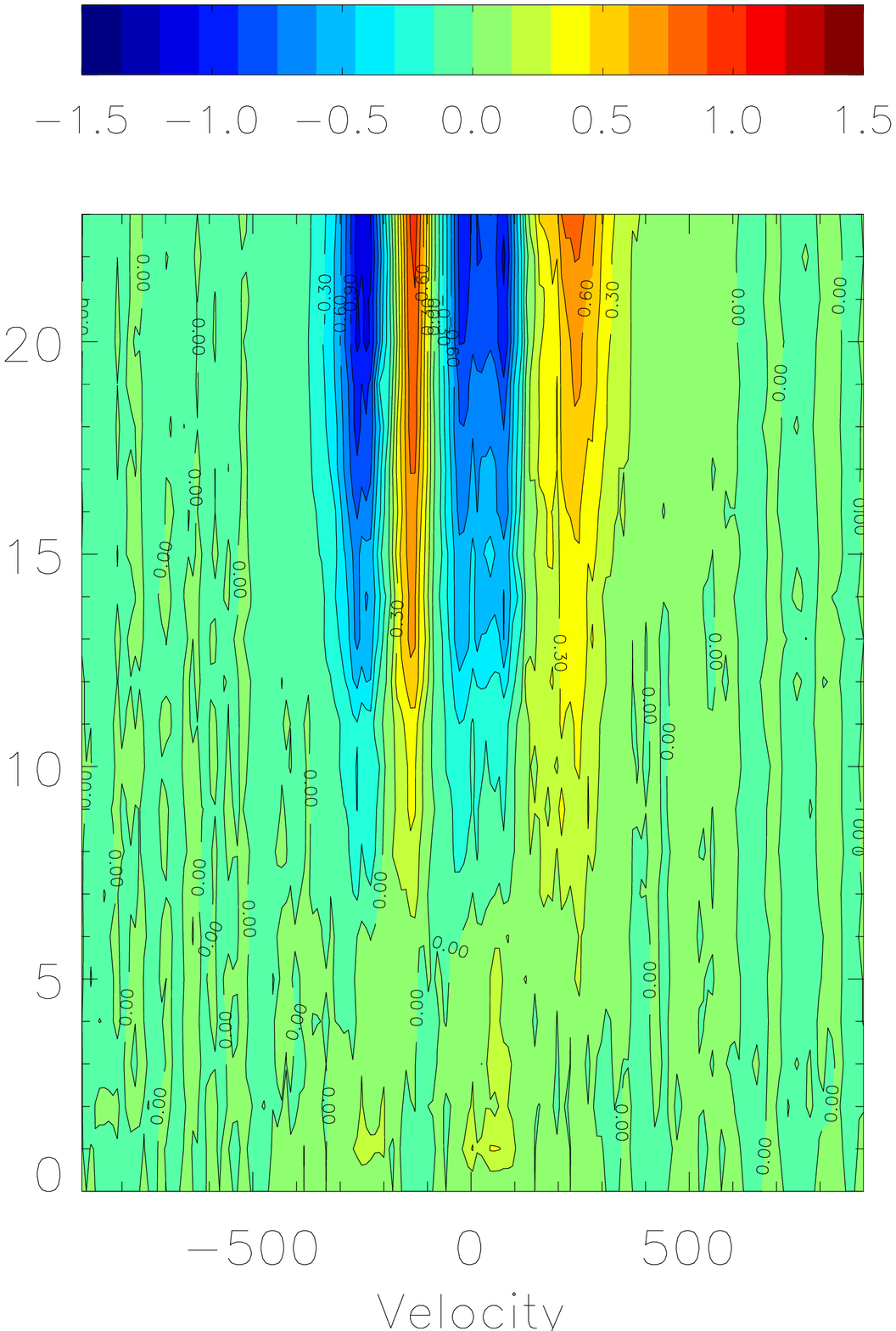} \\
 \includegraphics[scale=0.22]{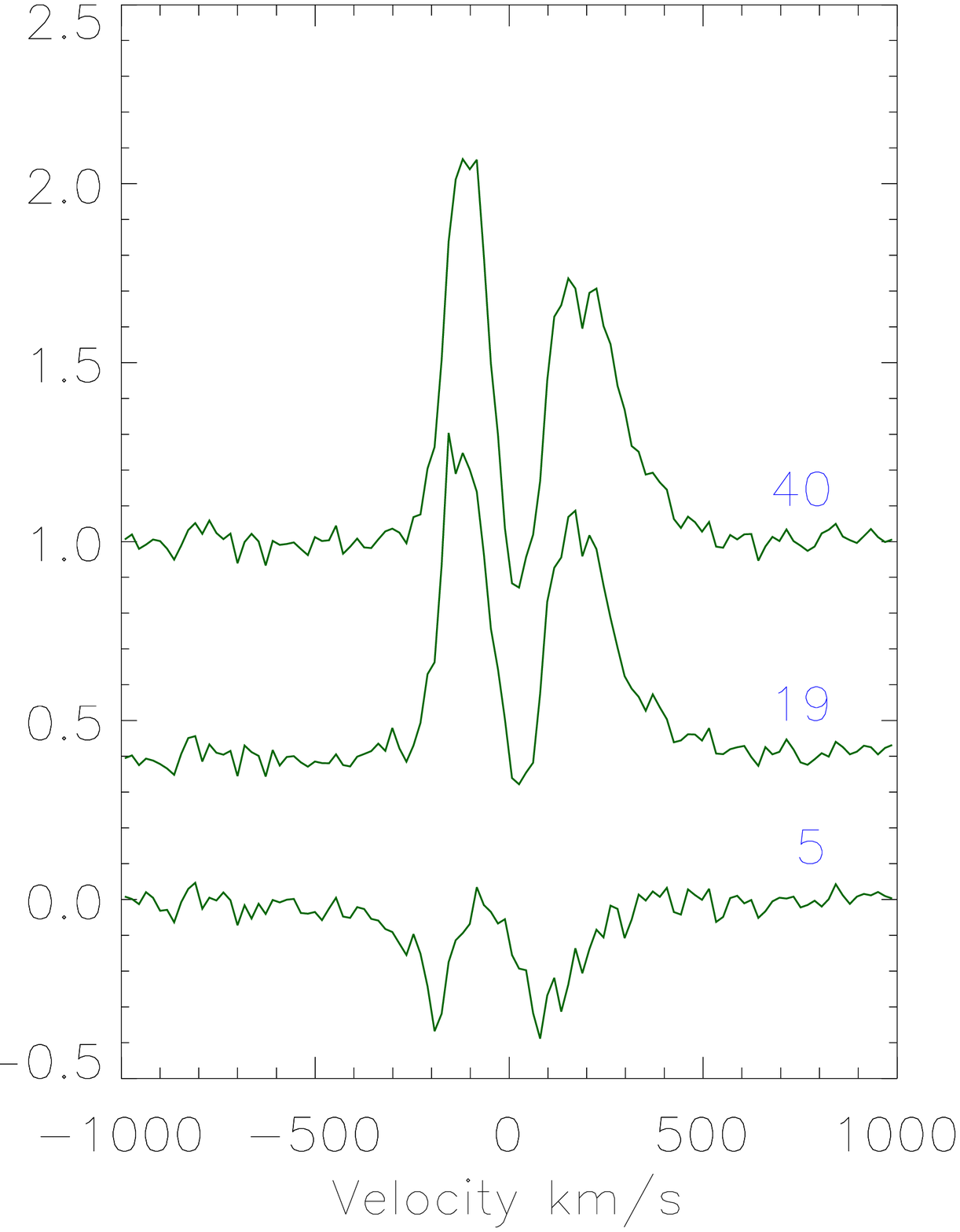} & \includegraphics[scale=0.22]{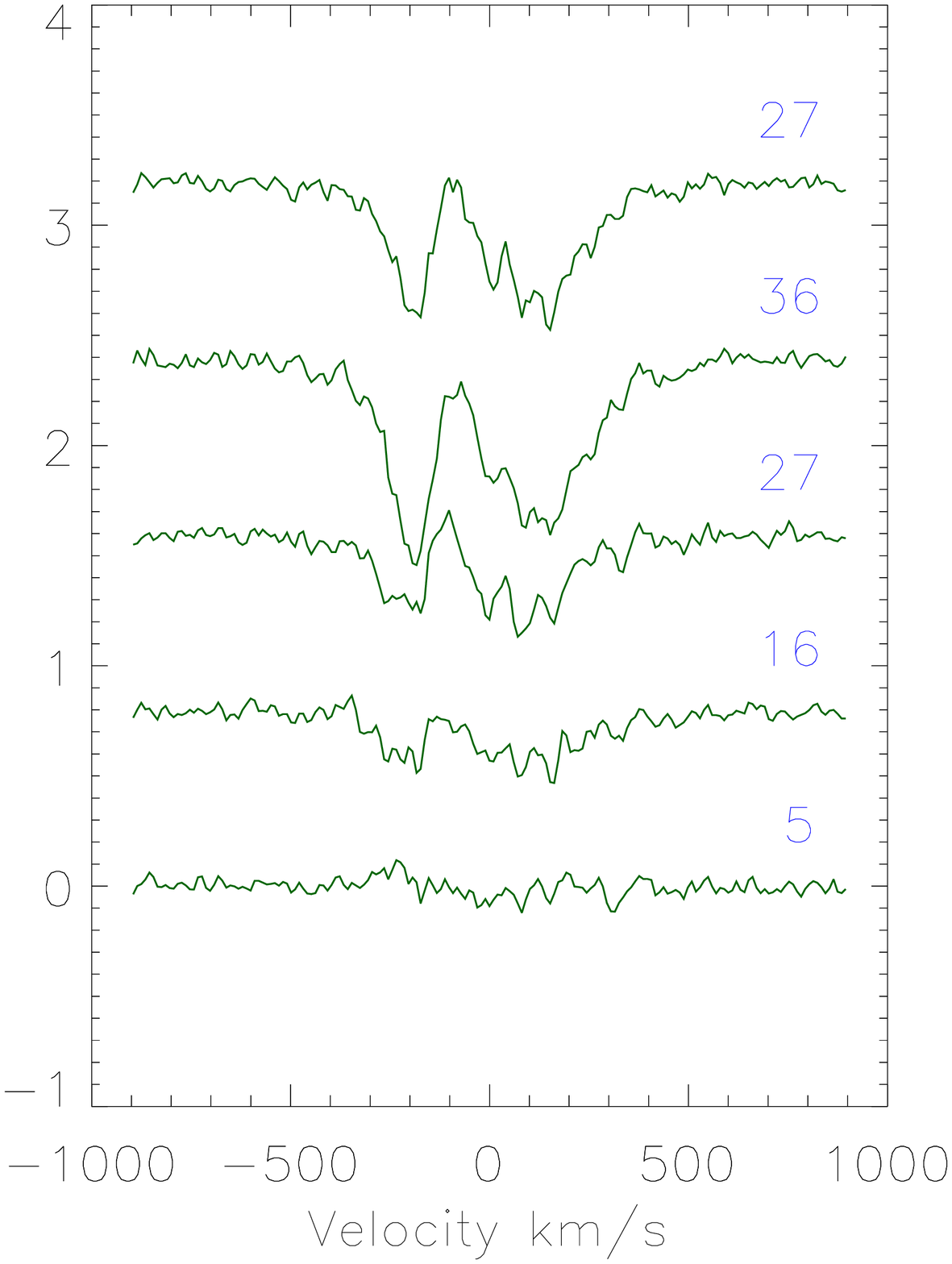} &  \includegraphics[scale=0.22]{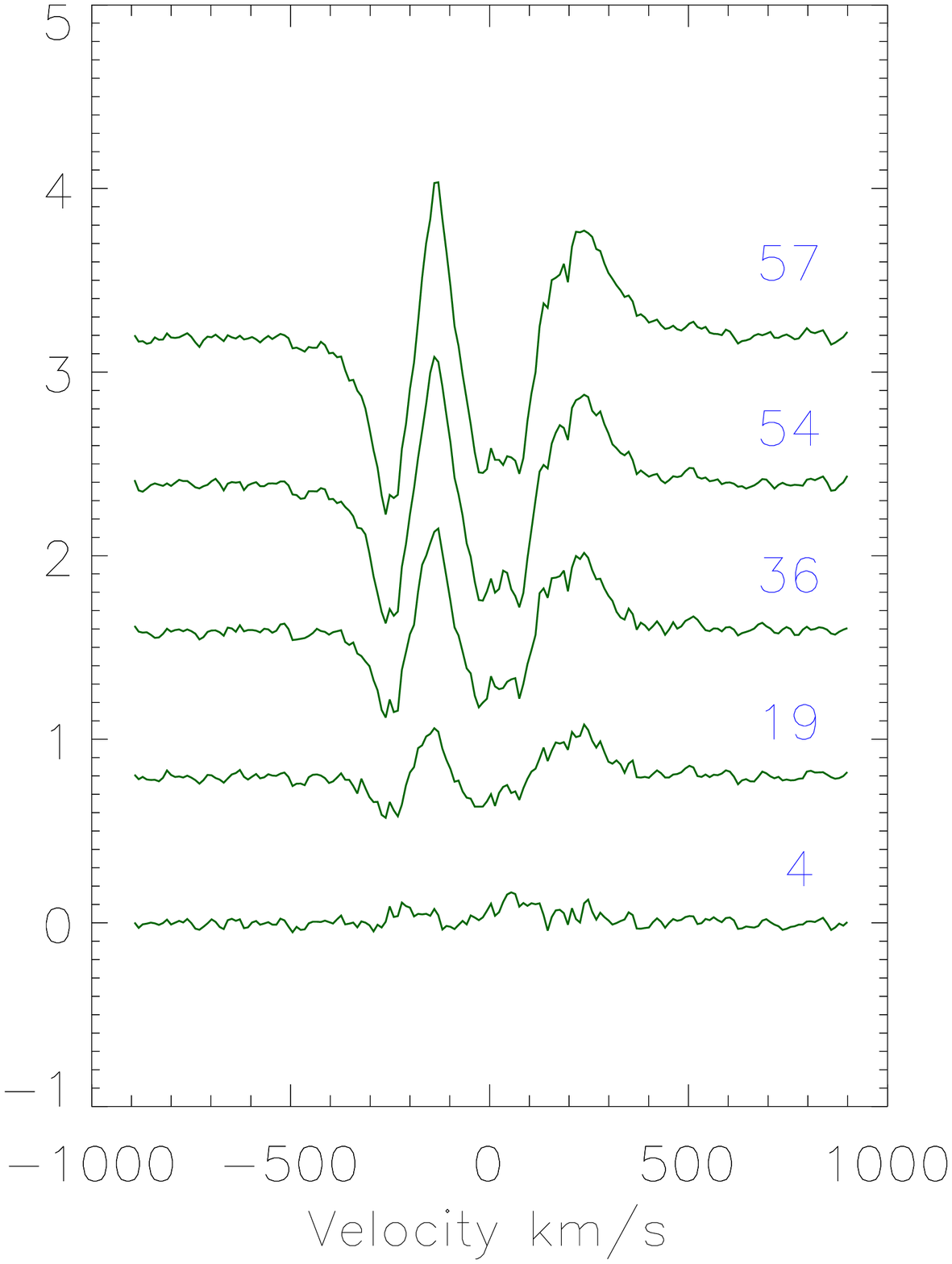} \\
\end{tabular}
\caption{RW Aur 2001 observations (left). RW Aur 2003 observations, Night 1 (middle) and Night 2 (right). }
\label{fig:RWAUR_2001}
\end{figure*}

\textbf{Stellar Properties:} RW Aur is a resolved triple system \citep{1993AJ....106.2005G}. The primary has an estimated stellar radius of 1.3\,-\,1.5 R$_{\odot}$ and a stellar mass of 1.1 M$_{\odot}$ \citep{2001A&A...369..993P}. A rotation period of 2.77 days has been reported for RW Aur A \citep{2001A&A...369..993P}. However, measurements of the longitudinal magnetic field in the star, found the changes in B$_{z}$ of -1.47 $\pm$ 0.15 to +1.10 $\pm$ 0.15 kG to be consistent with a rotation period of 5.6 days, and two rotating spots on the surface with opposite polarity. This is to be expected from an asymmetric accretion flow, where the two spots follow the base of the accretion flows. 

\noindent \textbf{Accretion Rates:} \citet{1989ApJ...341..340B} have reported an accretion rate of 2\,x\,10$^{-7}$ M$_{\odot}$.yr$^{-1}$ derived by fitting a boundary layer model to the observed emission.

\noindent \textbf{Outflows:} This object is associated with a very large asymmetric outflow which was found to contribute to the H$\alpha$ emission at $\sim$ -150\, to \,-180\,km\,s$^{-1}$ and at $\sim$ 109 km\,s$^{-1}$ \citep{1994ApJ...427L..99H}. \citet{2003A&A...405L...1L} found a bipolar jet with an inclination angle of 45$^{\circ}$, in this work it is assumed that the stellar system has the same inclination.  

\noindent \textbf{Disc Properties:} The inner disc radius is 2.7\,R$_{*}$ with a calculated co-rotation radius of 6.1\,R$_{*}$ \citep{2003ApJ...597..443G}, and an outer radius of $\leq$ 57 AU \citep{2006A&A...452..897C}. The disc inclination is thought to lie between 45\,-\,60${^\circ}$, which is in agreement with the jet inclination \citep{2006A&A...452..897C}.  Submillimeter observations have revealed  disc mass of  3\,x\,10$^{-4}$\,M$_{\odot}$ \citep{2006ApJ...653.1480W}. At the end of 2010, RW Aur was observed to dim by $\sim$ 2 magnitudes over the course of $\sim$ 180 days. This is attributed to an occultation by a large tidally disrupted trailing arm from the disc \citep{2013giec.conf40103R}, which most likely formed through an interaction with RW Aur B \citep{2006A&A...452..897C}.

\noindent \textbf{ISIS H$\alpha$ Observations:} Over the course of the observations, RW Aur shows a double peaked emission line with a central absorption and extended wings. During the single observation block in 2001 (Fig.\,\ref{fig:RWAUR_2001}) a double peaked profile is observed with a central absorption. Over the course of the observation block both peaks of emission strengthen with respect to the continuum. 

During the first night of observations in 2003 there is very little change in the profile (Fig.\,\ref{fig:RWAUR_2001}). A large change in the profile does take place between the first and second night's observations in 2003, where the red peak becomes much stronger the the blue emission peak. The changes in the profile on the second night take the form of increased emission in the red side of each of the emission peaks within the profile, while the central intensity decreases slightly.

\noindent \textbf{Previous H$\alpha$ Observations and Variations:} A similar H$\alpha$ profile to these was observed in 1976/77 with an EW of 128\AA~\citep{1979ApJS...41..369S}.

RW Aur was also observed in 2001 \citep{2013AJ....145..108C}, where EW measurements of 80.4, 65.6, 79.7, 72.9\,\AA~were found on November 17$^{th}$, 21$^{st}$, 25$^{th}$ and 27$^{th}$ respectively. These measurements are very close to what was observed in the ISIS sample on the 26$^{th}$ December 2001 (mean 74.04\,\AA). The same magnitude of variations (81.11 - 67.88\,\AA) was observed with ISIS during the half hour of observations, as \citet{2013AJ....145..108C} observed in one week. Their variations took the form of a fall in the red peak emission. 

\citet{1979BAAS...11..439S} found evidence for short term flares and changes while photometrically monitoring RW Aur, which they likened to the slow flare events on YZ Cmi. Relatively large photometric variations have also been reported, \citet{2001A&A...375..977P} found over periods of 3\,-\,4 days changes of 0.2\,-\,0.7 magnitudes in the V band. 

Observations of rapid line profile variability in the spectra of RW Aur, specifically the H$\gamma$ profile, have also been observed on time scales as short as 10 mins \citep{1982ApJ...256..156M}.

\subsection{DR Tau}

\begin{figure}
\centering
\begin{tabular}{cc}
\includegraphics[scale=0.22]{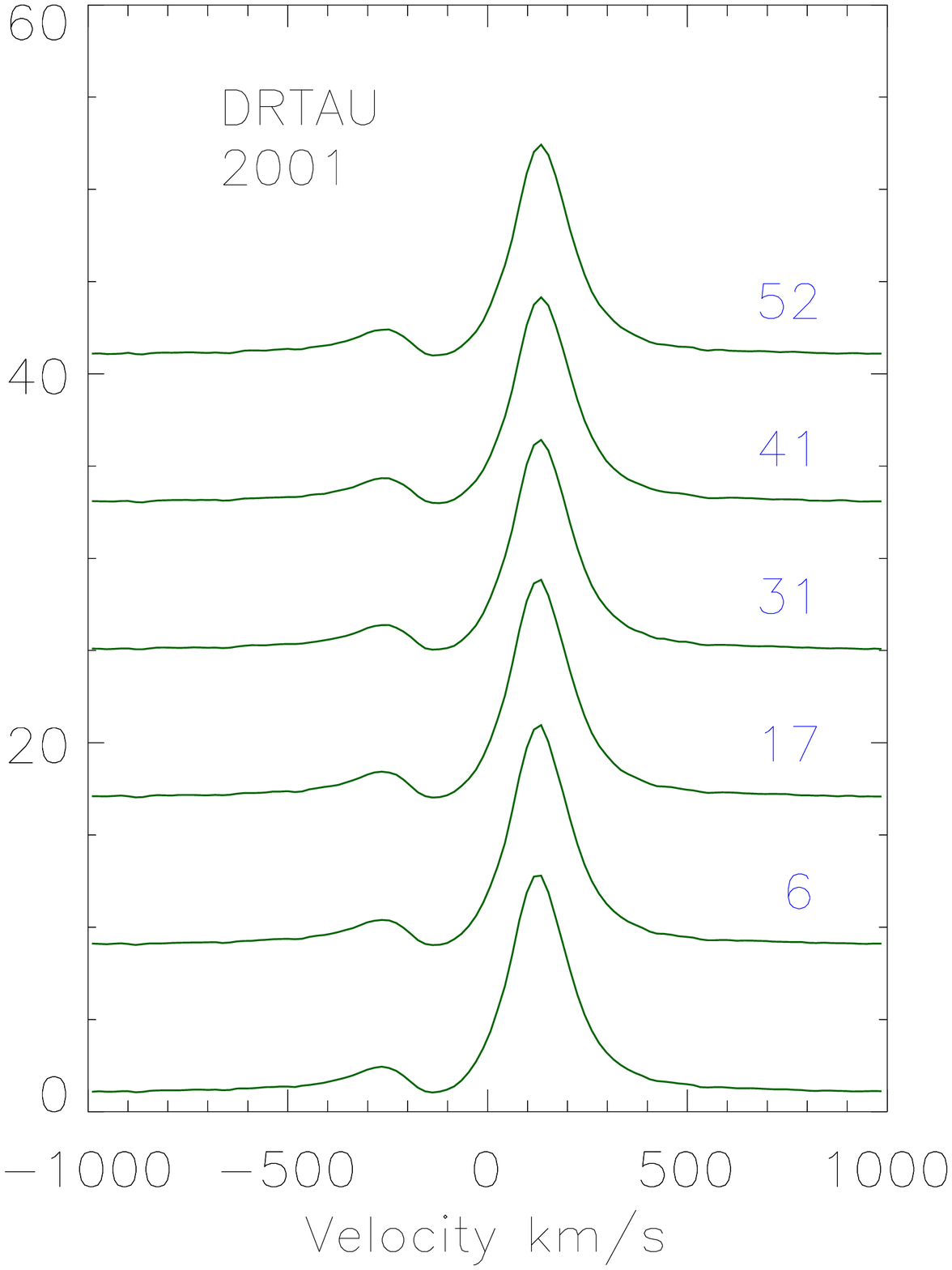} & \includegraphics[scale=0.22]{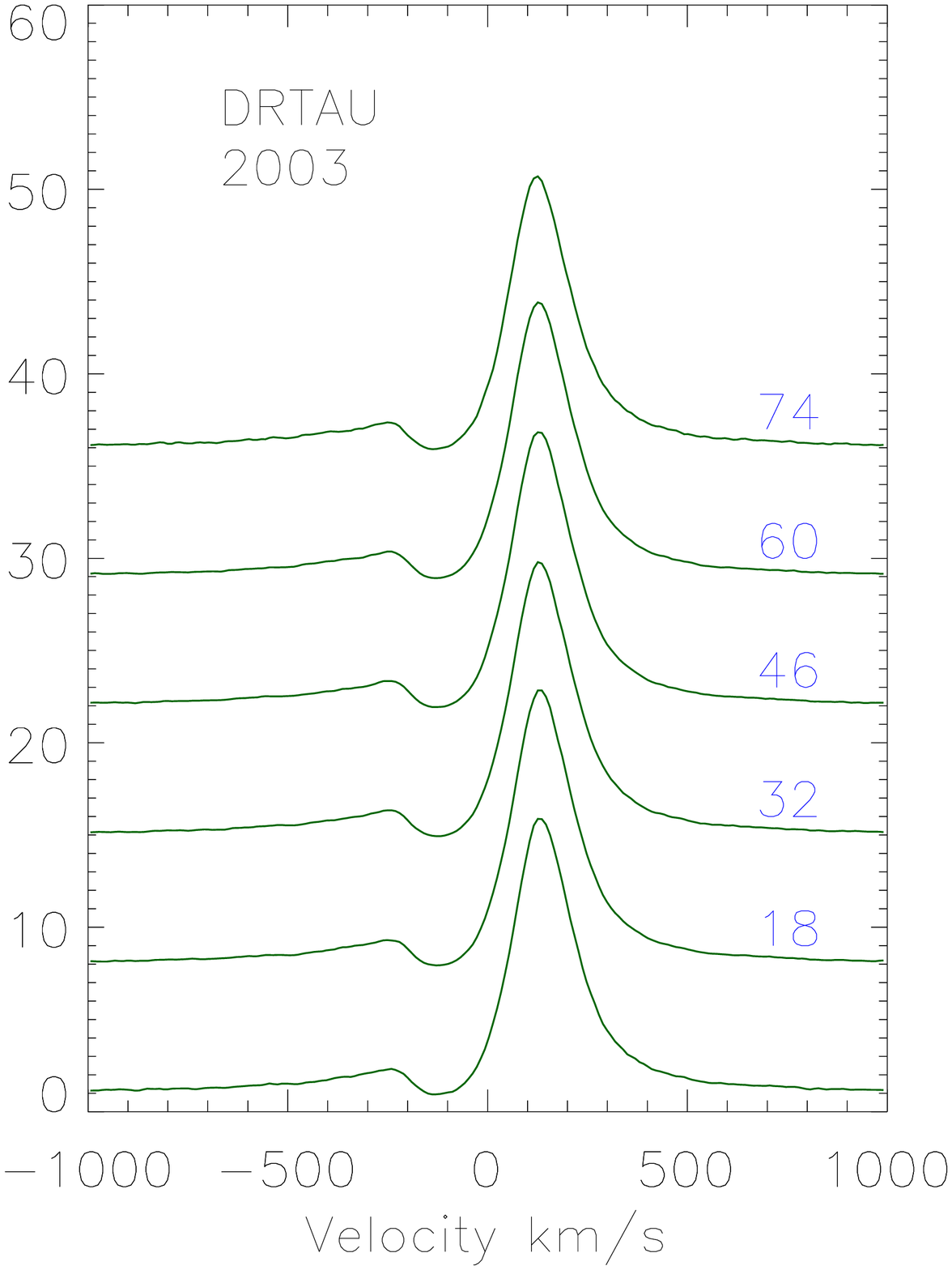} \\
\includegraphics[scale=0.22]{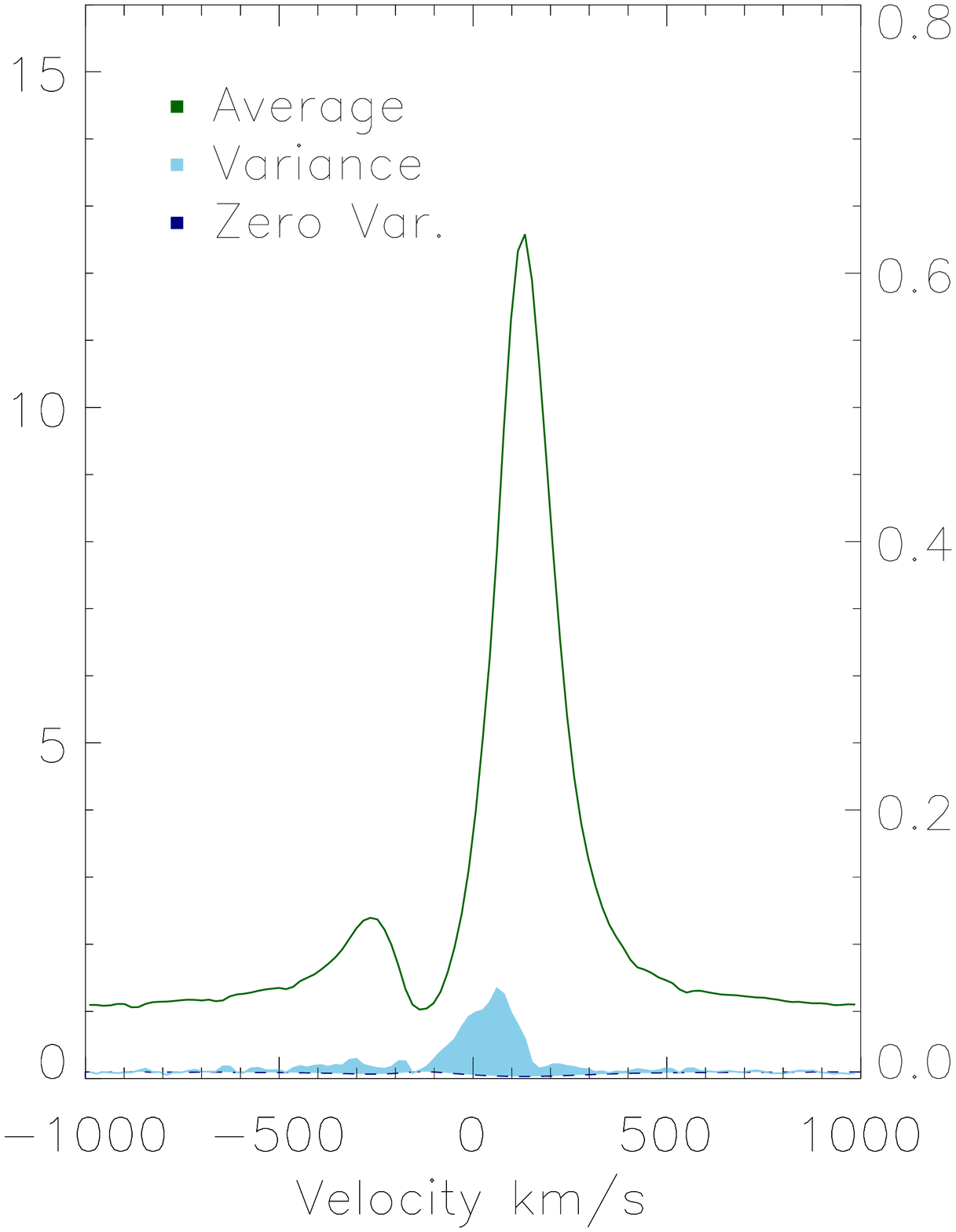} &  \includegraphics[scale=0.22]{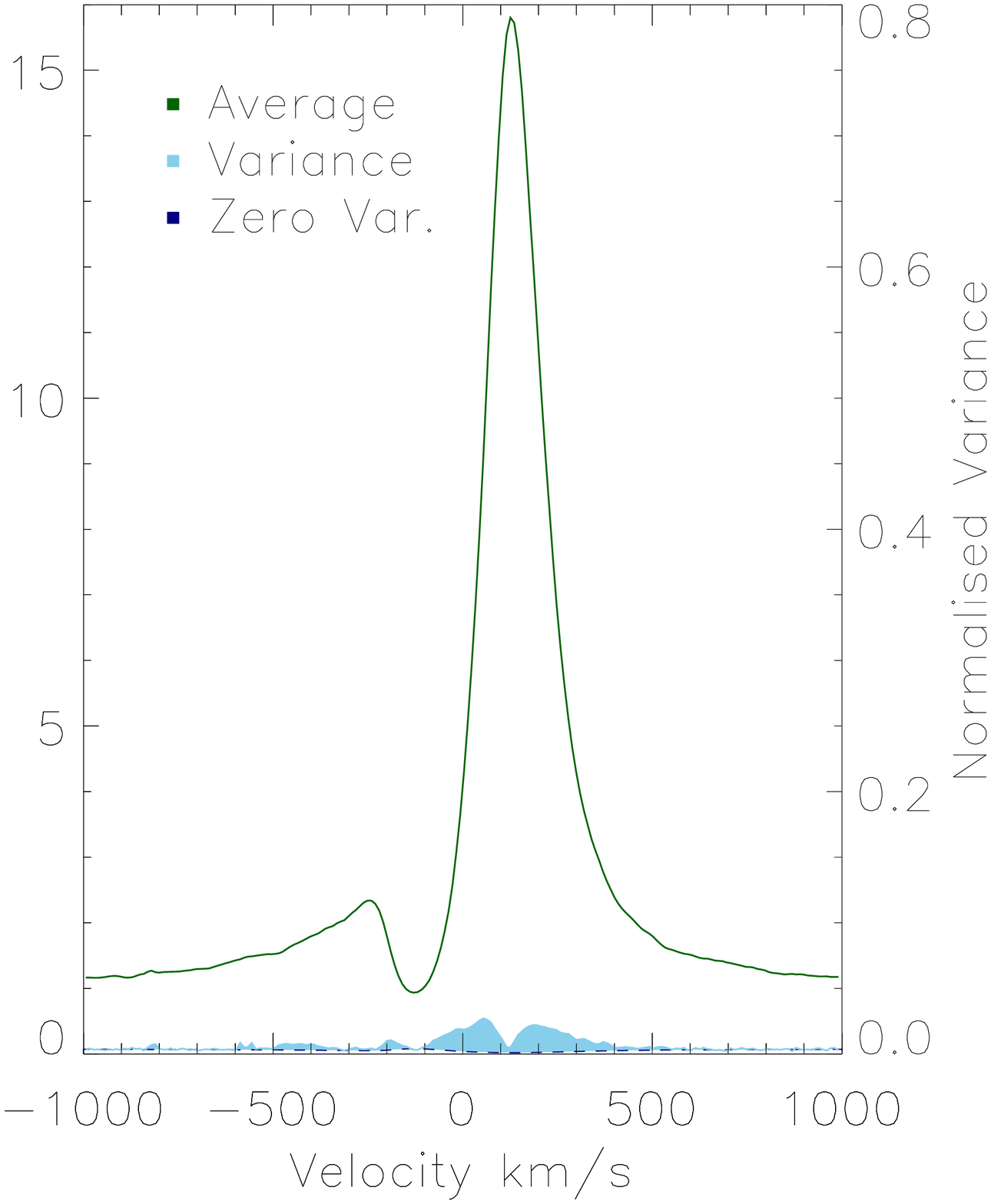} \\
 \includegraphics[scale=0.22]{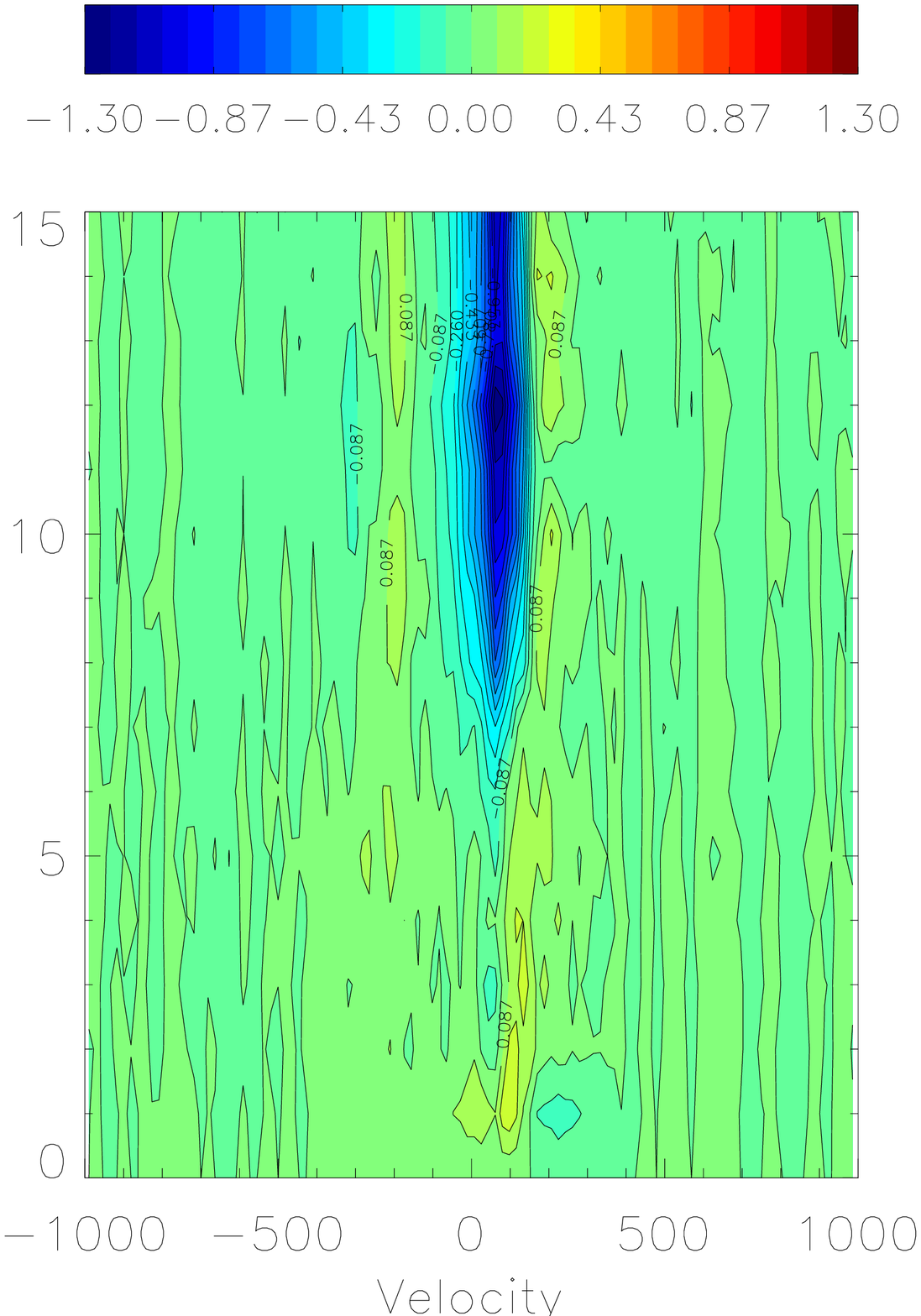} & \includegraphics[scale=0.22]{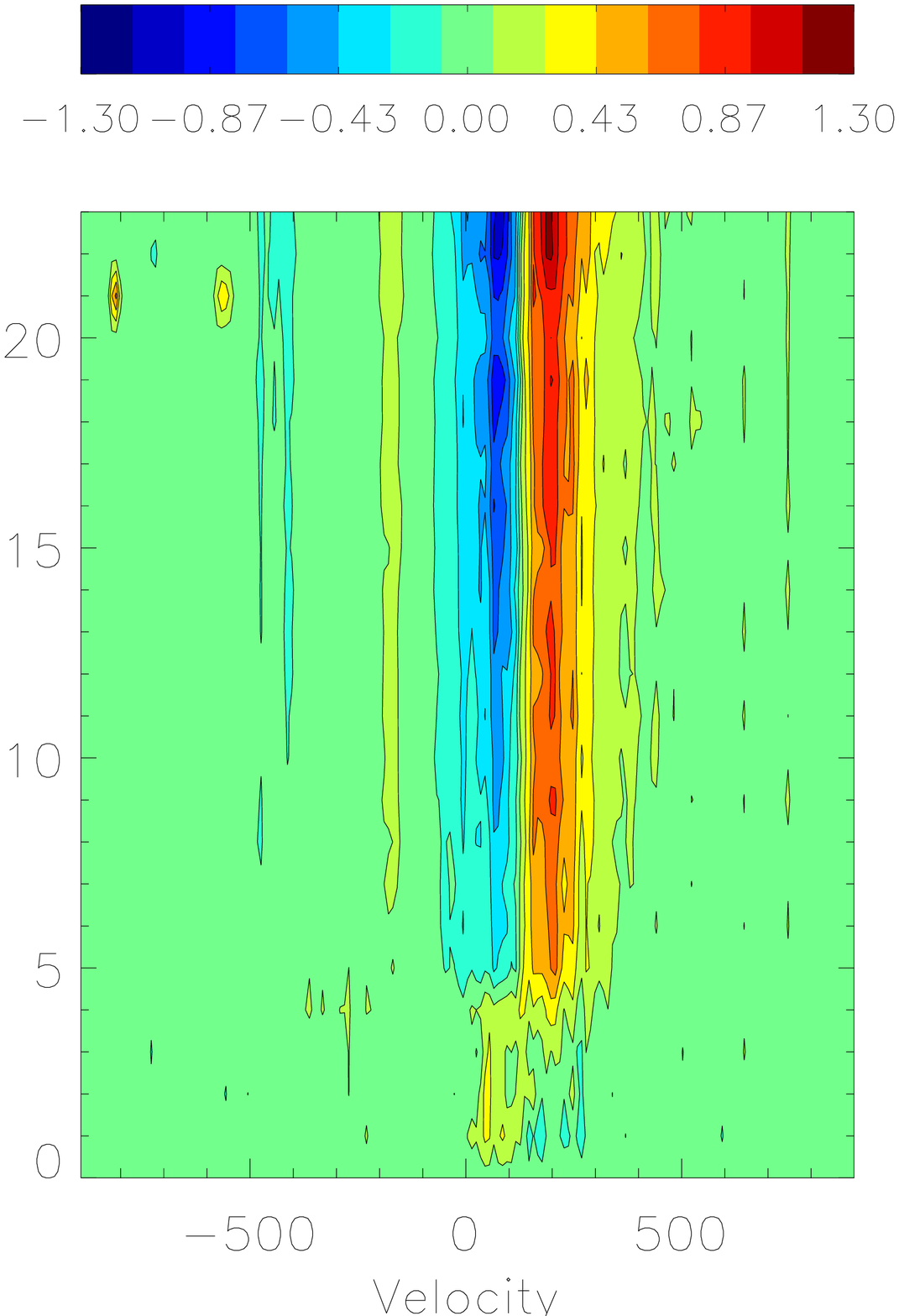}  \\
 \includegraphics[scale=0.22]{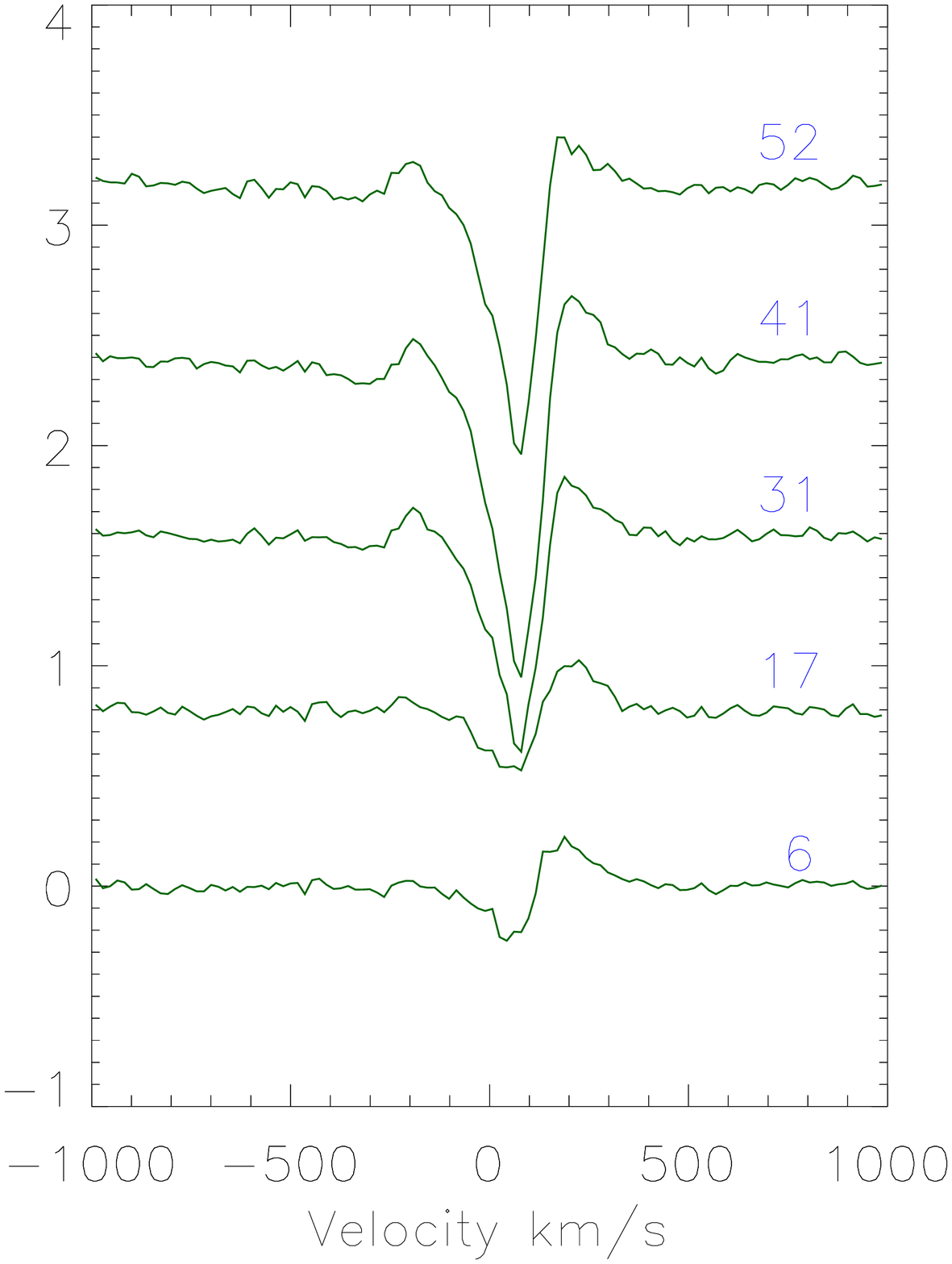} & \includegraphics[scale=0.22]{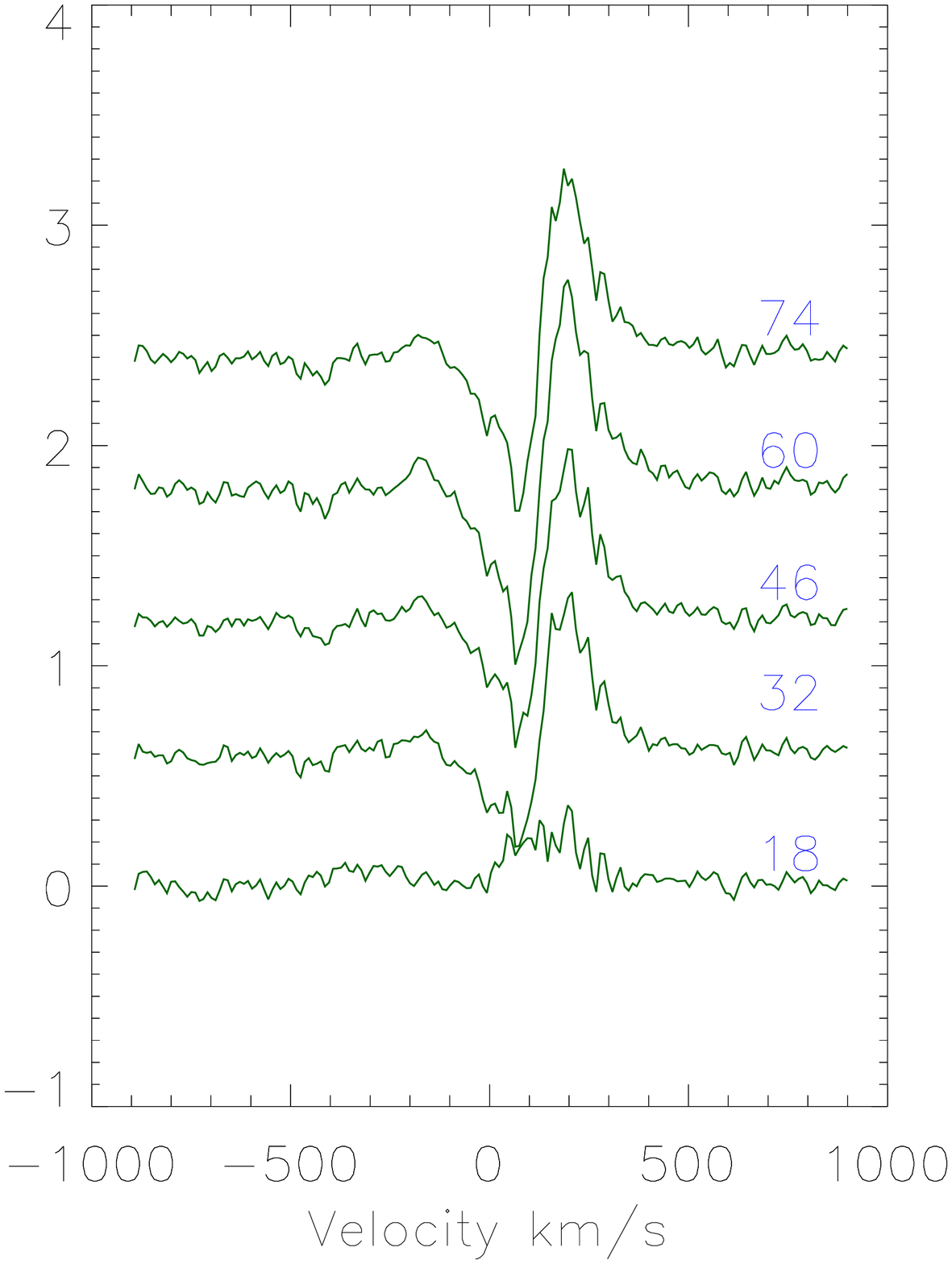} \\
\end{tabular}
\caption{DR Tau 2001 observations. Five spectra were observed on the previous day to these observations, but are not shown here (left). DR Tau 2003 Observations (right). }
\label{fig:DRTAU_plots_2001}
\end{figure}

\textbf{Stellar Properties:} DR Tau has a stellar mass of 1.0\,M$_{\odot}$ and a radius of 5.1\,R$_{\odot}$ \citep{1988ApJ...330..350B,1998ApJ...492..323G}. The period is not very well constrained, \citet{1995AJ....109.2800J} found signatures in the emission lines of DR Tau with periods of 5.1 and 7.9 days. 

\noindent \textbf{Disk properties}: A disc mass of 0.1 $M_{\odot}$ has been estimated with R$_{in}$ of 0.05\,AU \citep{2002ApJ...581..357K}. There have been a number of published inclinations for this system, which vary from edge on orientation to a pole on orientation \citep{2001AJ....122.3335A,2001ApJ...550..944M,2002ApJ...581..357K}. However, based on the low rotational velocity and a period of 5.1 days \citep{1995AJ....109.2800J}, \citet{2005MNRAS.359.1049V} suggest a pole-on orientation to be more plausible. 

\noindent \textbf{Accretion:} The lack of photospheric lines in the observed spectra of DR Tau is attributed to high levels of veiling in the continuum \citep{1990ApJ...363..654B,1994PhDT........16V}. \citet{1989ApJ...341..340B} have reported an accretion rate of 5\,x\,10$^{-7}$ M$_{\odot}$yr$^{-1}$ derived by fitting a boundary layer model to the observed emission.

\noindent \textbf{Outflows:} It has been suggested that most of the H$\alpha$ emission is from a stellar wind \citep{2001ApJ...550..944M}. The forbidden emission line [OI] $\lambda$6300 has been observed with profiles composed of two peaks, one at the central wavelength and the second blue-shifted. This is a strong suggestion for a collimated outflow from this source \citep{1995ApJ...452..736H,1997A&AS..126..437H}.

\noindent \textbf{ISIS H$\alpha$ Observations:} DR Tau shows very similar profiles in both 2001 and 2003 (Fig.\,\ref{fig:DRTAU_plots_2001}). The profile takes the form of a strong red-shifted emission line with extended wings, and a strong blue absorption that extends to below the continuum. In 2001 the blue-shifted `peak' is more pronounced, suggesting a slightly stronger high-velocity (\textgreater \,400\,km\,s$^{-1}$) blue-shifted absorption. The variations in 2001 are contained to a single wavelength range centred at 100\,km\,s$^{-1}$, which is close to the large peak of the profile.  No significant variations occur over the time-scale of the observations. 

In 2003 the variance profile is different, showing that the changes in the line take the form of a change in emission strength either side of the emission peak (Right most panels Fig.\,\ref{fig:DRTAU_plots_2001}). The first three spectra in this observation block differ slightly from the rest of the spectra, resulting in a difference of $\sim$ 20\AA~in EW but no significant change in the 10\%w.

\noindent \textbf{Previous H$\alpha$ Observations and Variations:} \citet{2001AJ....122.3335A} decomposed the H$\alpha$ emission line for DR Tau into three parts (1) a strong red-shifted emission peak (like the emission line in this work) (2) a blue-shifted wind absorption component and (3) a relatively low amplitude component with a FWHM centred at rest velocity. Most of the variability was found in the blue side of the profile, which was found to be incoherent with changes in the red emission. (Interestingly in the case of the ISIS observations, most variability is found on the red side in emission peak). They attribute the red emission to be a result of emission from a hot inflow. 

The large photometric variations of 1\,-\,3 magnitudes that have been observed have been attributed to hot and cool spots on the surface of DR Tau \citep{1977A&A....61..737B,1993A&AS..101..485B,1994AJ....107.2153K} and high veiling (see above).

\subsection{GW Ori}

\begin{figure}
\centering
\begin{tabular}{cc}
\includegraphics[scale=0.22]{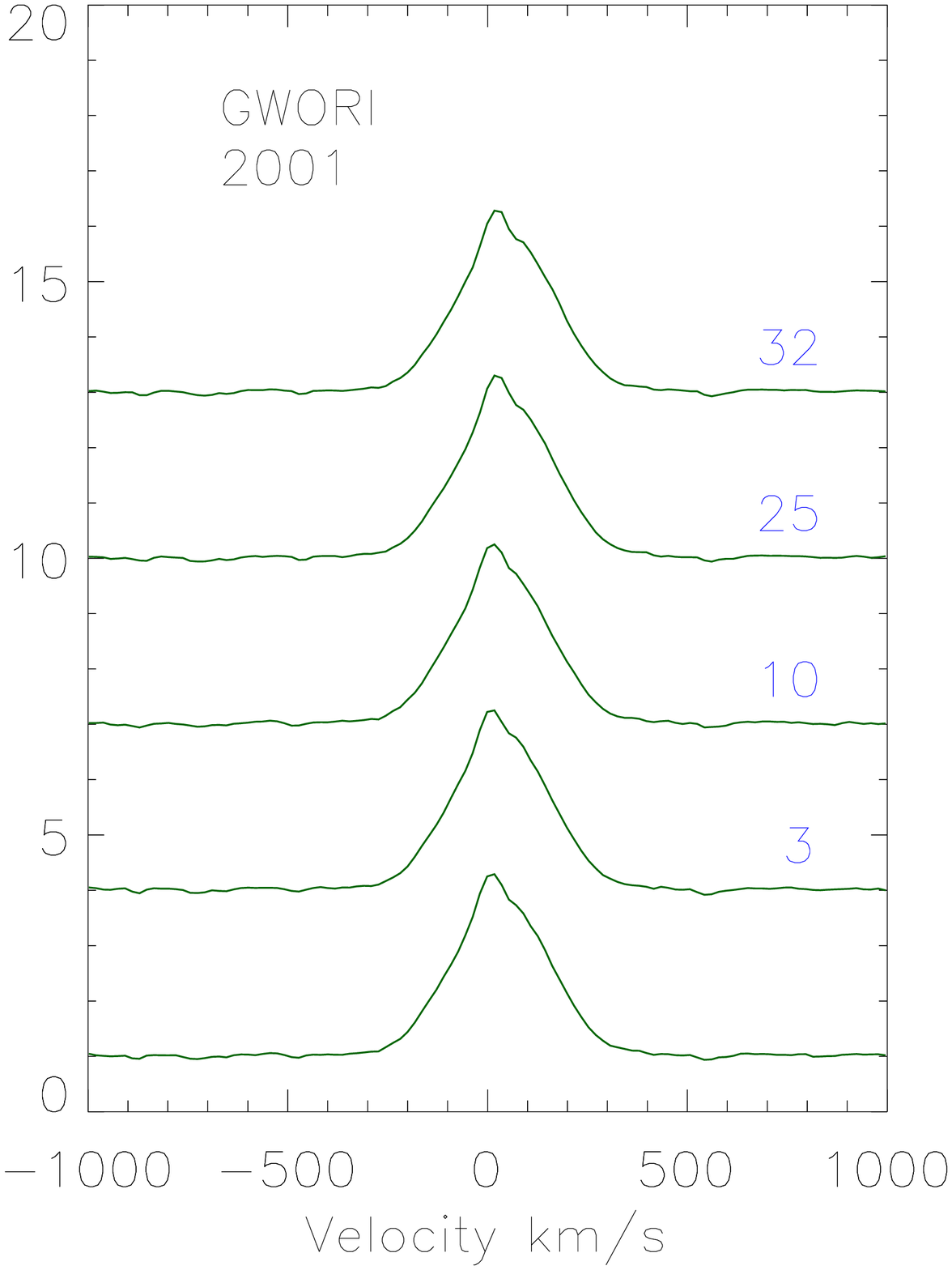} & \includegraphics[scale=0.22]{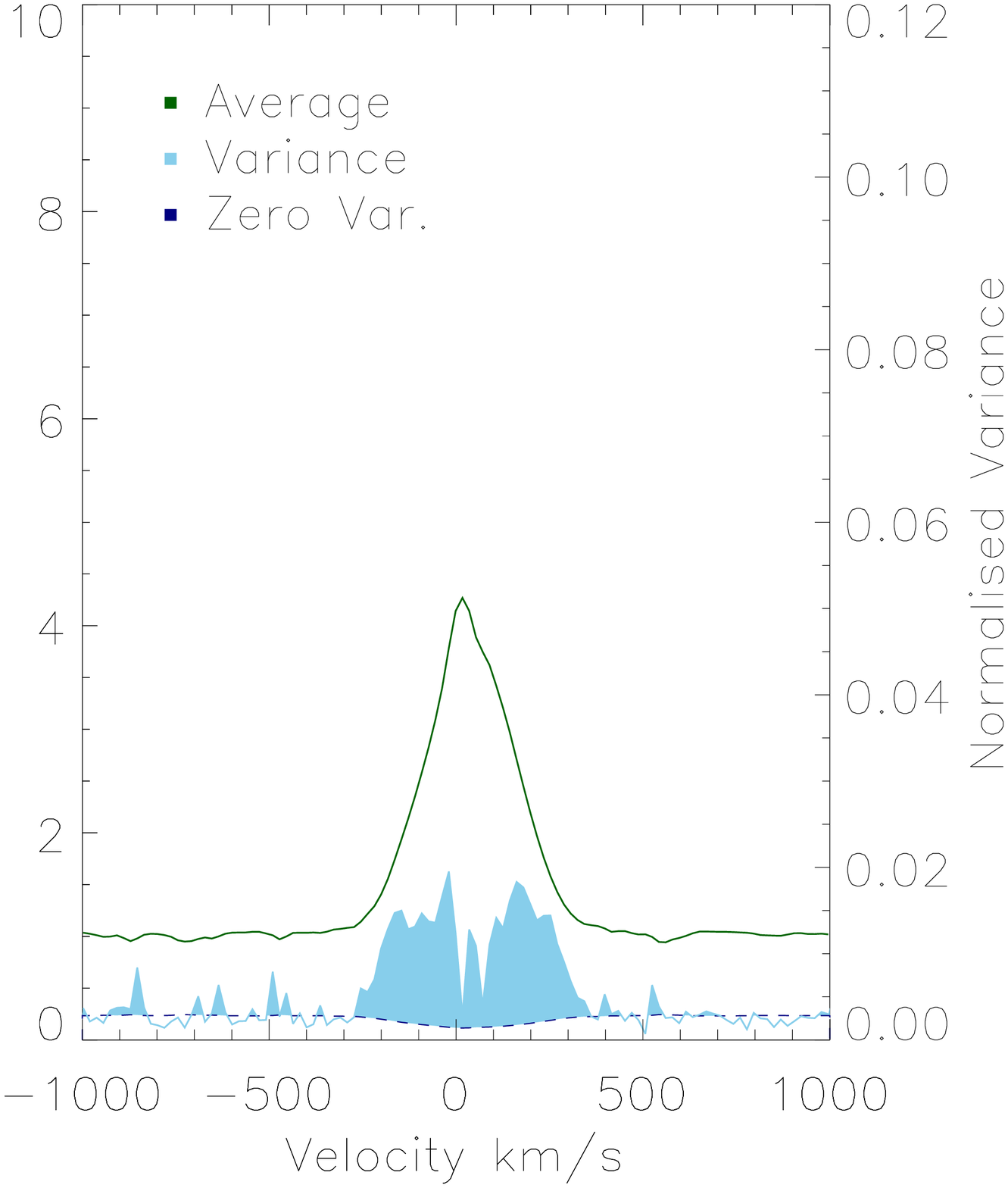} \\
\includegraphics[scale=0.22]{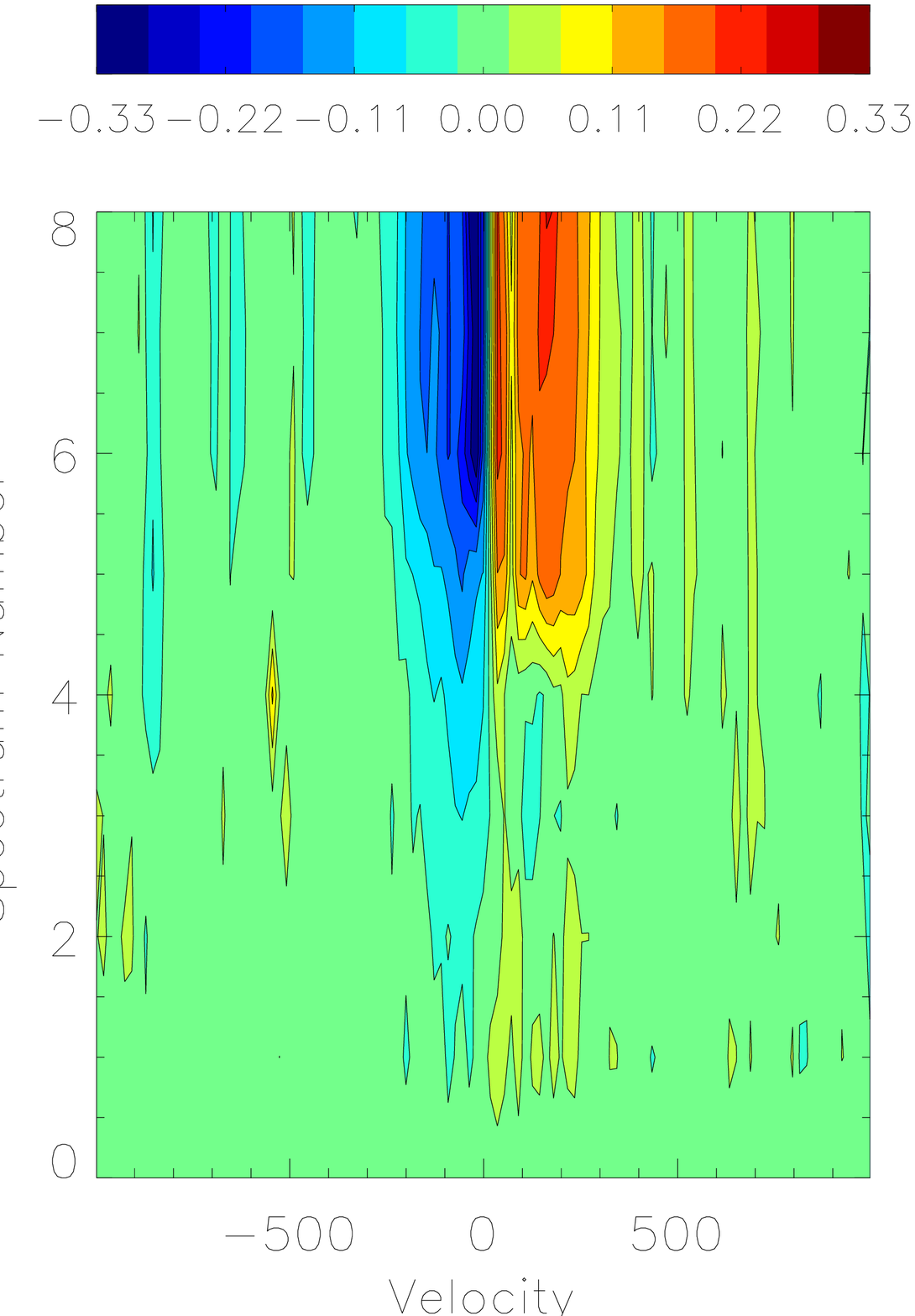} & \includegraphics[scale=0.22]{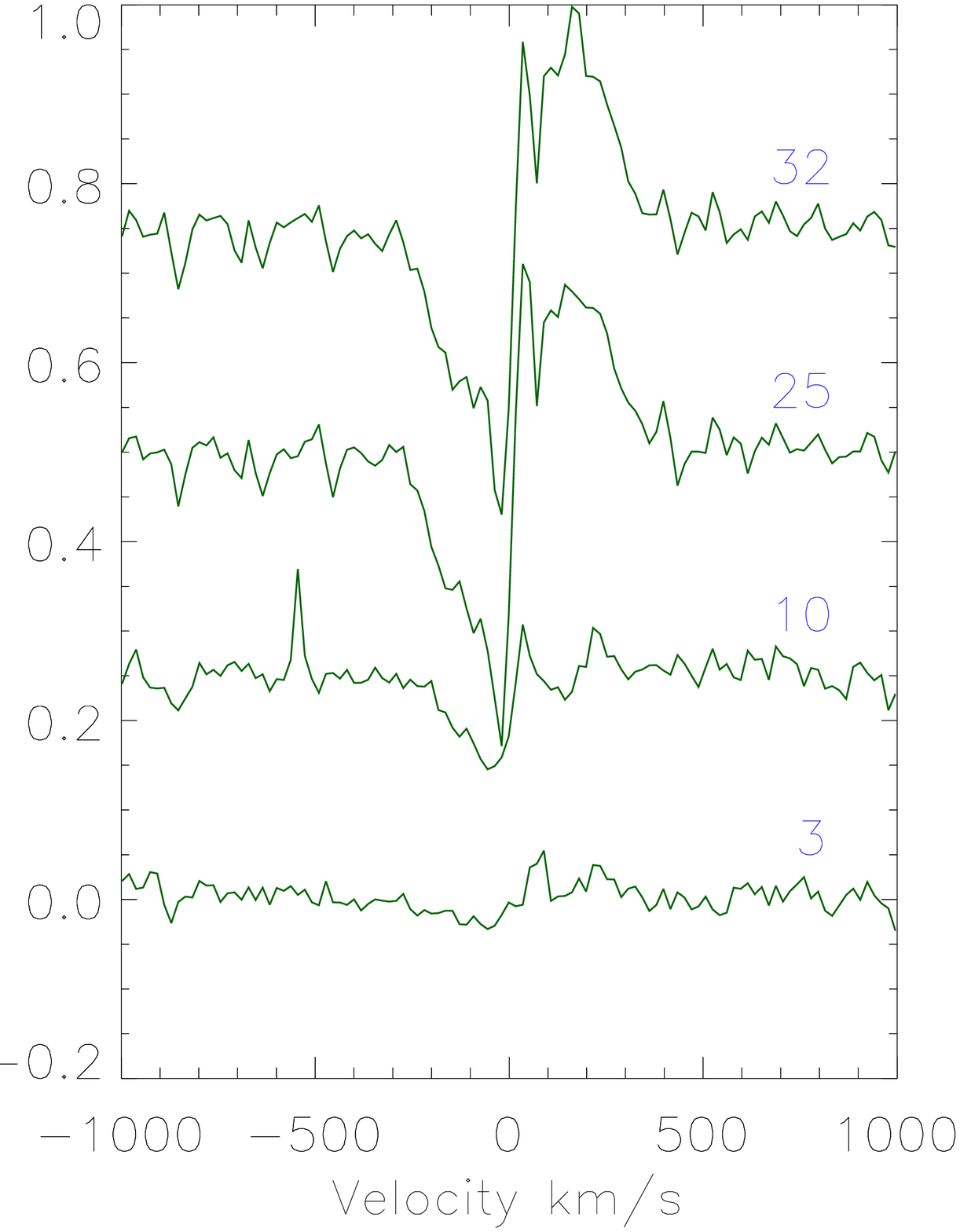} \\
\end{tabular}
\caption{GW Ori 2001 observations.  }
\label{fig:GWORI_plots_2001}
\end{figure}

\begin{figure}
\centering
\begin{tabular}{cc}
\includegraphics[scale=0.22]{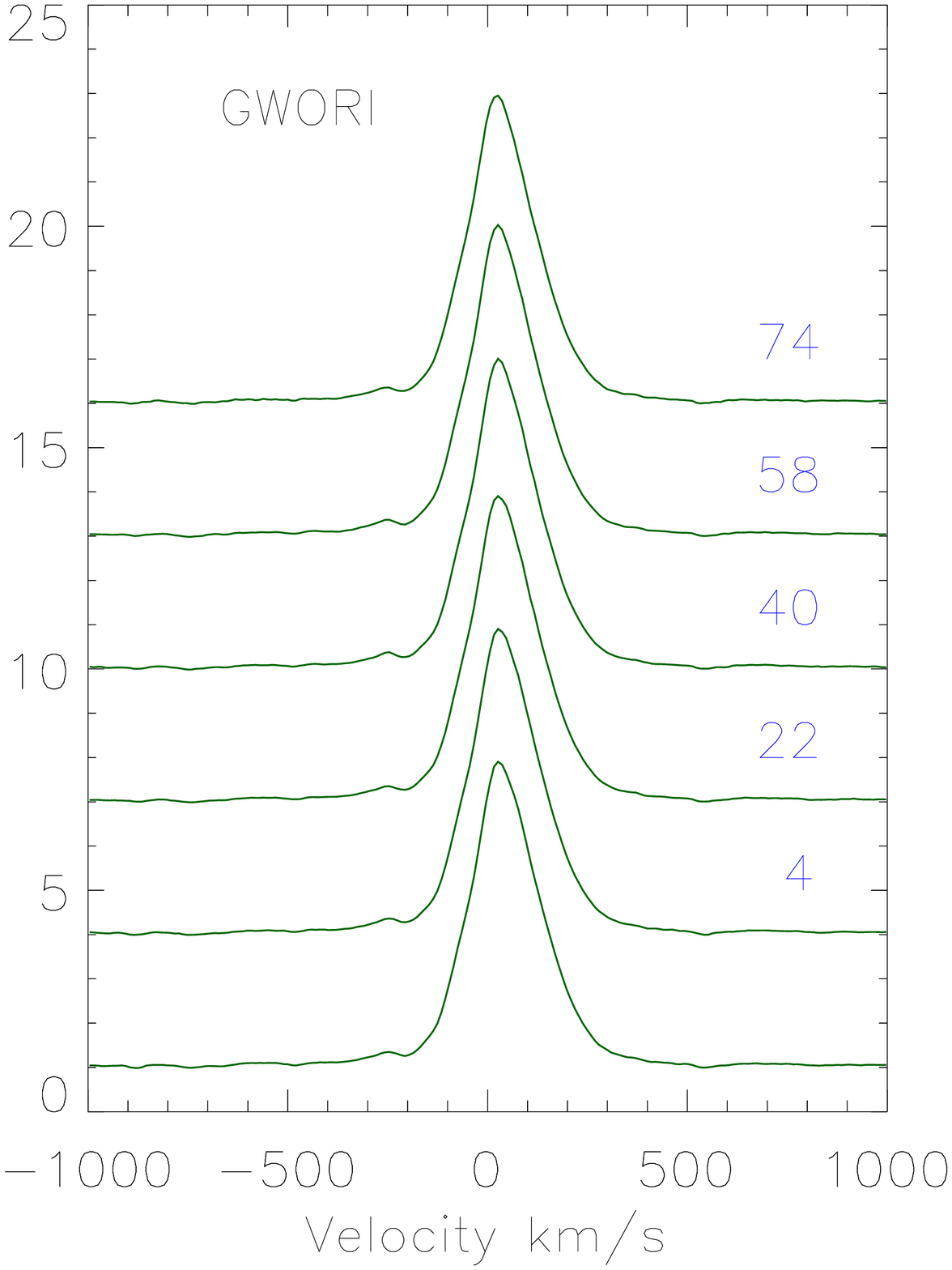} & 
\includegraphics[scale=0.22]{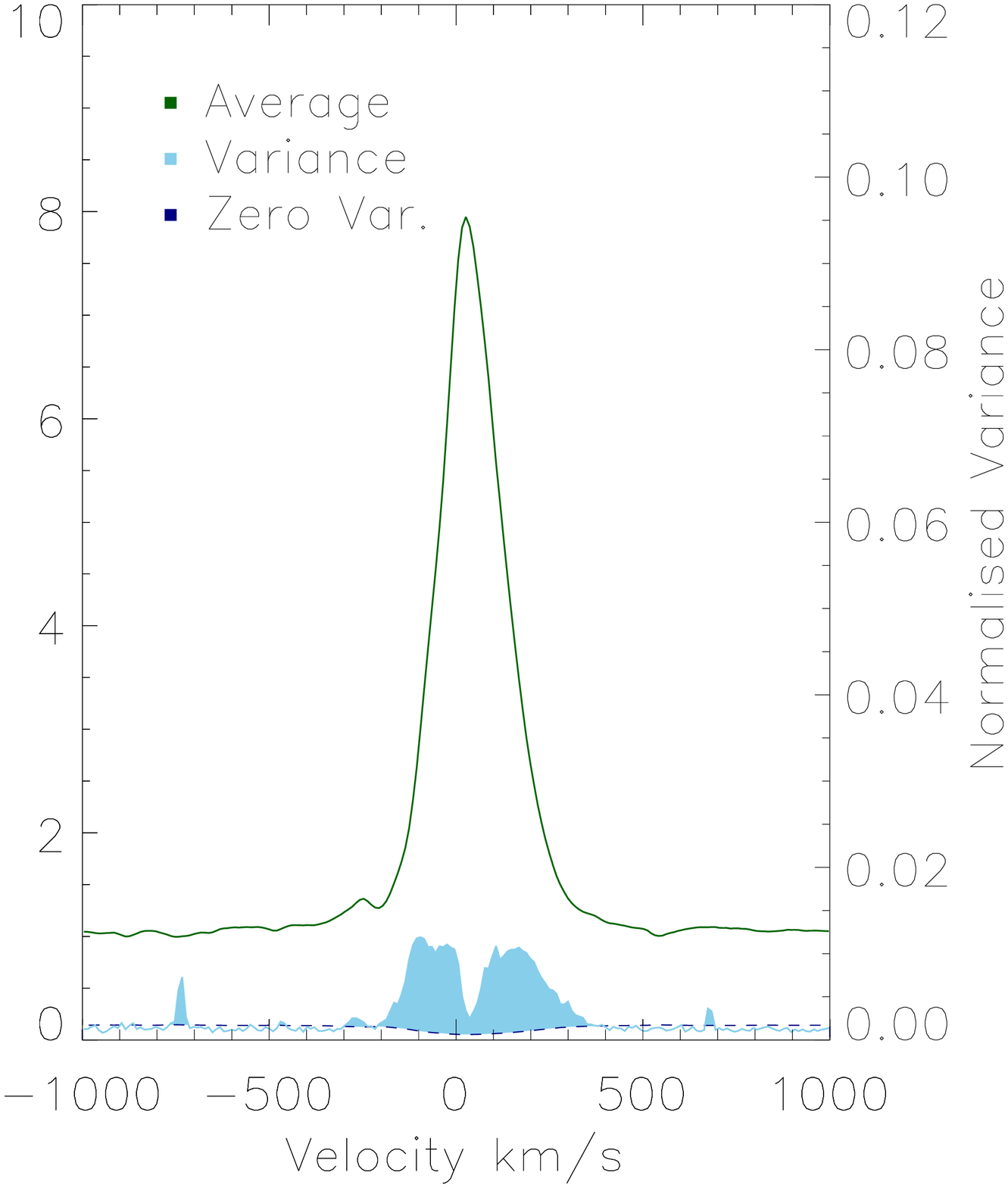}  \\
\includegraphics[scale=0.22]{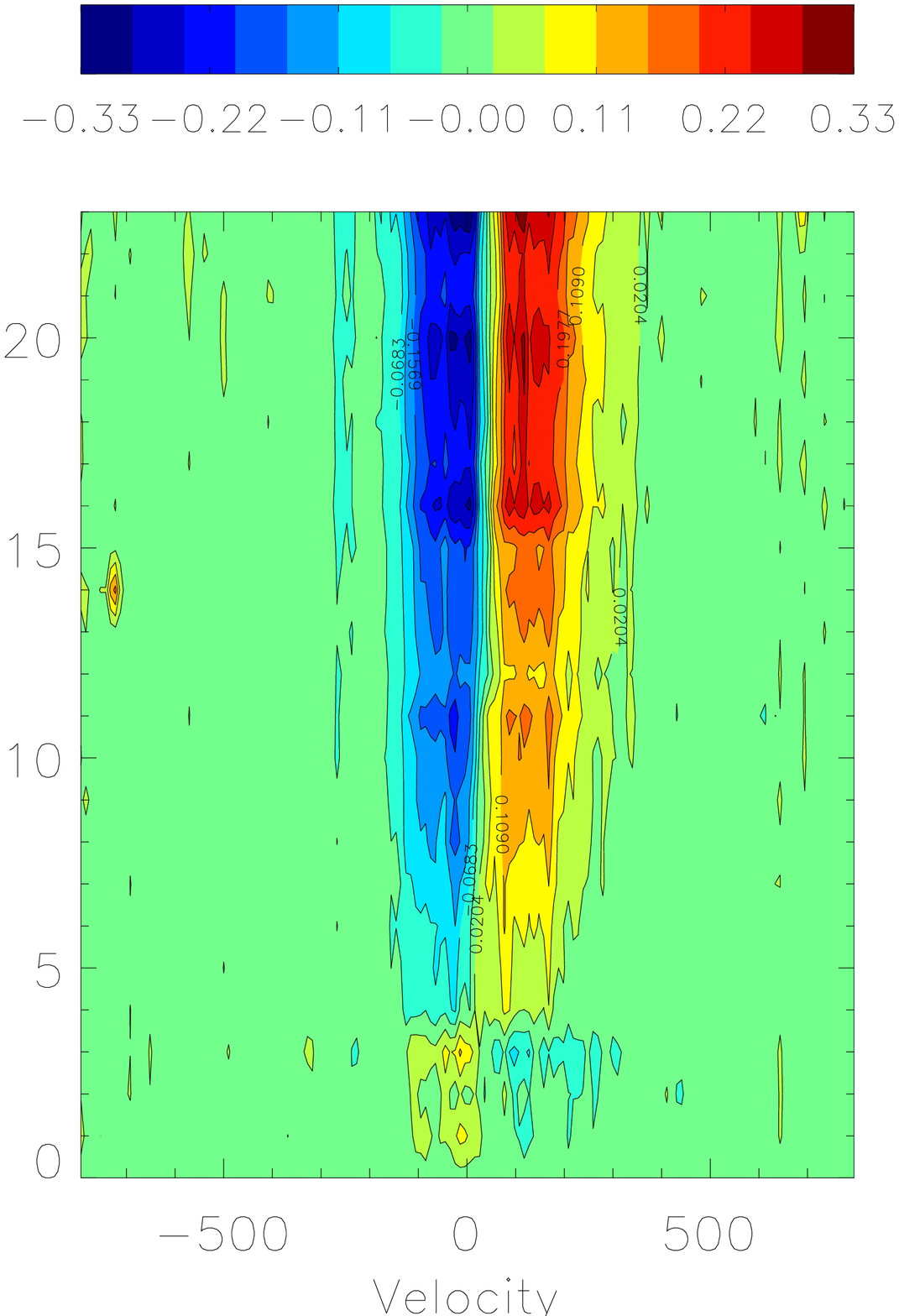} &
\includegraphics[scale=0.22]{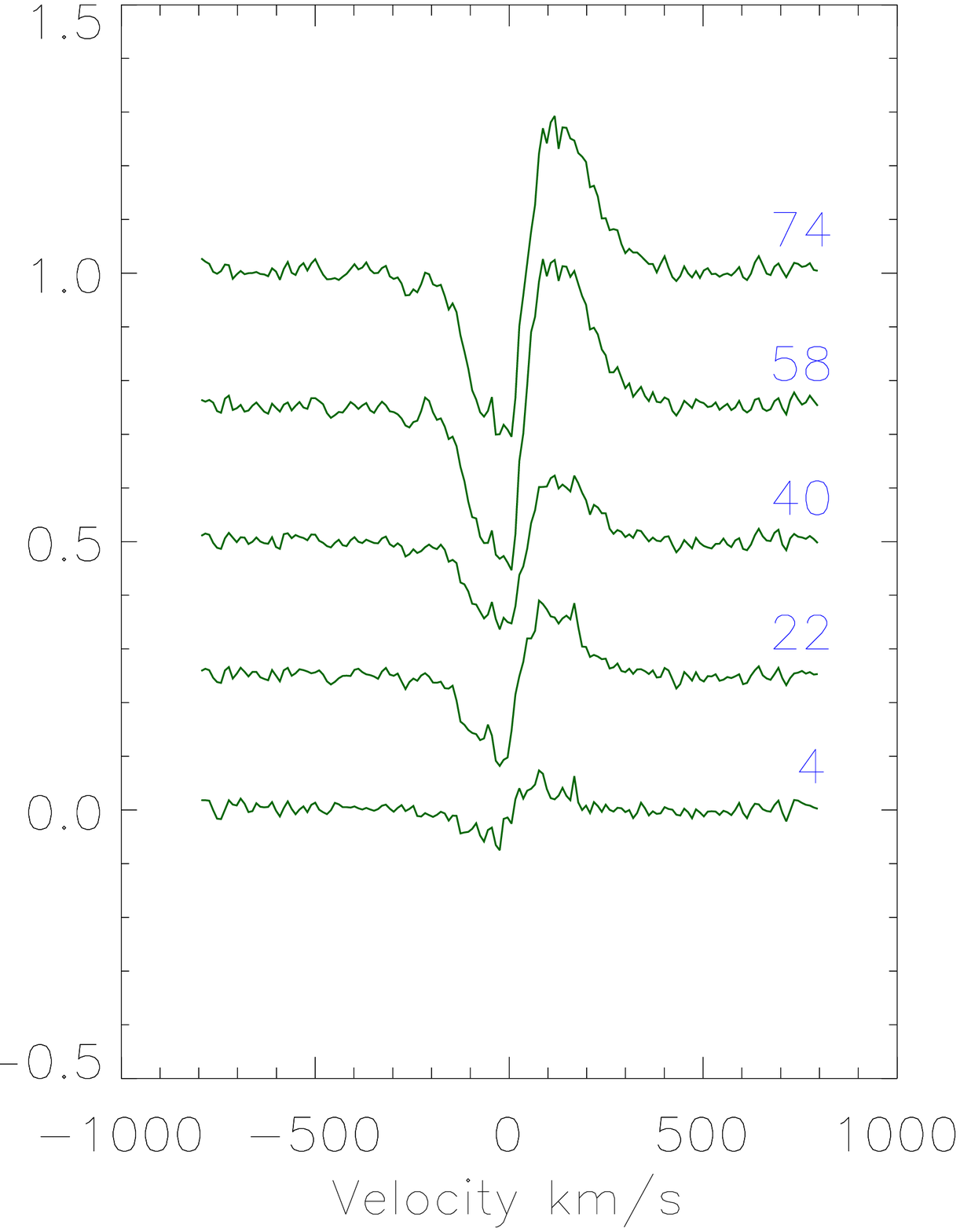} \\
\end{tabular}
\caption{GW Ori 2003 observations.}
\label{fig:GWORI_plots_2003}
\end{figure}

\textbf{Stellar Properties:} GW Ori is a binary system with a period of 242 days \citep{1977MNRAS.181..657M}. GW Ori itself has a period of 3.2 days \citep{1986A&A...165..110B}. Signatures of magnetic activity have been observed through X-ray emission from GW Ori \citep{1981ApJ...243L..89F}.

\noindent \textbf{Disc Properties:} GW Ori shows a very large IR excess \citep{1980MNRAS.191..499C} and the strongest sub-millimeter emission found for a T Tauri star \citep{1995AJ....109.2655M} originating in a circumbinary disc of mass 1.5\,M$_{\odot}$. The inclination of the disc is 15\,-\,27 $^{\circ}$ \citep{1989A&A...211...99B,1991AJ....101.2184M}. The disc contains a gap at $\sim$ 0.17 \,-\,3.3 AU \citep{1991AJ....101.2184M} in which the secondary lies.

\noindent \textbf{ISIS H$\alpha$ Observations:} The observations show a roughly symmetric H$\alpha$ emission profile, that shows small variations in both wings. In 2001 an asymmetry is seen in the wings, with the red wing showing slightly stronger emission (Fig.\,\ref{fig:GWORI_plots_2001}). Across the observation block this wing grows in strength compared to the blue wing. 

In 2003, the two emission wings are more symmetric, and the line is stronger (Fig.\,\ref{fig:GWORI_plots_2003}). The changes in the emission take a similar form as in 2001. Low amplitude waves are seen in the 2003 time series of H$\alpha$ EW measurements.

\noindent \textbf{Previous H$\alpha$ Observations and Variations:} Photometric monitoring for flares was carried out over one night for this object, but no flare activity was found \citep{1981ApJ...244..520W}. A very similar H$\alpha$ profile was observed in 1977 with an EW of 45\,\AA~\citep{1979ApJS...41..369S}.

\subsection{AB Aur}
 \textbf{Stellar Properties:} AB Aur is the brightest Herbig star in the sky, and is a very well studied object. It is found to have a stellar mass of 2.5\,M$_{\odot}$ and a stellar radius of 2.5\,R$_{\odot}$ \citep{1993A&AS..101..629B,1986ApJ...303..311P}. \citet{1999A&A...345..884C} found a period of 34 hours from the monitoring of photospheric lines. X-ray emission has been observed from AB Aur, however the origin of it is not clear but it is possibly from a magnetic corona \citep{2007A&A...468..541T}

\noindent \textbf{Disc Properties:} Emission from the disc has been observed to extend out to 580\,AU, with a double spiral structure within the disk at radii of 200\,-\,450\,AU \citep{2004ApJ...605L..53F}. A large 100\,AU inner hole in the dust continuum has also been observed in the sub-millimeter emission \citep{2005ASPC..344..168O}. There has been many contradictory reports of inclination angles, which has lead to the belief that the disc is warped. The inclination angle has been reported to be 76$^{\circ}$ \citep{1997ApJ...490..792M}, \textless 45$^{\circ}$ \citep{1999ApJ...523L.151G}, 27$^{\circ}$\,-\,35$^{\circ}$ \citep{2003ApJ...588..360E} which was later revised to 8$^{\circ}$\,-\,16$^{\circ}$ \citep{2004ApJ...613.1049E}. More recently \citet{2004ApJ...605L..53F} found an inclination angle of 30$^{\circ}$ $\pm$ 5 $^{\circ}$ and large spiral structures within the disc. Modelling of the outer mm emission from the disc suggests a inclination of \textless 30$^{\circ}$ \citep{2001A&A...371..186N,2005ApJ...622L.133C}.

\begin{figure*}
\begin{tabular}{cccc}
\includegraphics[scale=0.22]{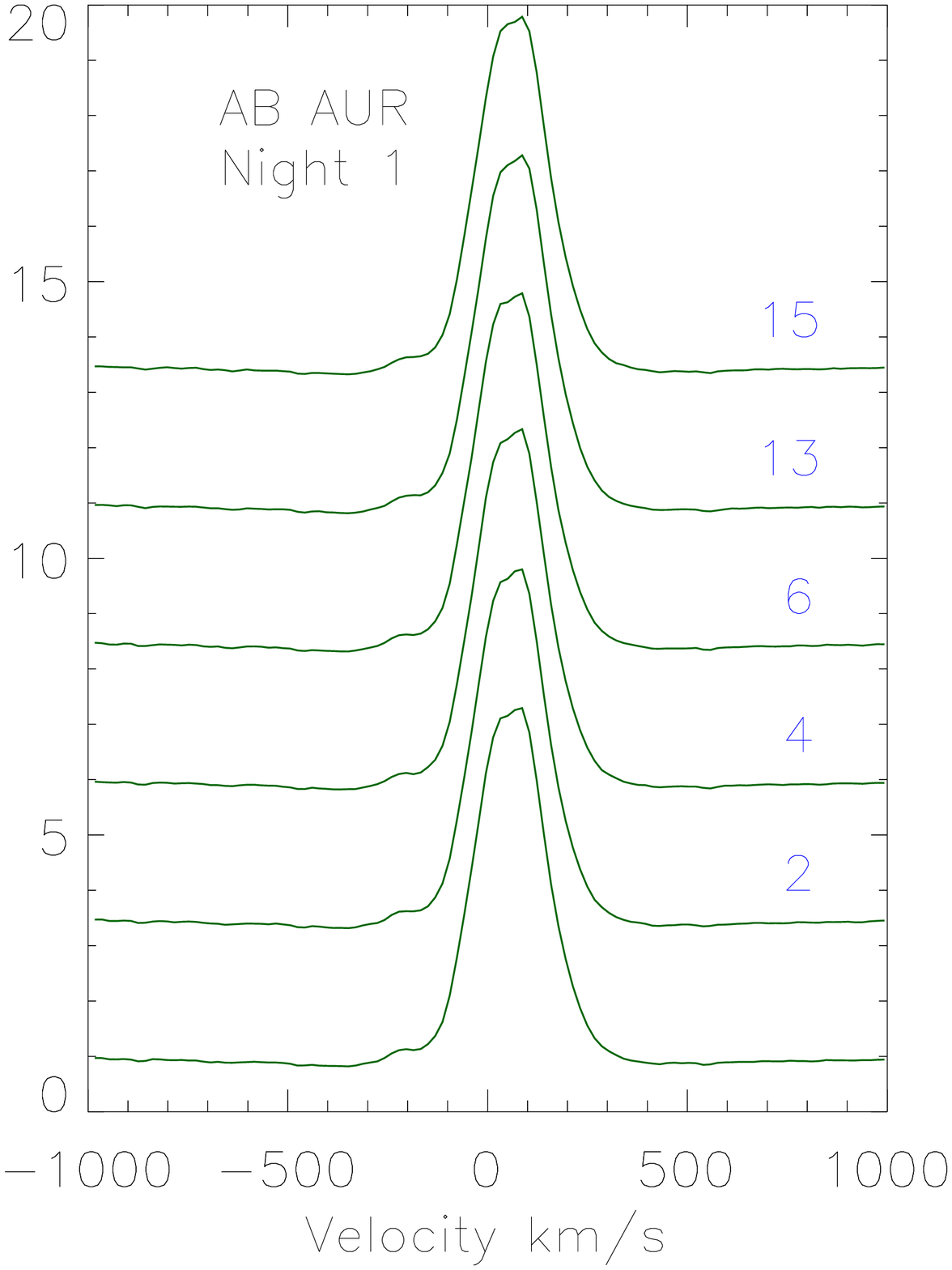} & \includegraphics[scale=0.22]{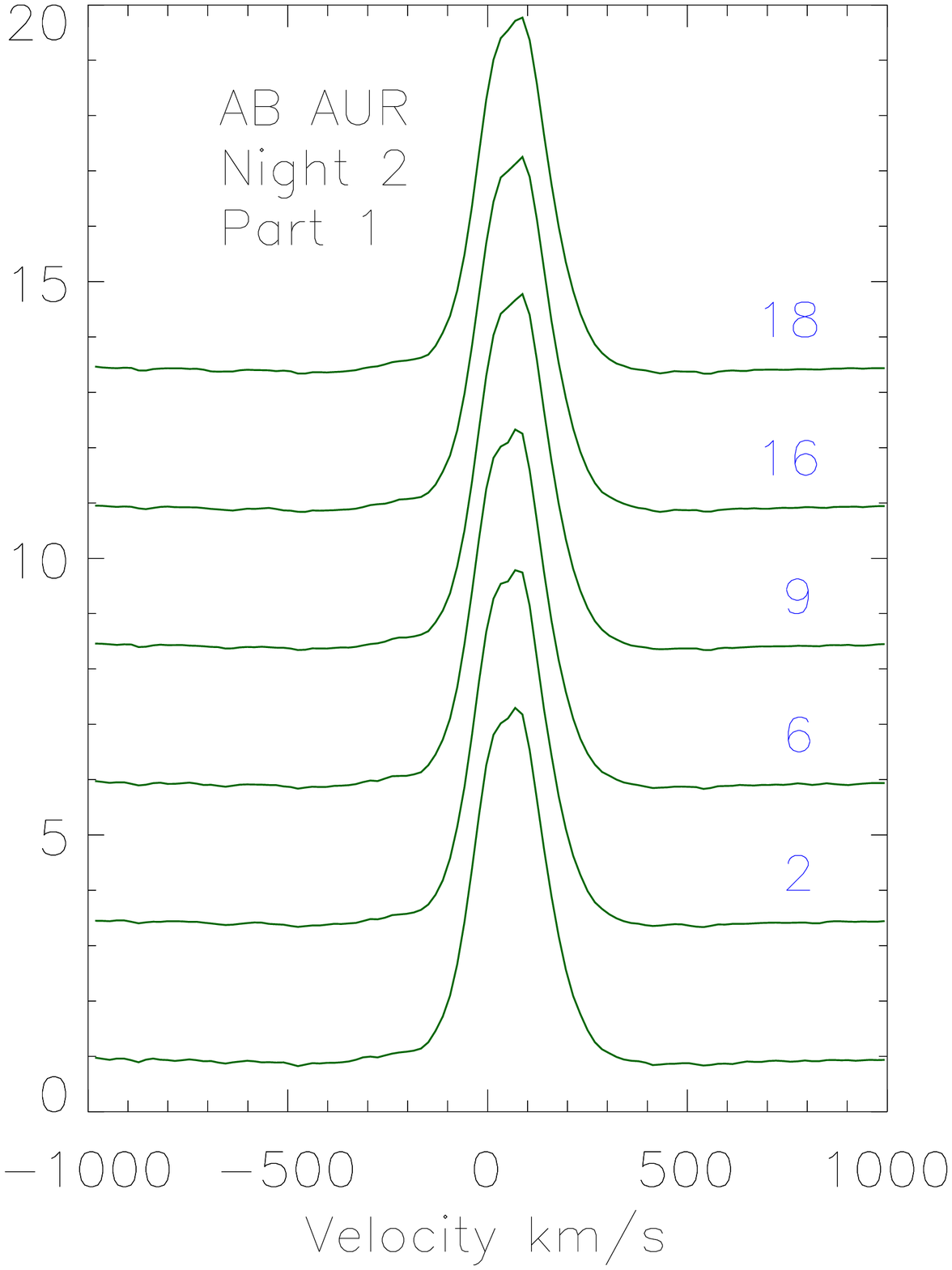} &\includegraphics[scale=0.22]{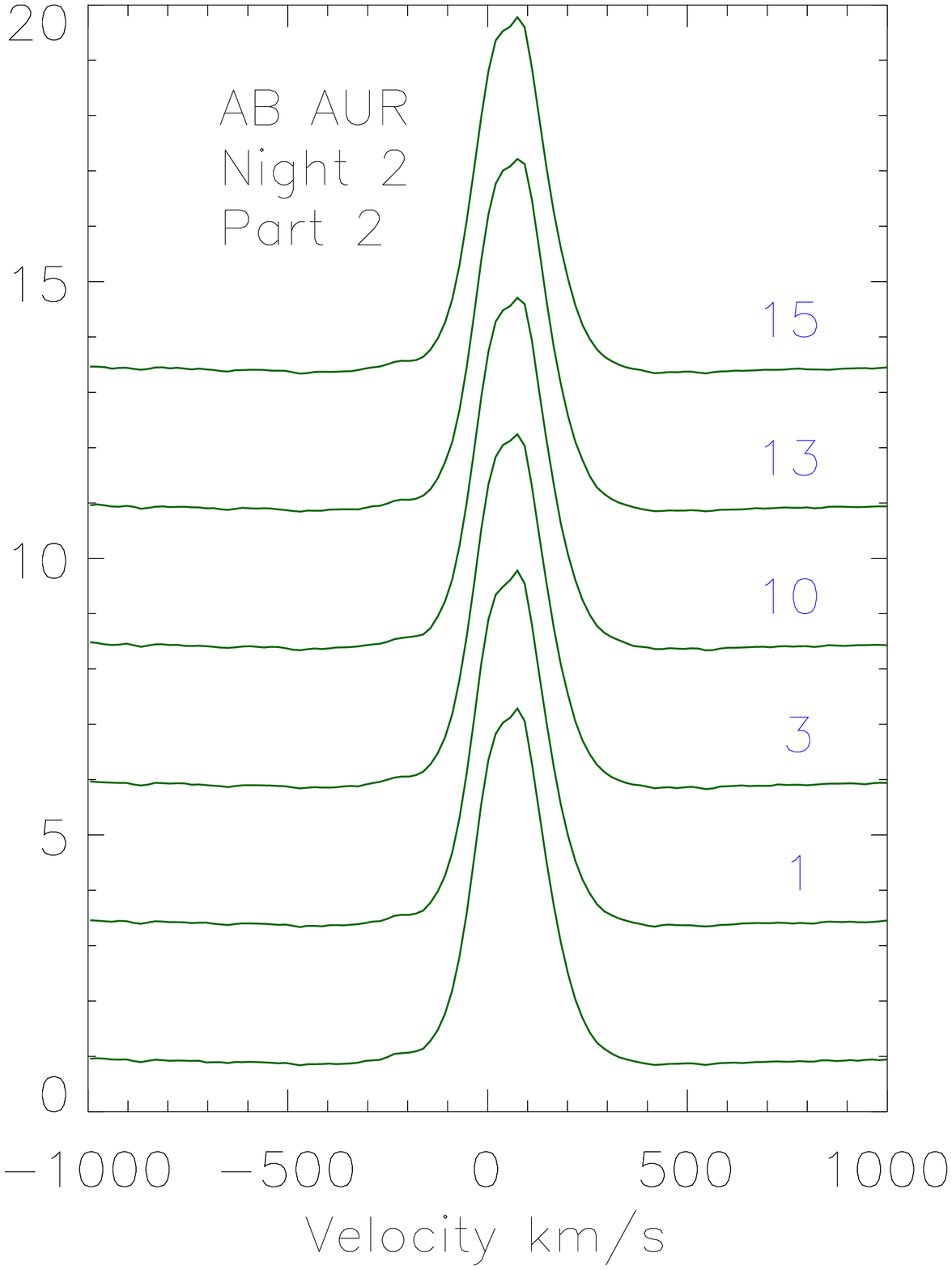} &  \includegraphics[scale=0.22]{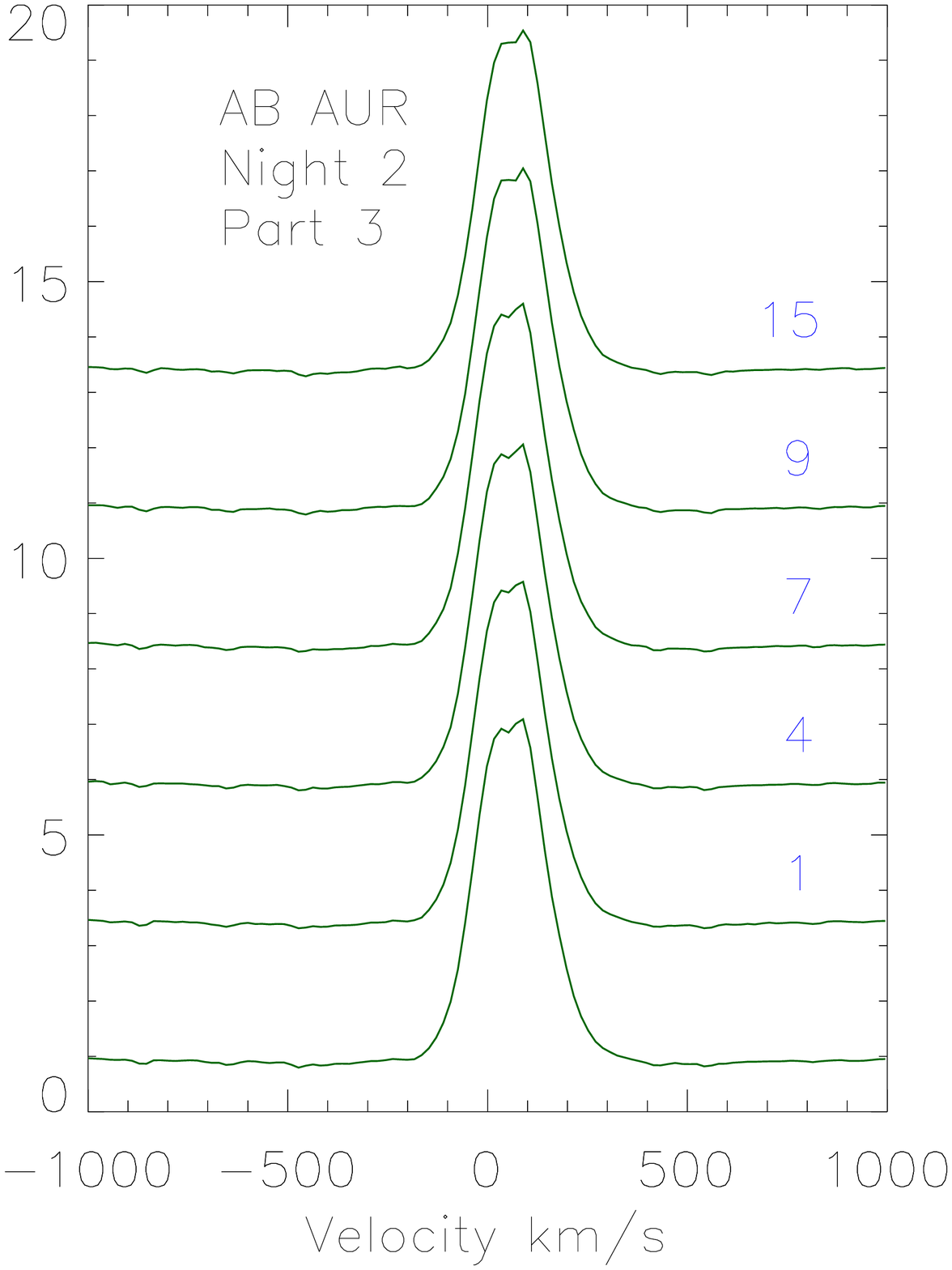}\\
\includegraphics[scale=0.22]{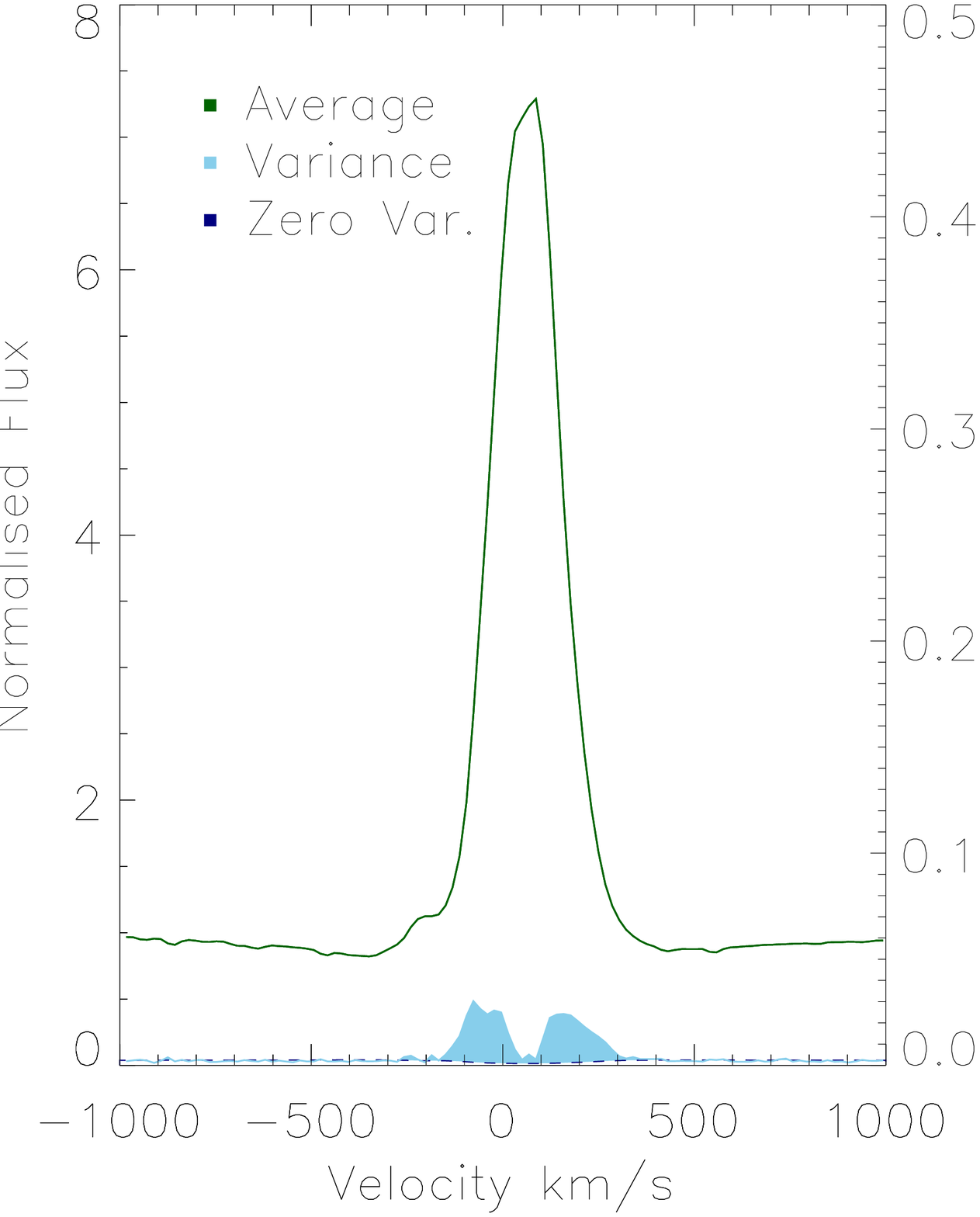} &\includegraphics[scale=0.22]{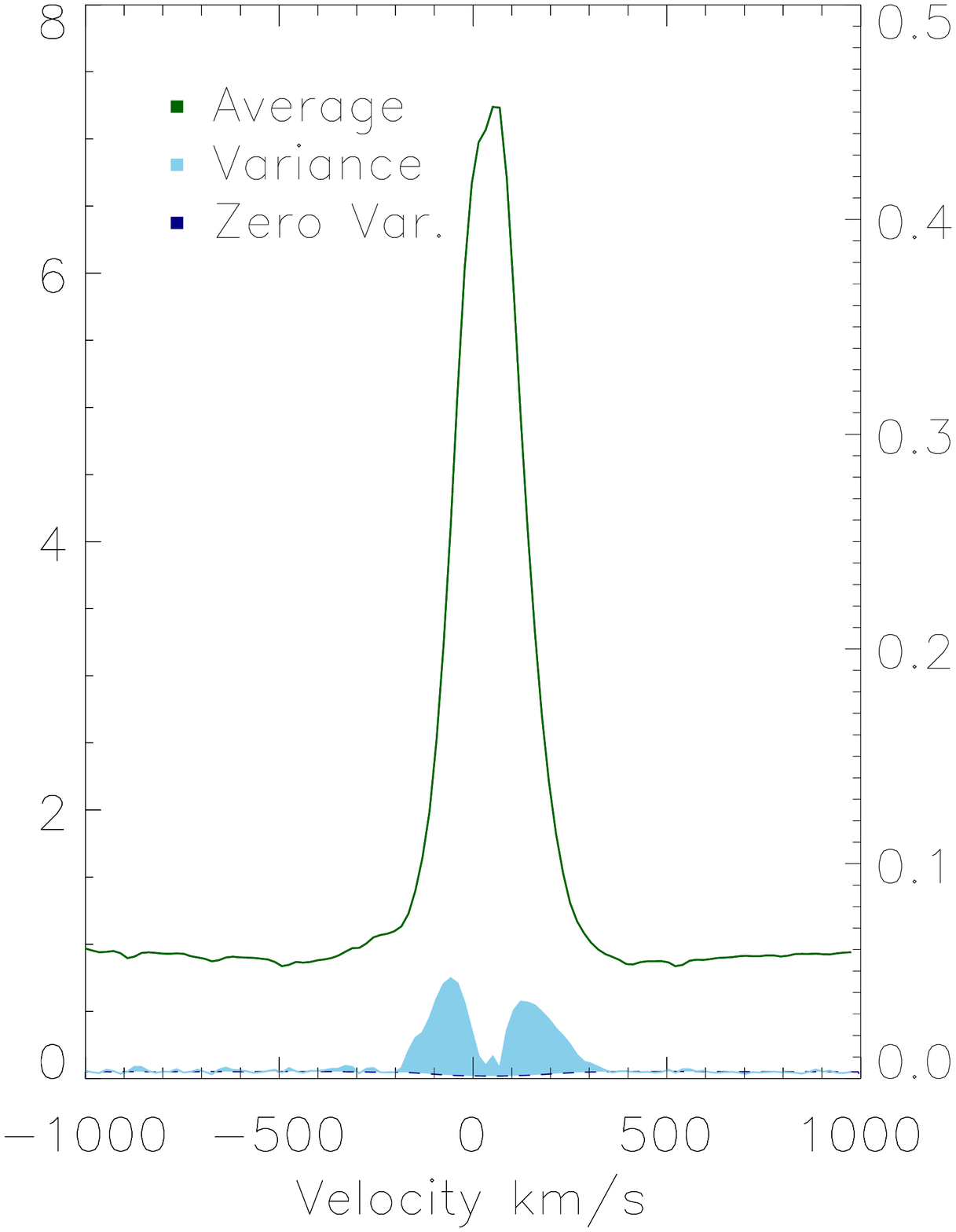} & \includegraphics[scale=0.22]{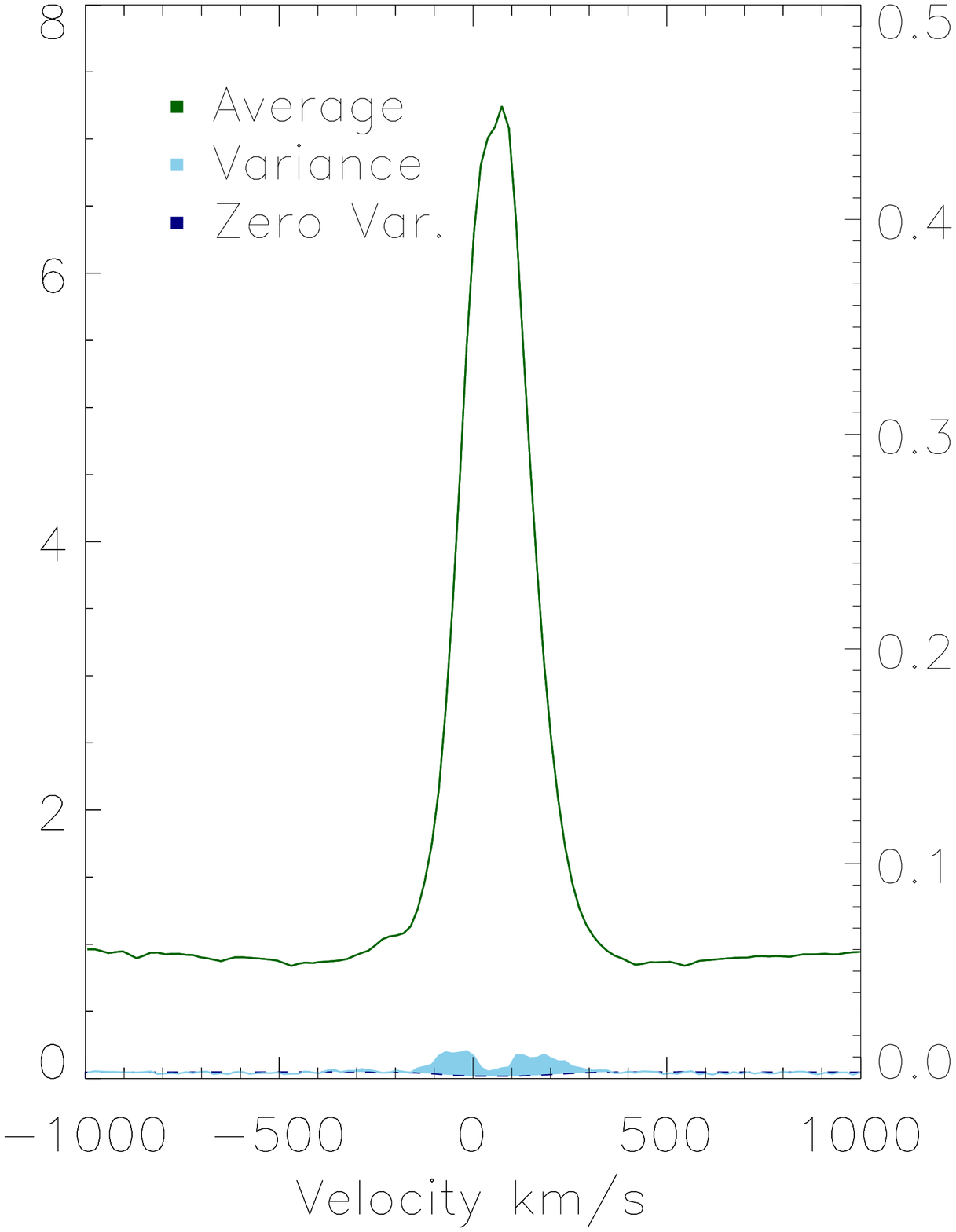} &  \includegraphics[scale=0.22]{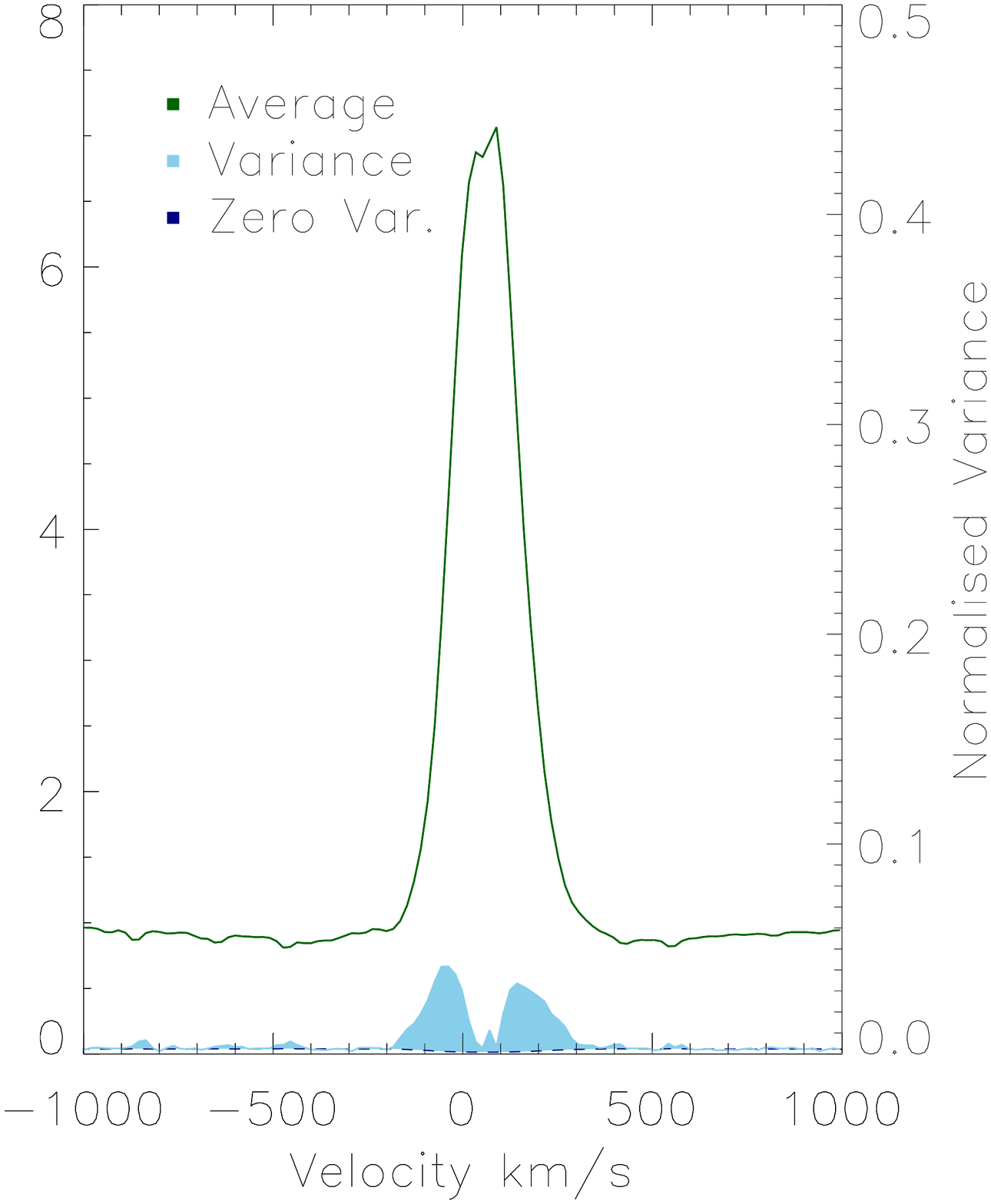}\\
\includegraphics[scale=0.22]{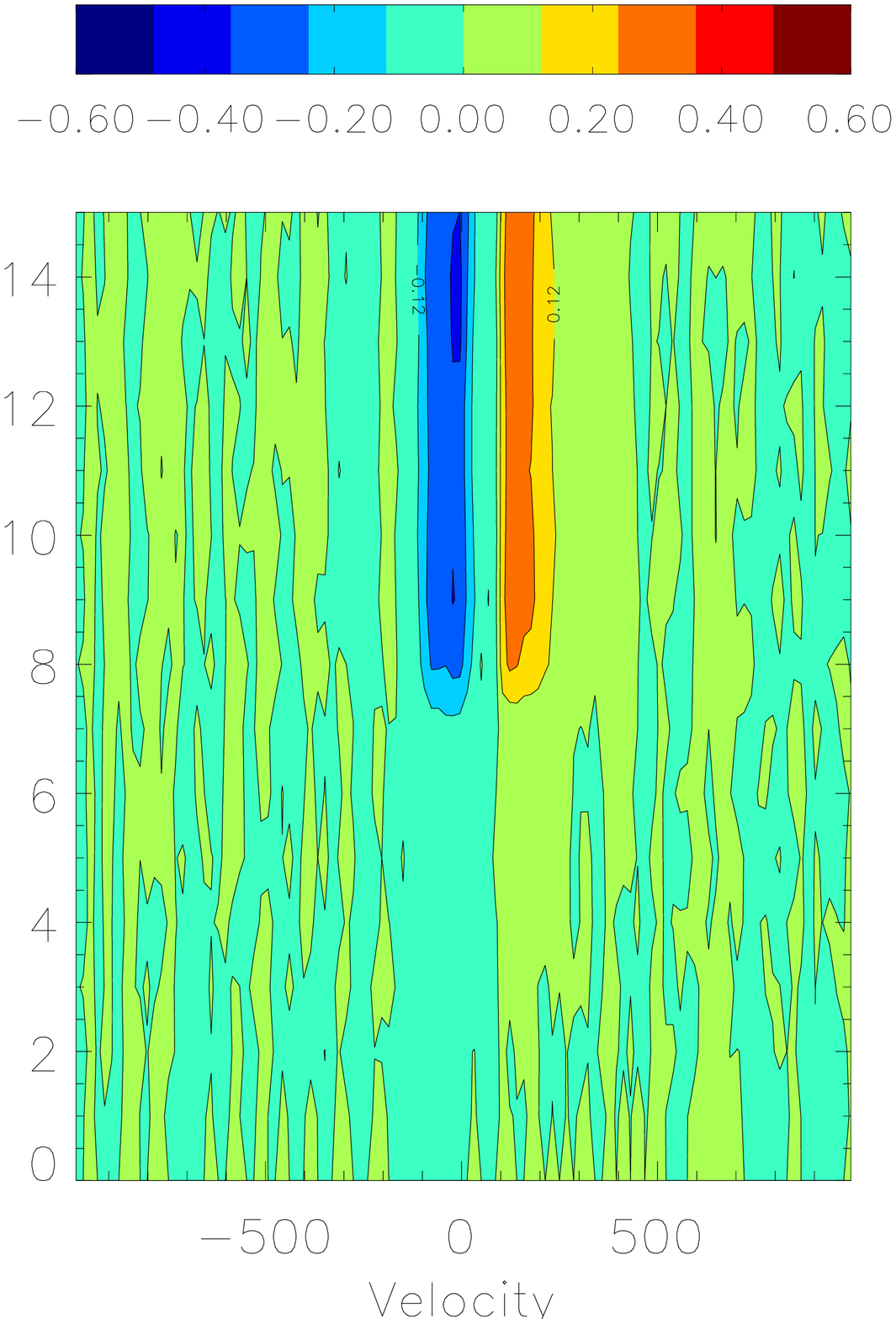} & \includegraphics[scale=0.22]{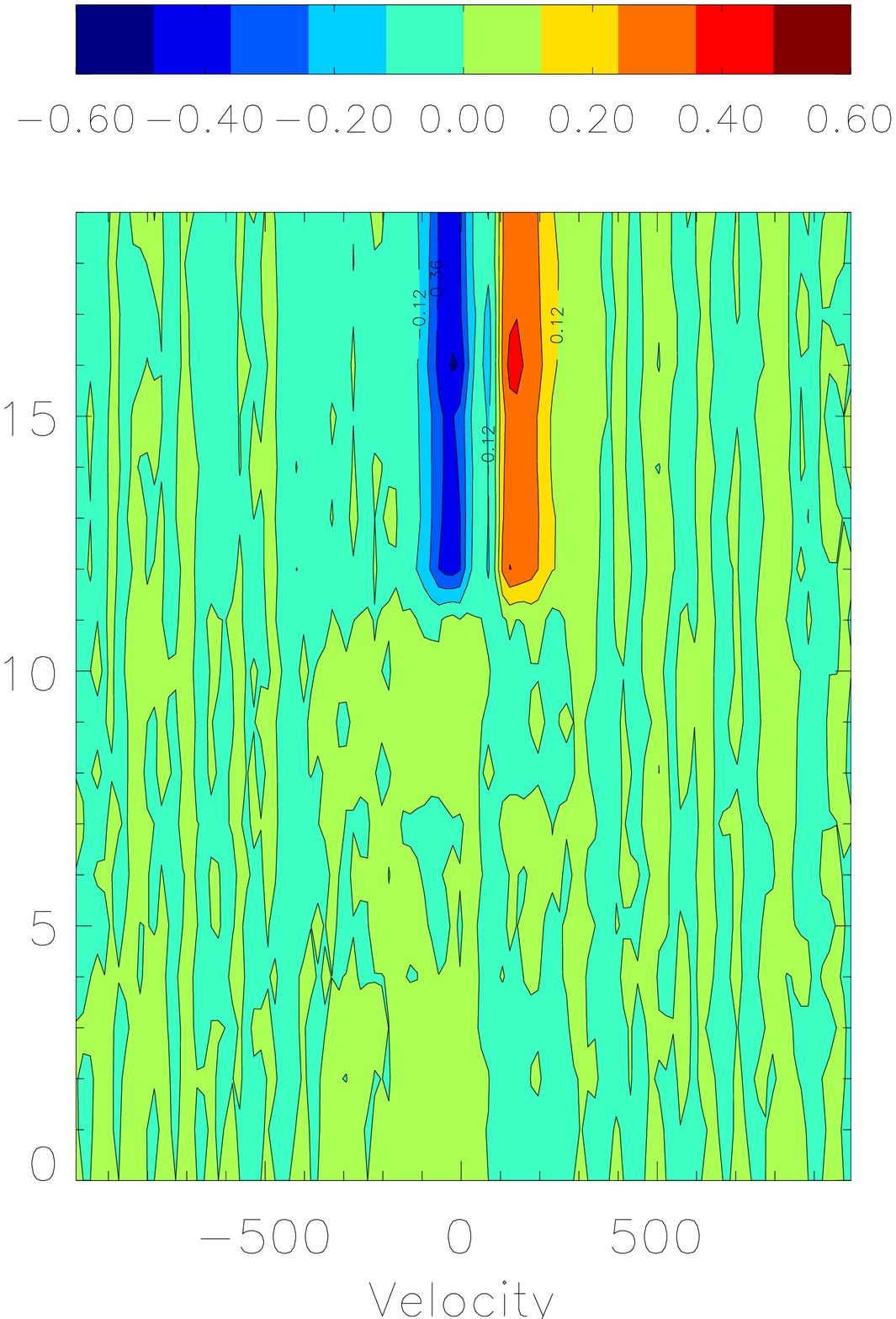} & \includegraphics[scale=0.22]{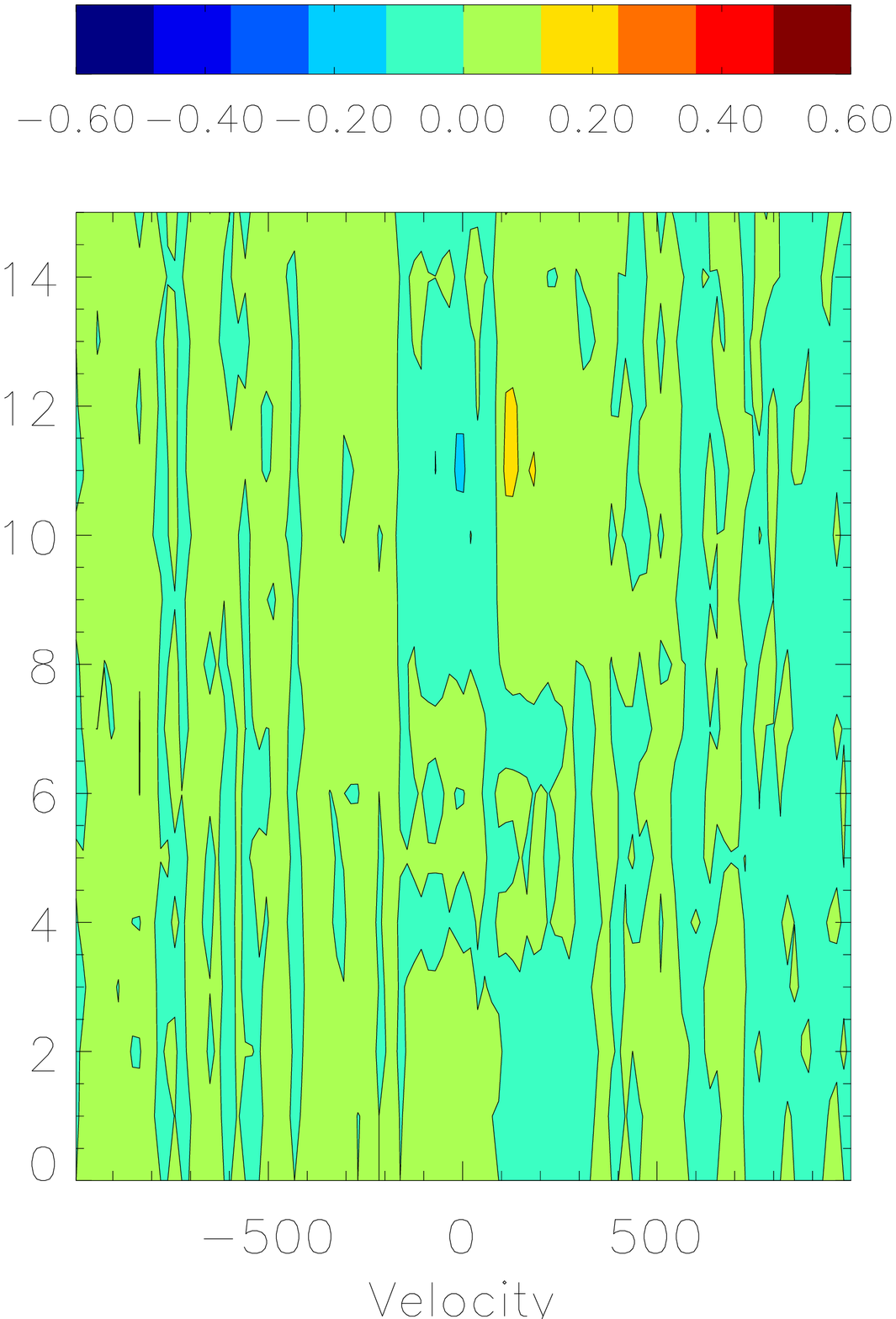} &\includegraphics[scale=0.22]{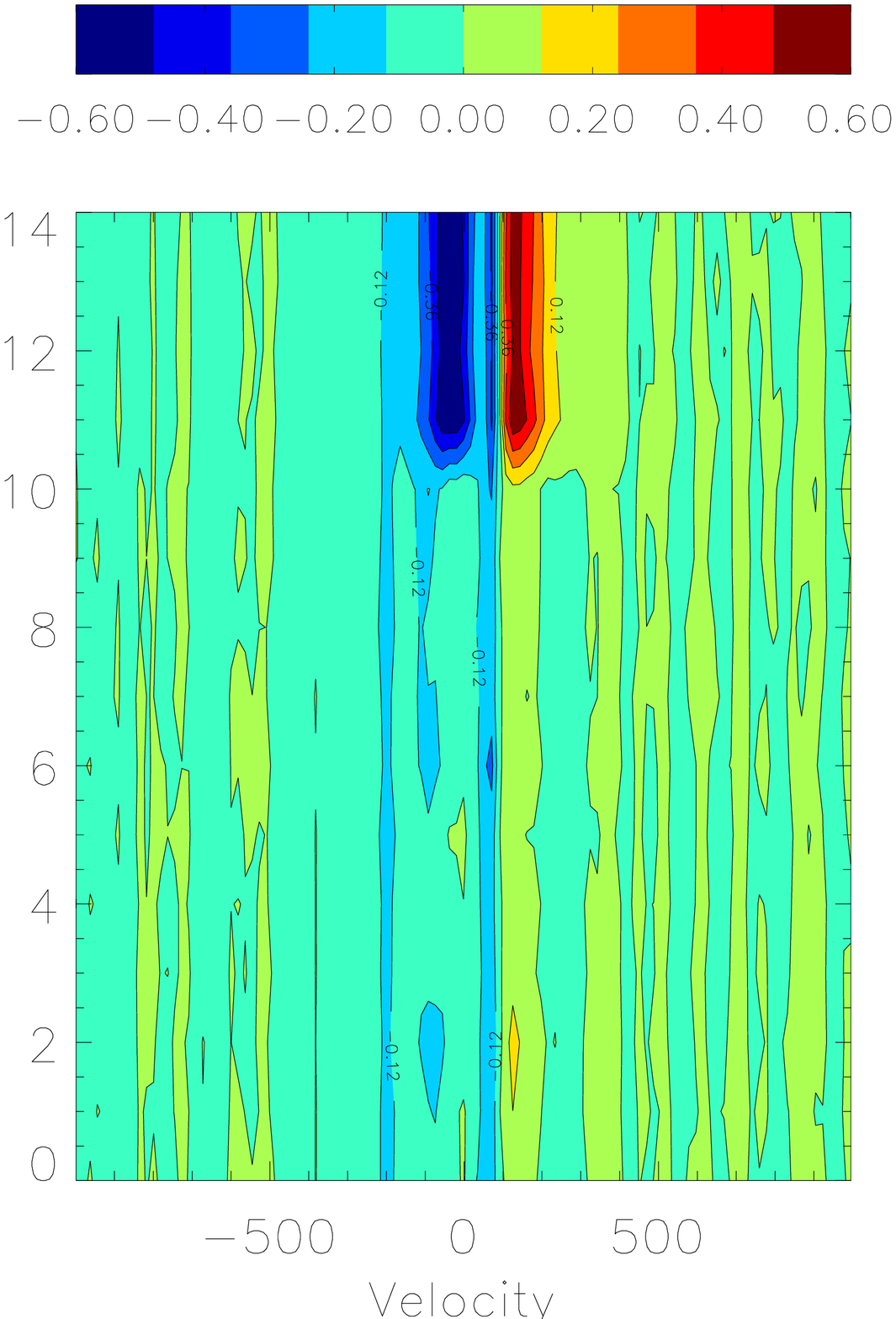} \\
\includegraphics[scale=0.22]{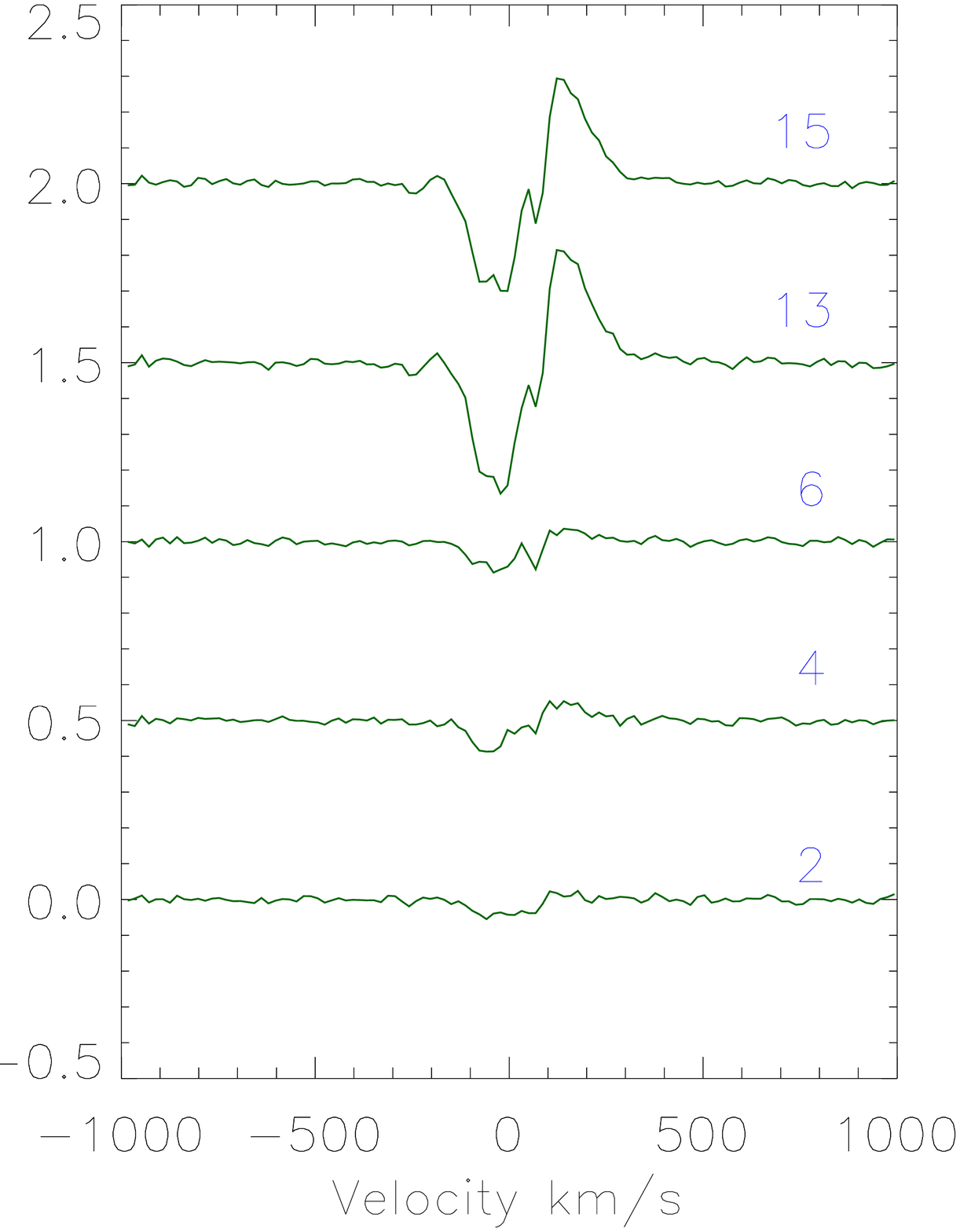} & \includegraphics[scale=0.22]{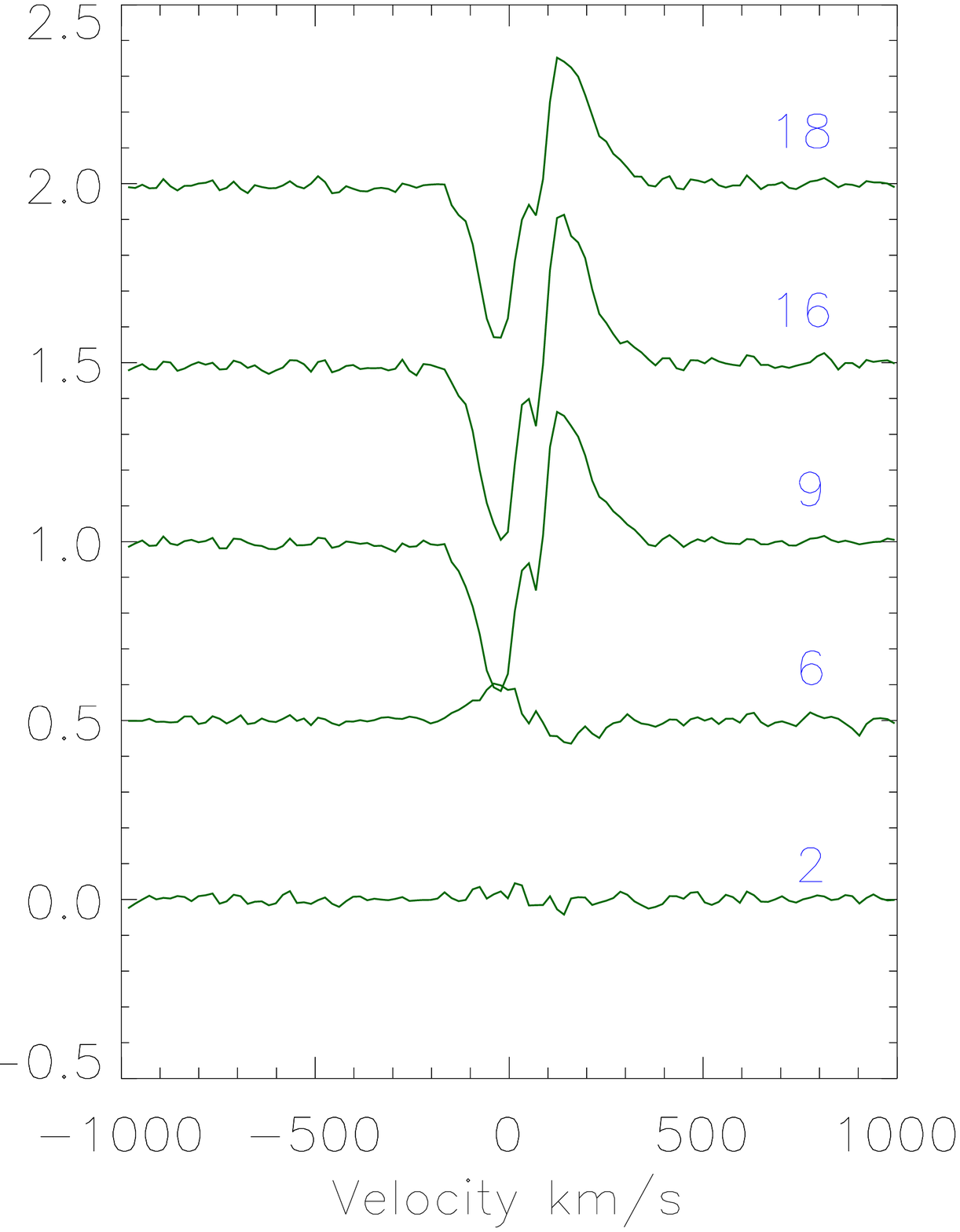} & \includegraphics[scale=0.22]{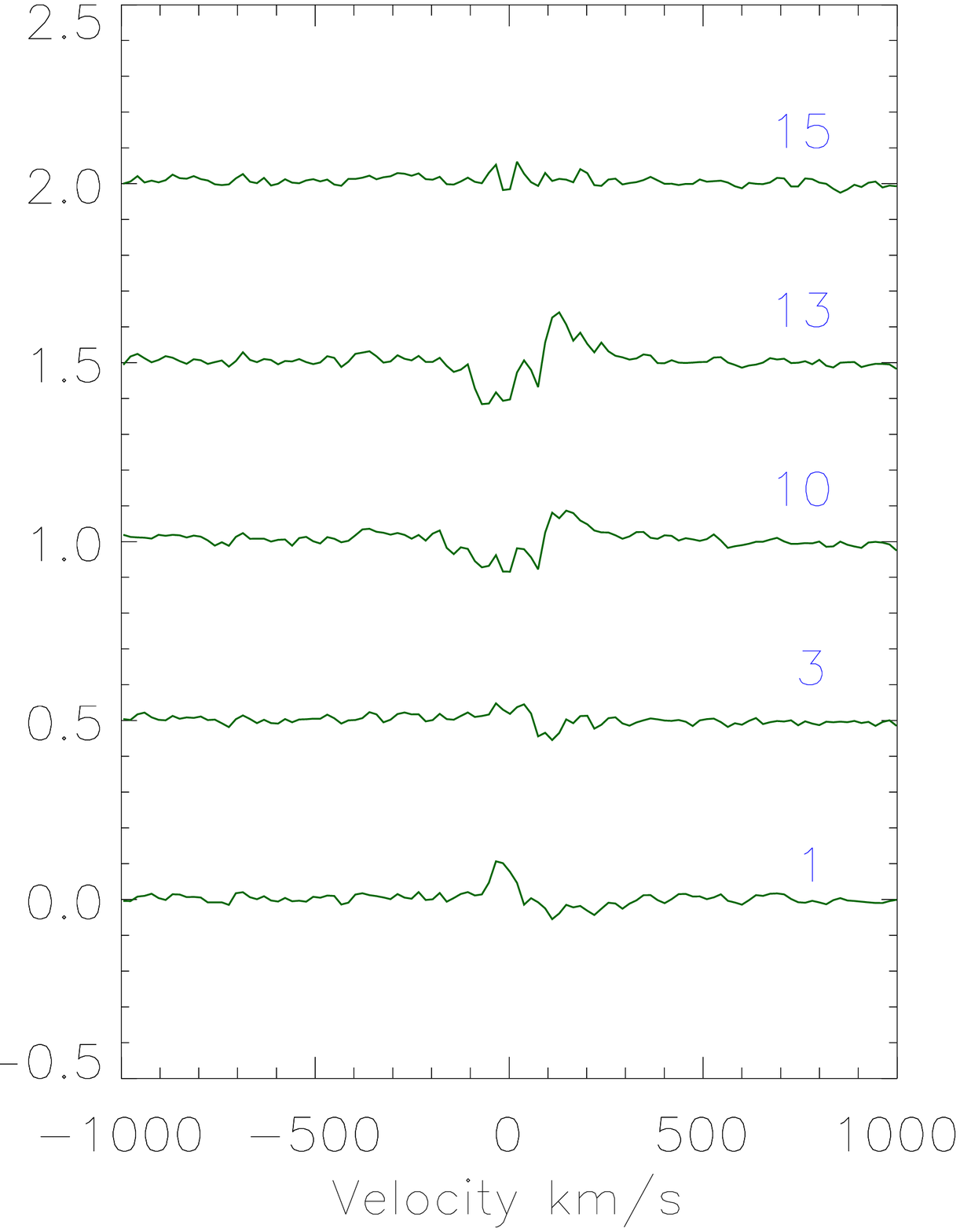} & \includegraphics[scale=0.22]{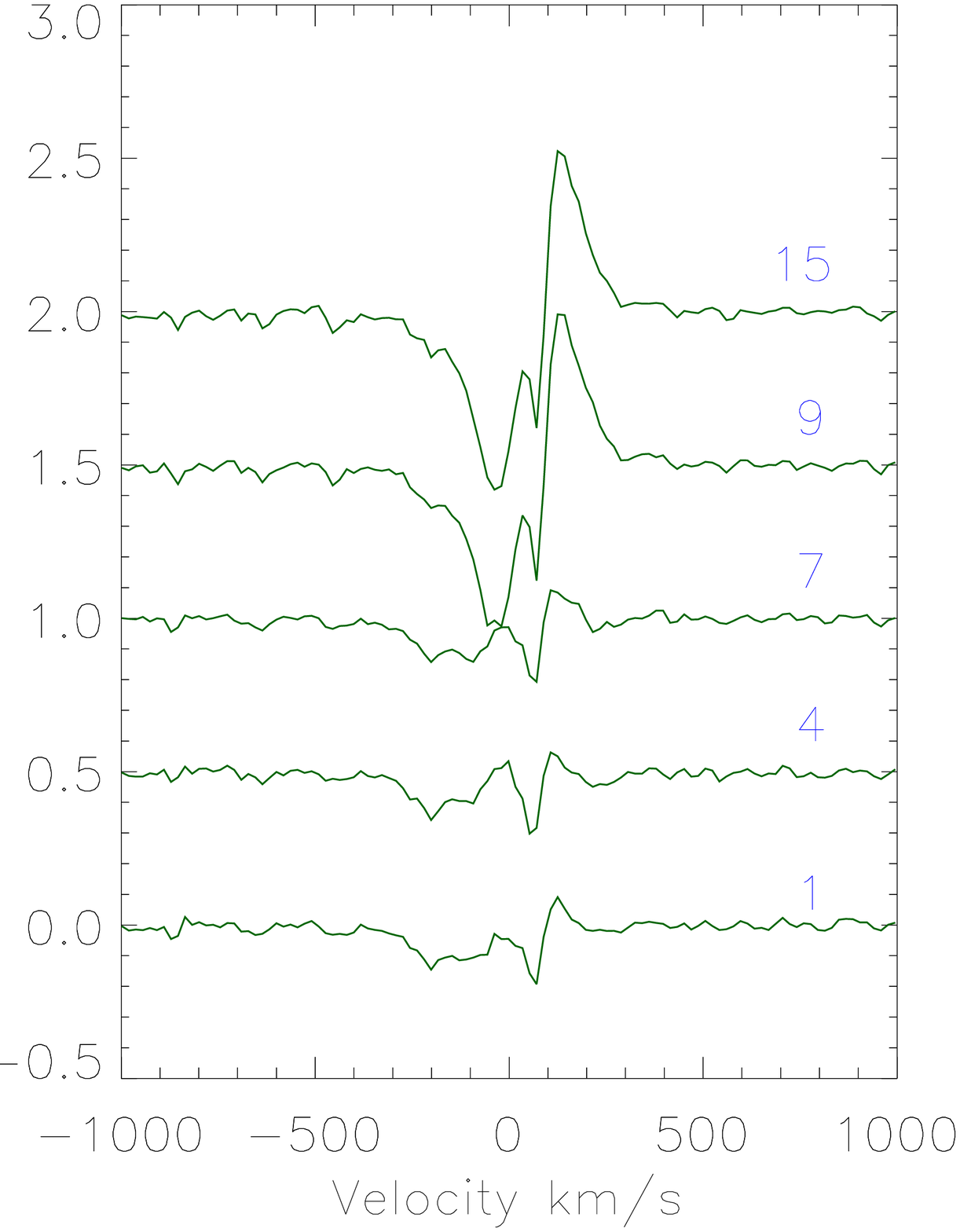}\\
\end{tabular}
\caption{AB Aur 2001 observations. Night 1 (left). First block of observations Night 2 (middle left). Second Block of observations Night 2 (middle right).Third block of observations Night 2 (right).}
\label{fig:ABAUR_profiles_2001_1}
\end{figure*}

\noindent \textbf{ISIS H$\alpha$ Observations:} Across the four nights of observations in 2003, AB Aur shows a strong P-Cygni emission profile. (see Fig.\,\ref{fig:ABAUR_profiles_2003_1}). Studying the variance profiles for the first and second night of observations, the changes in the emission line are concentrated in the wings of the profile. On the first night the variations are entirely in the blue wing, and on the second night both wings show variations. No significant variations occur during the fourth night. Within the blue-shifted absorption, a small emission feature can be seen. This changes position between each nights observation. From the first to the second night, the shift is to the blue side, and from the second to the third night it shifts further back to the red side. It cannot be distinguished from the wing or continuum on the fourth night. 

On the third night of 2003 large distinctive variations in the H$\alpha$ emission are seen (Fig.\,\ref{fig:ABAUR_profiles_2003_1}). The profile decreases in strength and expands a number of times during the observation block. When it increases again it is narrower. These variations manifest themselves as changes in three distinct regions of variations in the variance profiles. This results in not very large changes in the H$\alpha$ EW of $\sim$ 220\,-\,234\,\AA. However during this night there are much more significant variations in the 10\%w, $\sim$ 345\,-\,549\,km\,s$^{-1}$, which is an change of of 60\% in 10\%w. This bi-modal behaviour can clearly be seen in the surface plot for AB Aur in Fig.\,\ref{fig:ABAUR_profiles_2003_1}. This is unusual behaviour within this sample: no other object shows such strong, distinct peaks in the variance profile. 

During the 2001 observations, AB Aur, does not show the same P-Cygni profile, rather a small emission peak in the blue wing can be seen (Fig\,\ref{fig:ABAUR_profiles_2001_1}). The emission line does not show any drastic changes as it later does in 2003, the changes are small in 2001, and take place in two velocity ranges one centred at $\pm$ 200\,km\,s$^{-1}$. 

\noindent \textbf{Previous H$\alpha$ Observations:} As mentioned in the main text in Chapt.\,4, Sect.\,\ref{sec:rapid_events}, \citet{1995A&A...298..585B} have previously observed striking night-to-night variations in H$\alpha$ profile both in intensity and profile shape. The time sampling of these variations were on longer time-scales than those in the ISIS sample (hours rather than minutes). The authors argue that these P-Cygni variations are connected with the motion of circumstellar inhomogeneities. Large changes in the H$\alpha$ emission were again recorded between 1987 and 1996, where the line changes more than 100\% which is attributed to changes in the wind structure \citep{1998A&A...340..163B}.

\subsection{BP Tau}

\begin{figure}
\centering
\begin{tabular}{cc}
\includegraphics[scale=0.22]{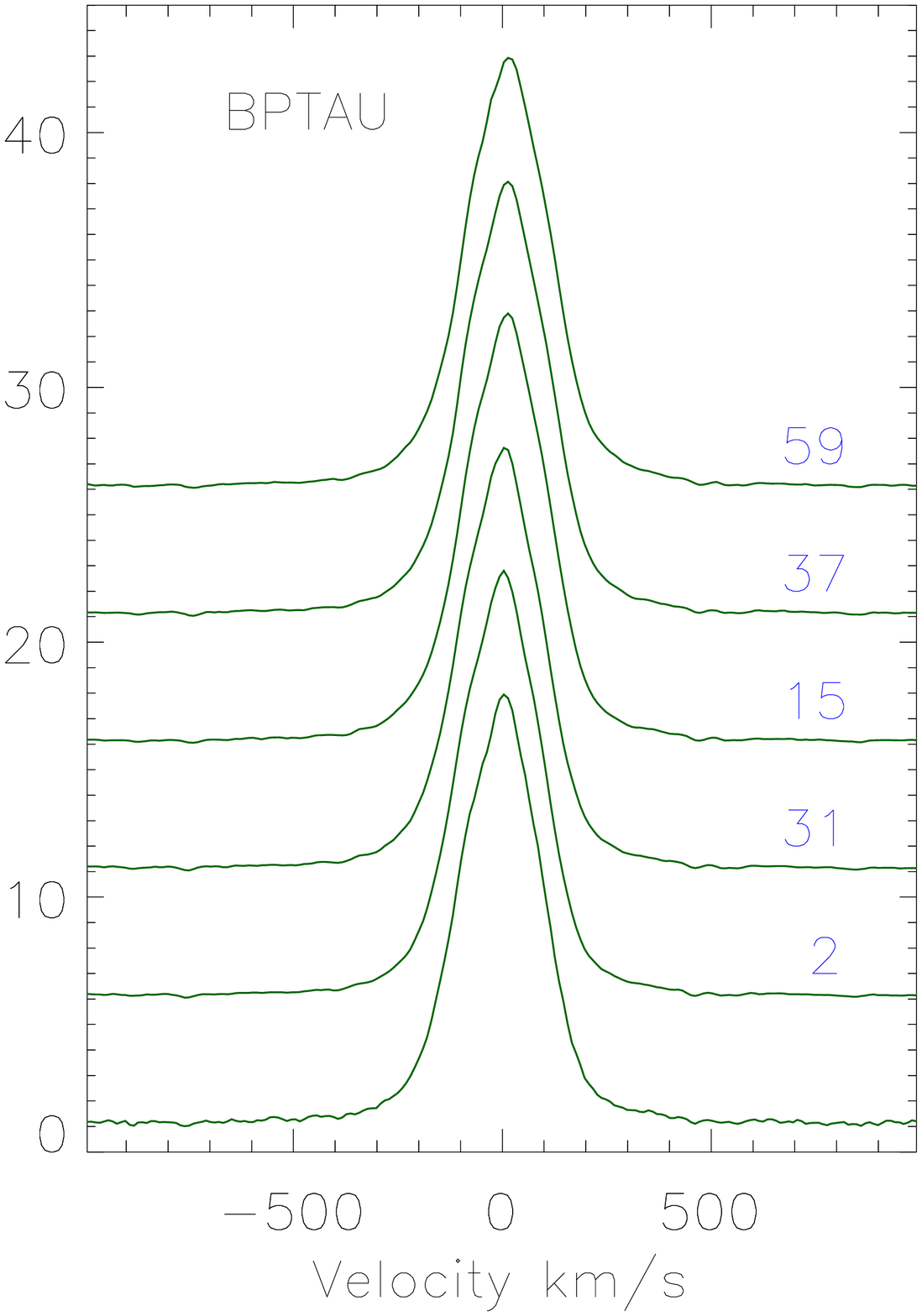} &\includegraphics[scale=0.22]{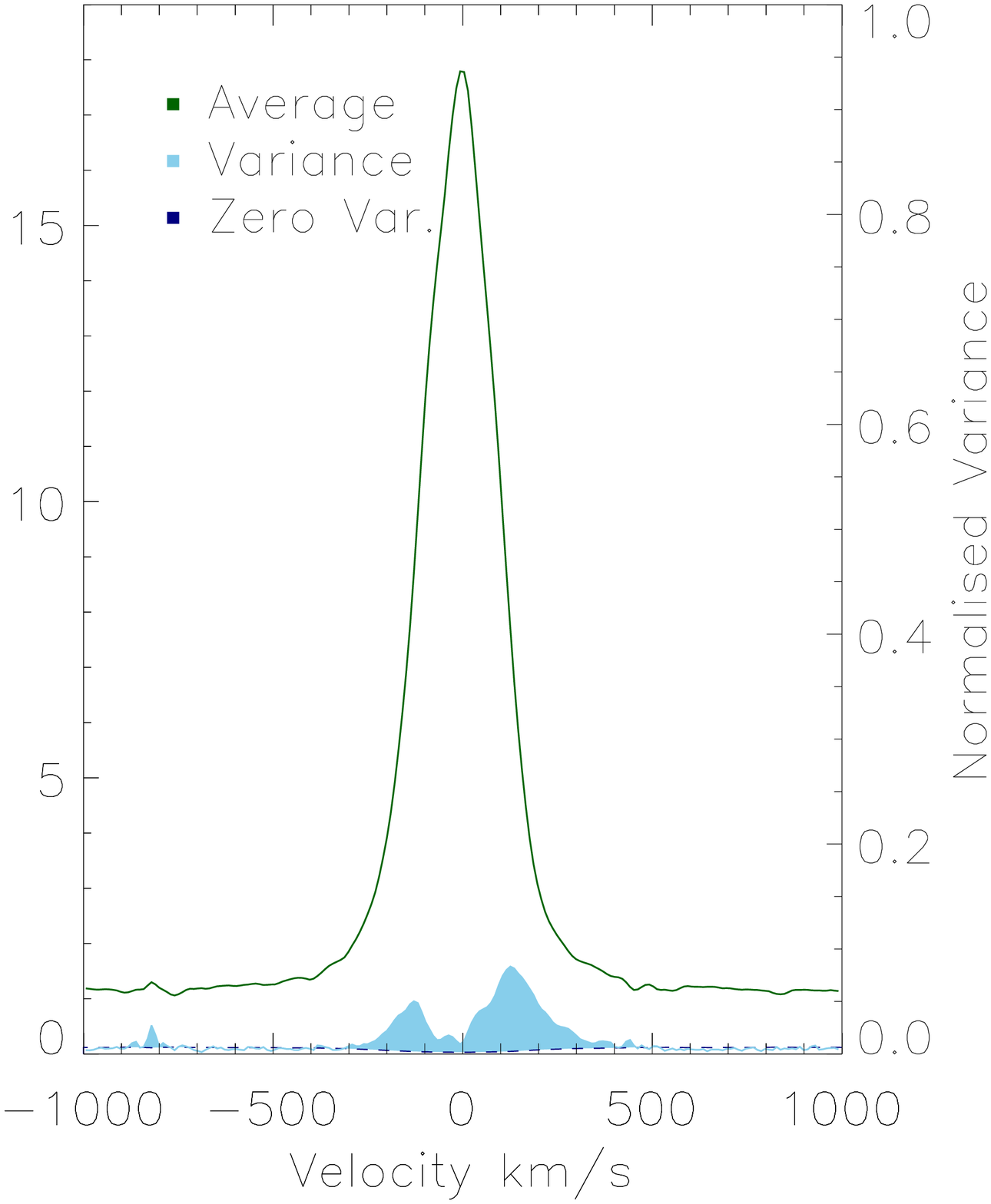}  \\
\includegraphics[scale=0.22]{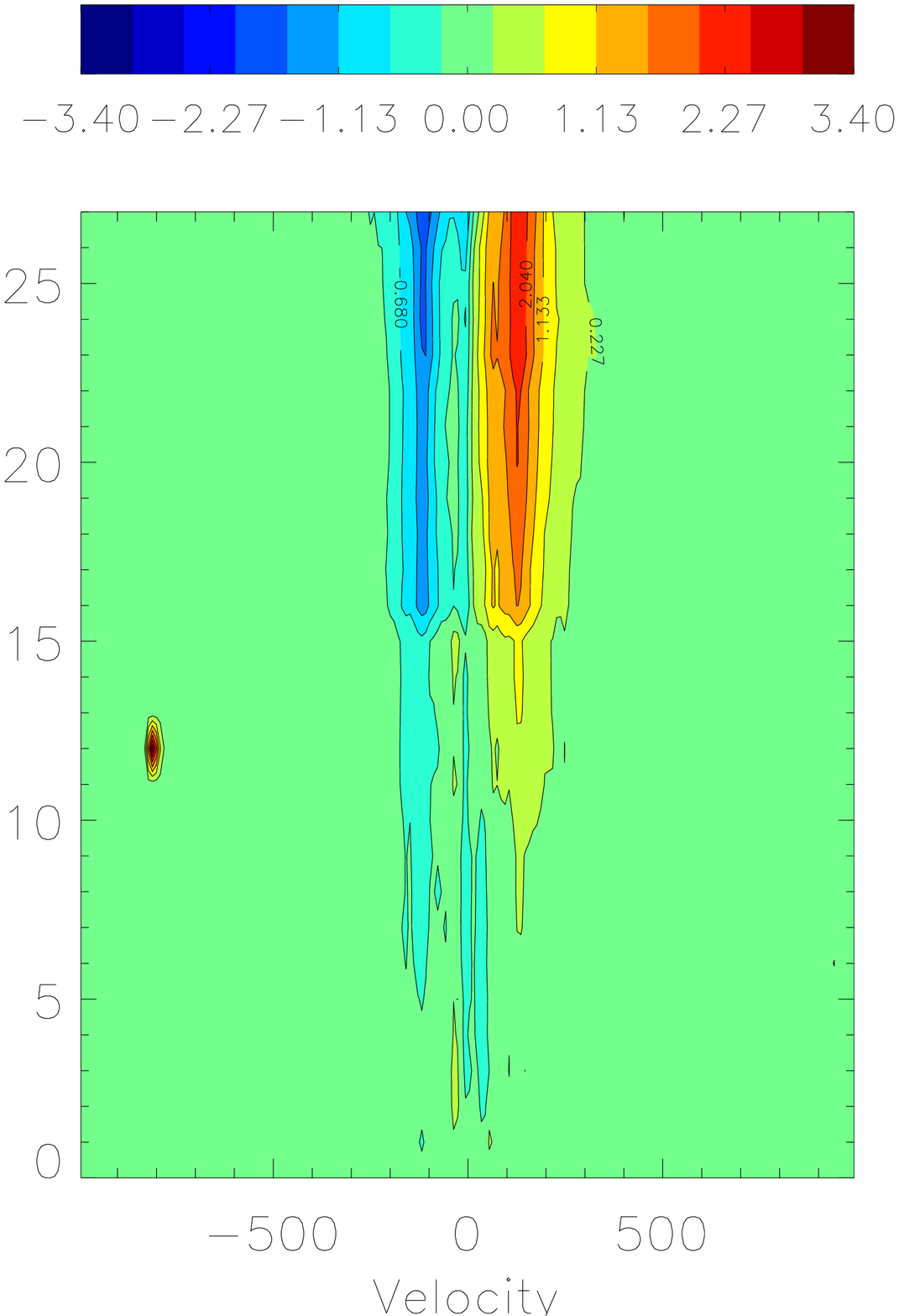}& \includegraphics[scale=0.22]{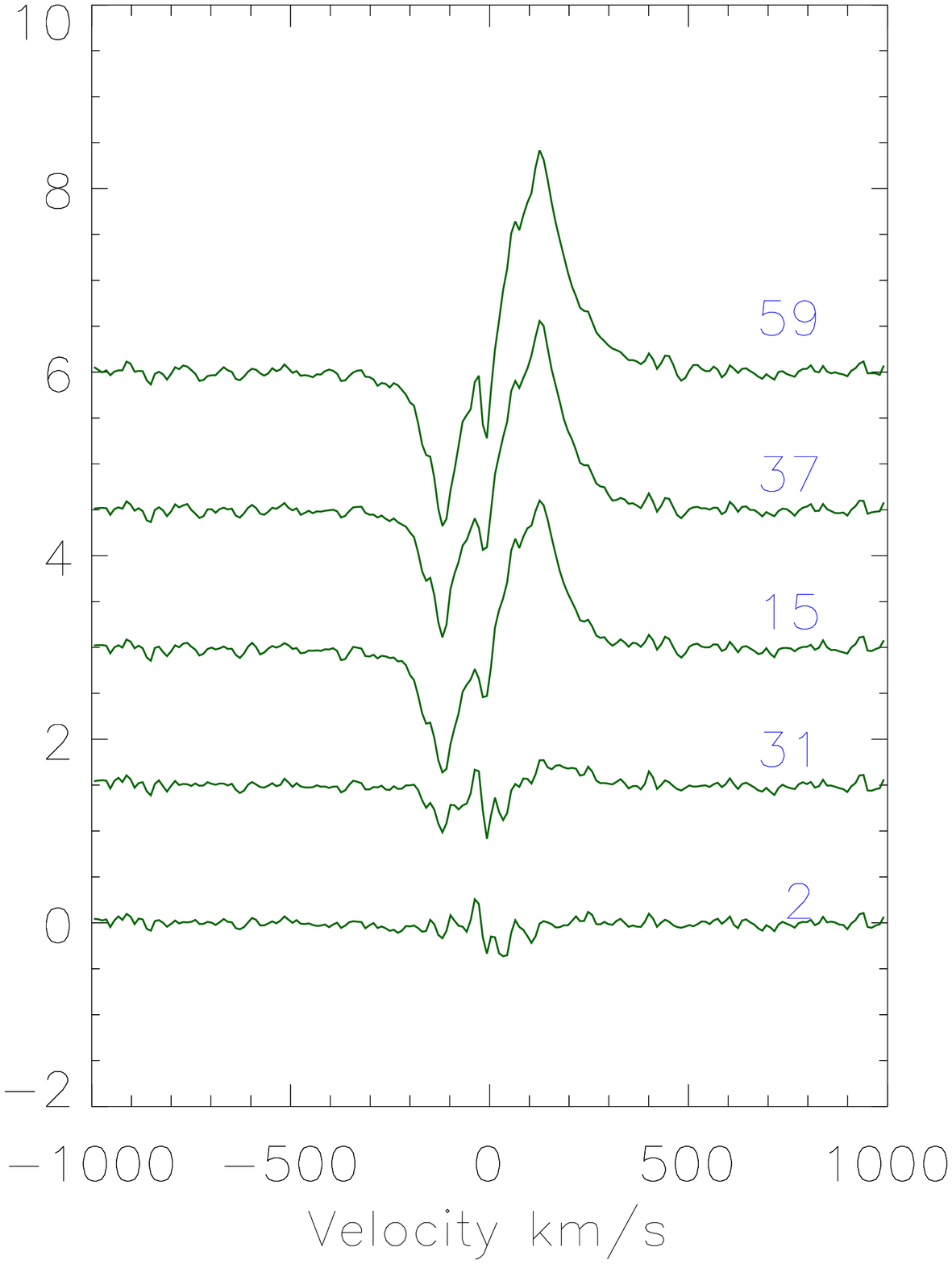}\\
\end{tabular}
\caption{ BP Tau 2003 observations.}
\label{fig:BPTAU_plots}
\end{figure}
\textbf{Stellar Properties:} Observations of BP TAU have revealed that it contains strong magnetic field, both a dipole (1.2 kG) and an octupole (1.6 kG) component, and both are slightly shifted with respect to the rotation axis (20$\degree$ and 10$\degree$, \citealt{2011MNRAS.413.1061L}). The accretion spots coincide with the two main high-latitude octupole poles and overlap with dark photospheric spots which each cover about 2\% of the surface.
X-ray emission has also been observed from BP Tau \citep{1995A&A...297..391N}.

\noindent \textbf{Disc Properties:} BP Tau has a reported inclination of $\sim$ 30$^{\circ}$ \citep{2003A&A...402.1003D,2000ApJ...545.1034S}. The disc is small with an outer radius of 120 AU, and there are suggestions that BP Tau is in the process of clearing its disc \citep{2003A&A...402.1003D}.

\noindent \textbf{Accretion:} An accretion rate of 3\,x\,10$^{-8}$M$_{\odot}$yr$^{-1}$ has been estimated from the UV excess \citep{1998ApJ...492..323G}.

\noindent \textbf{ISIS H$\alpha$ Observations:} BP Tau shows a roughly symmetric H$\alpha$ emission profile across the course of the nights observations (Fig.\,\ref{fig:BPTAU_plots}). The variance profile shows two peaks of variations in the blue and red wings, where the changes in the red side are more pronounced. The time series of the H$\alpha$ EW measurements shows what looks like a period across the hour of observation, during which the EW changes by about 58\,\AA. The 10\%w does not show the same kind of period as the EW, increasing slightly by $\sim$ 30\,km\,s$^{-1}$.
For BP Tau, four spectra were also observed from the night before the main observations. There is a large difference in EW (70\AA) between these two nights observations, with an increase in 10\%w over these two nights of $\sim$ 20\,km\,s$^{-1}$. 

\noindent \textbf{Previous H$\alpha$ Observations and Variations:} \citet{2008MNRAS.386.1234D} have suggested that rotational modulation dominates the observed variability in BP Tau reaching about $\pm$20\,-\,25 \% for H$\beta$ and $\pm$10 \% for H$\alpha$.

\citet{1981ApJ...244..520W} attributed observed photometric variations on the time-scales of minutes to flare activity. However a later study that carried out simultaneous observations of the optical and X-ray emission, found no correlation between the changes in the two \citep{1997A&A...324..155G}. This led the authors to conclude that there was no flare activity in BP Tau over the course of the observations, and suggested that the interaction between stellar field and circumstellar disc could act to quench any flare activity. They also attributed the variations observed in the optical to changes in the accretion rate \citep{1996A&A...307..791G}. 

UV continuum flares have been observed in BP Tau, on time-scales of hours superimposed on a longer variability pattern which is attributed to rotational modulation of the stellar flux by hot spots on the surface \citep{1997ApJ...482..465G}.

\subsection{RY Tau}
\textbf{Stellar Properties:} RY Tau has a stellar mass 2.2\,M$_{\odot}$ of and a radius 2.7 R$_{\odot}$ \citep{1988ApJ...330..350B}. A period of 5.7 days has been found for RY Tau \citep{1987AJ.....94..150H}, and it is a confirmed X-ray emitter \citep{1995ApJ...446..331D}

\noindent \textbf{Disc Properties:} This system has an inclination angle of $\sim$30$^{\circ}$  \citep{2003SPIE.4838.1037A} with a disk mass 0.2\,$M_{\odot}$. A large cavity has been found in the disc, where the gap is $\sim$ 20\,AU, with the outer disc wall lying at 18\,AU, and inner wall at 0.42\,AU. The disc is classified as a transitional disc \citep{2010ApJ...710..265P}.

\begin{figure*}
\begin{tabular}{cccc}
\includegraphics[scale=0.22]{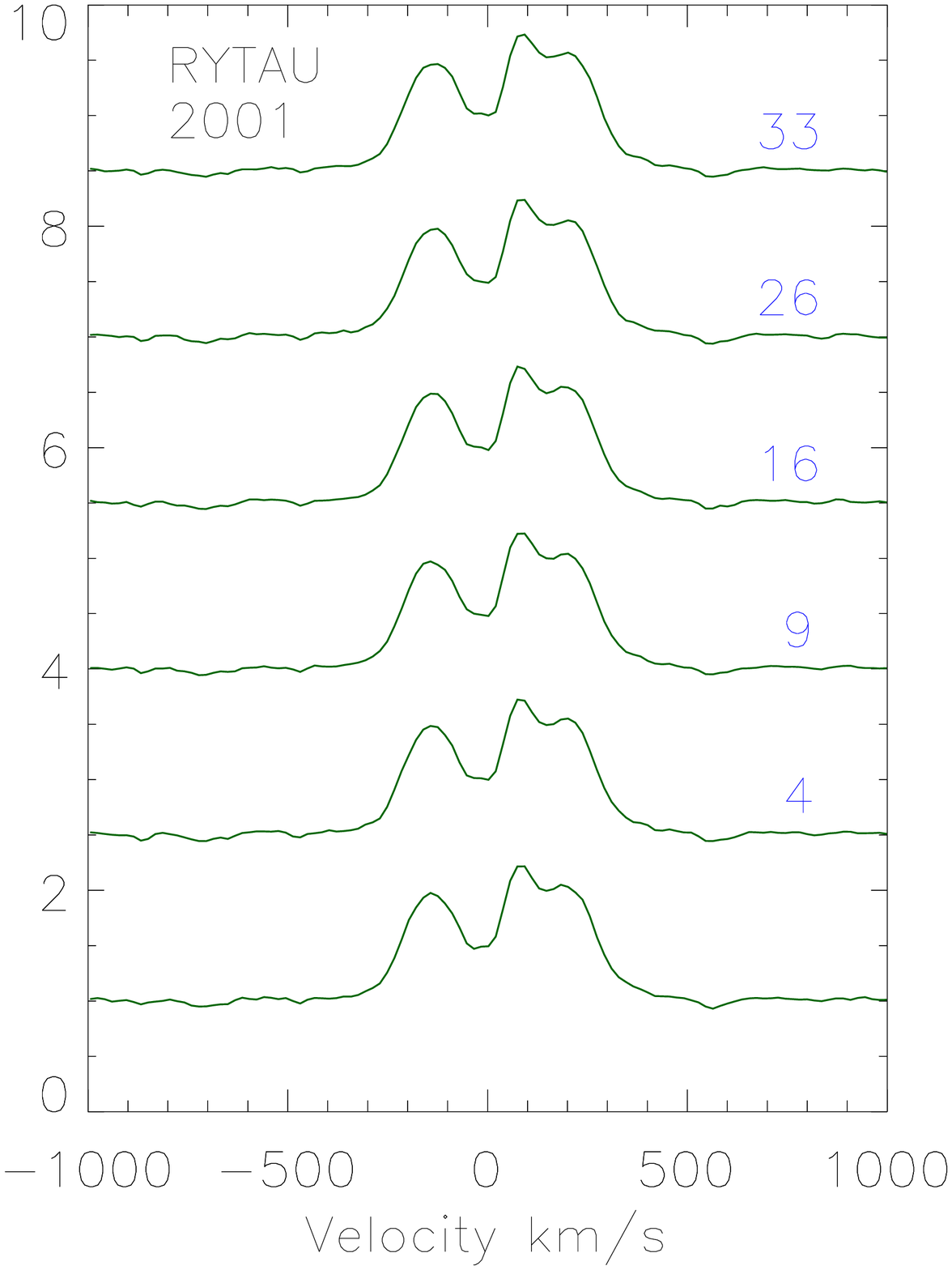} & \includegraphics[scale=0.22]{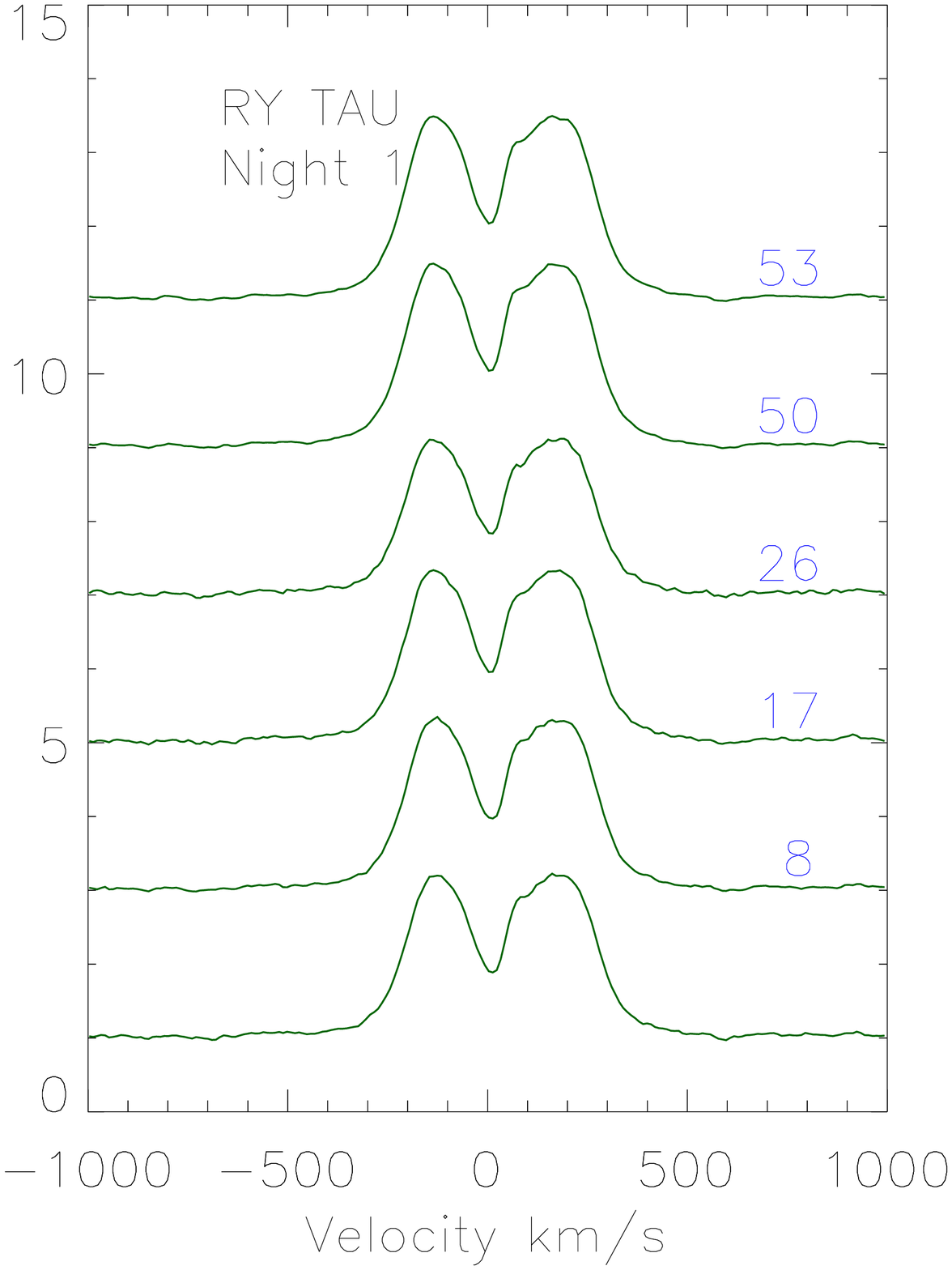} & \includegraphics[scale=0.22]{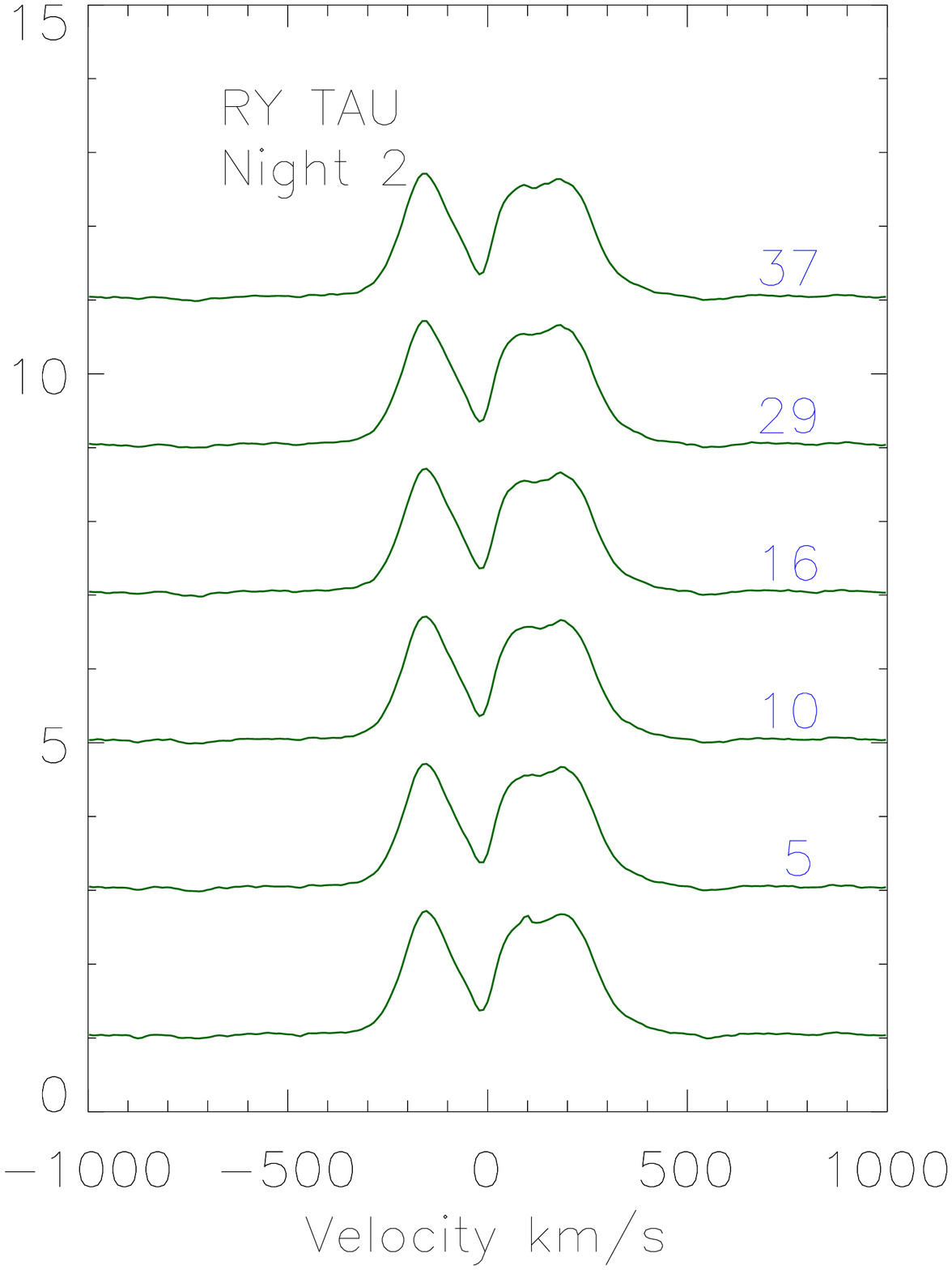} & \includegraphics[scale=0.22]{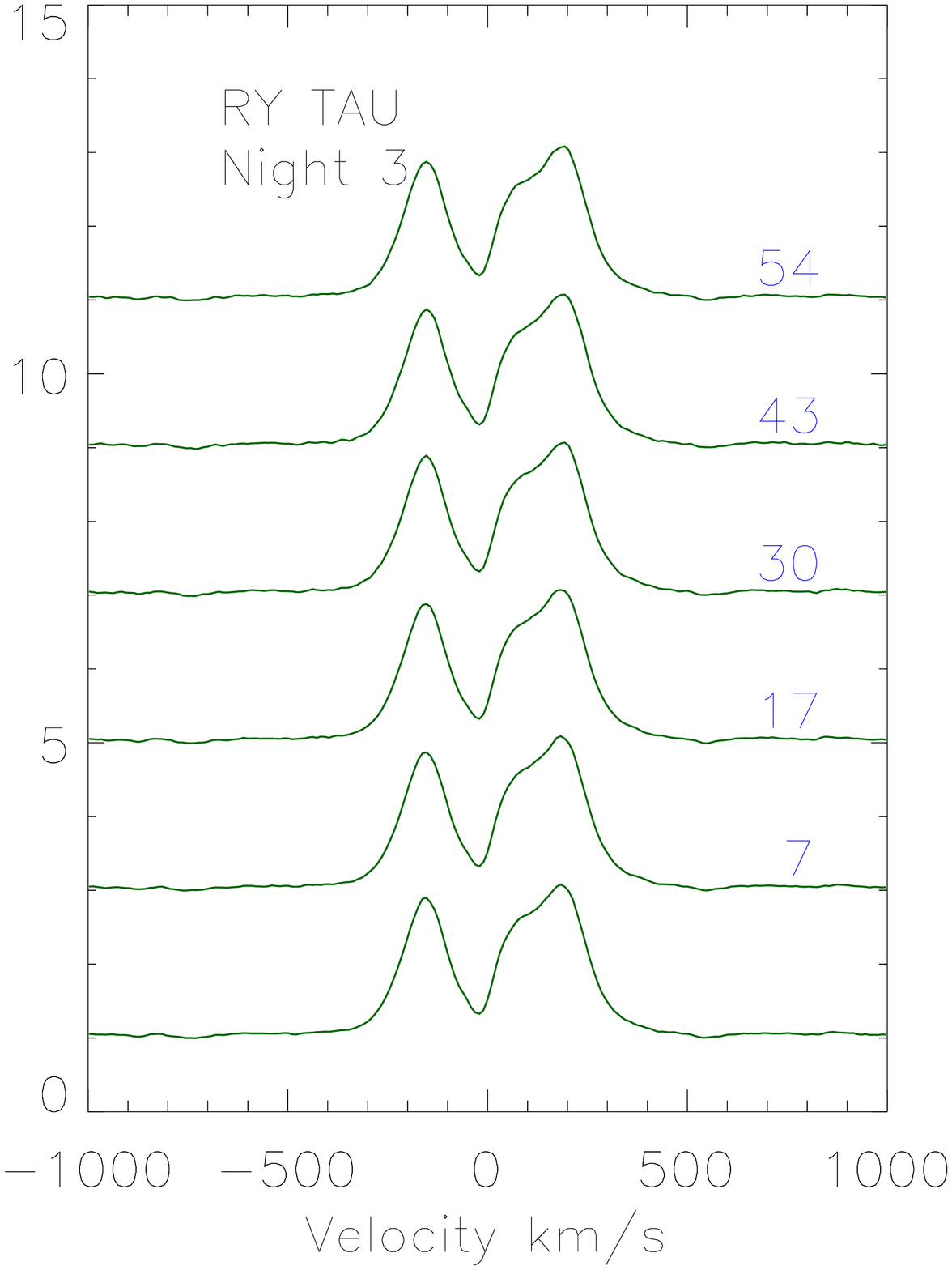}\\
 \includegraphics[scale=0.222]{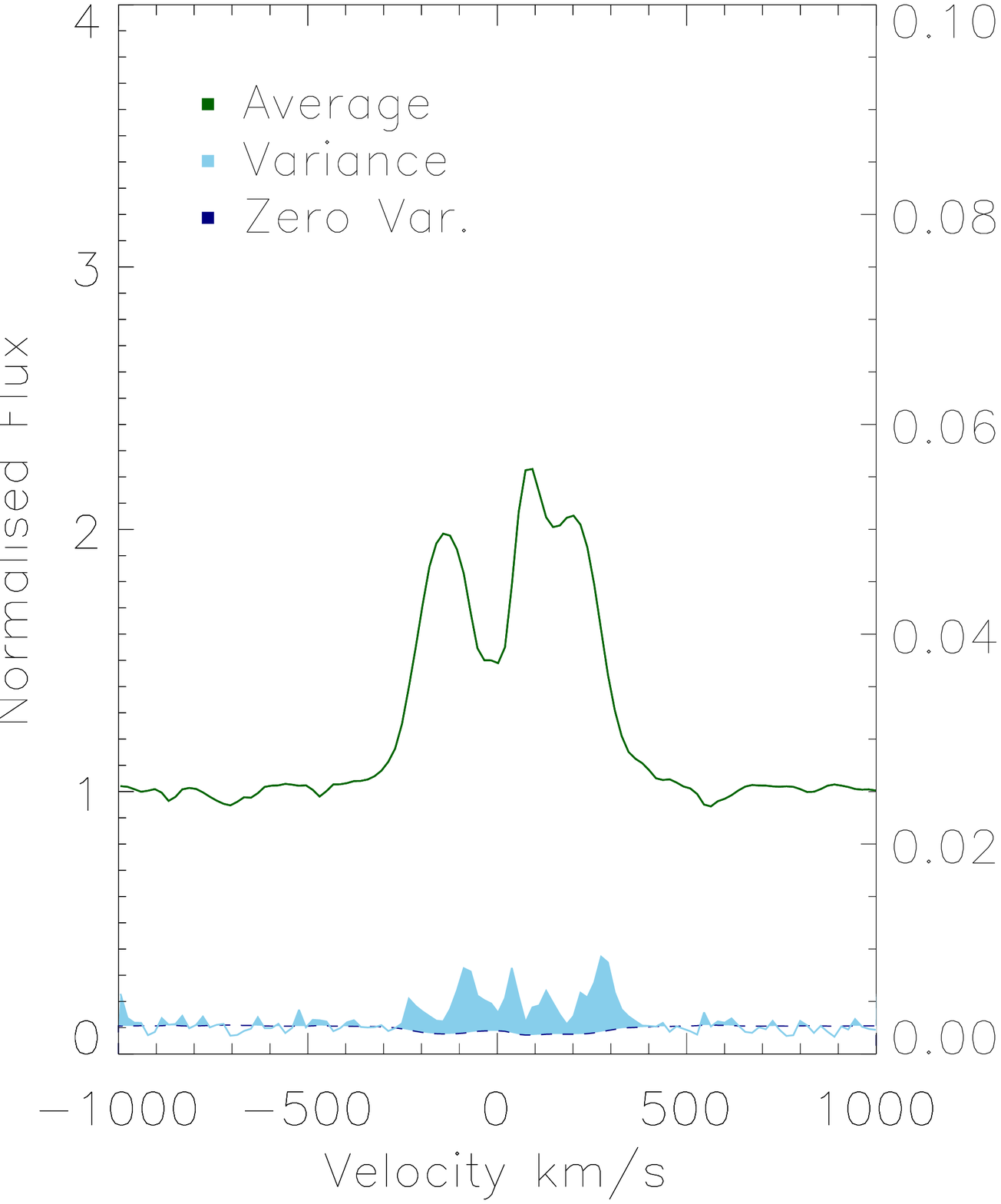}  & \includegraphics[scale=0.22]{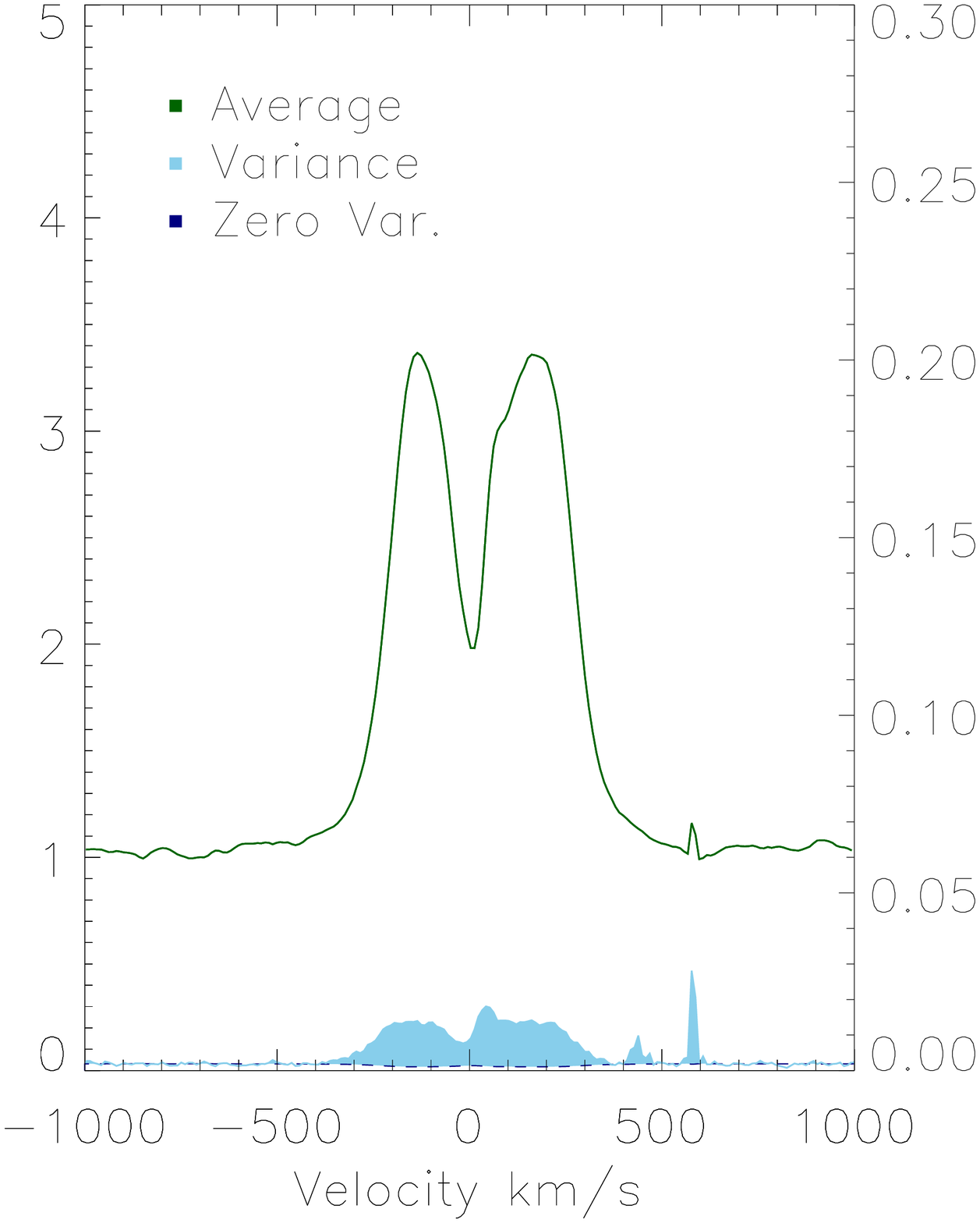}  & \includegraphics[scale=0.22]{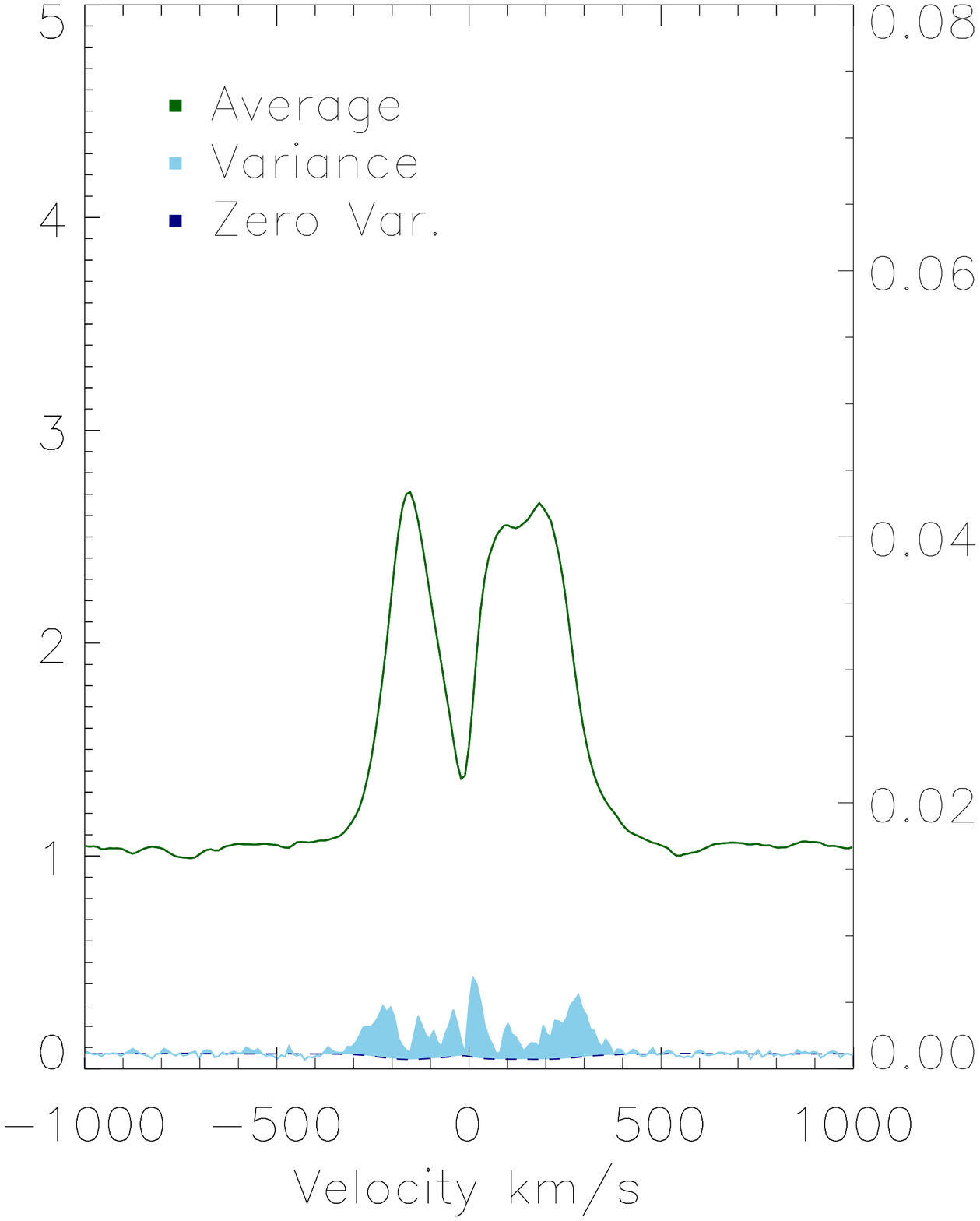}  & \includegraphics[scale=0.22]{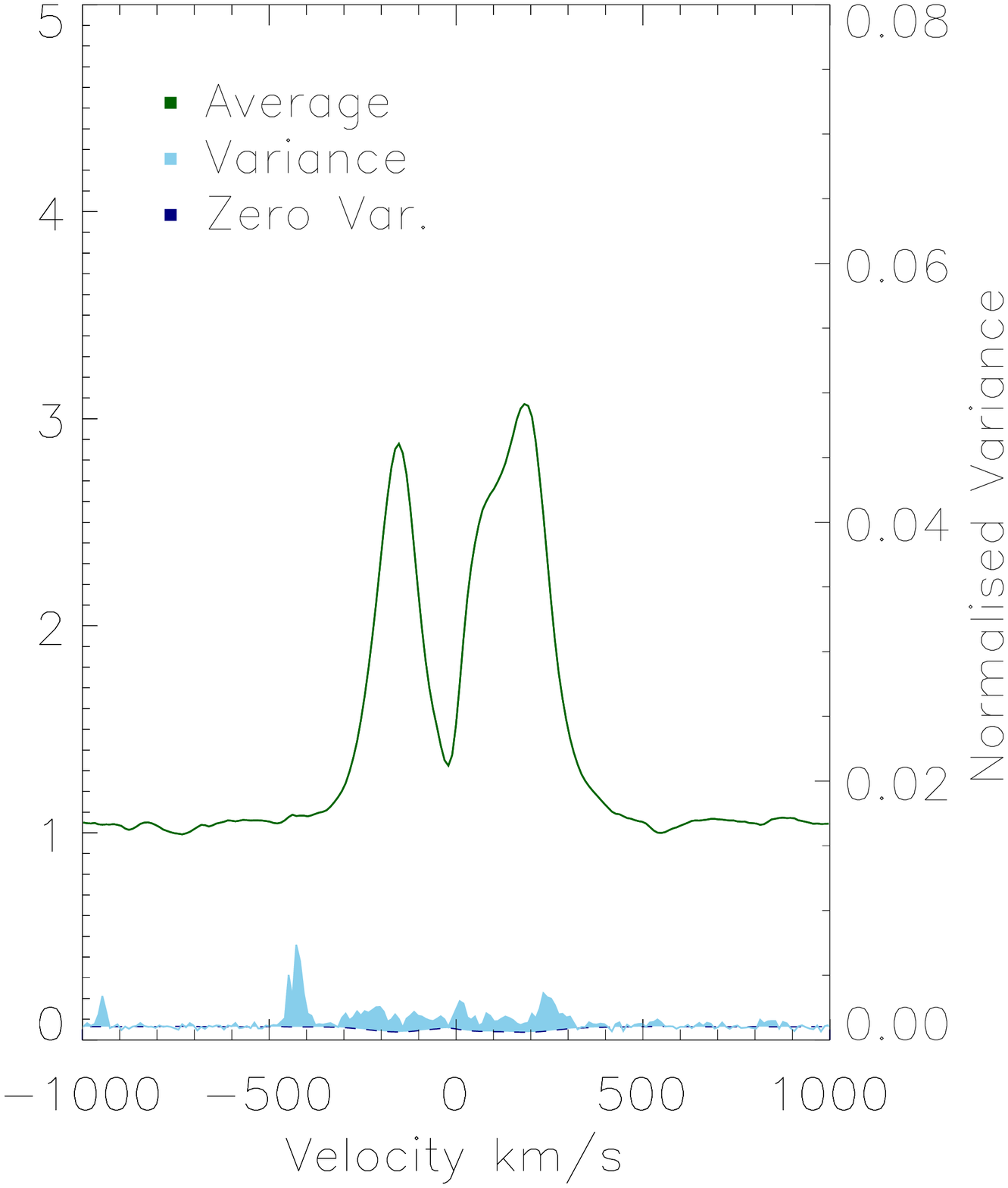} \\
 
\includegraphics[scale=0.22]{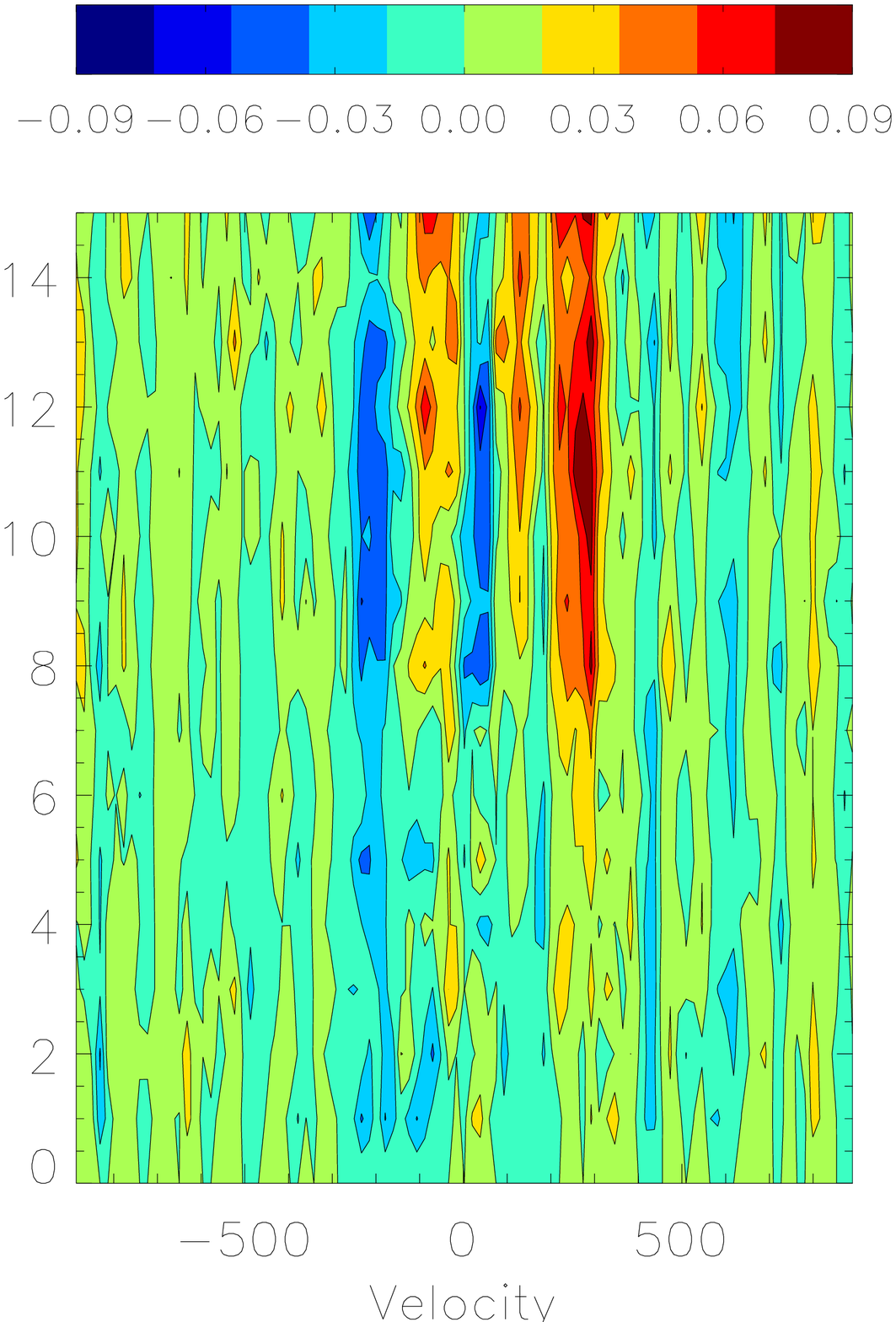}  & \includegraphics[scale=0.22]{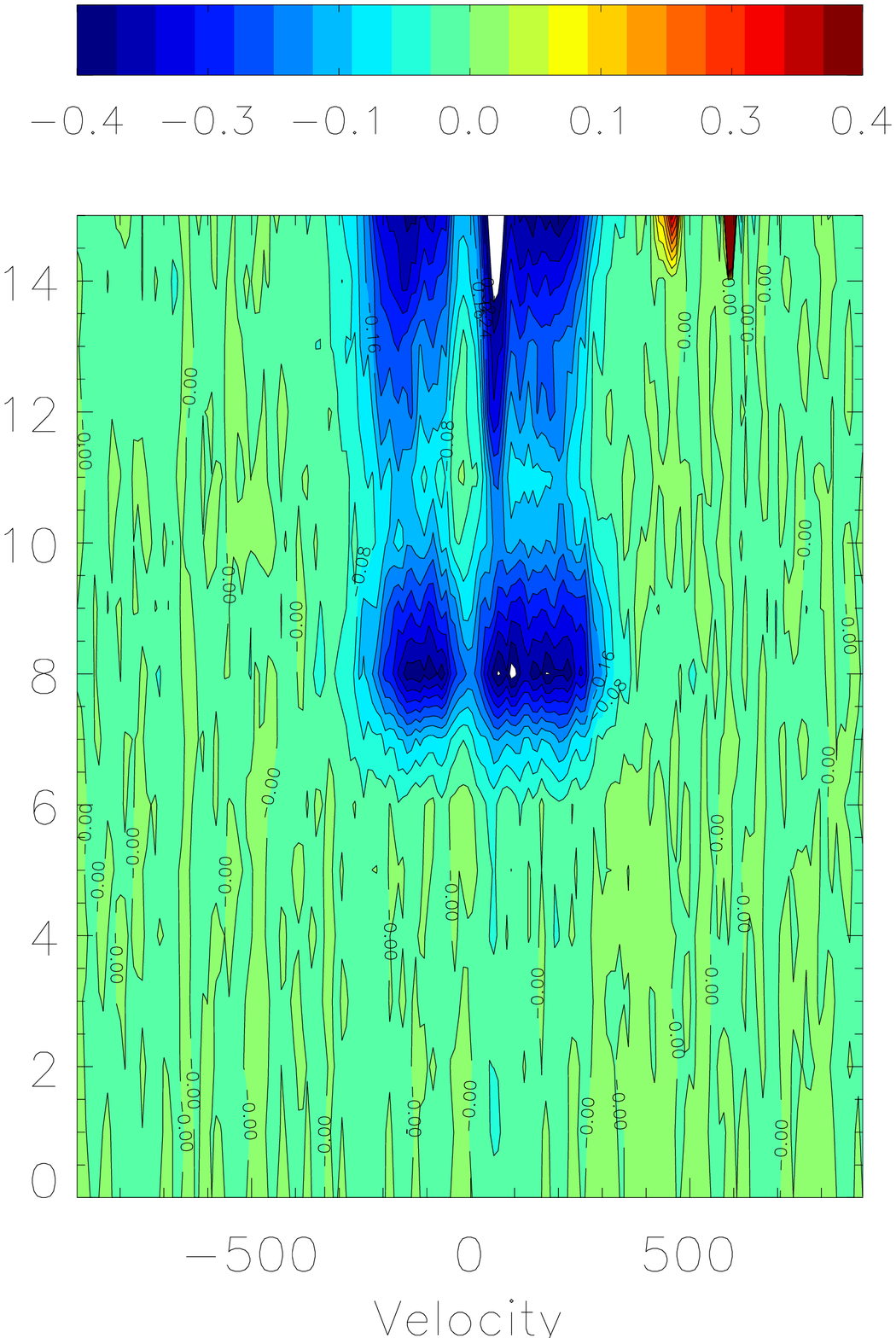}  & \includegraphics[scale=0.22]{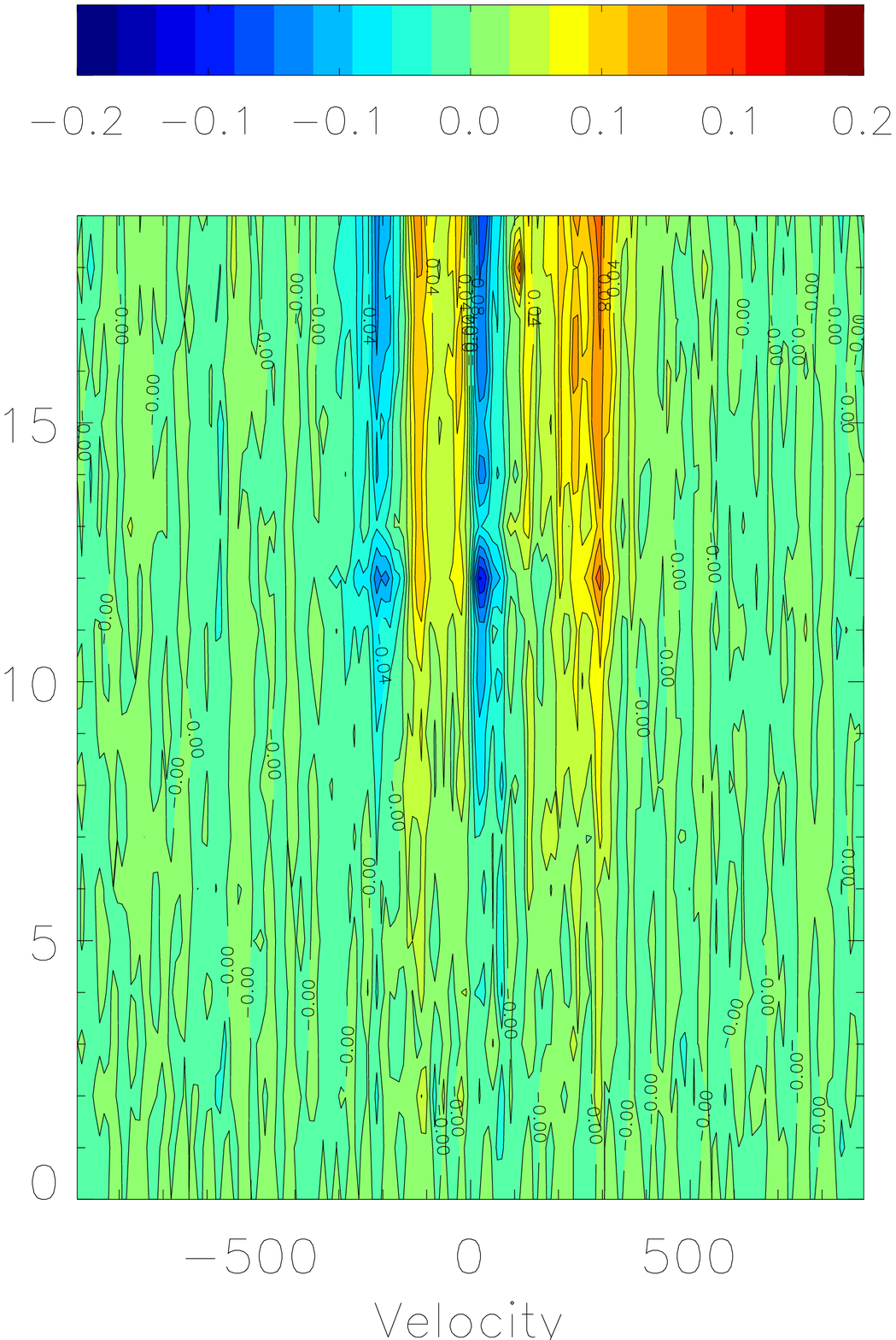} & \includegraphics[scale=0.22]{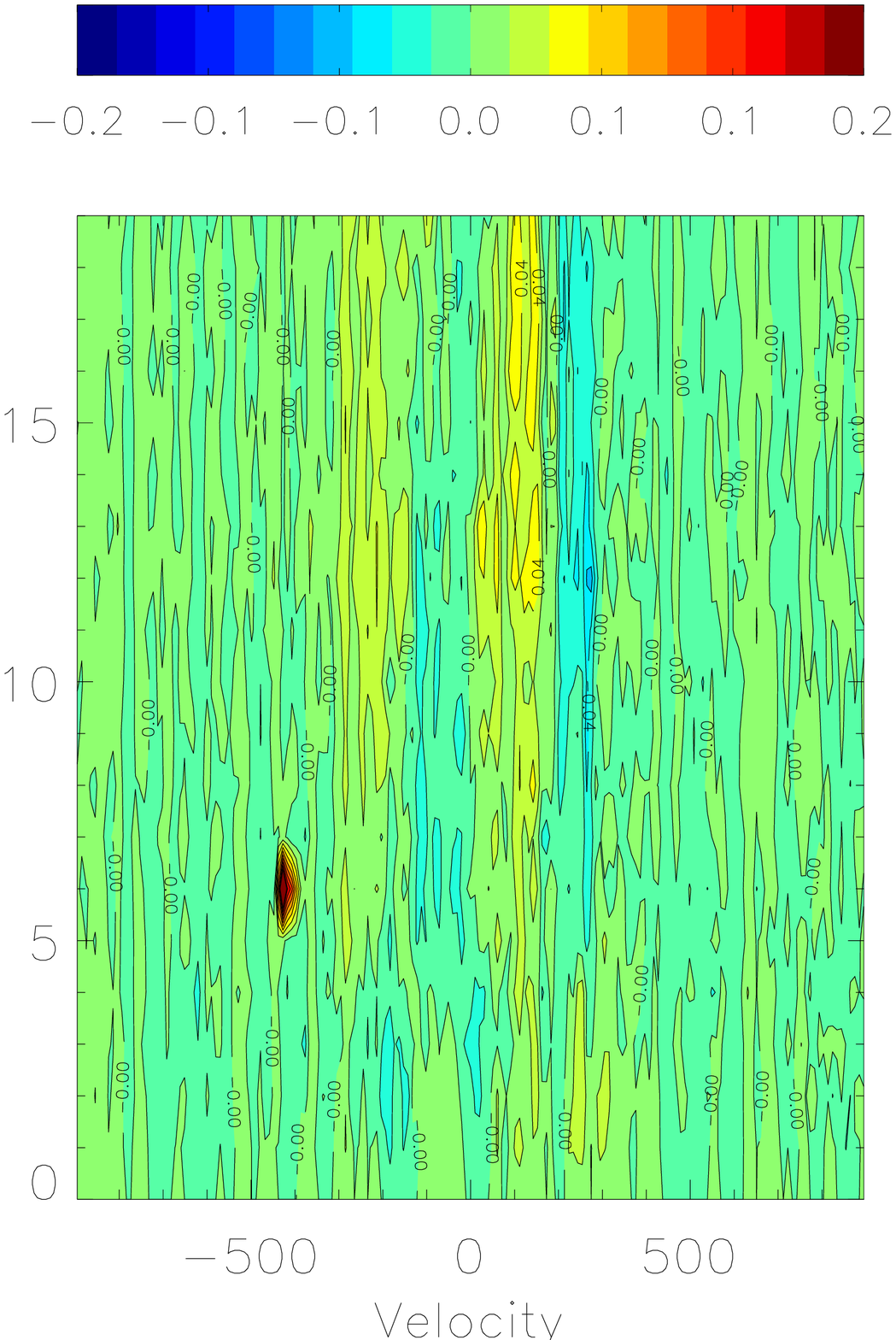}\\

 \includegraphics[scale=0.22]{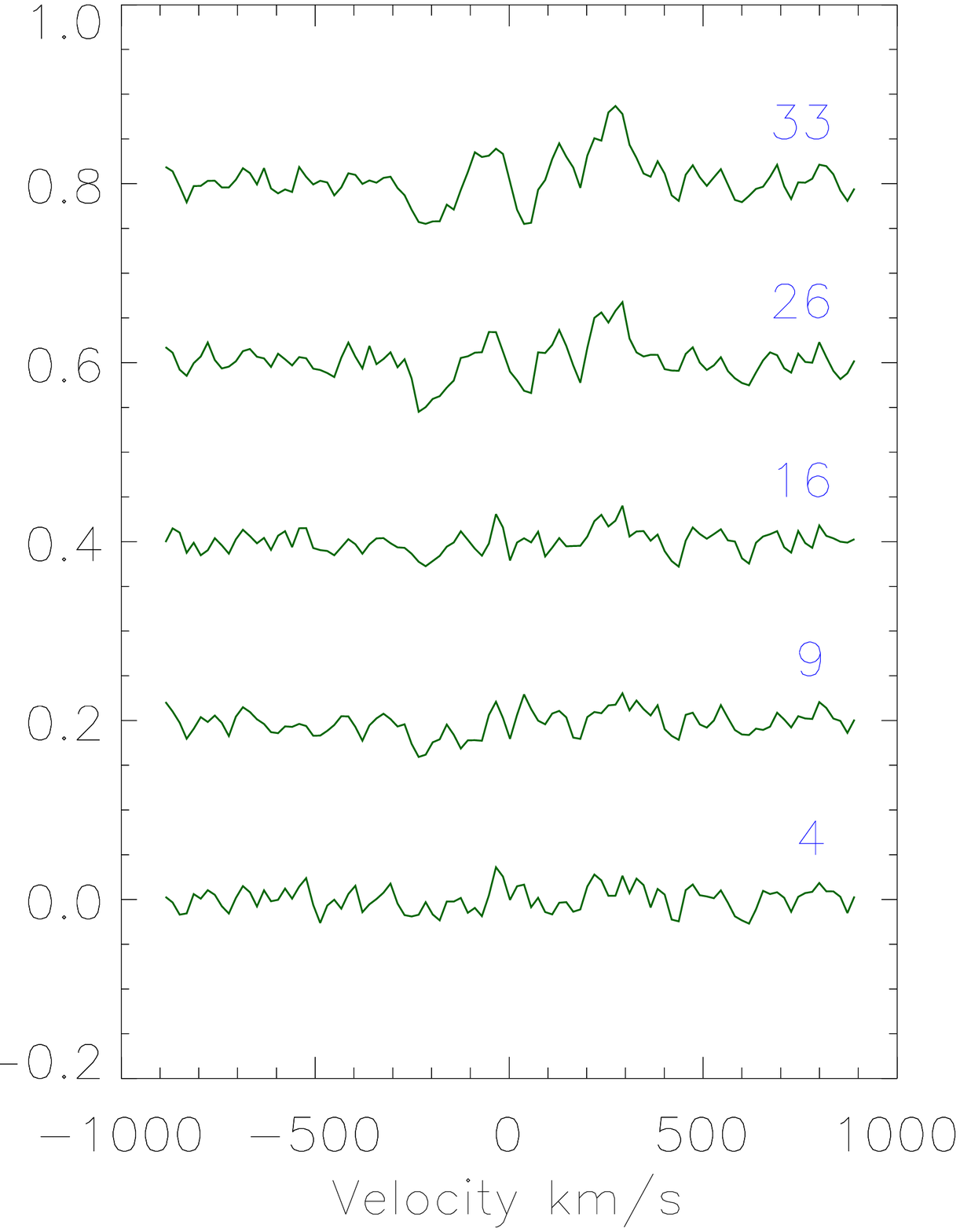} & \includegraphics[scale=0.22]{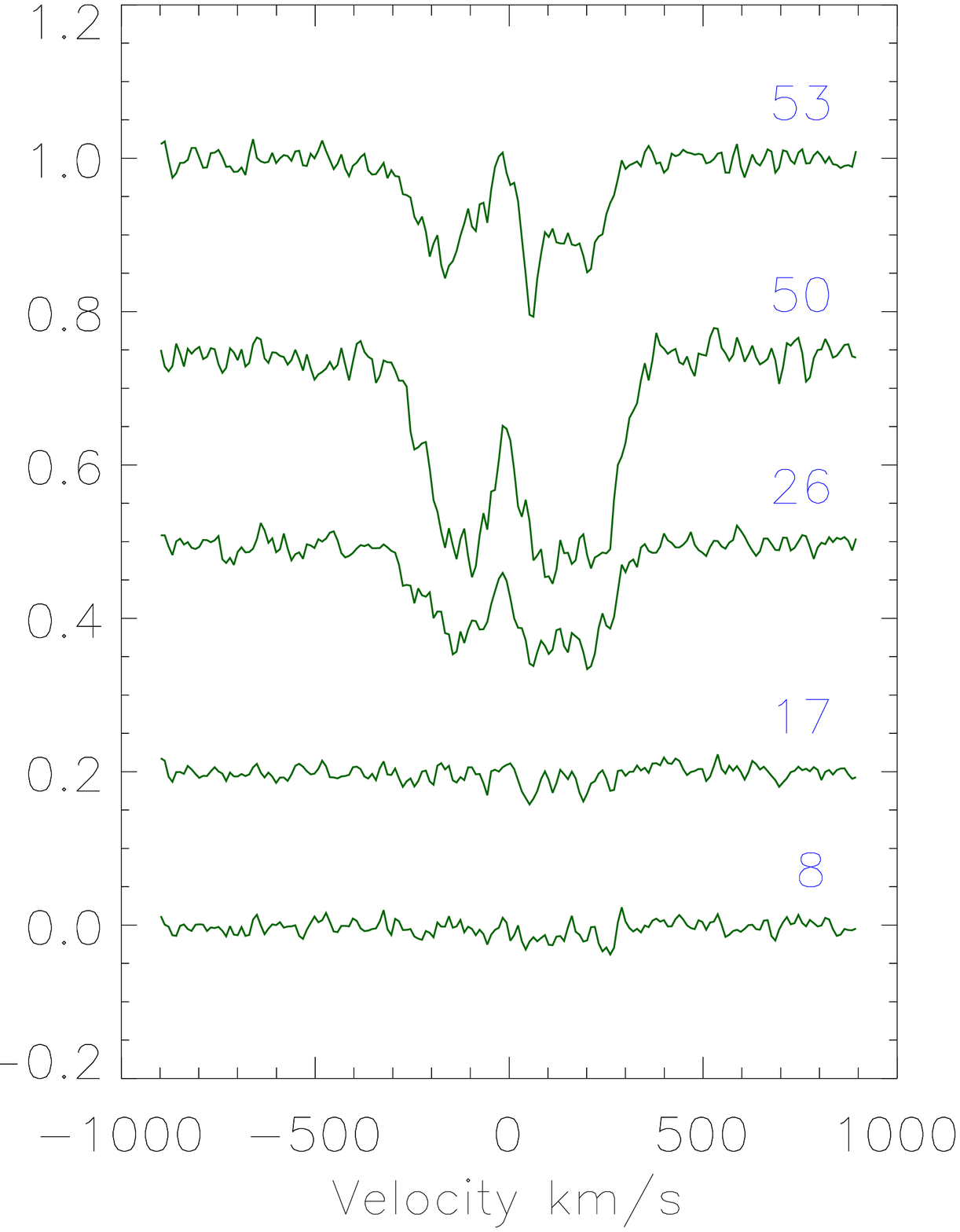}  & \includegraphics[scale=0.22]{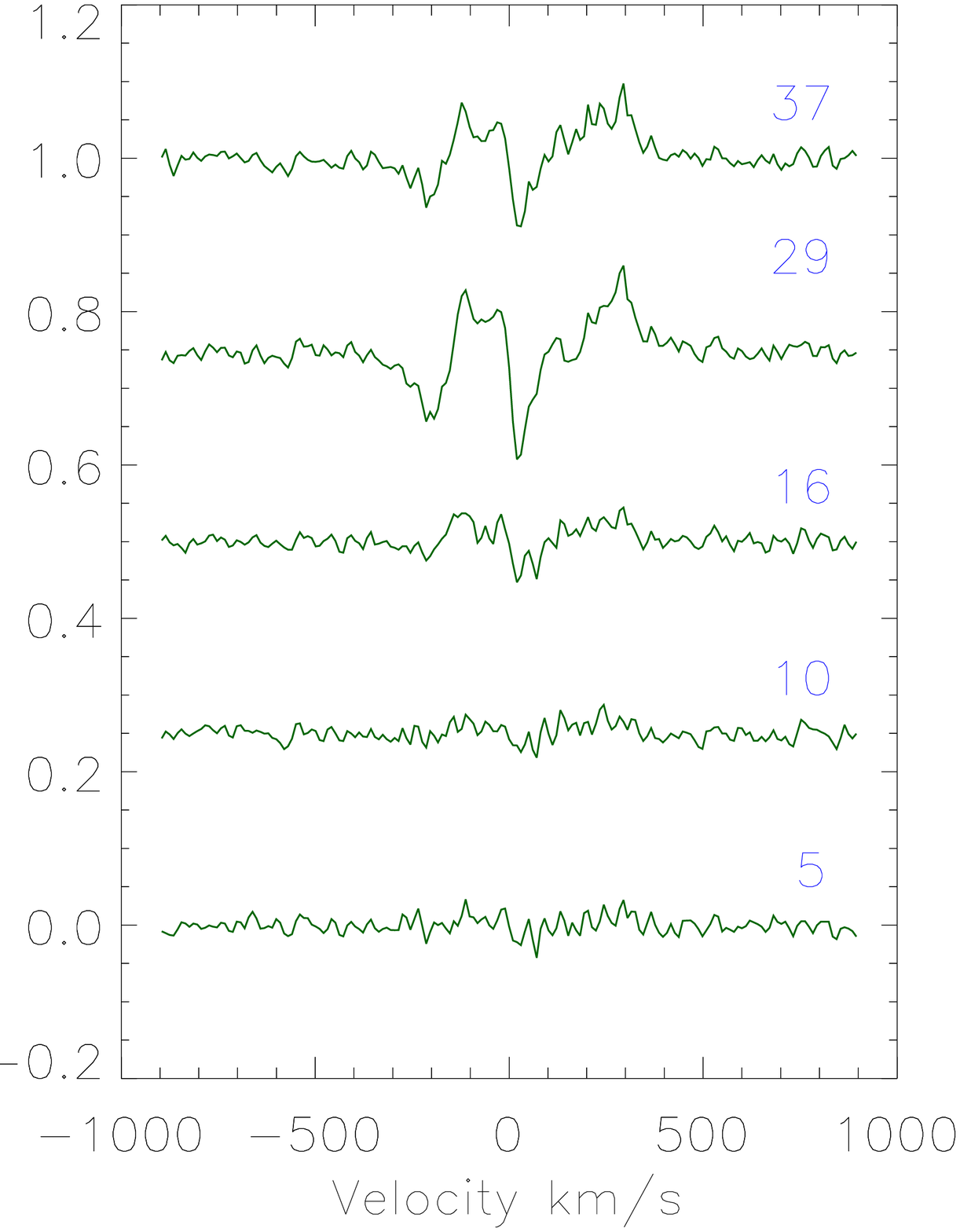} &  \includegraphics[scale=0.22]{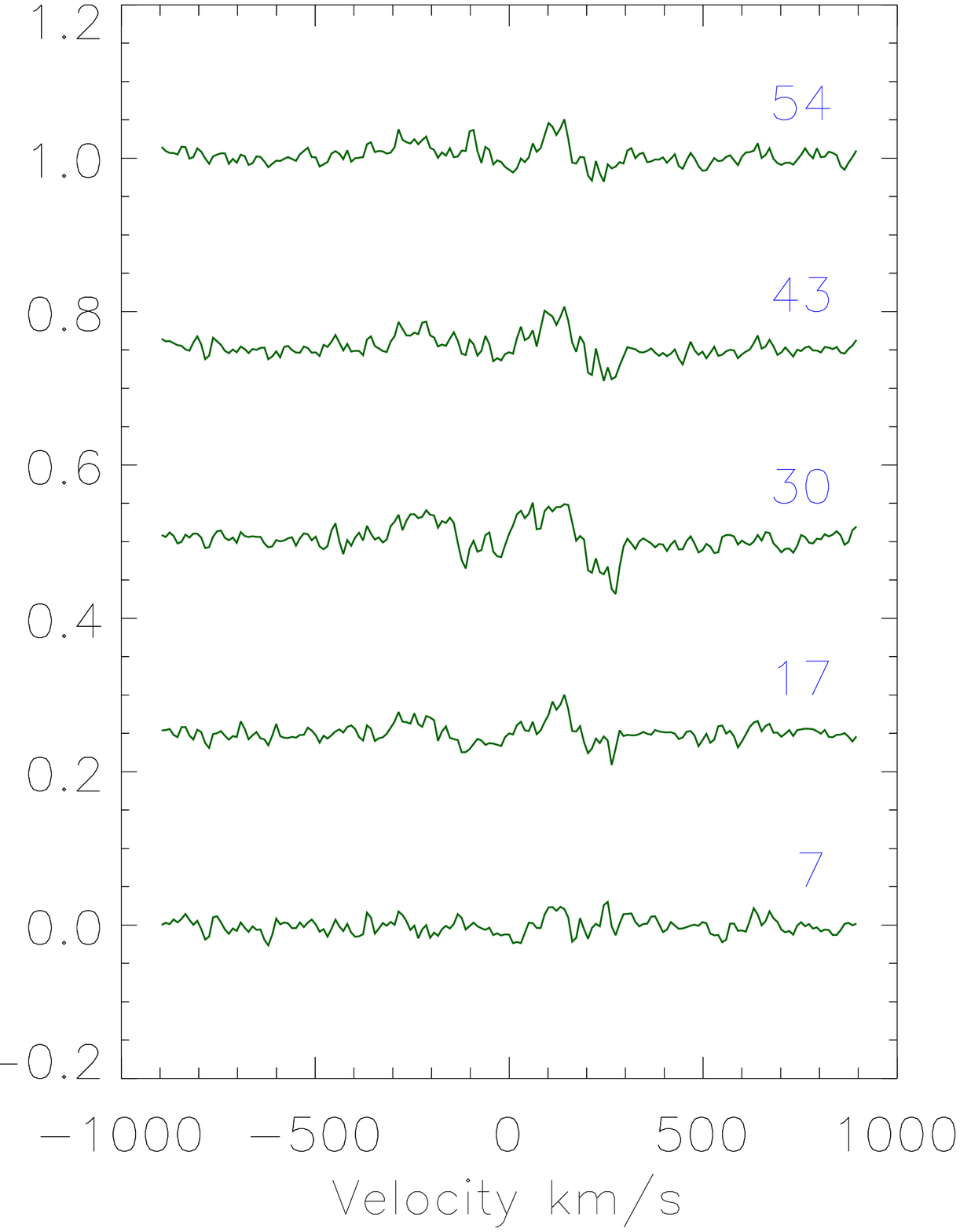} \\
\end{tabular}
\caption{RY Tau 2001 observations (left). RY Tau 2003 observations, Night 1 (middle left), Night 2 (middle right) and Night 2 (right).}
\label{fig:RYTAU_plots_2001}
\end{figure*}

\noindent \textbf{Accretion:} Observations have suggested that RY Tau has low levels of veiling \citep{1993AJ....106.2024V}. \citet{1989ApJ...341..340B} derived an accretion rate of 7.5\,x\,10$^{-8}$ M$_{\odot}$yr$^{-1}$.

\noindent \textbf{ISIS H$\alpha$ Observations:} The H$\alpha$ emission takes the form of a double peaked profile with a central absorption. In the single observation block of 2001, both peaks are of equal strength (Fig.\,\ref{fig:RYTAU_plots_2001}). The red-shifted peak shows an asymmetry, with an excess emission in the wing close to the line centre. 

In the 2003 observations the strength of the emission line has dropped, and the excess emission in the red-shifted emission peak becomes more developed (Panels on right of Fig.\,\ref{fig:RYTAU_plots_2001}). Comparing the three nights the red emission peak shows more variations. It changes in strength with respect to the blue peak, and also there is a slight change in the morphology of the peak.  On the third night the red emission peak becomes a double peak. 
The changes in the H$\alpha$ EW on night 1 mainly takes the form of a drop in flux across the emission line. However there are some pronounced changes seen in the variance profile in the red side of the absorption feature. The second and third night have no significant changes, but there are some small changes in the red side of absorption again. From first to second night the H$\alpha$ EW drops from 245\,\AA~to 203\,\AA, a change of $\sim$ 17\% of original EW.

\noindent \textbf{Previous H$\alpha$ Observations:}  
RY Tau was also observed in October and November 2010 \citep{2013AJ....145..108C}, with EW measurements of 10.3, 14.9, 15.5, 14.3, 11.8\,\AA~recorded on October 21$^{st}$, November 17$^{th}$, 21$^{st}$, 25$^{th}$ and 27$^{th}$ respectively. Over the month time-scales of these observations, the profile is similar to what is observed in the ISIS sample. The red peak also shows similar variations in the rise and fall, and also the appearance of an extra absorption feature in the red peak (as seen in our 2001 Dec. 26$^{th}$ observation).  

Variations in H$\alpha$ profile have also been found on the time-scales of 10\,-\,20 mins without variations in the star's brightness in \citep{1975PZ.....20..153K}. Three separate nights observations took place covering time scales of $\sim$6.4hrs and 3 hrs. The profiles changed between nights, but in two out of three occasions they were stable through out the night. One night however the profile did go through changes, the EW varying between 19 to 22\AA, and absorption features changed in strength, the blue absorption becoming stronger and weaker again, while a red absorption feature appears and disappears. Most spectra where taken with 15 min. exposures and the brightness of the star was lower on the night of the variations than on the other two nights \citep{1975PZ.....20..153K}.
The emission profile in this case is similar to what is observed in the ISIS observations, a double peaked profile, however in this case the blue peak is significantly smaller.
Interestingly when this object was observed again in 1976/77 multiple absorption components were observed in H$\alpha$ emission profile \citep{1979ApJS...41..369S}.

\subsection{SU Aur} 

\textbf{Stellar Properties:} SU Aur has as stellar mass of 2\,M$_{\odot}$ and a radius of 3.1 R$_{\odot}$ \citep{1988ApJ...330..350B}. The period of SU Aur has not been well established, probably lying between 1.7 \citep{2003ApJ...590..357D} and 2.7 days \citep{1987AJ.....94..150H}. 

\noindent \textbf{Outflows:} Evidence has been found of a bipolar outflow from SU Aur \citep{2001AAS...199.6015G}.

\noindent \textbf{Accretion:} An accretion rate of 6\,x\,10$^{-8}$ M$_{\odot}$yr$^{-1}$ was found for this object by \citet{1989ApJ...341..340B}.

\noindent \textbf{Disc Properties:}  Interferometric observations found the system inclination to be close to 63$^{\circ}$ \citep{2002ApJ...566.1124A} with an inner disc hole of radius 0.05\,-\,0.08 AU.

\begin{figure}
\begin{tabular}{cc}
\includegraphics[scale=0.22]{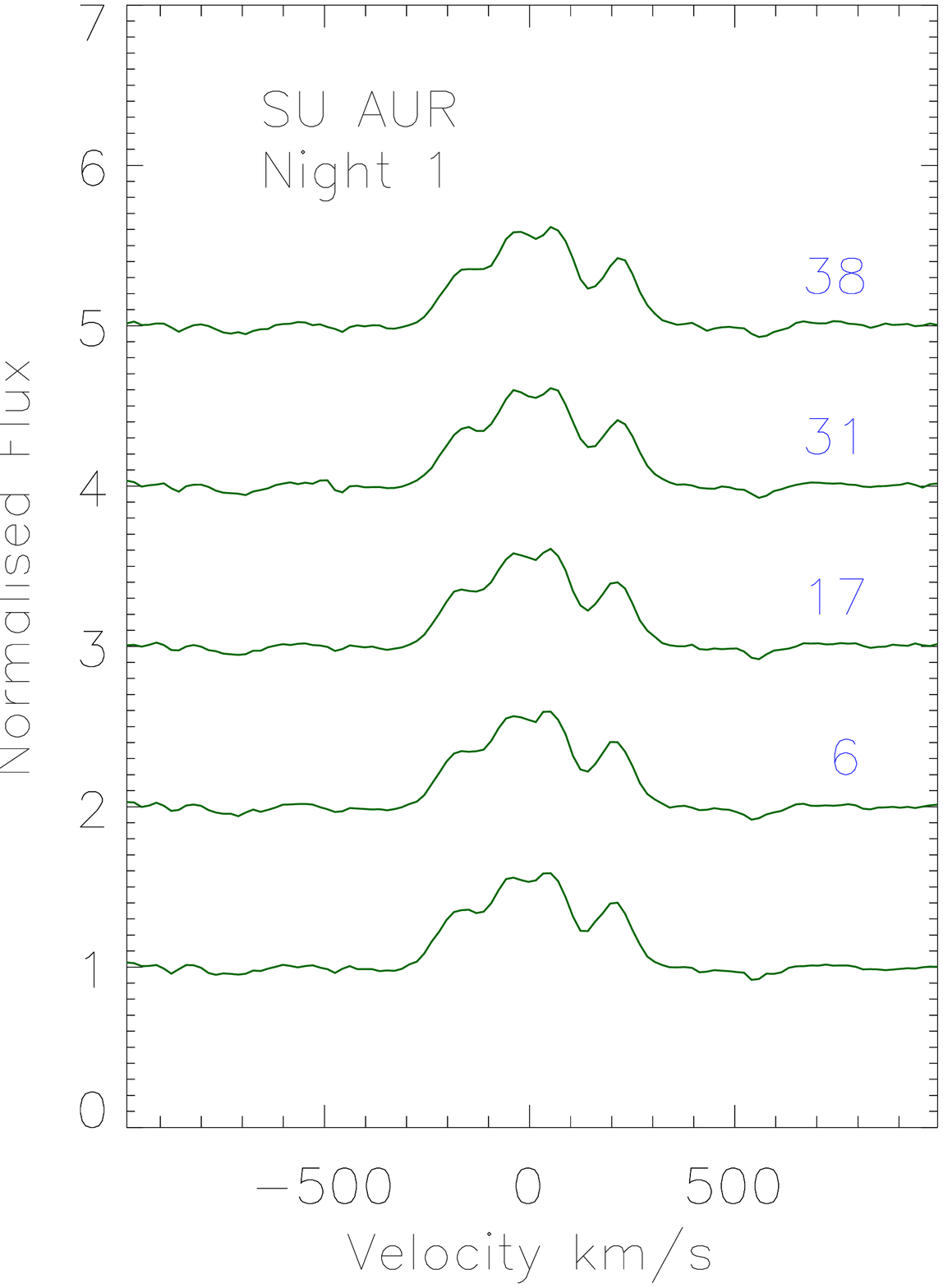} & \includegraphics[scale=0.22]{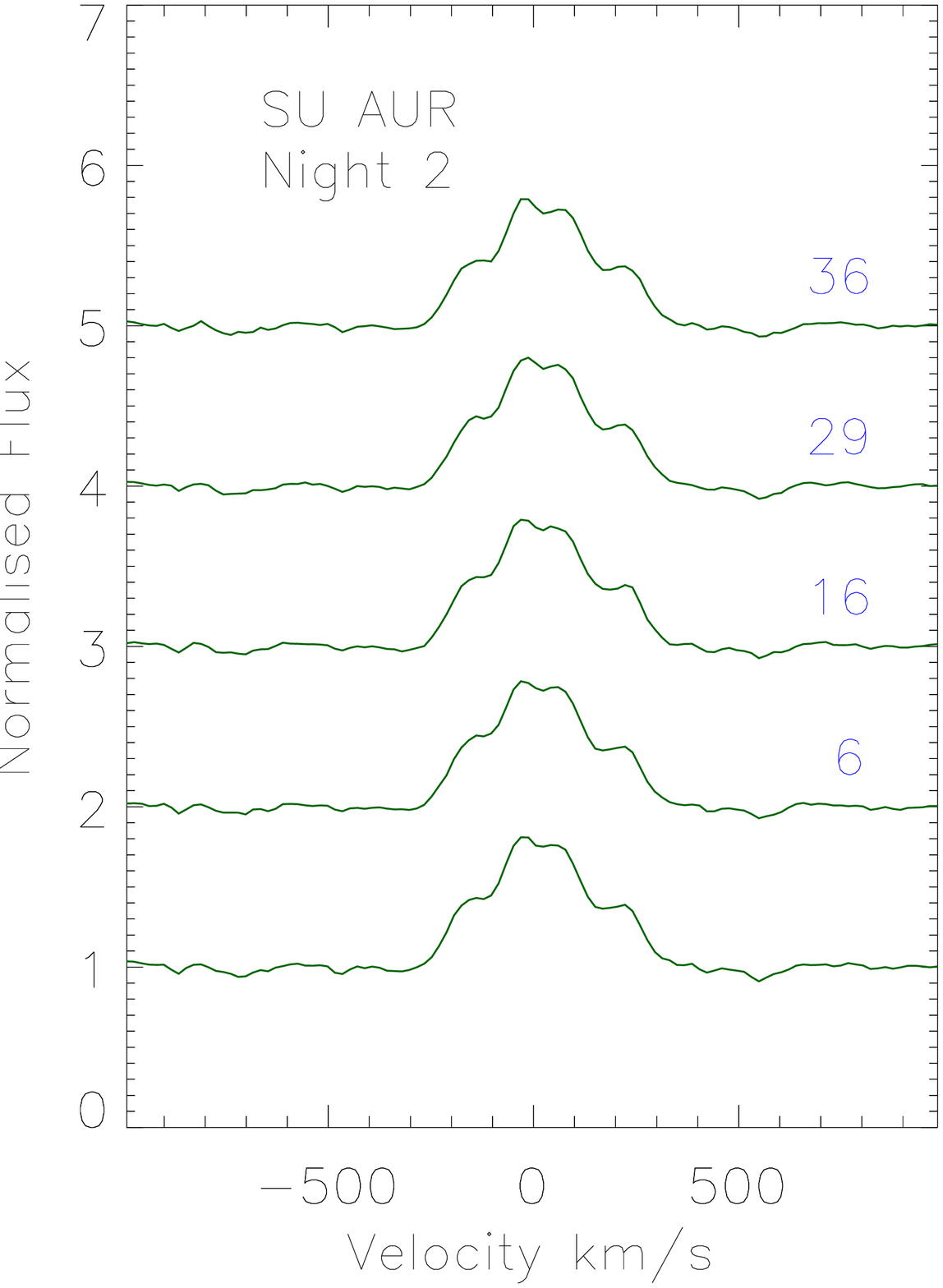}\\
\includegraphics[scale=0.22]{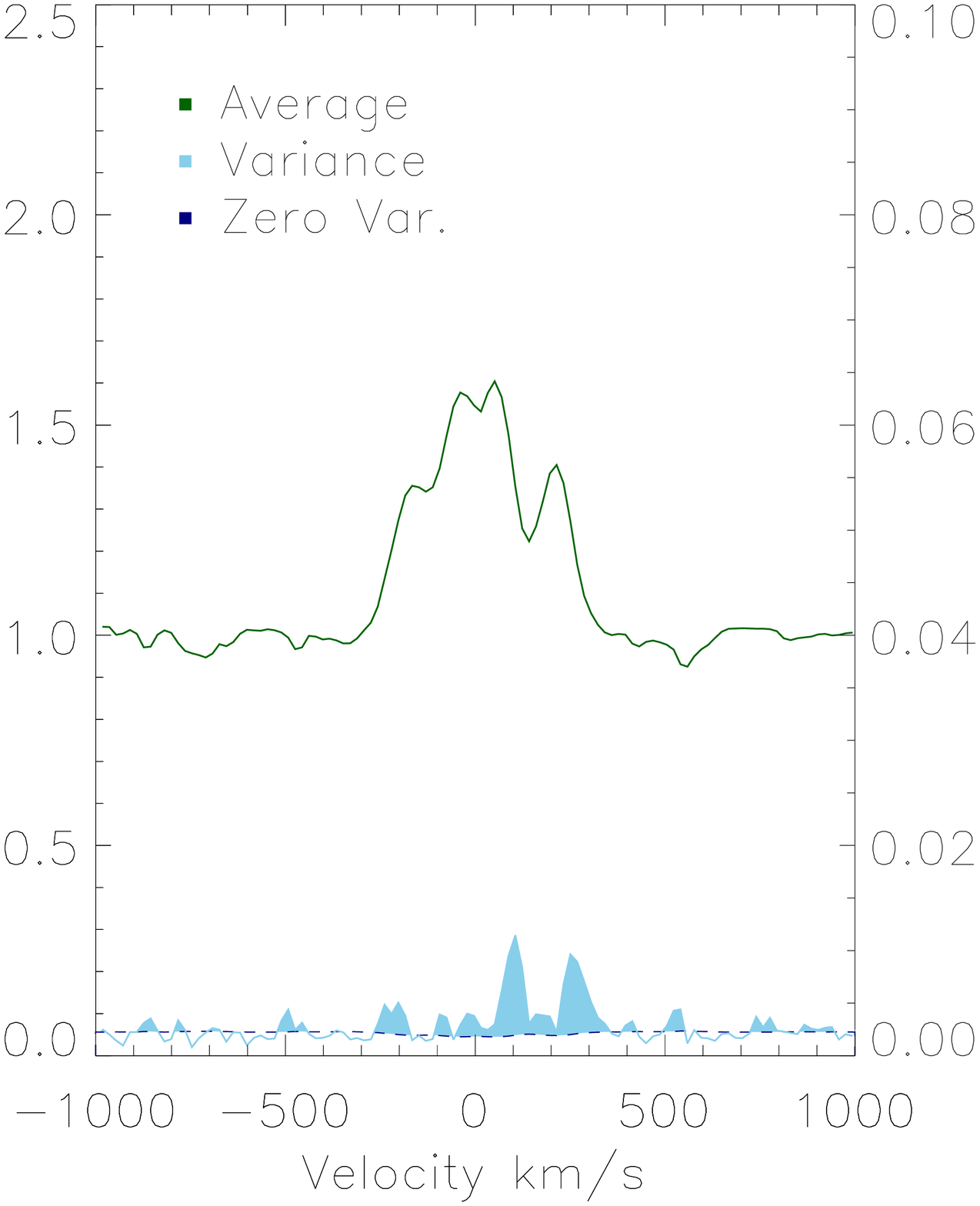} & \includegraphics[scale=0.22]{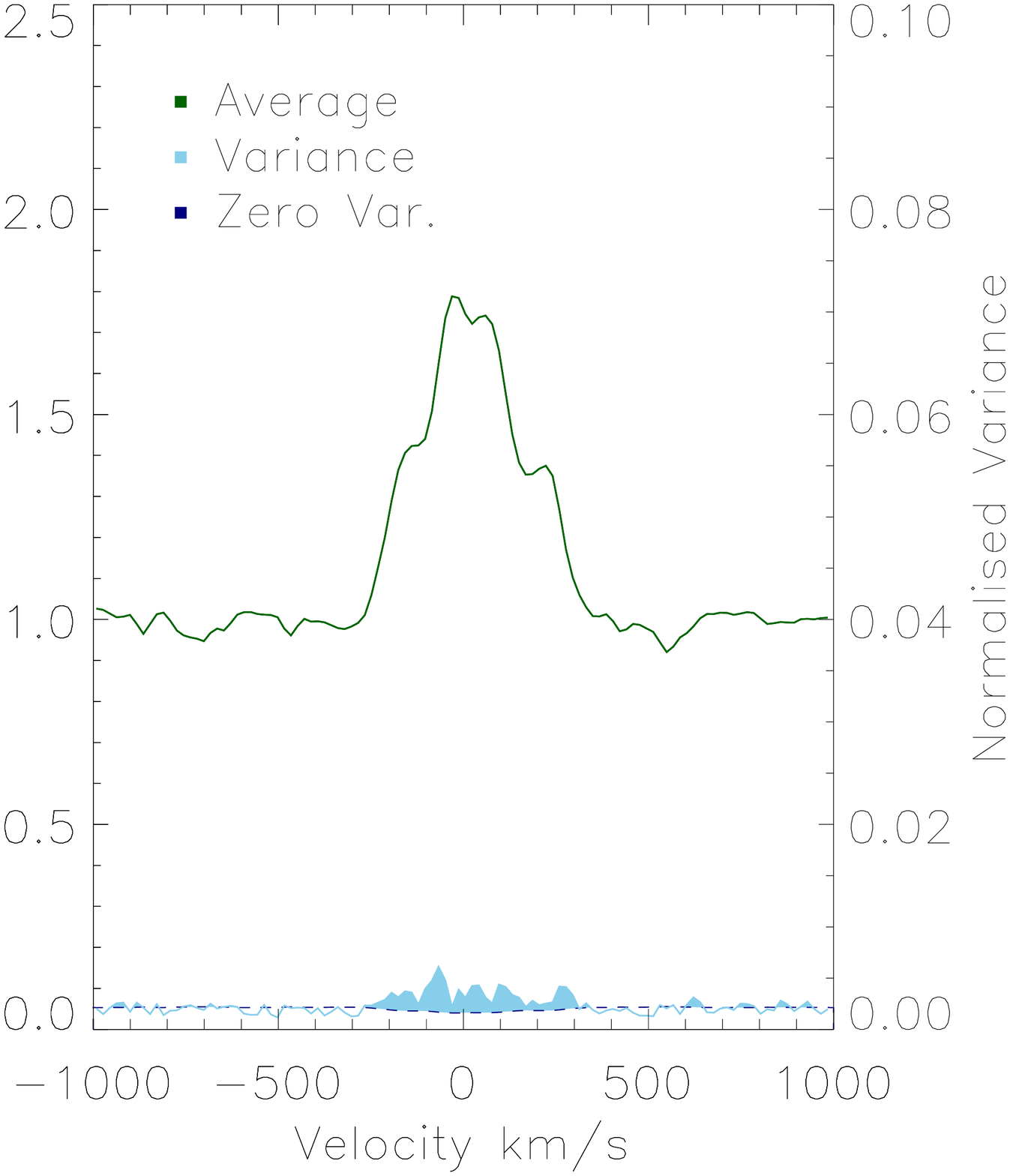} \\
\includegraphics[scale=0.22]{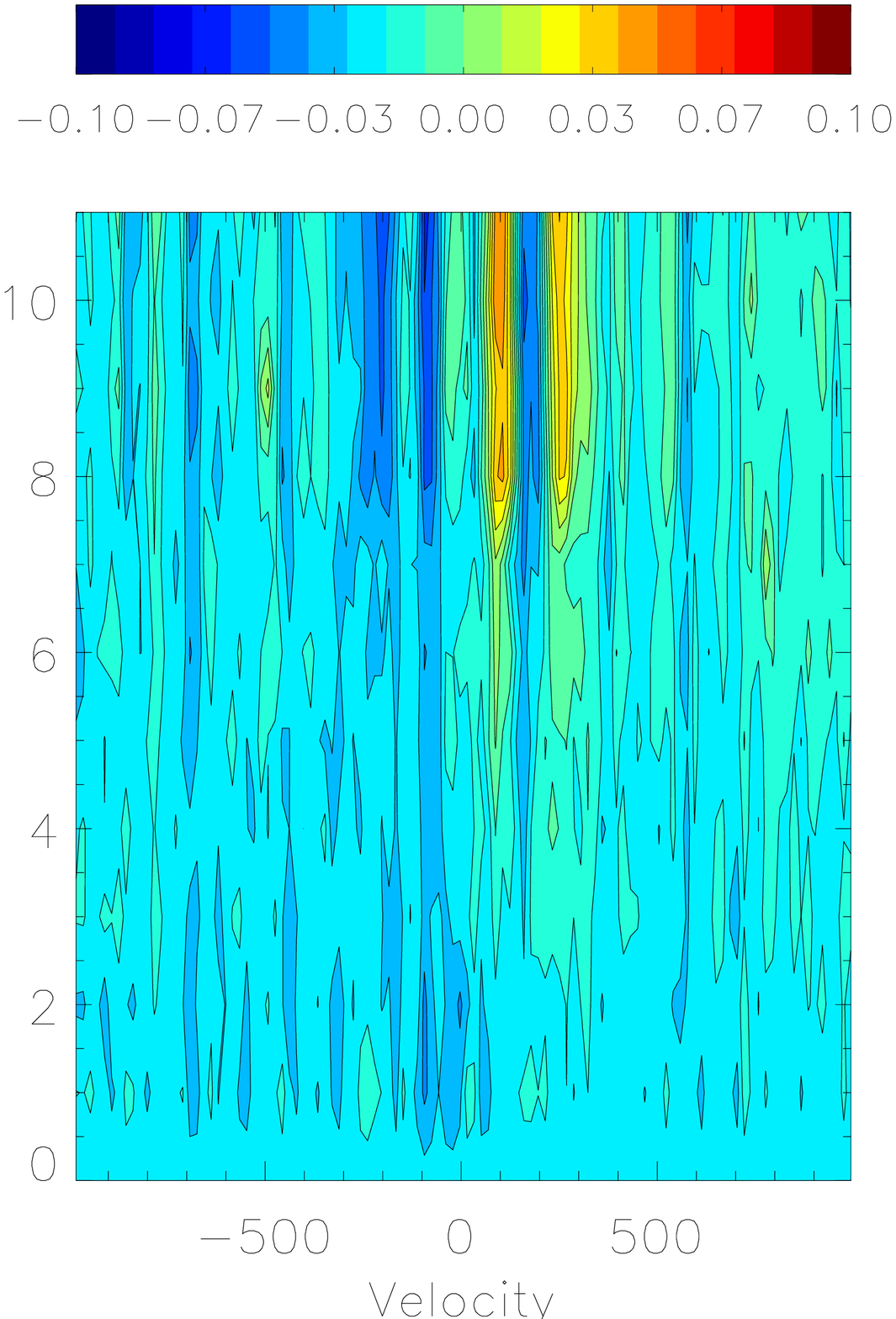} & \includegraphics[scale=0.22]{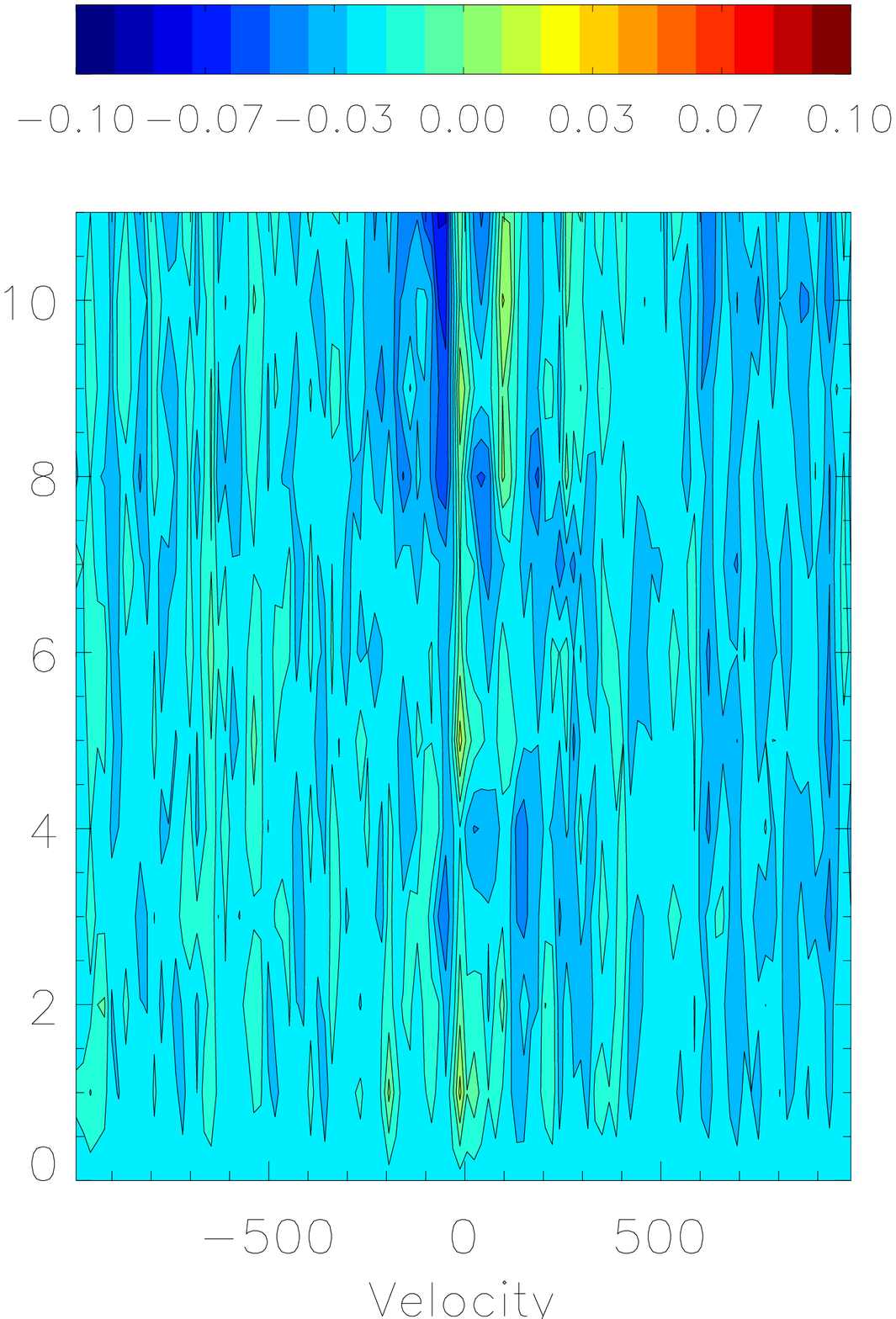}\\
\includegraphics[scale=0.22]{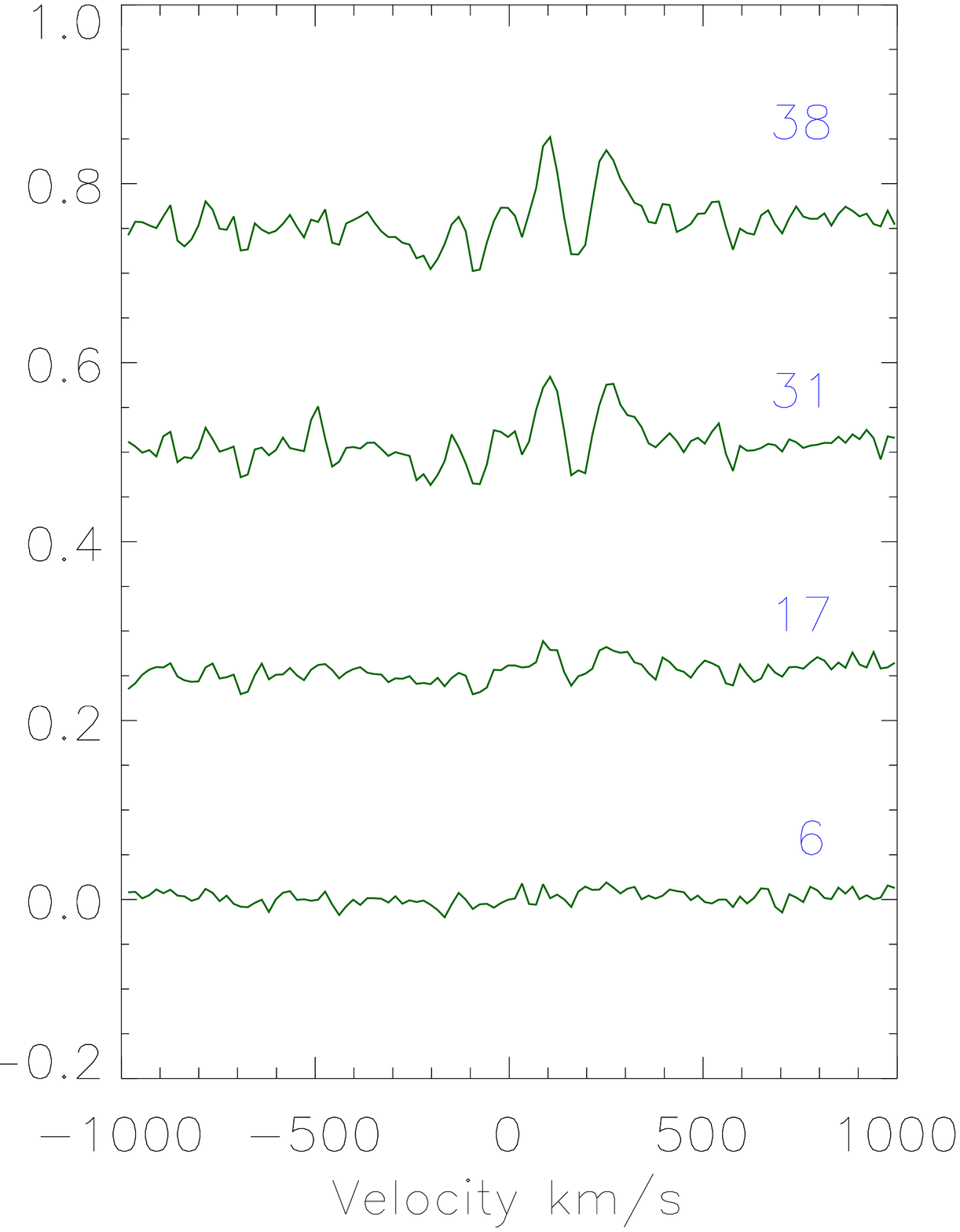} & \includegraphics[scale=0.22]{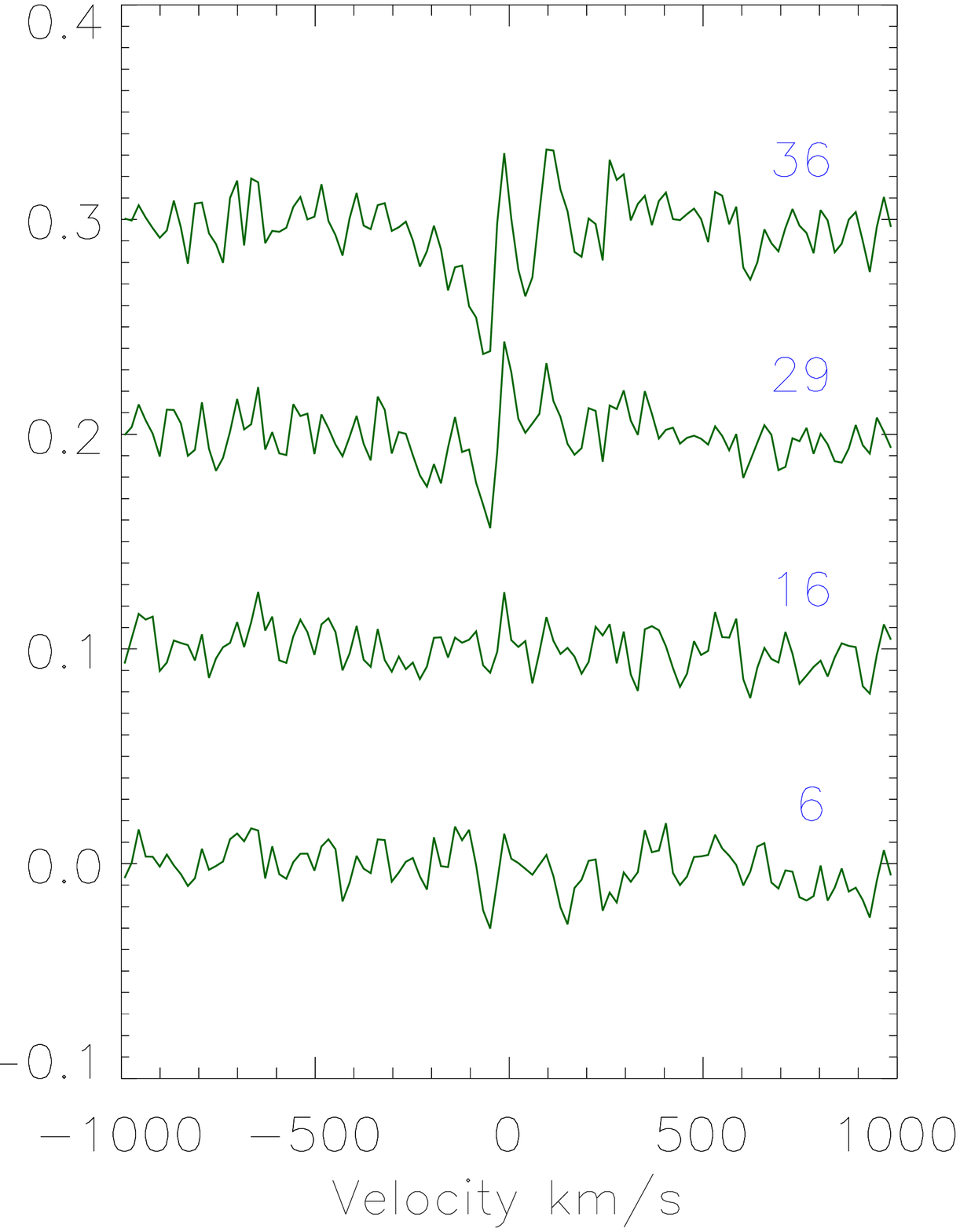} \\
\end{tabular}
\caption{SU Aur 2001 observations, Night 1 (left) and Night 2 (right). }
\label{fig:SUAUR_plots_2001_1}
\end{figure}

\begin{figure*}
\begin{tabular}{cccc}
\includegraphics[scale=0.22]{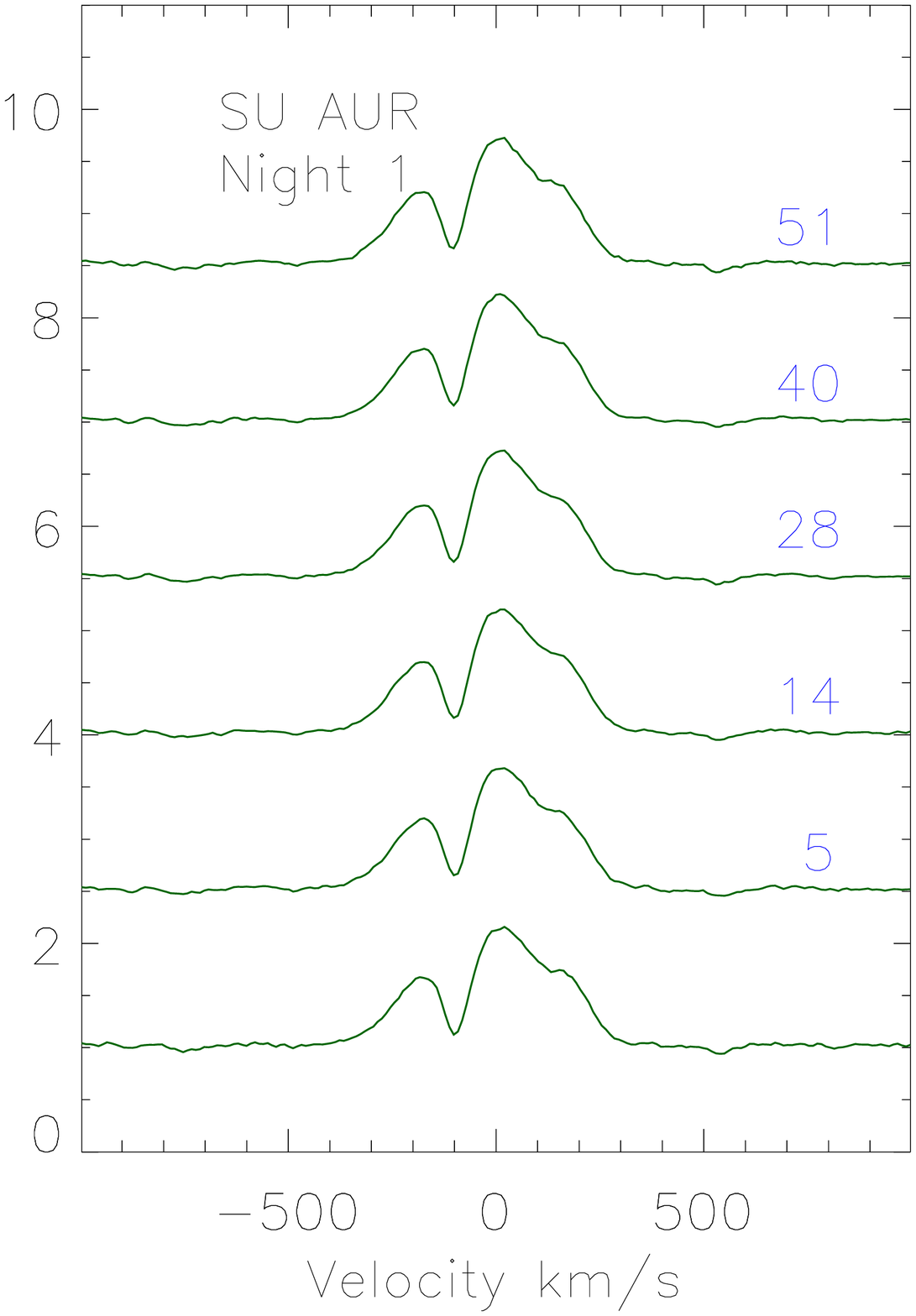}& \includegraphics[scale=0.22]{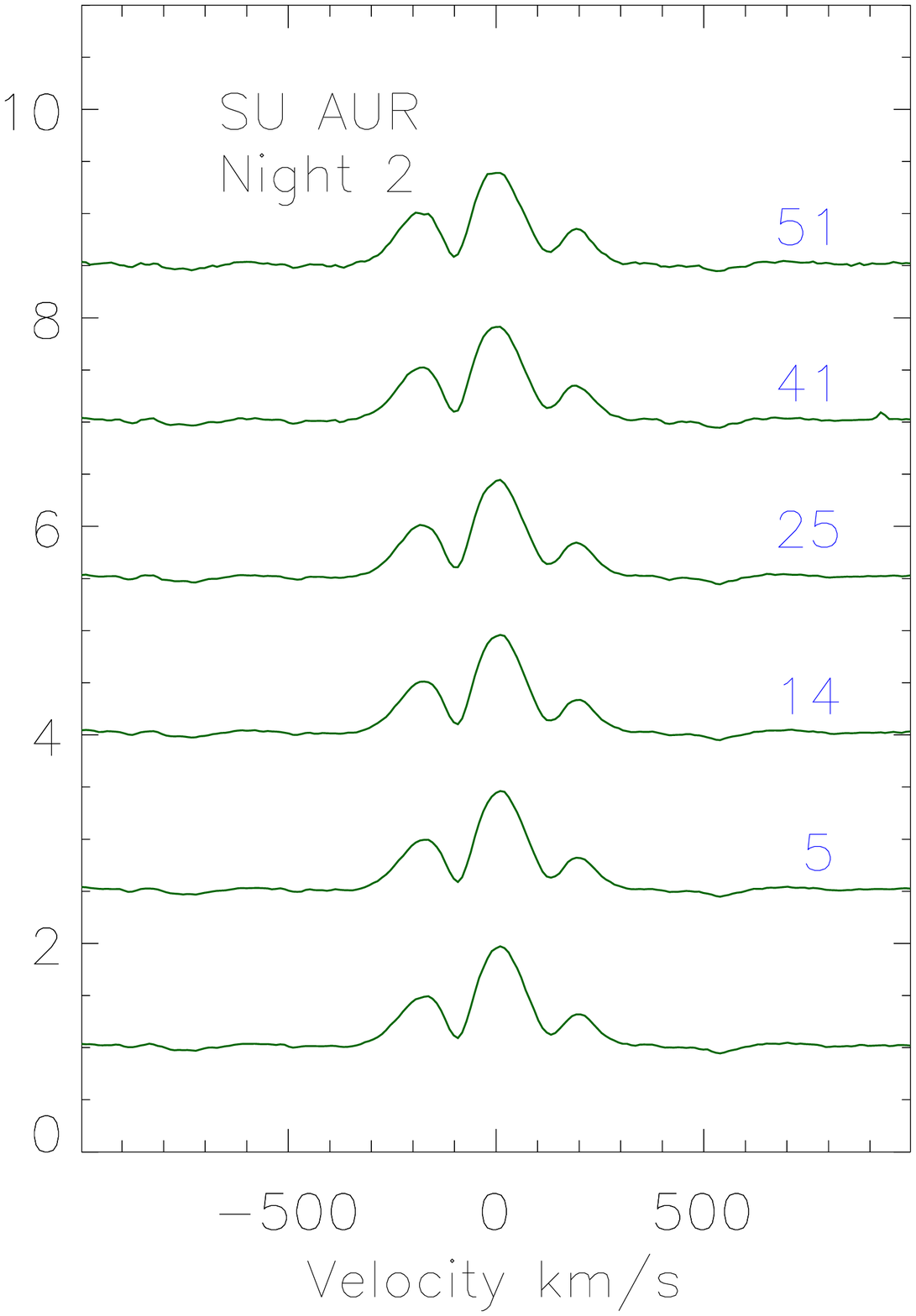} & \includegraphics[scale=0.22]{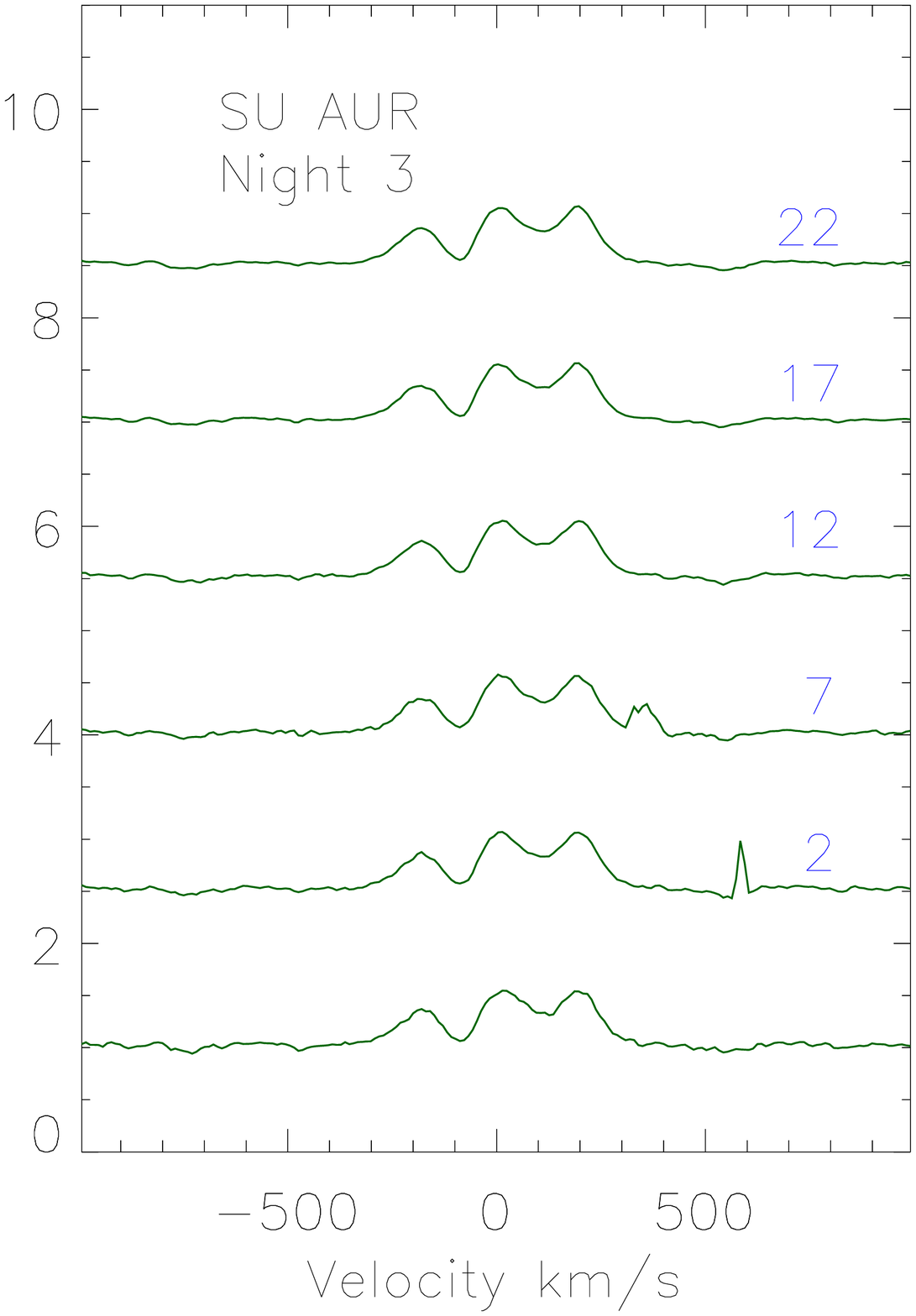} \\
\includegraphics[scale=0.22]{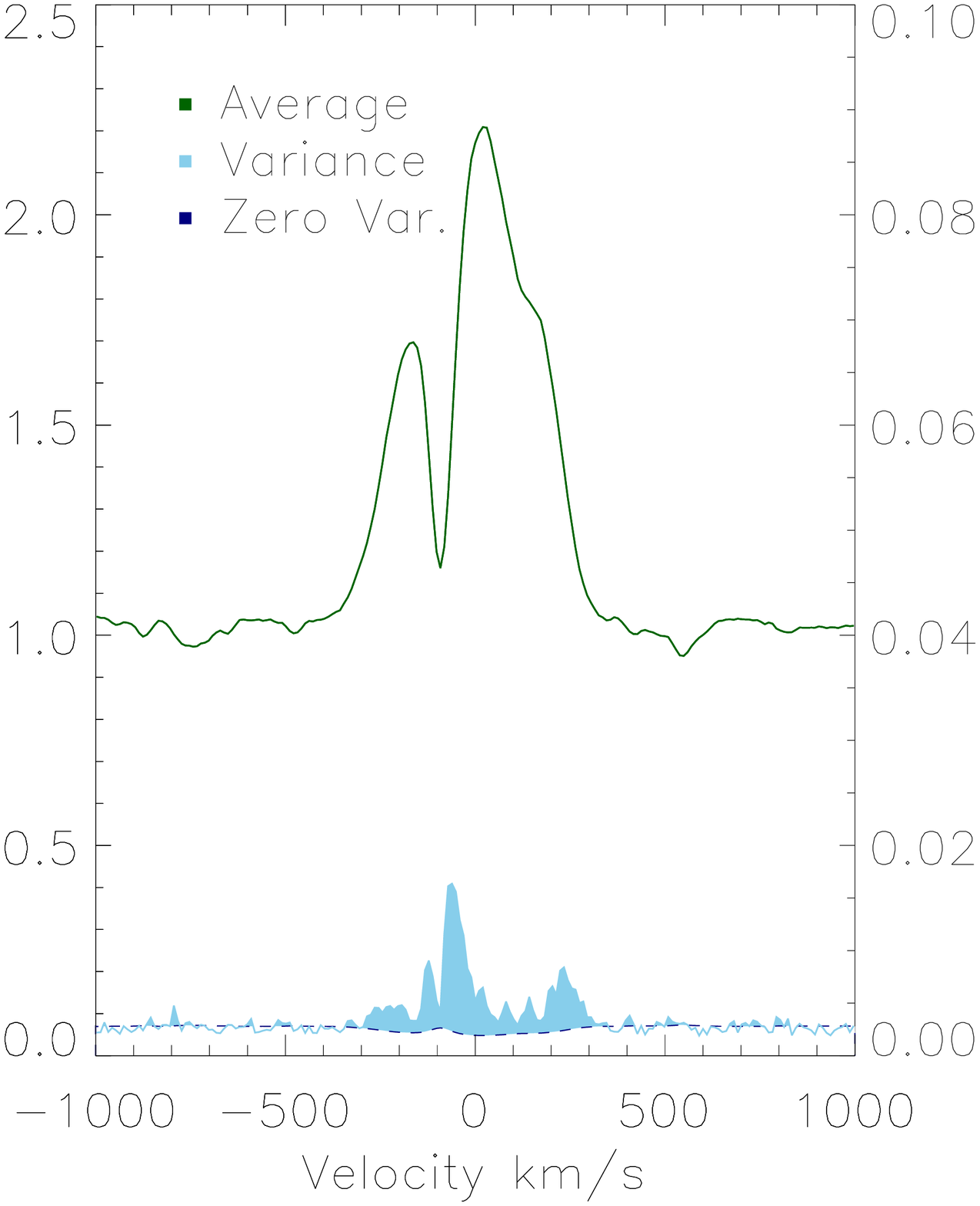} & \includegraphics[scale=0.22]{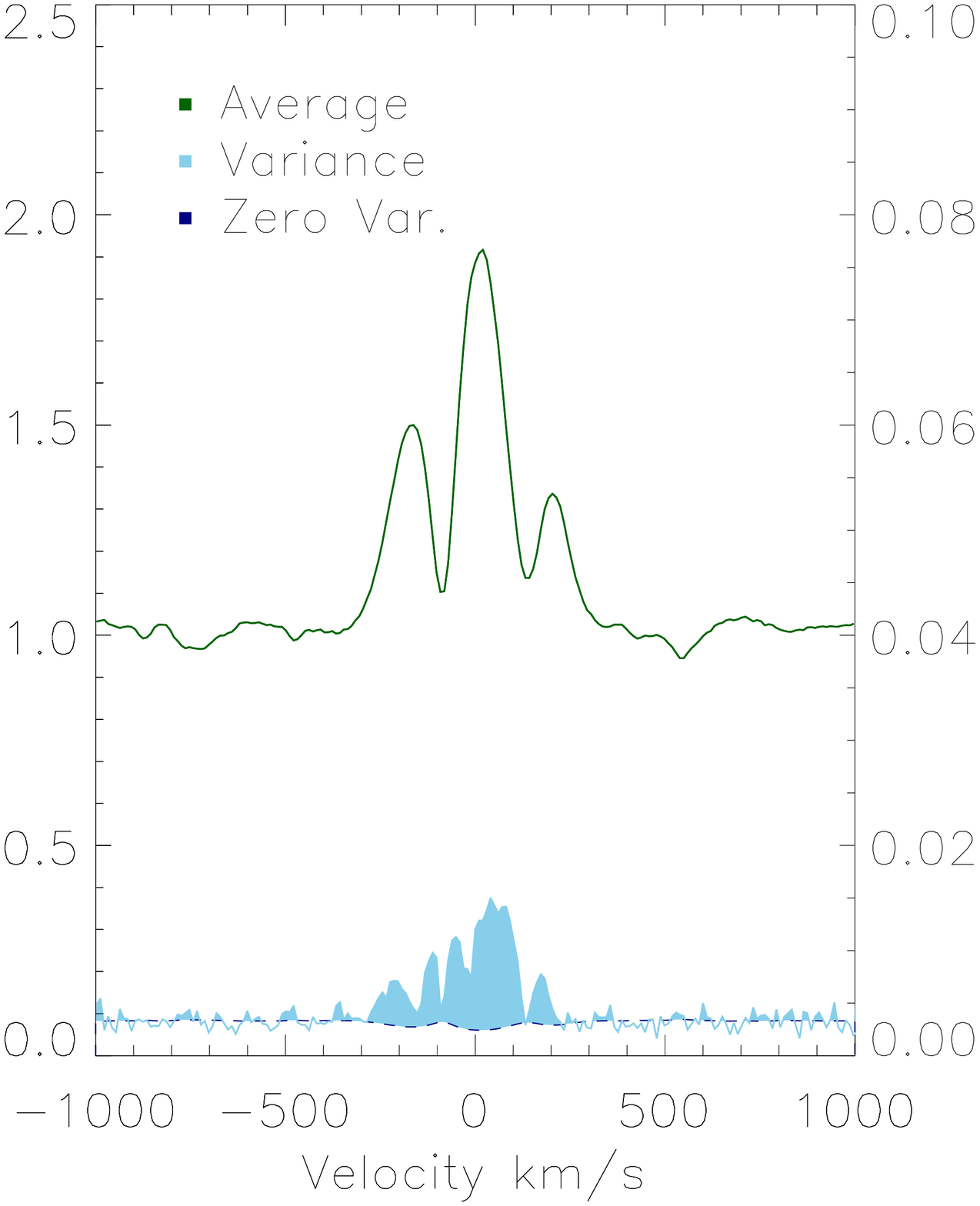} & \includegraphics[scale=0.22]{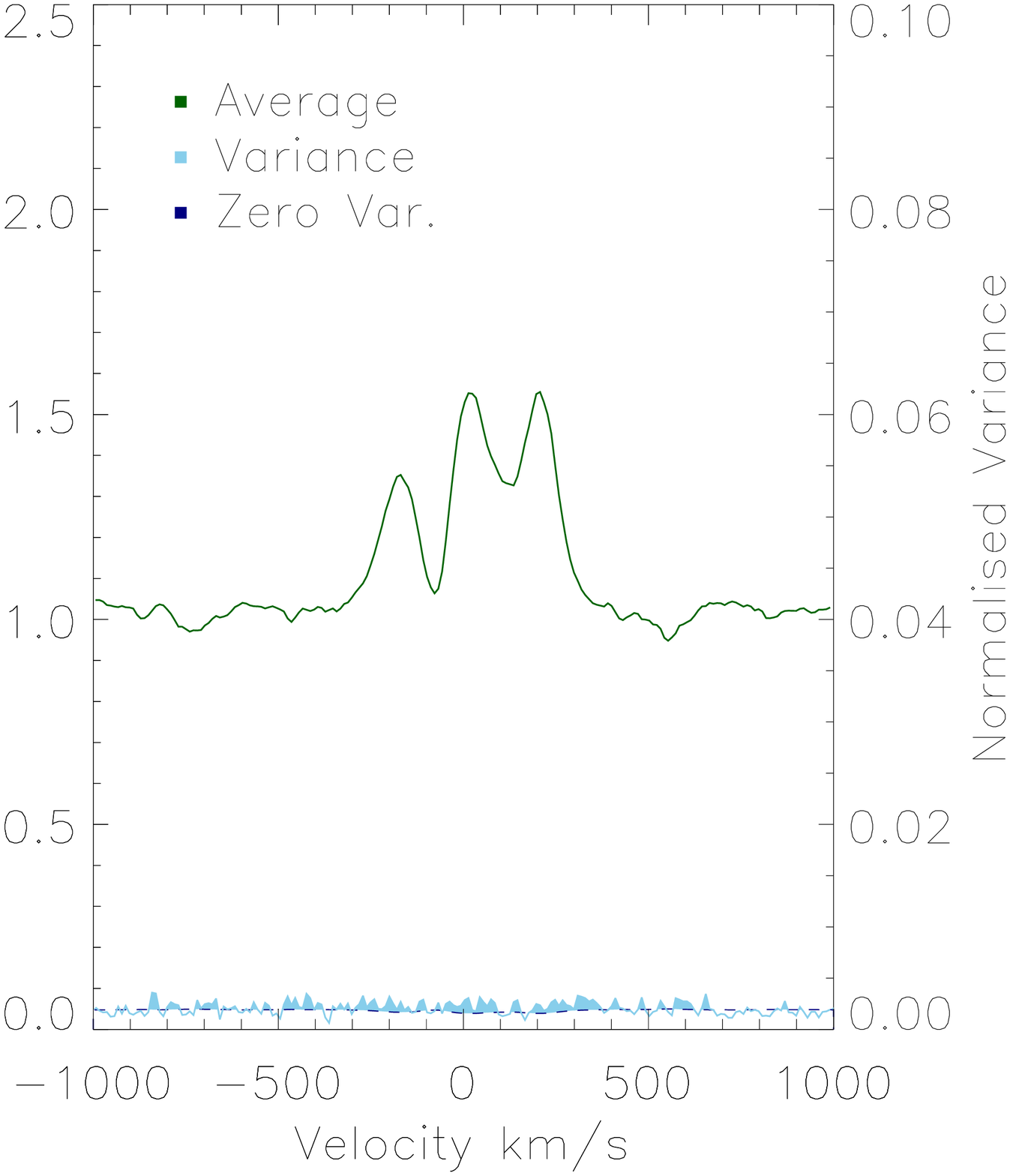} \\
\includegraphics[scale=0.22]{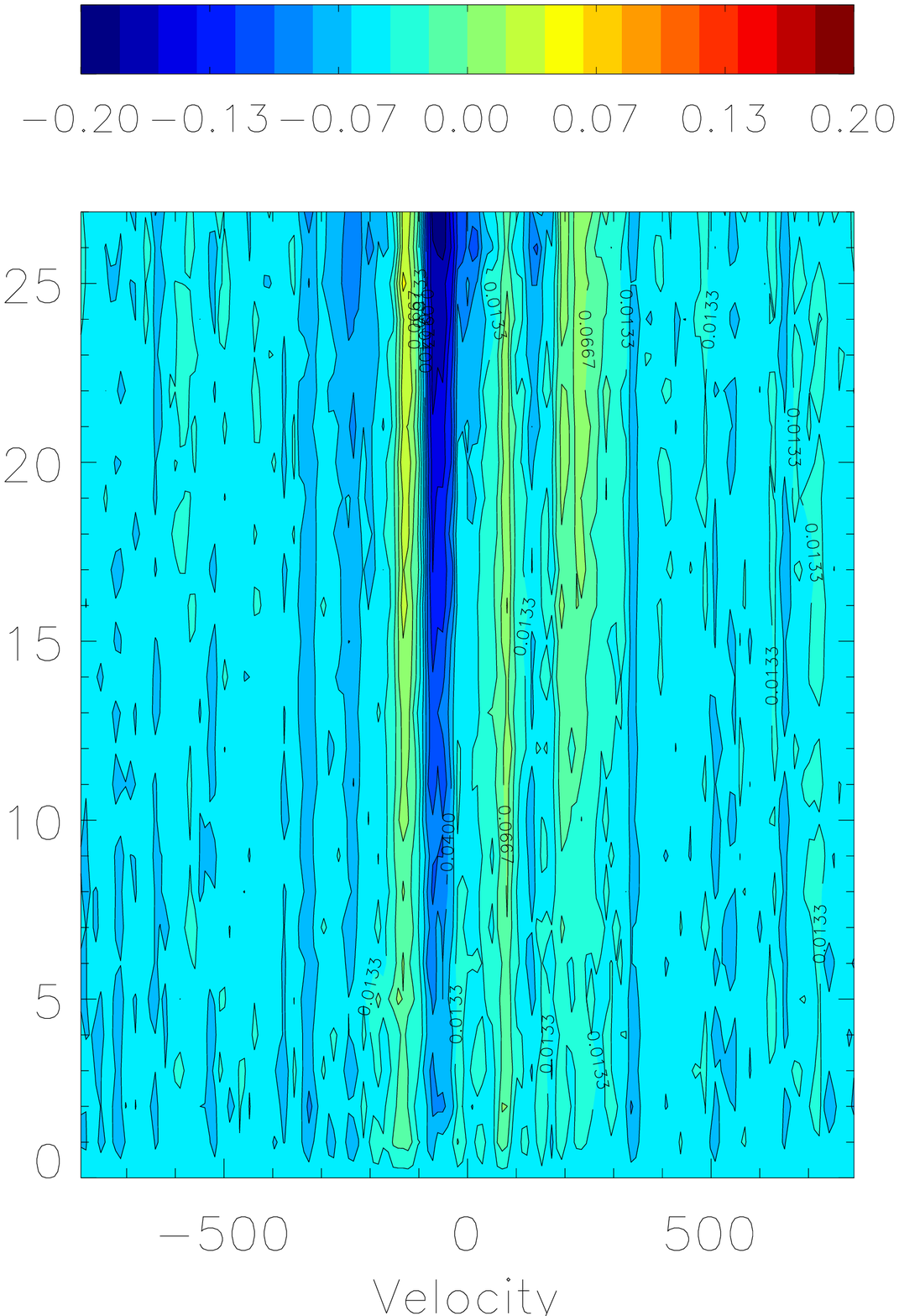} &\includegraphics[scale=0.22]{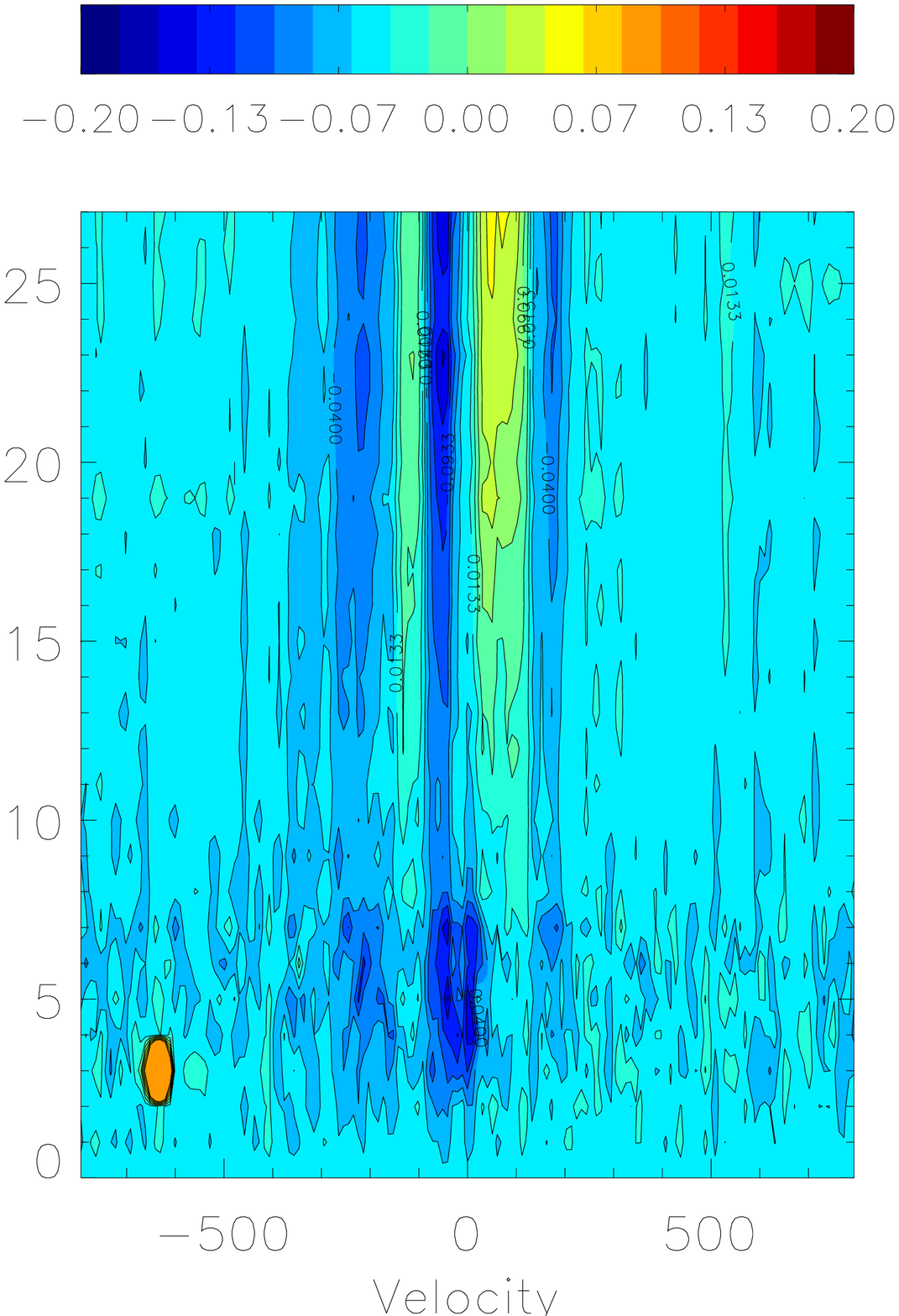} & \includegraphics[scale=0.22]{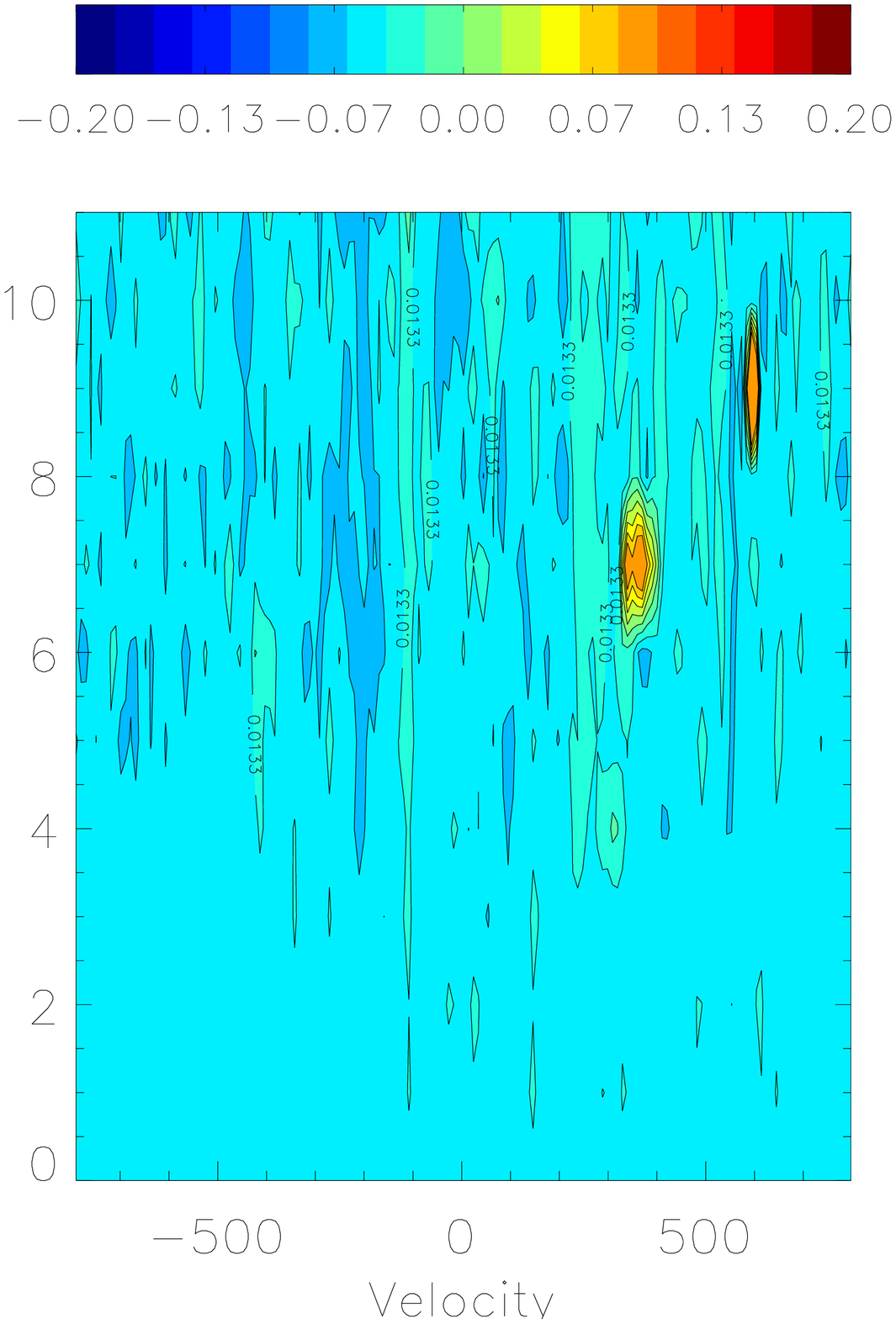} \\
\includegraphics[scale=0.22]{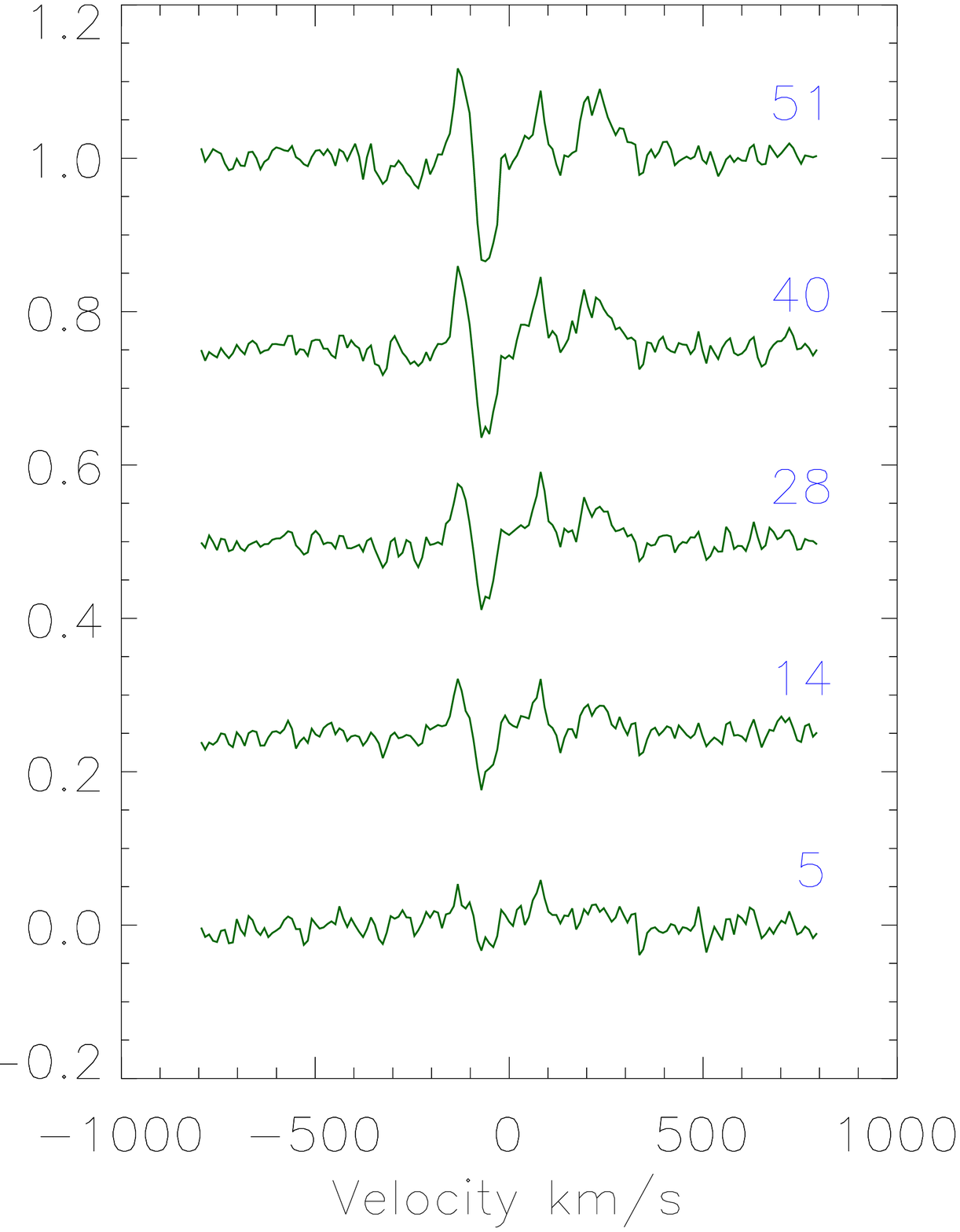} & \includegraphics[scale=0.22]{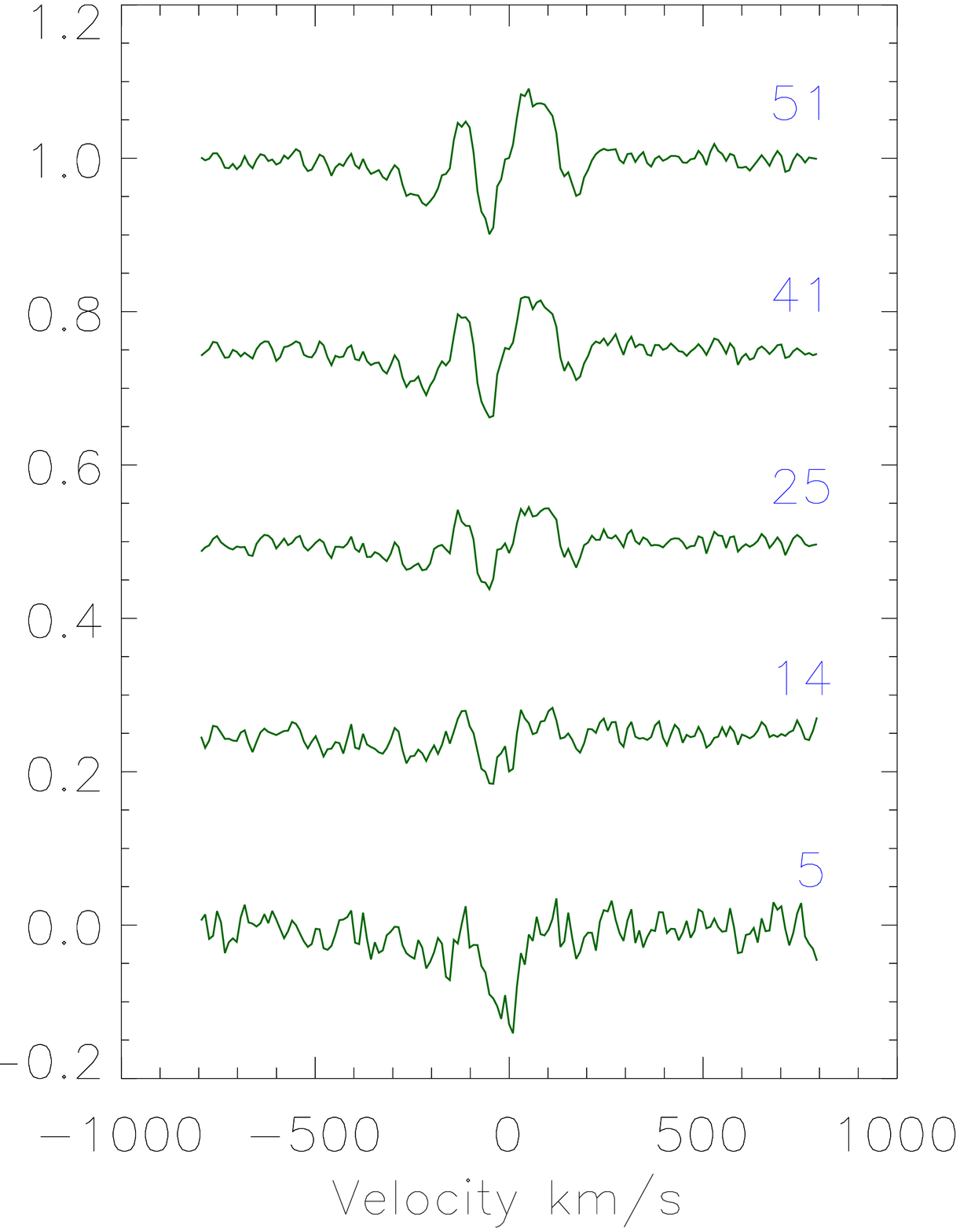} & \includegraphics[scale=0.22]{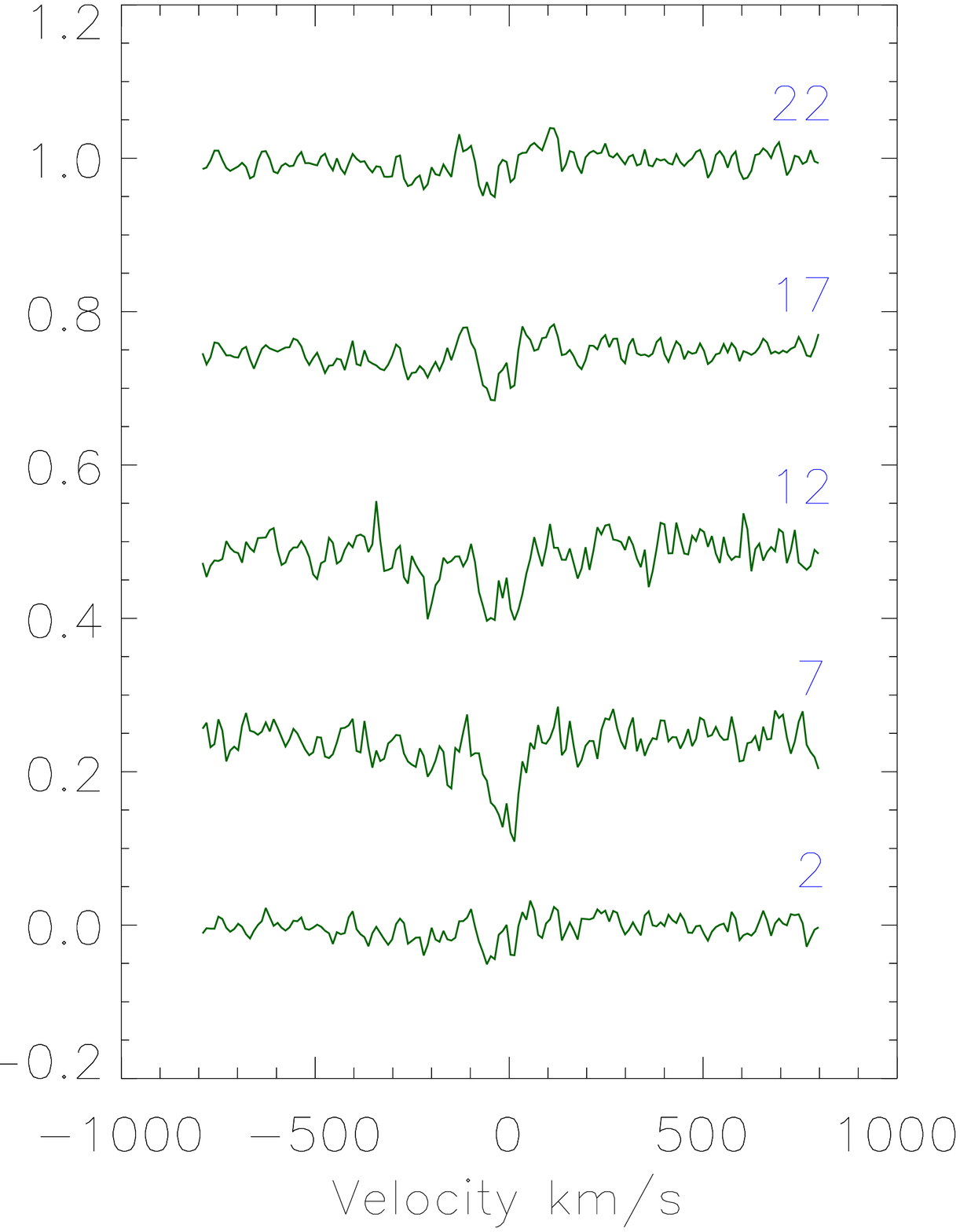} \\
\end{tabular}
\caption{SU Aur 2003 observations, Night 1 (left), Night 2 (middle) and Night 3 (right). }
\label{fig:SUAUR_plots_2003_1}
\end{figure*}

\noindent \textbf{ISIS H$\alpha$ Observations:} SU Aur shows one of the most complex profiles in the sample.
On the first night of the 2001 observations, the emission line is made up of four peaks (Fig.\,\ref{fig:SUAUR_plots_2001_1}). Two are very close together in the line centre, and two lie in either wing at about 200\,km\,s$^{-1}$. The peak in the red wing is the most prominent as it is slightly separated from the main emission. During the second night of 2001, this red shifted peak of emission shifts to lower wavelengths and merges with the main emission line. The small variations that do occur in 2001 seem to be concentrated on the red side of the line. 

On the first night of 2003 the line is composed of two emission peaks, and a blue-shifted central absorption  (Fig.\,\ref{fig:SUAUR_plots_2003_1}). The emission peak on the red side is asymmetric on the first night of observations and separates into a central and red-shifted emission peak for the second and third nights observations.
Over the course of the observations, no large variations are seen. On the first night, the changes seem to be concentrated in the blue side of the absorption feature (which is unusual for this sample) and also the red wing of the main emission. The H$\alpha$ EW falls by half from the first to the second night, from 80\AA~to 40\AA. On the second night of 2003 observations, it is the central peak in the emission line that grows relative to the other two. On the third night the profile shows unchanging steady emission.

\noindent \textbf{Previous H$\alpha$ Observations:} \citet{1995ApJ...449..341J} found evidence for unsteady accretion in SU Aur in the form of red displaced absorption feature in H$\beta$ with a period of 3 days, (where each observation was taken within a few hours or few days of each other). The H$\alpha$ profile shows some similarities to ours. It has two overlying absorption features than grow and weaken over course of observations. However, in the case of Johns \& Basri, the blue absorption feature is always the stronger, and below the continuum. In the ISIS observations it is the stronger in two out of three nights in 2003, on the second night it is equal to the depth of the red absorption. Again in 1977 a similar profile to what we observe on the first night of our observations is seen \citet{1979ApJS...41..369S}.

In 1992, SU Aur was observed again showing a very similar profile to our first nights observations in 2003 \citep{1992ASPC...26..441J}. In this case the blue-shifted absorption is shown to vary over the rotation period, and is attributed to an outflow originating in the inner disc and co-rotating with the star.

\subsection{T Tau}

\begin{figure*}
\centering
\begin{tabular}{ccc}
\includegraphics[scale=0.22]{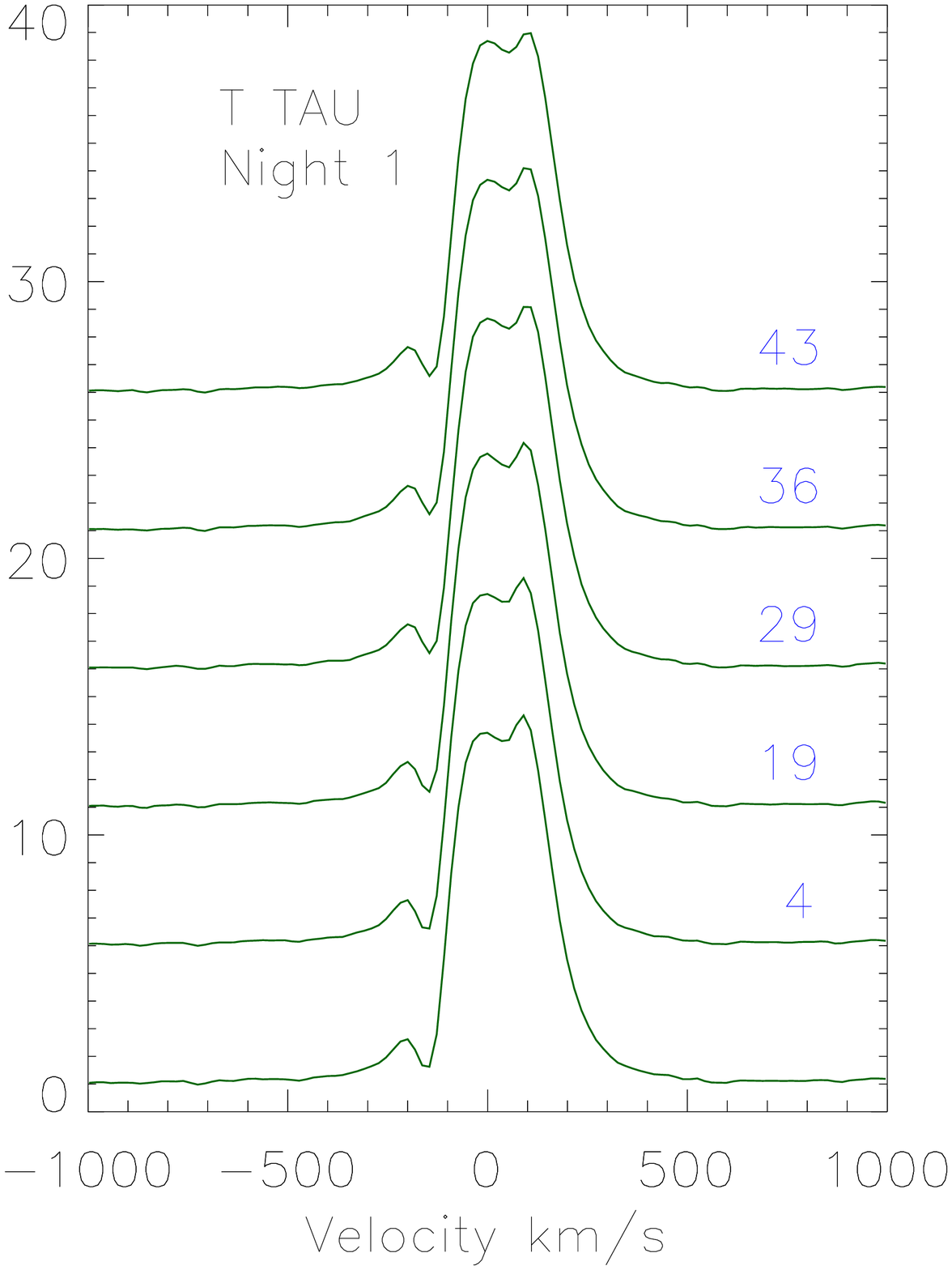} & \includegraphics[scale=0.22]{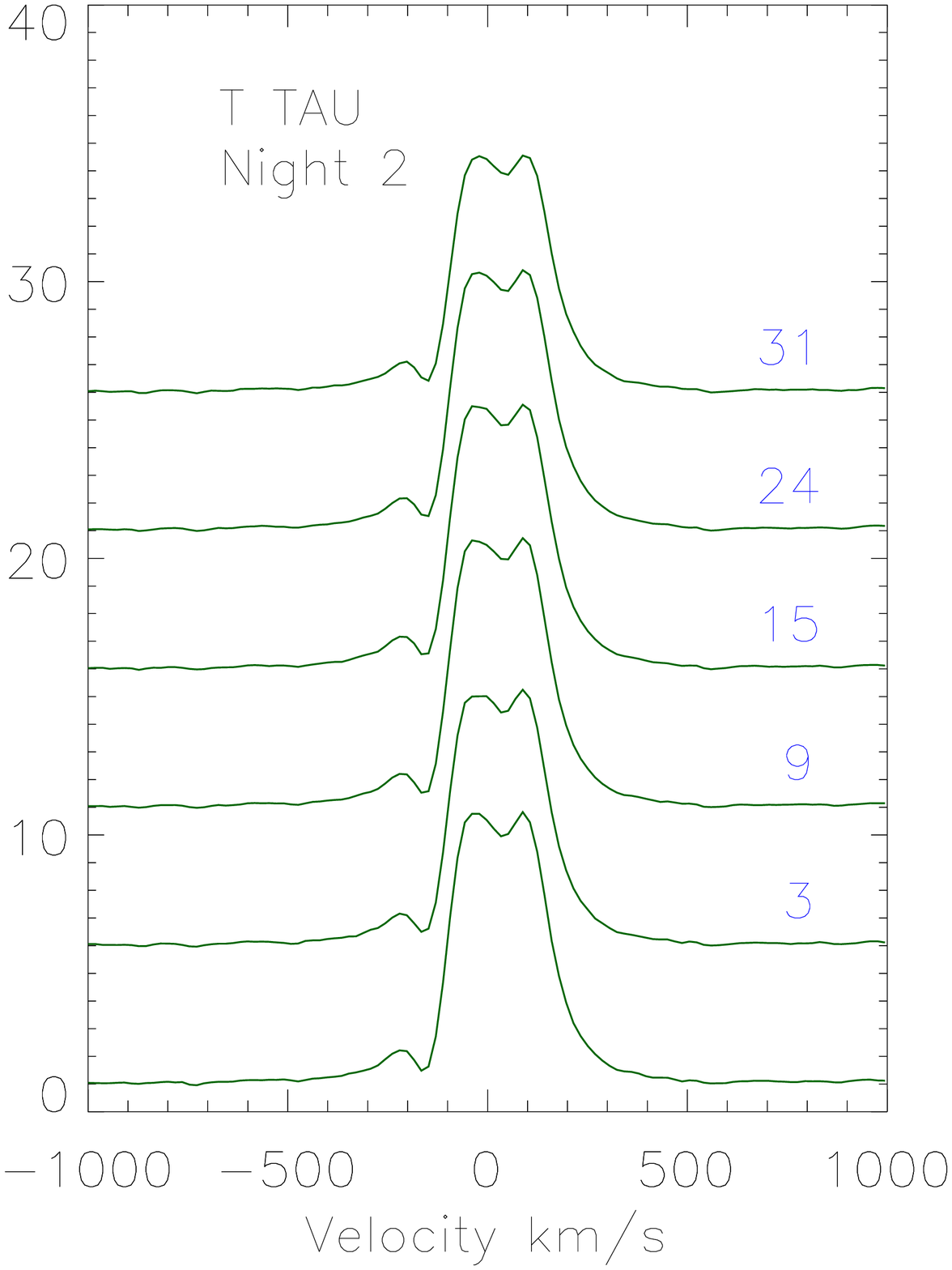} & \includegraphics[scale=0.22]{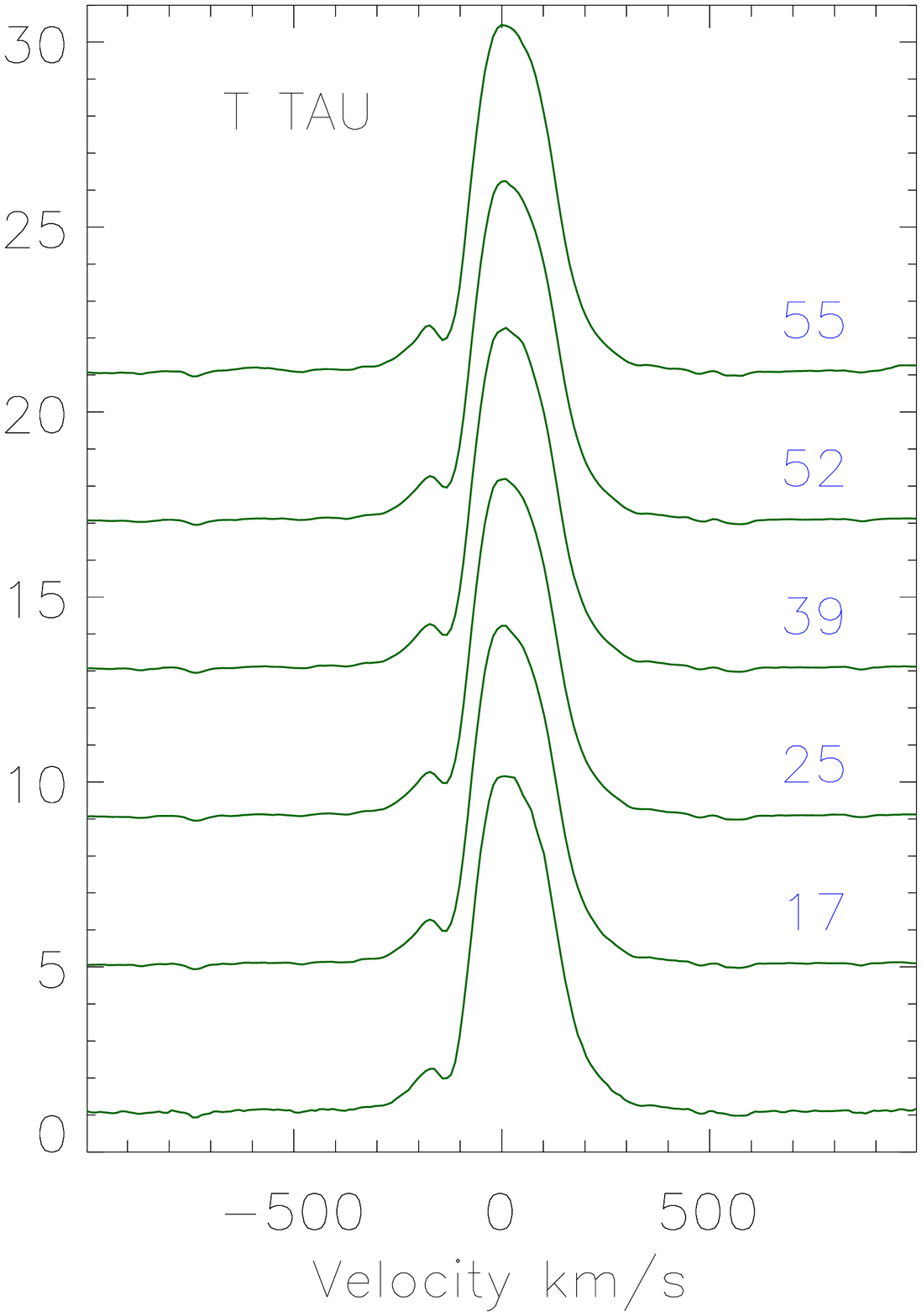} \\
\includegraphics[scale=0.22]{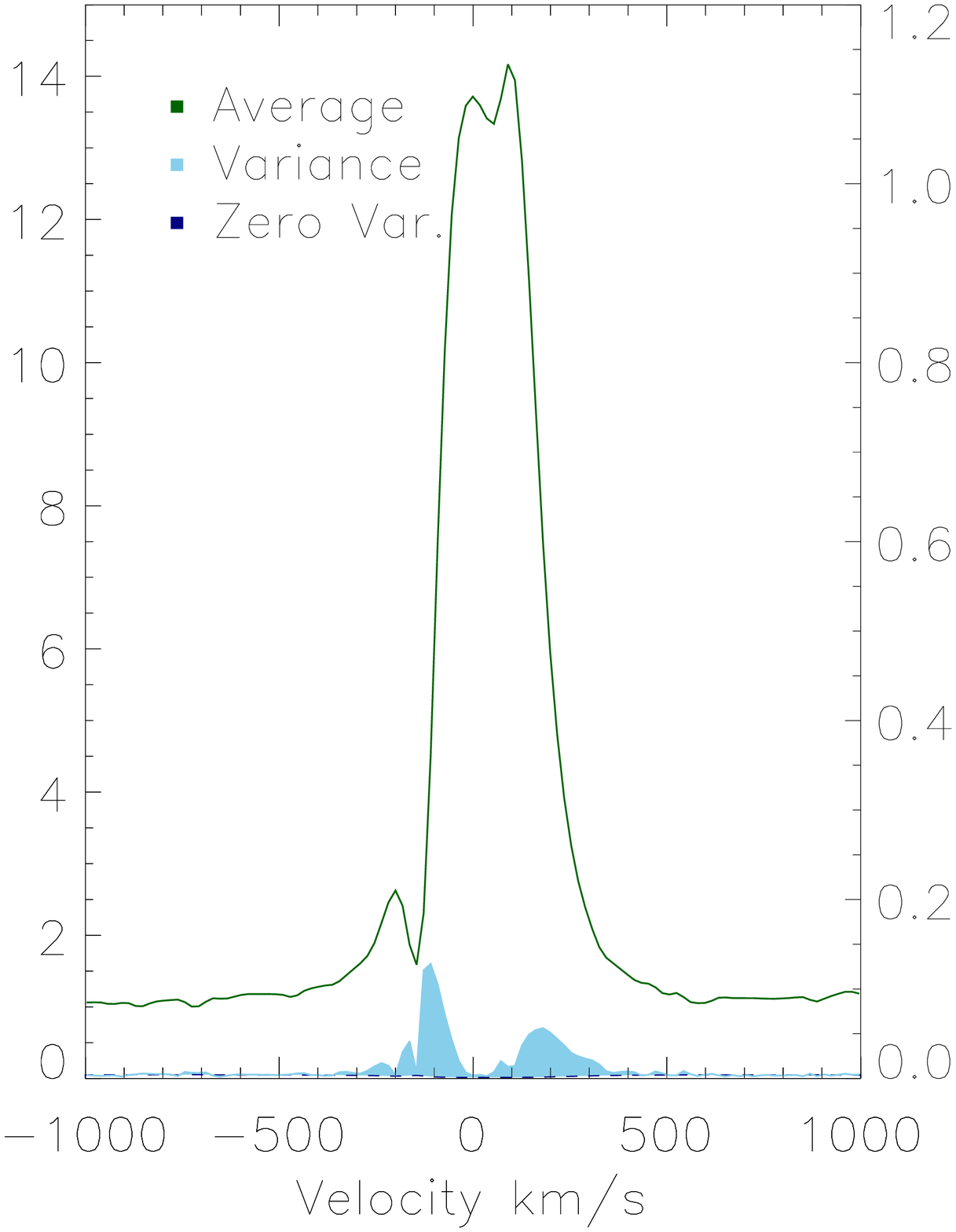} & \includegraphics[scale=0.22]{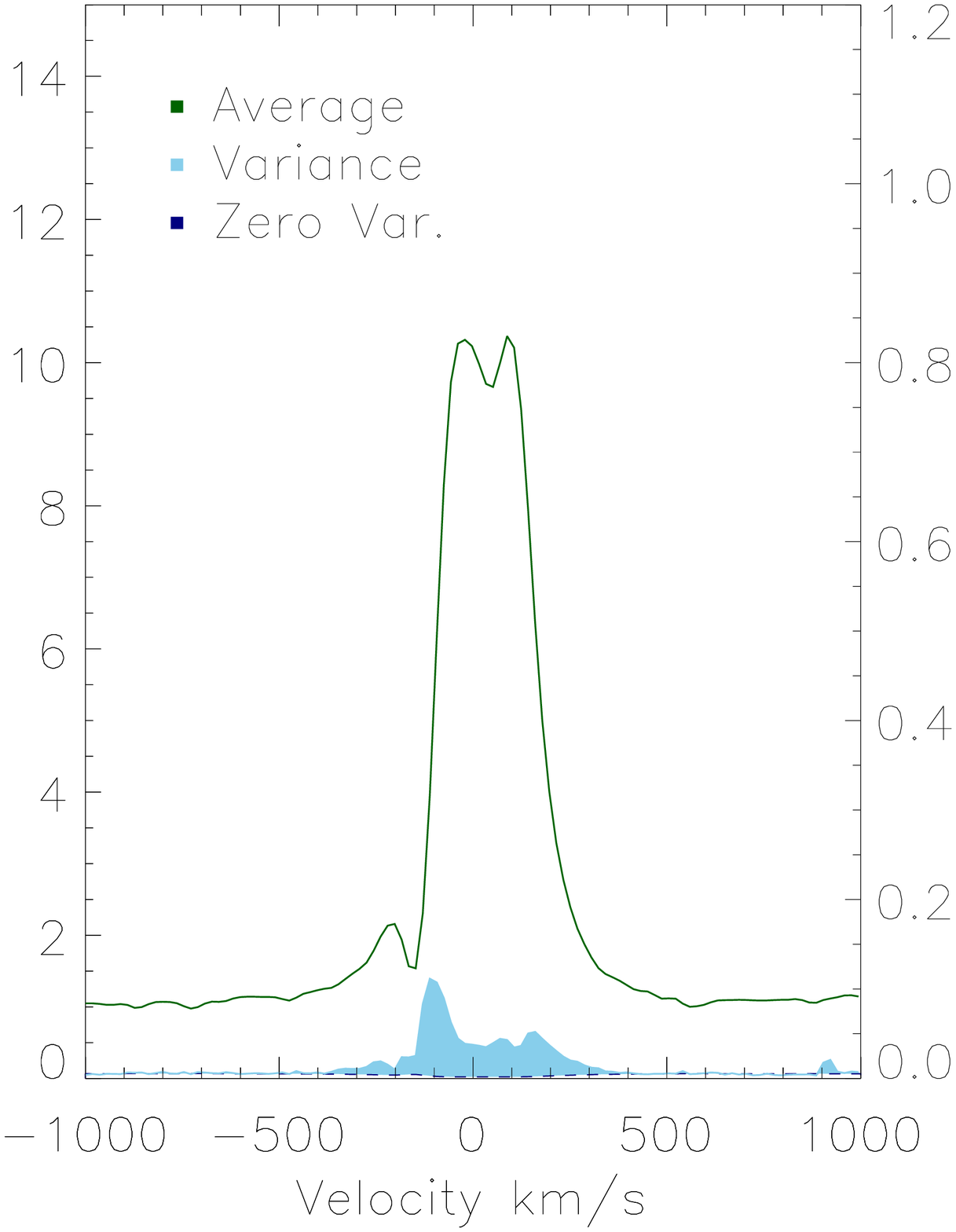} & \includegraphics[scale=0.22]{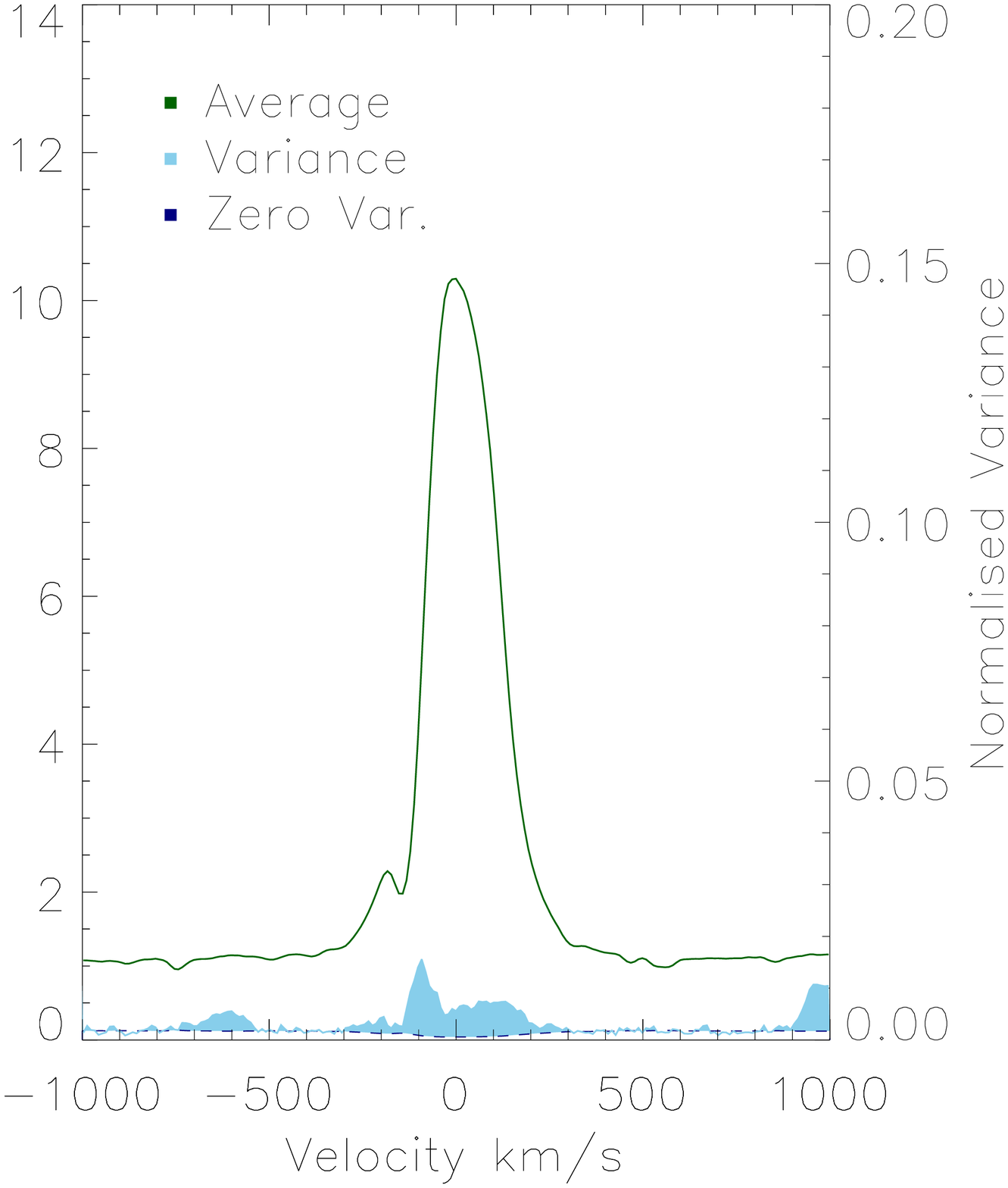} \\
\includegraphics[scale=0.22]{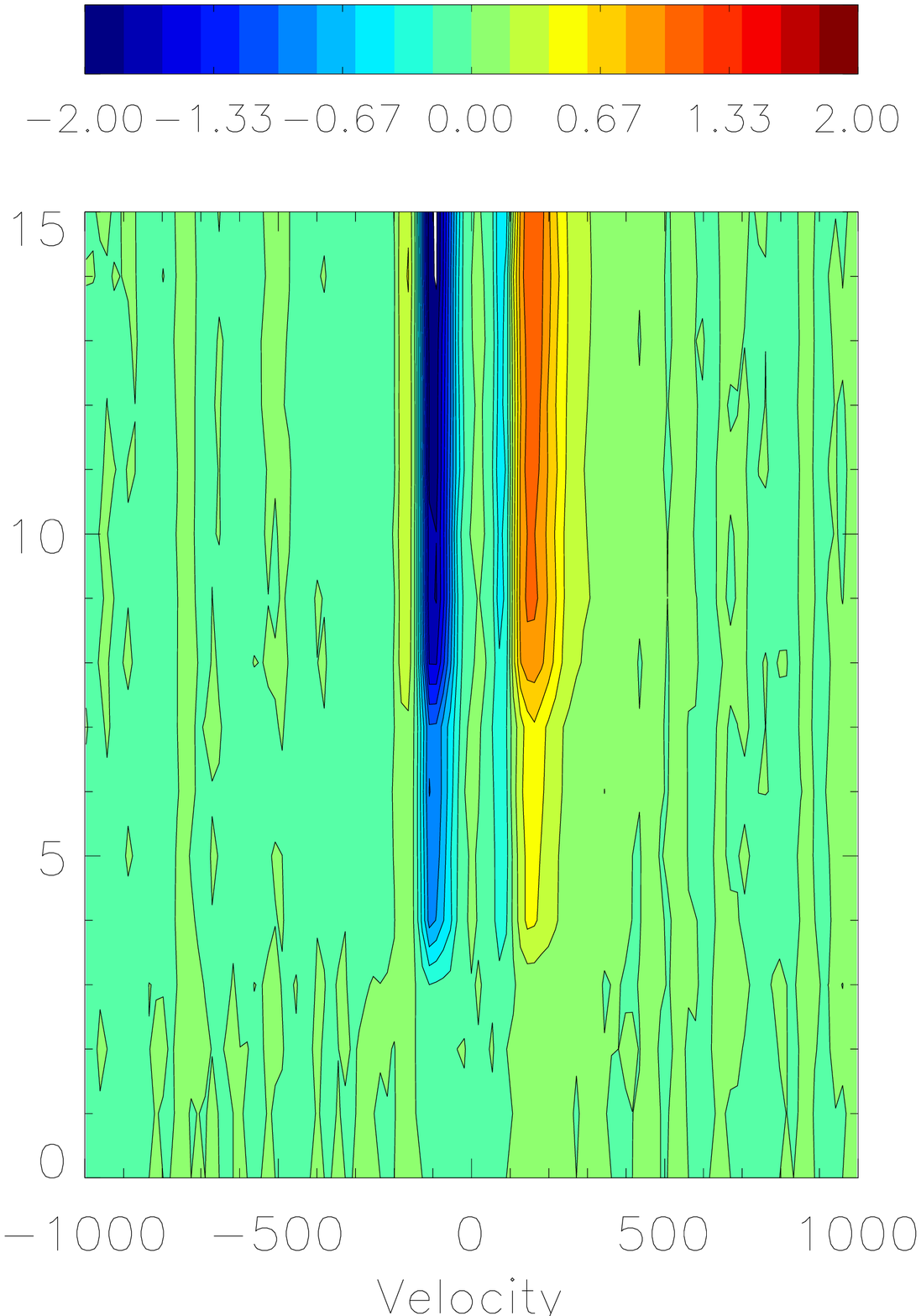}  & \includegraphics[scale=0.22]{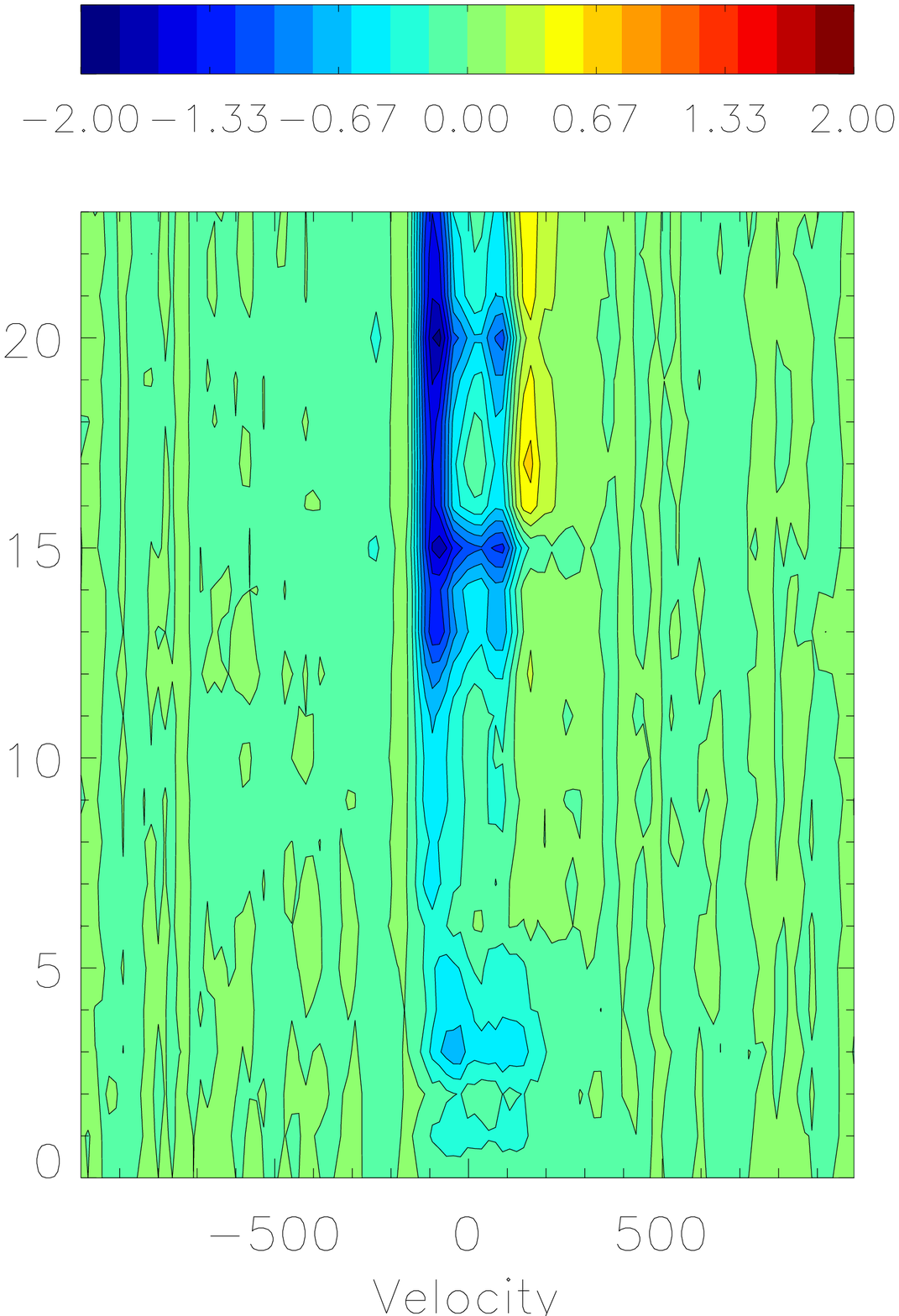} & \includegraphics[scale=0.22]{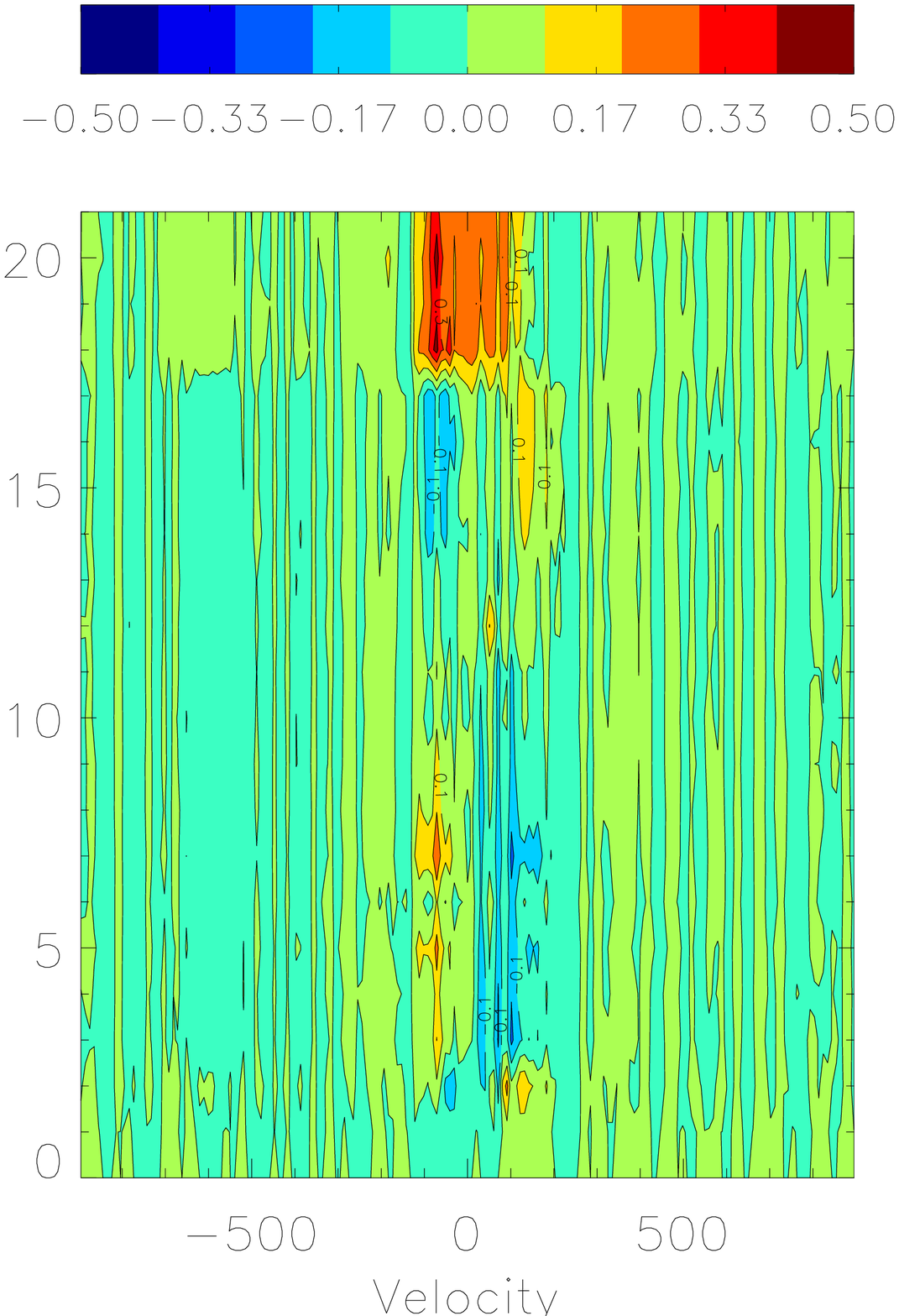} \\
\includegraphics[scale=0.22]{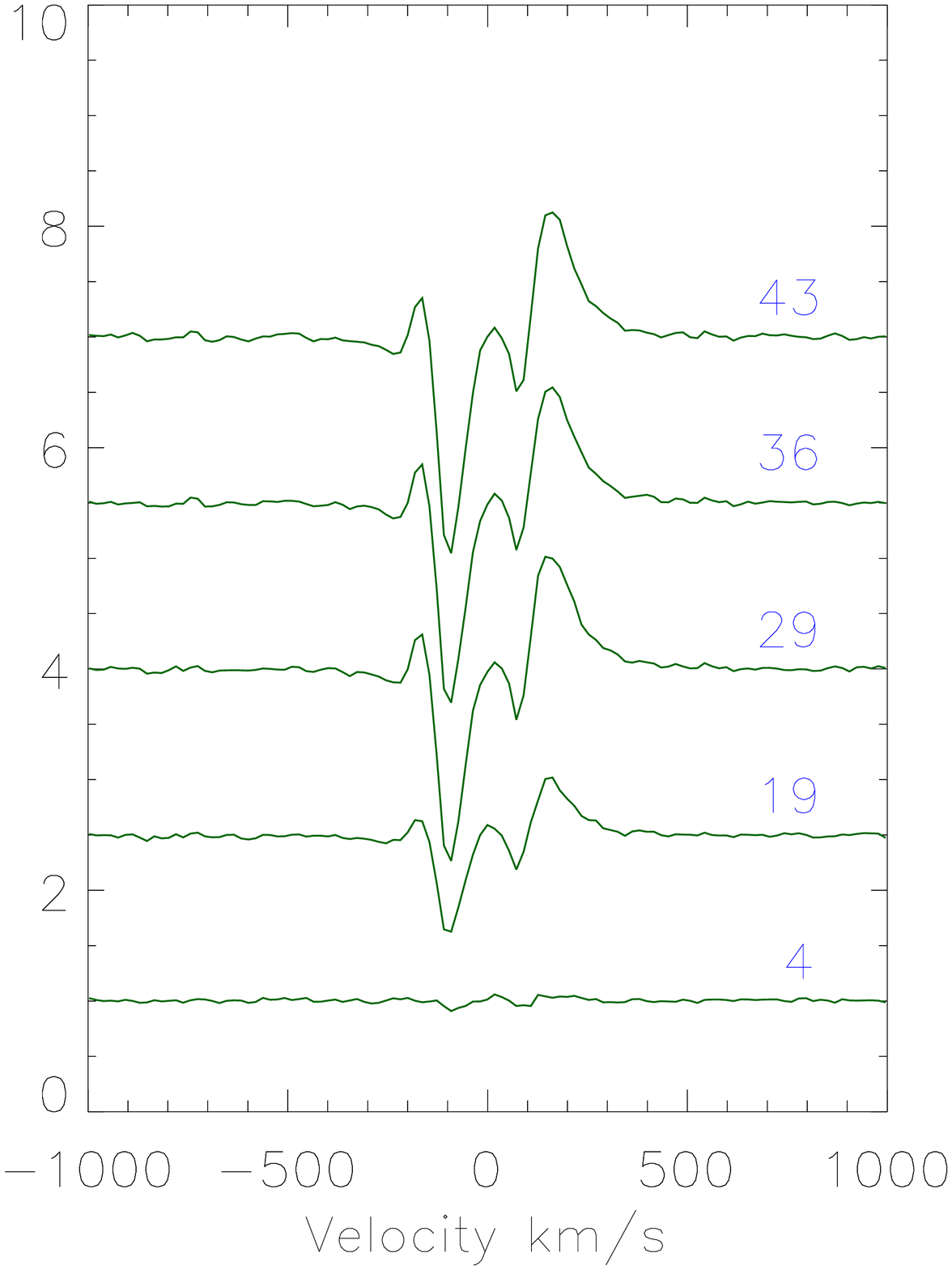} & \includegraphics[scale=0.22]{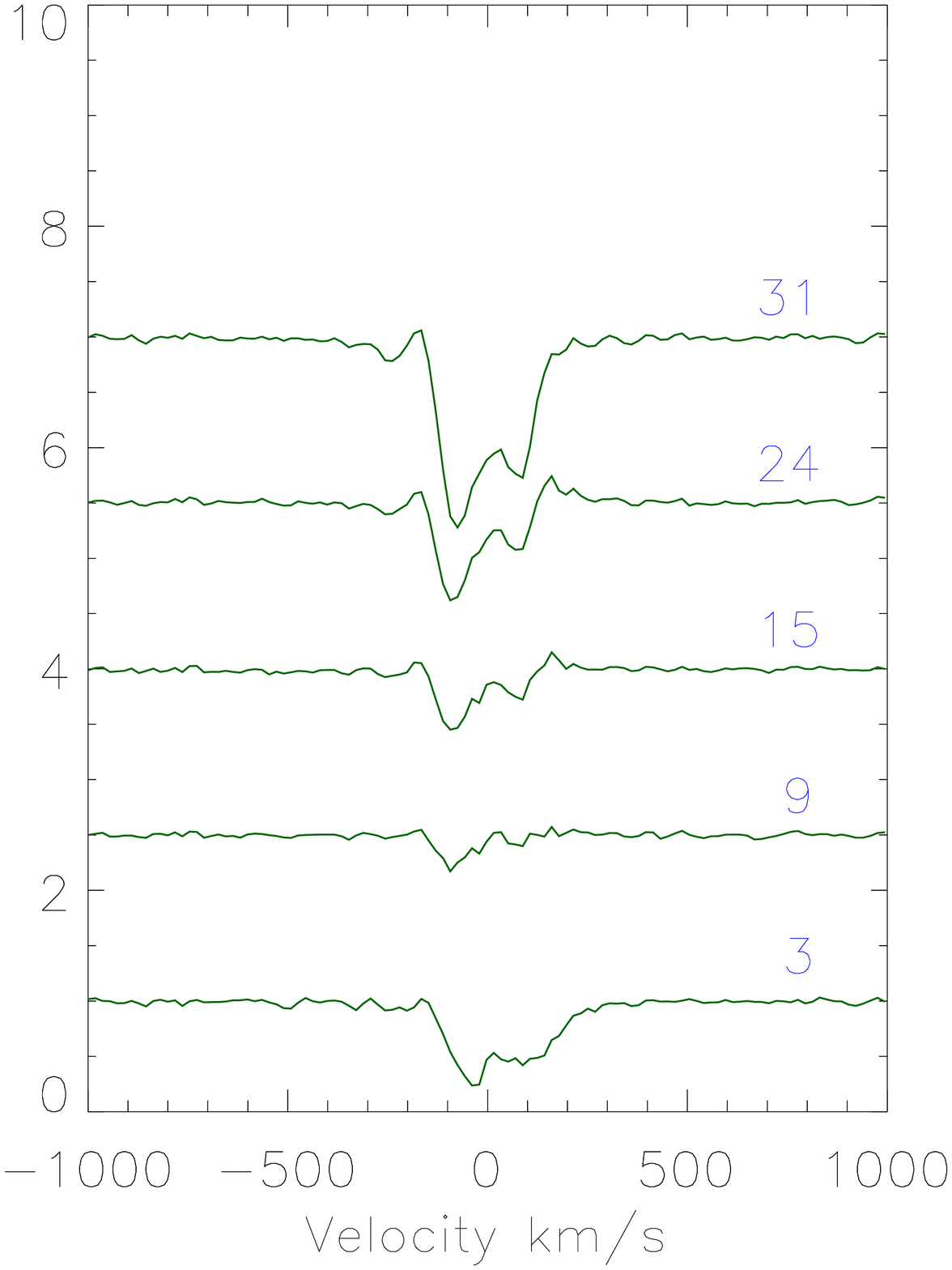} & \includegraphics[scale=0.22]{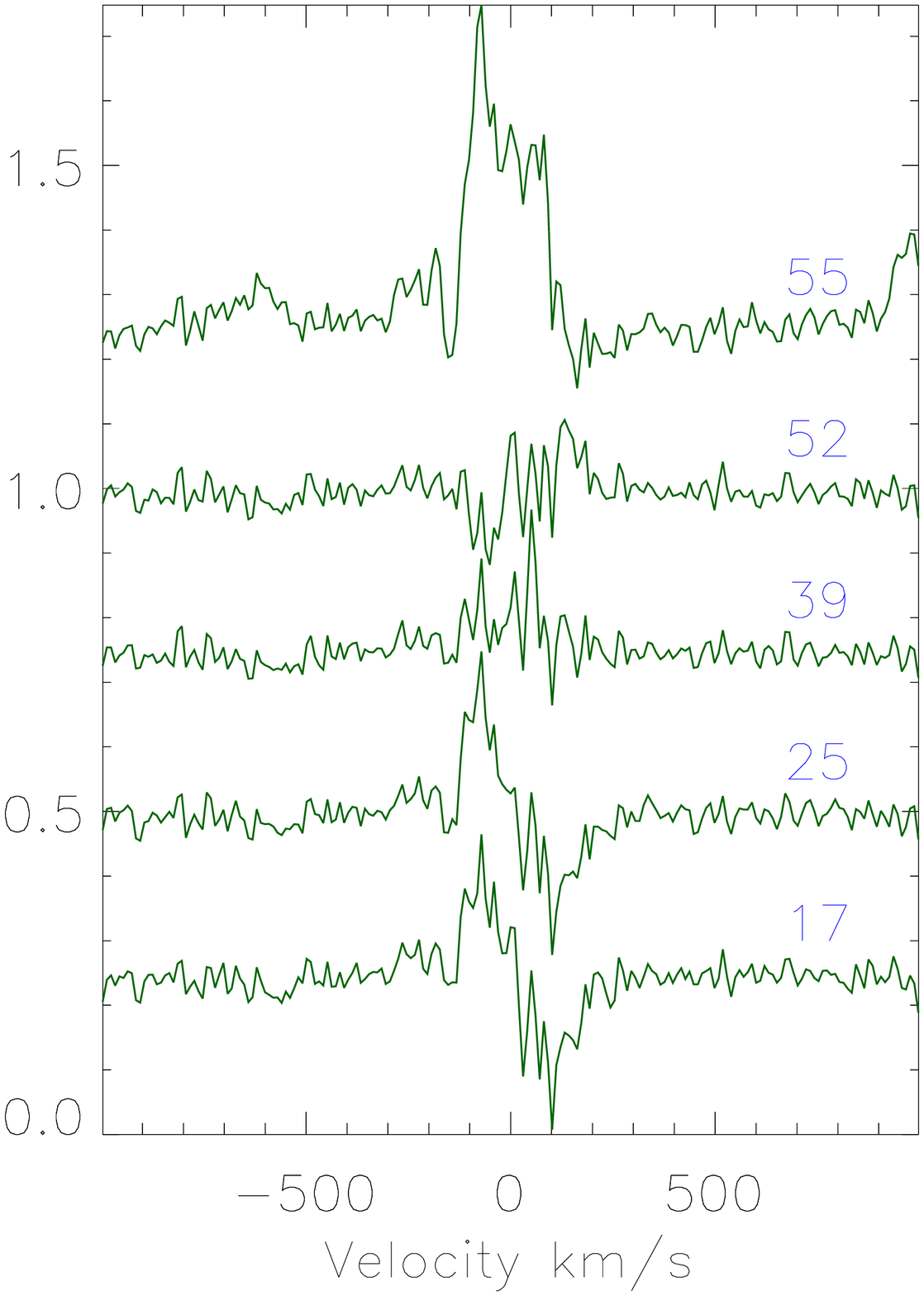}\\
\end{tabular} 
\caption{T Tau 2001 observations, Night 1 (left) and Night 2 (middle). 2003 observations (right)  }
\label{fig:TTAU_plots_2001_1}
\end{figure*}

\textbf{Stellar Properties:} The T Tauri system is one of the best studied pre-main sequence systems. Even though it gave the name to the class of T Tauri stars, it is not a typical one. It is a triple system, where T Tau N is visually bright (V $\sim$ 10), with a strongly IR variable companion, T Tau S which is now known to be a binary. These observations are of the T Tau N, the optical visible T Tauri star in the system. It has  stellar mass of 2.0 M$_{\odot}$ and a radius of 3.3 R$_{\odot}$ \citep{1988ApJ...330..350B}. A period of 2.8 days has been found for T Tau N \citep{1986ApJ...310L..71H}. X-ray emission has been observed in T Tau, however \citet{2007A&A...468..529G} suggest the accretion processes suppresses the emission, and it is lower that what is expected.

\noindent \textbf{Disc Properties:} T Tau has a disc mass of 0.008 M$_{\odot}$ \citep{2005ApJ...631.1134A}. \citet{2002ApJ...566.1124A} report a system inclination angle of 29$^{\circ}$ based on interferometric observations.

\noindent \textbf{Accretion:}  There have been a number of accretion rate estimates found for T Tau.  \citet{1989ApJ...341..340B} report an accretion rate of 1.1\,x\,10$^{-8}$M$_{\odot}$yr$^{-1}$, in rough agreement with that found by \citet{2004AJ....128.1294C} and \citet{2001ApJ...556..265W}.

\noindent \textbf{Outflows:}  T Tau has been observed in the radio, with emission consistent with both an extended outflow and non-thermal emission \citep{1993ApJ...403L..43P}. \citep{1994AJ....107.1461S} found the flux to be consistent with a mass loss rate of 3.7\,x\,10$^{-8}$M$_{\odot}$yr$^{-1}$.

\noindent \textbf{ISIS H$\alpha$ Observations:} In 2001, T Tau's emission profile is composed of a central emission peak with a small absorption feature in the line centre and with a larger blue-shifted absorption feature in the wing (Fig.\,\ref{fig:TTAU_plots_2001_1}). During both nights, the changes are mostly in the blue wing, very close to the centre of the blue-shifted absorption feature. A smaller strengthening of the emission in the red wing occurs at the same time.

In 2003, T Tau shows weaker emission, and also the central absorption feature has disappeared (Fig.\,\ref{fig:TTAU_plots_2001_1}). The variance profile again shows a peak near the blue shifted absorption feature, but the small change that does occur takes the form of a strengthening across the line.

\noindent \textbf{Previous H$\alpha$ Observations:} Multiple observations were taken of T Tau's H$\alpha$ emission in 1976/77. These observations show a similar profile to what is observed in the ISIS sample, but with a small red-shifted emission feature, that sometimes appears as a knee in the red wing, and sometimes as a distinguished emission feature. During these observations the EW was seen to change by 10\% over a three day period \citep{1979ApJS...41..369S}.

\subsection{UX Tau}

\begin{figure}
\centering
\begin{tabular}{cc}
\includegraphics[scale=0.22]{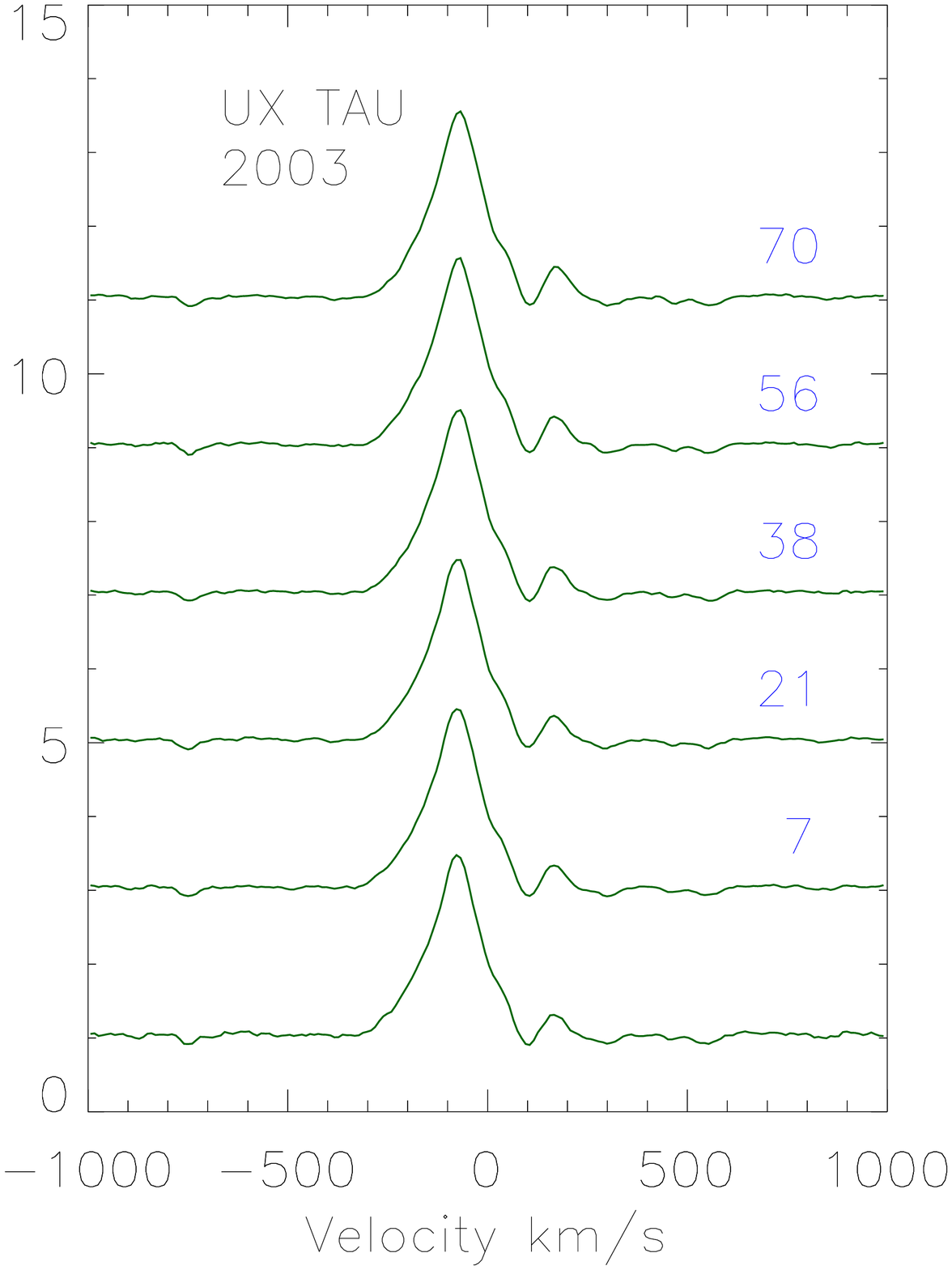} & \includegraphics[scale=0.22]{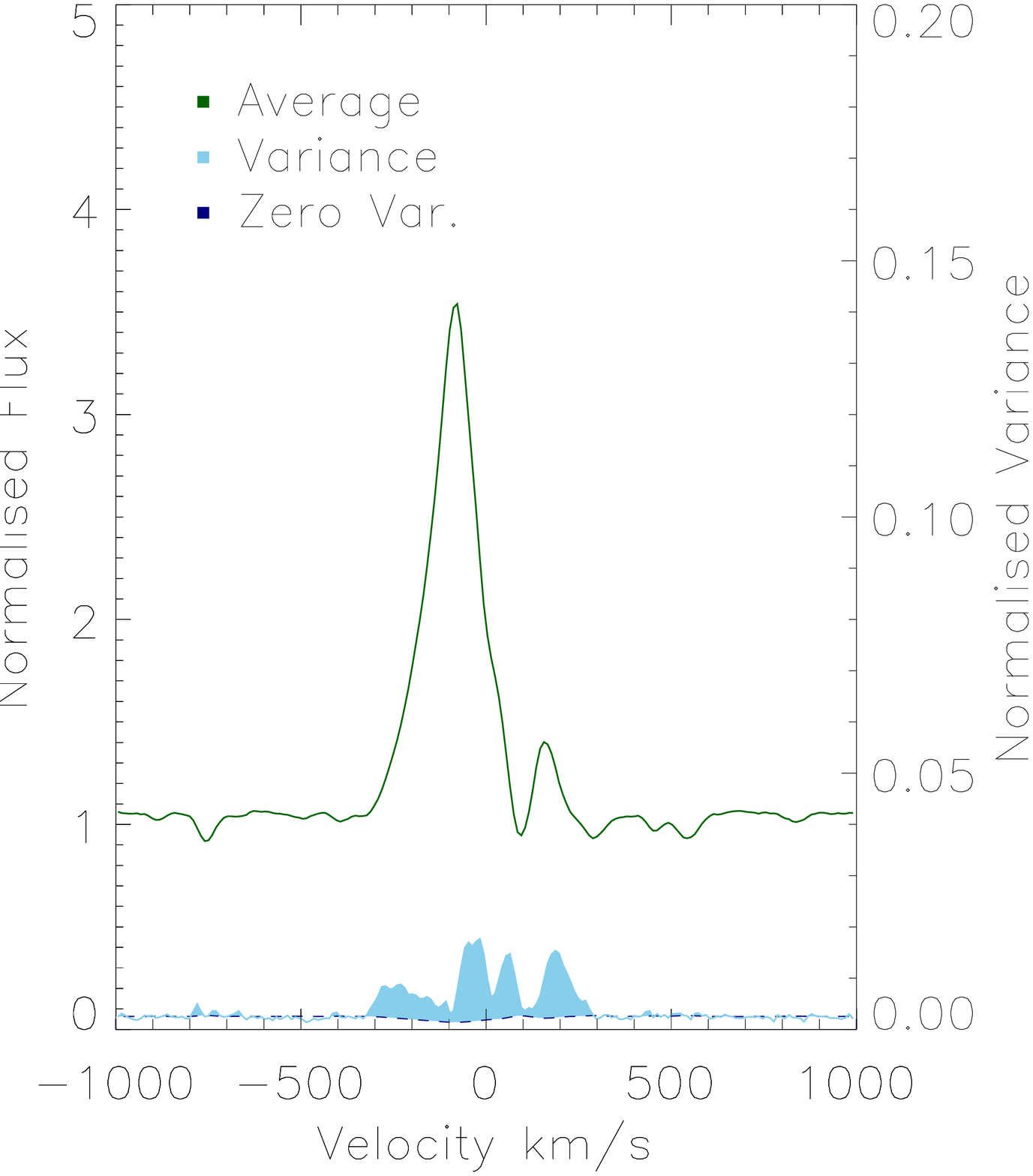} \\
\includegraphics[scale=0.22]{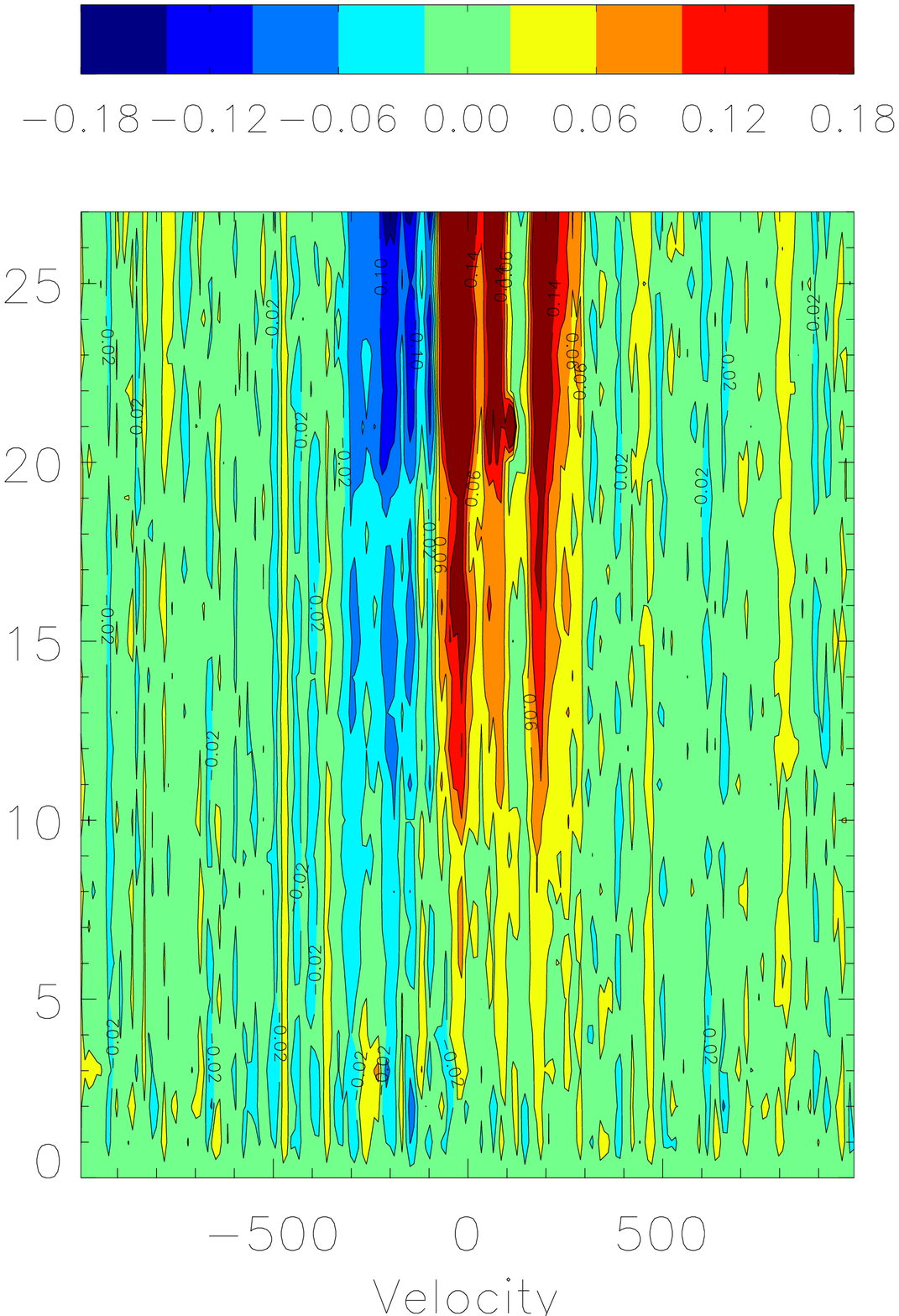}  & \includegraphics[scale=0.22]{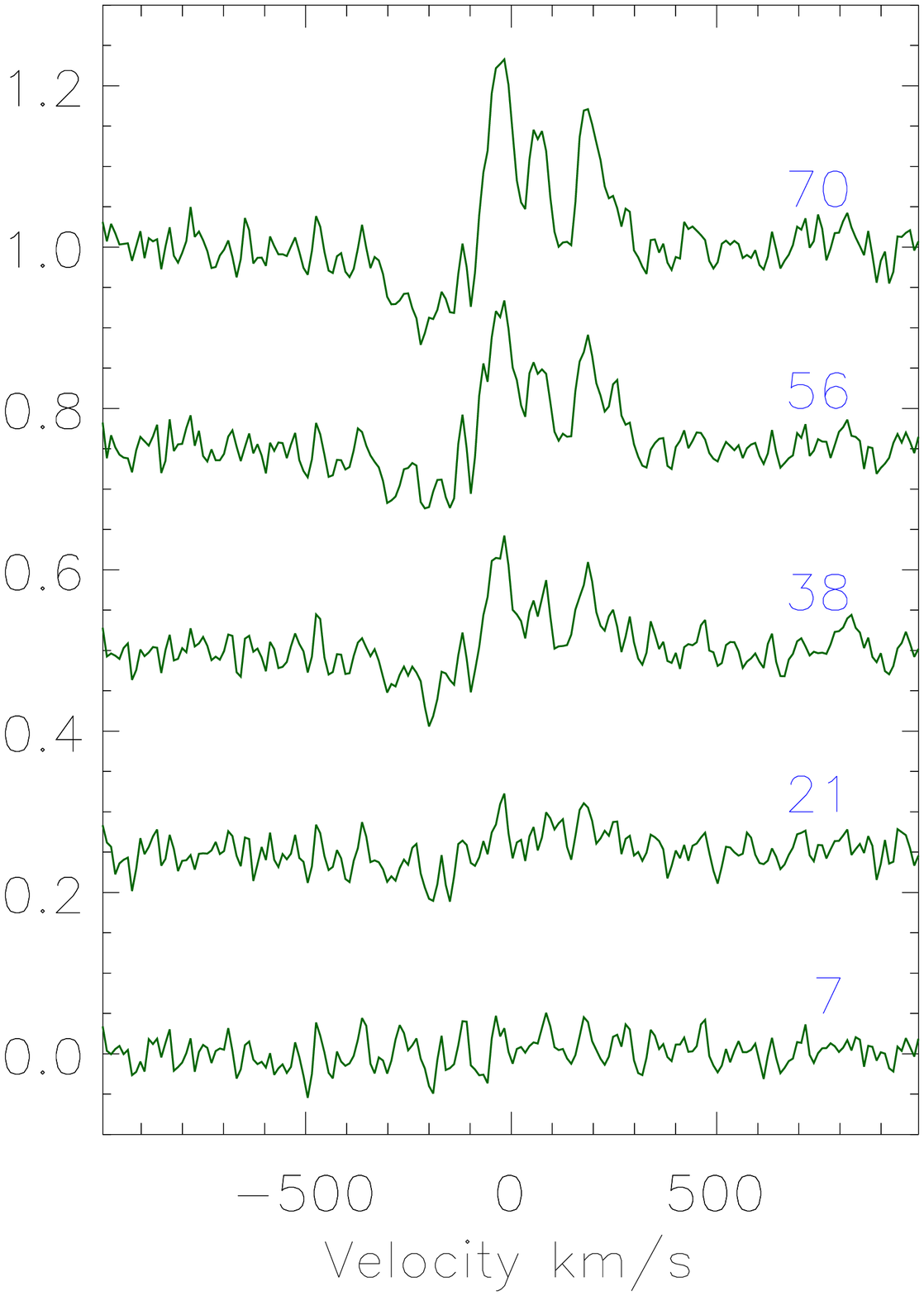} \\
\end{tabular}
\caption{ UX Tau 2003 observations.}
\label{fig:UXTAU_plots}
\end{figure}

\textbf{Stellar Properties:} UX Tau is a multiple system \citep{1979AJ.....84.1872J}, where the primary star UX Tau A is separated from  the binary UX Tau b by 5.86" and UX Tau C by 2.63". The primary has a mass of 1.3 M$_{\odot}$ and a radius of  2.0 R$_{\odot}$ \citep{2009ApJ...704..531K,2007ApJ...670L.135E}.

\noindent \textbf{Disc Properties:} The circumstellar disc has been classified as a transitional disc \citep{2010ApJ...710..265P} and takes the form of an optical thin inner disc separated by a gap from an optically thick outer disc, where the inner wall lies at 0.21 AU and the outer wall at 30 AU. The disc has a mass of 0.005 M$\odot$ \citep{2005ApJ...631.1134A} and a tilt angle of $\sim$ 46$^{\circ}$ $\pm$ 2 $^{\circ}$. 

\noindent \textbf{ISIS H$\alpha$ Observations:} UX Tau shows a reverse P-Cgyni profile in these observations (Fig.\,\ref{fig:UXTAU_plots}). Across the hour observation, both emission peaks grow in size. The H$\alpha$ EW gradually increases by $\sim$ 10\AA~(12\%). The time-series of H$\alpha$ EW measurements show it increases in a wave like manner. The changes in the profile occur on the blue side of absorption (closer to main peak), and in the wings. This is the only profile with red absorption in this sample.

\noindent \textbf{Previous H$\alpha$ Observations:} Over the course of this work no previous H$\alpha$ observations of this target were found.

\subsection{V773 Tau}

\begin{figure}
\centering
\begin{tabular}{cc}
\includegraphics[scale=0.22]{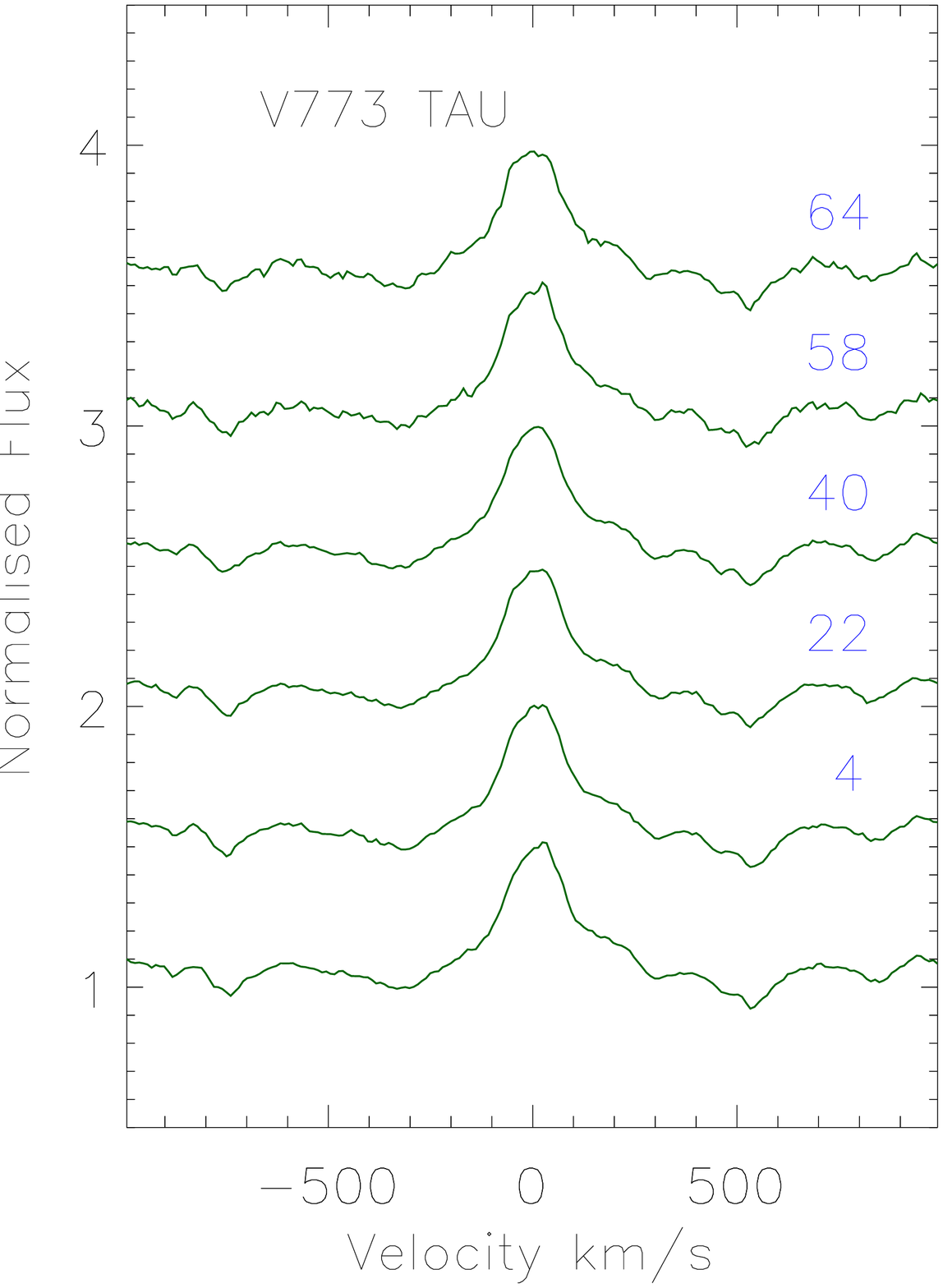} & \includegraphics[scale=0.22]{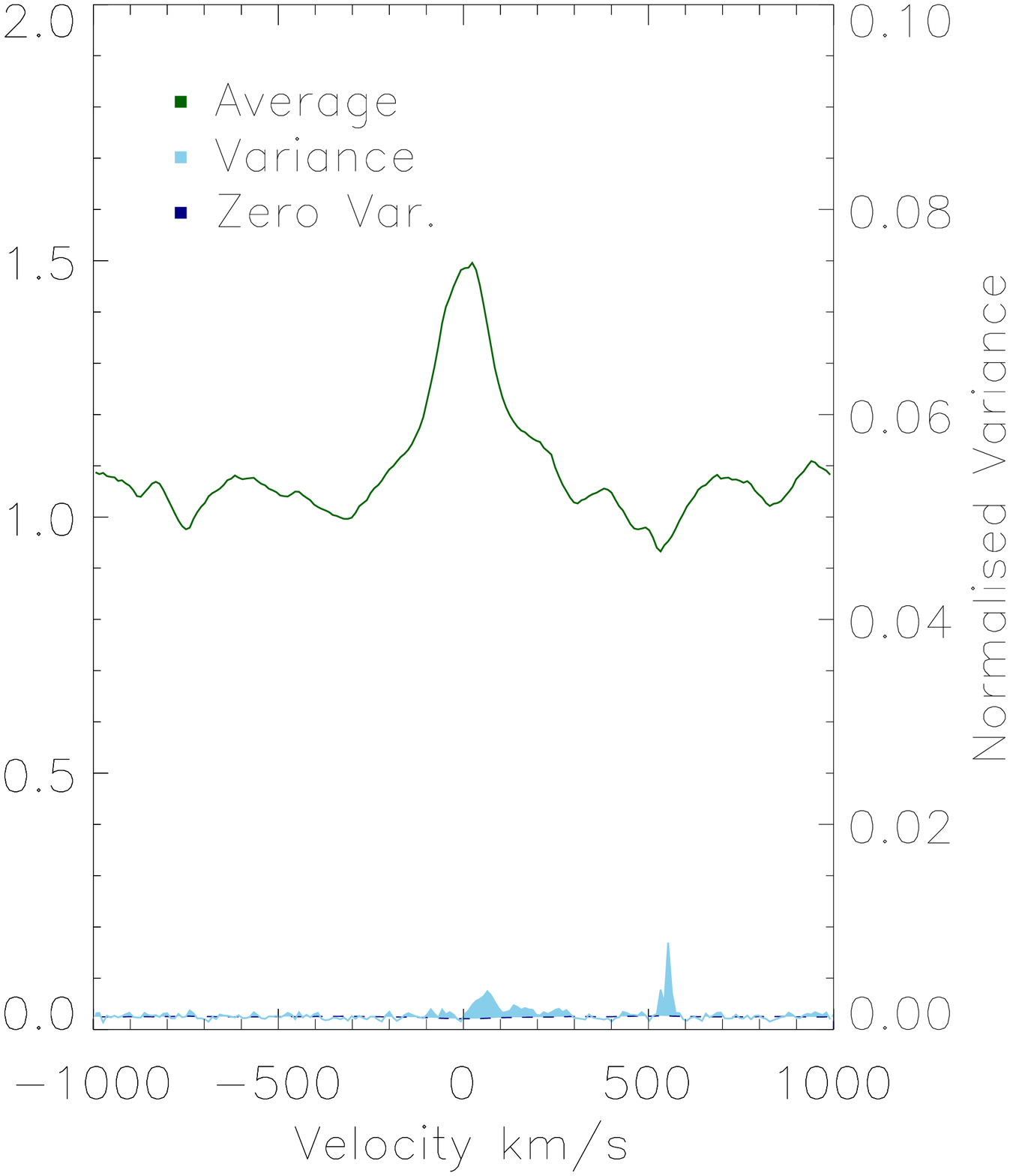}\\
\includegraphics[scale=0.22]{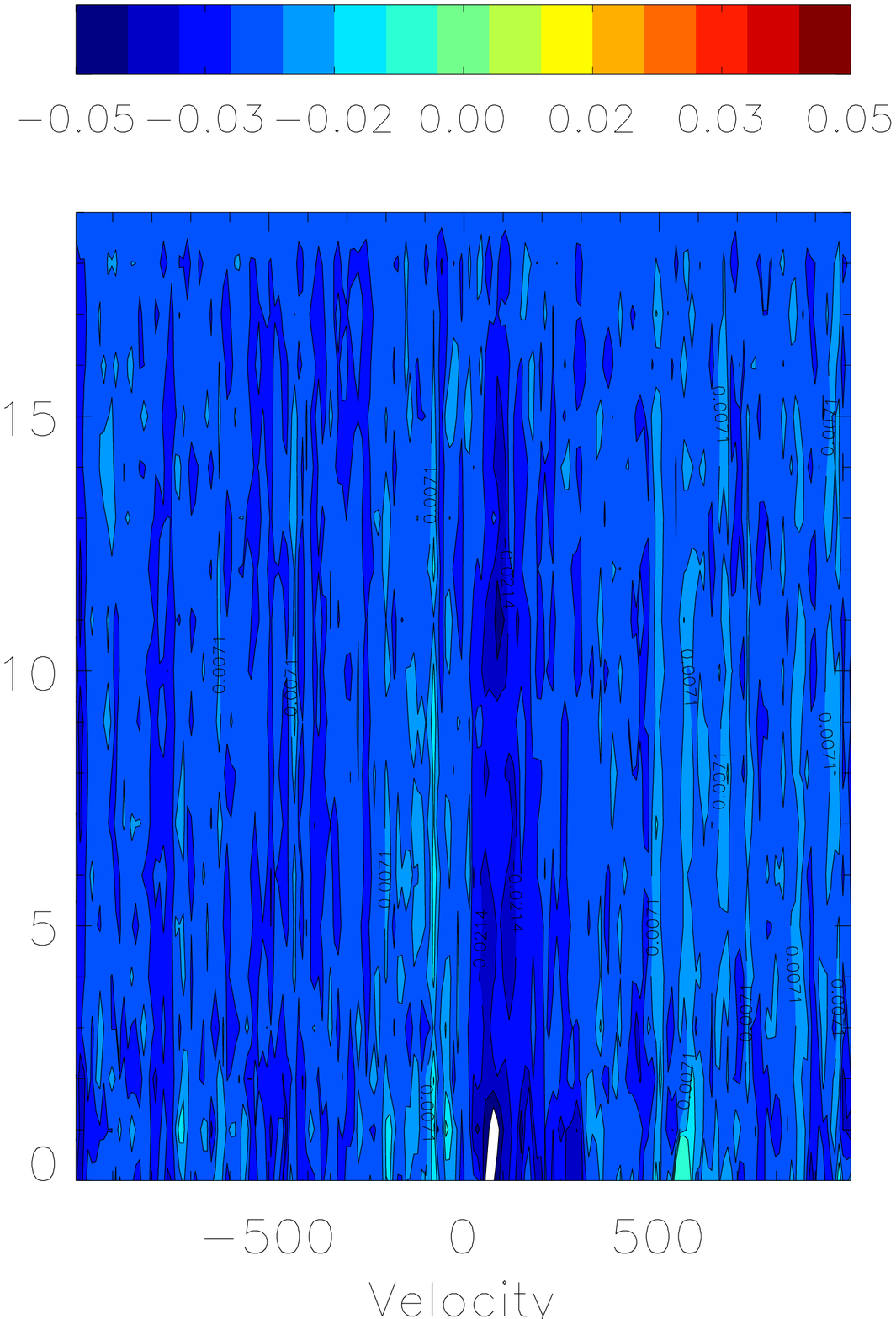} & \includegraphics[scale=0.22]{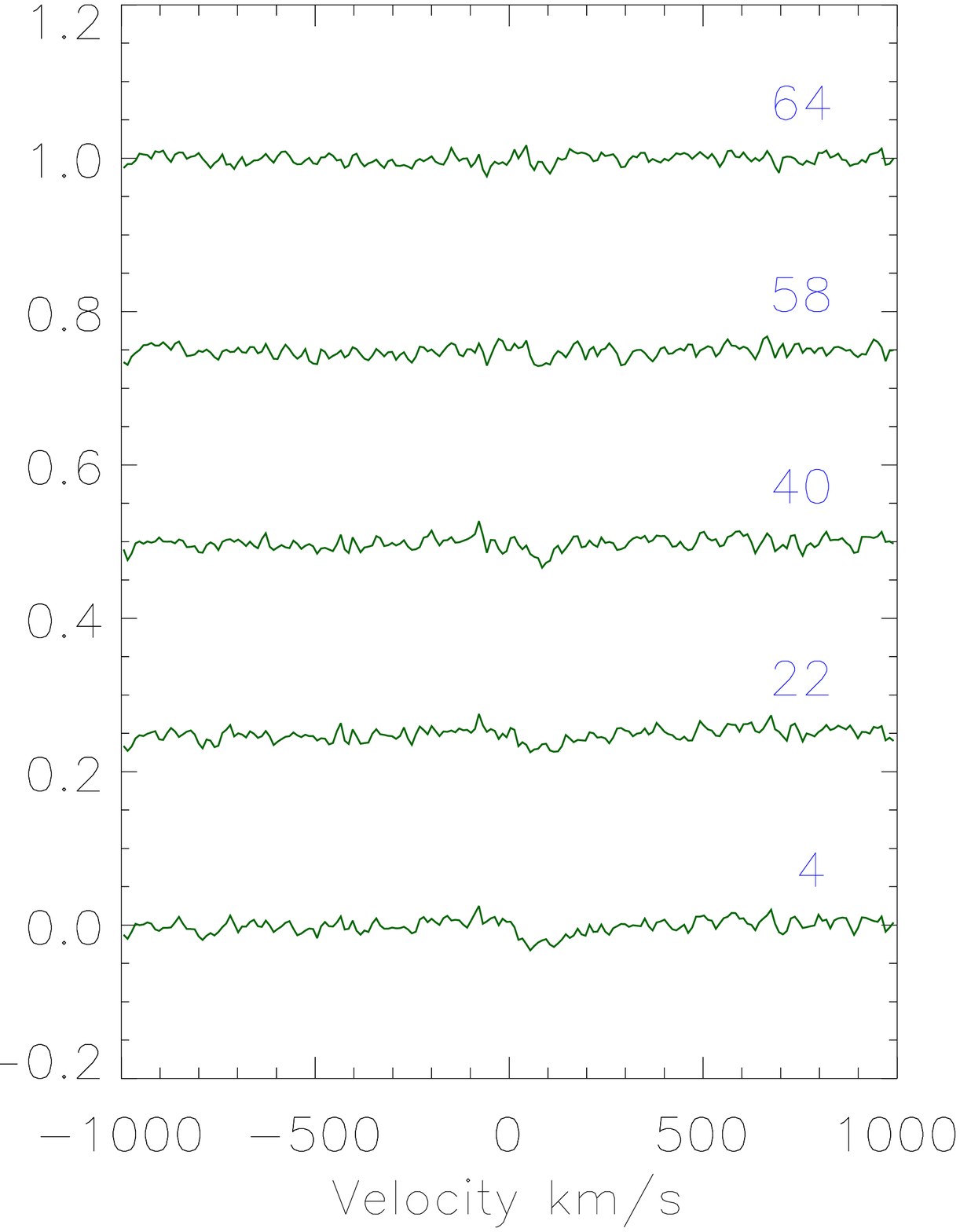}\\
\end{tabular}
\caption{V773 Tau 2003 observations. }
\label{fig:V773_plots}
\end{figure}

\textbf{Stellar Properties:} V773 Tau is a spectroscopic binary \citep{2009ApJ...704..531K} with an orbit period of 51 days \citep{2007ApJ...670.1214B}, with both components having periods of 3.43 days \citep{1983ApJ...267..191R}. It is highly variable with luminous non-thermal radio emission 
\citep{1990AJ....100.1610O,1991ApJ...382..261P,1996A&A...309..493D} and very bright, highly variable X-ray emission \citep{2000A&A...357..206G}. It is well known for its flaring, particularly around periastron passage, suggesting an interaction of the magnetic fields of the binary \citep{2002A&A...382..152M}.

\noindent \textbf{Disc Properties:} V773 Tau has been found to have a circumstellar disc of mass 0.0005 M$_{\odot}$ \citep{2005ApJ...631.1134A}

\noindent \textbf{ISIS H$\alpha$ Observations:} V773 Tau shows the weakest emission line within the sample, consisting of an asymmetric emission line with enhanced emission in the red wing (Fig.\,\ref{fig:V773_plots}). Across the observations the emission shows no significant variations.

\noindent \textbf{Previous H$\alpha$ Observations:} \citet{2000A&A...357..206G} observed V773 Tau in the optical and X-ray simultaneously and found a loose correlation between the H$\alpha$ EW and X-ray emission strength.  

\subsection{BF Ori}

\begin{figure}
\centering
\begin{tabular}{ccc}
 \includegraphics[scale=0.22]{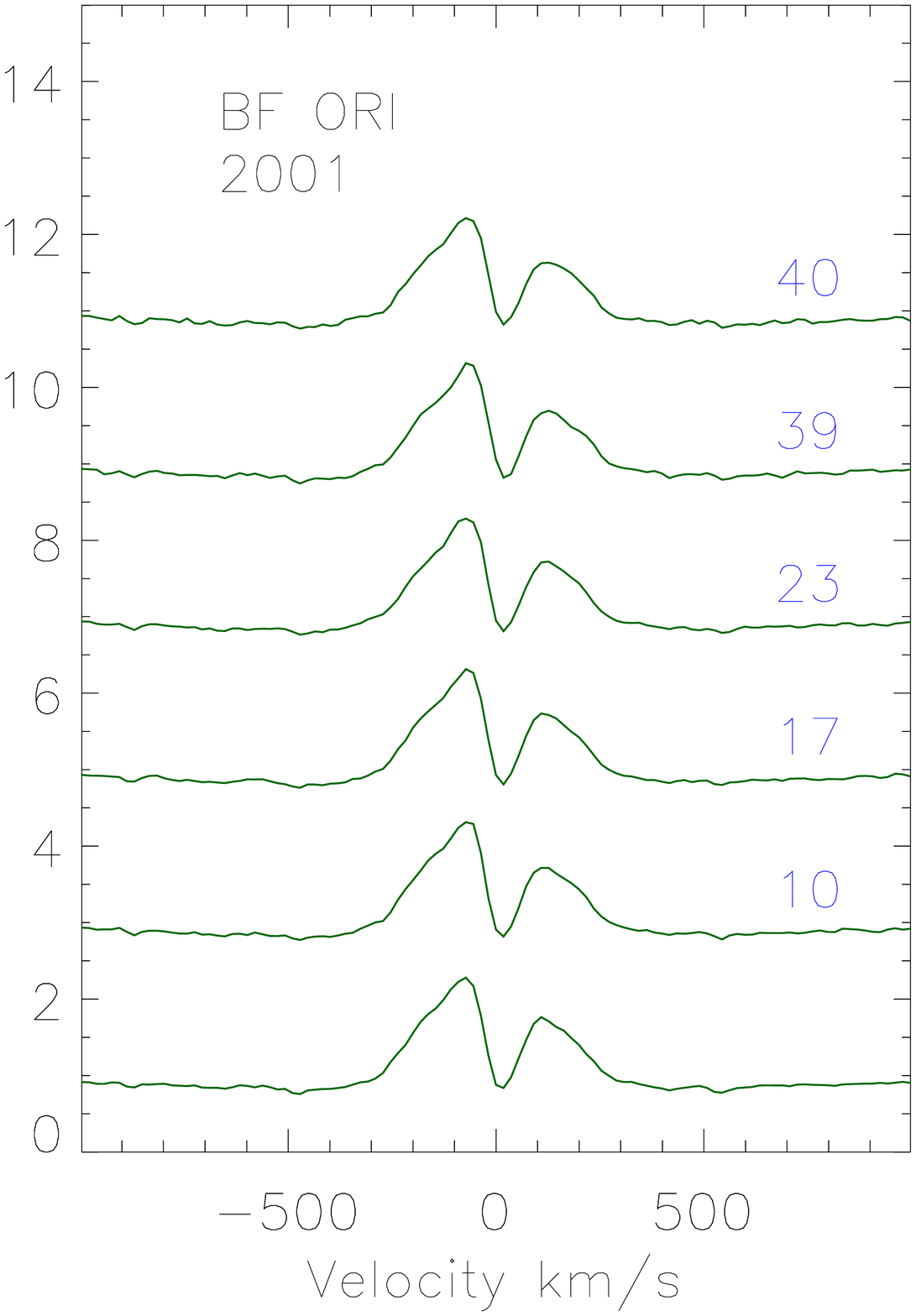} & 
 \includegraphics[scale=0.22]{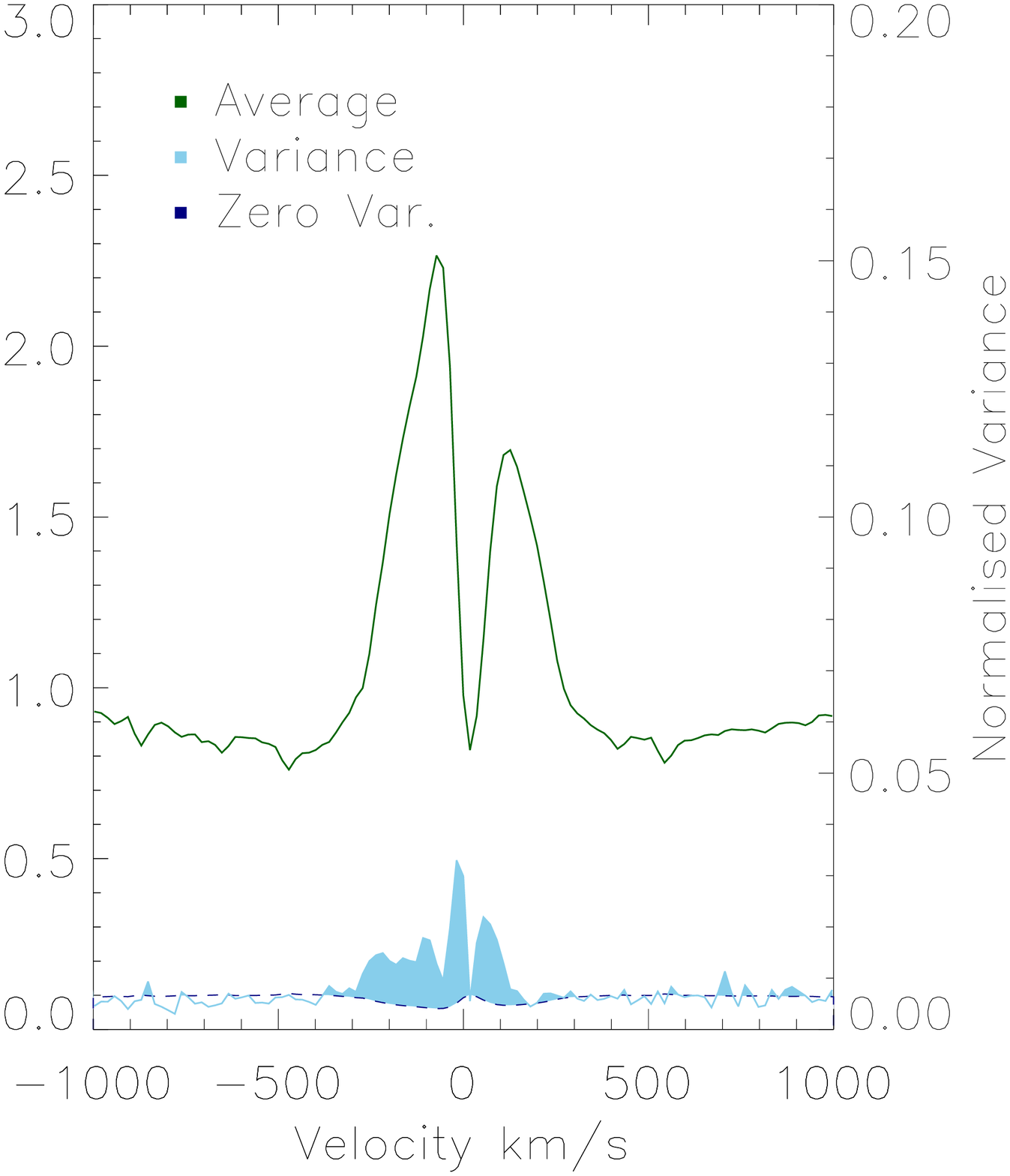}  \\
 \includegraphics[scale=0.22]{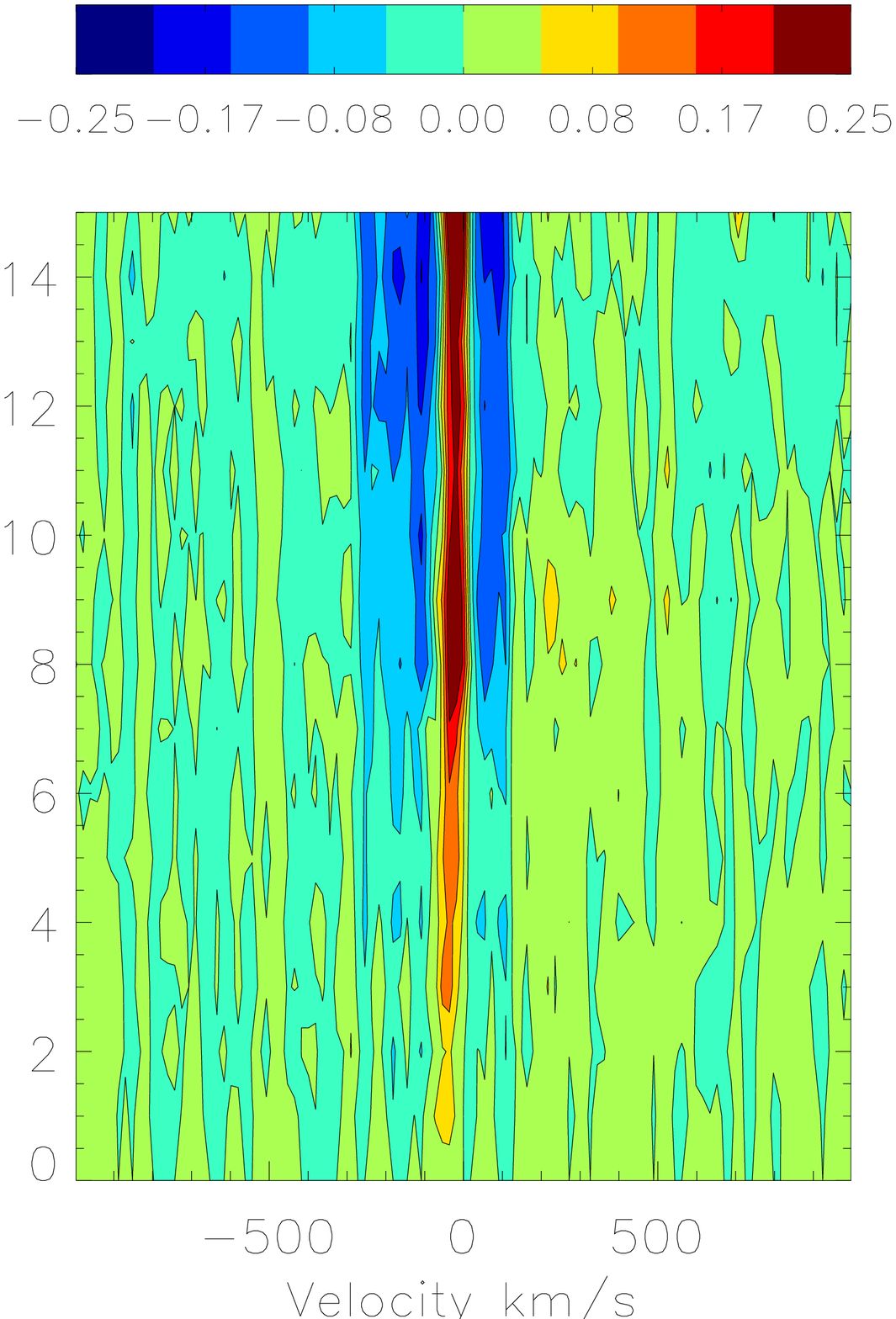}& 
 \includegraphics[scale=0.22]{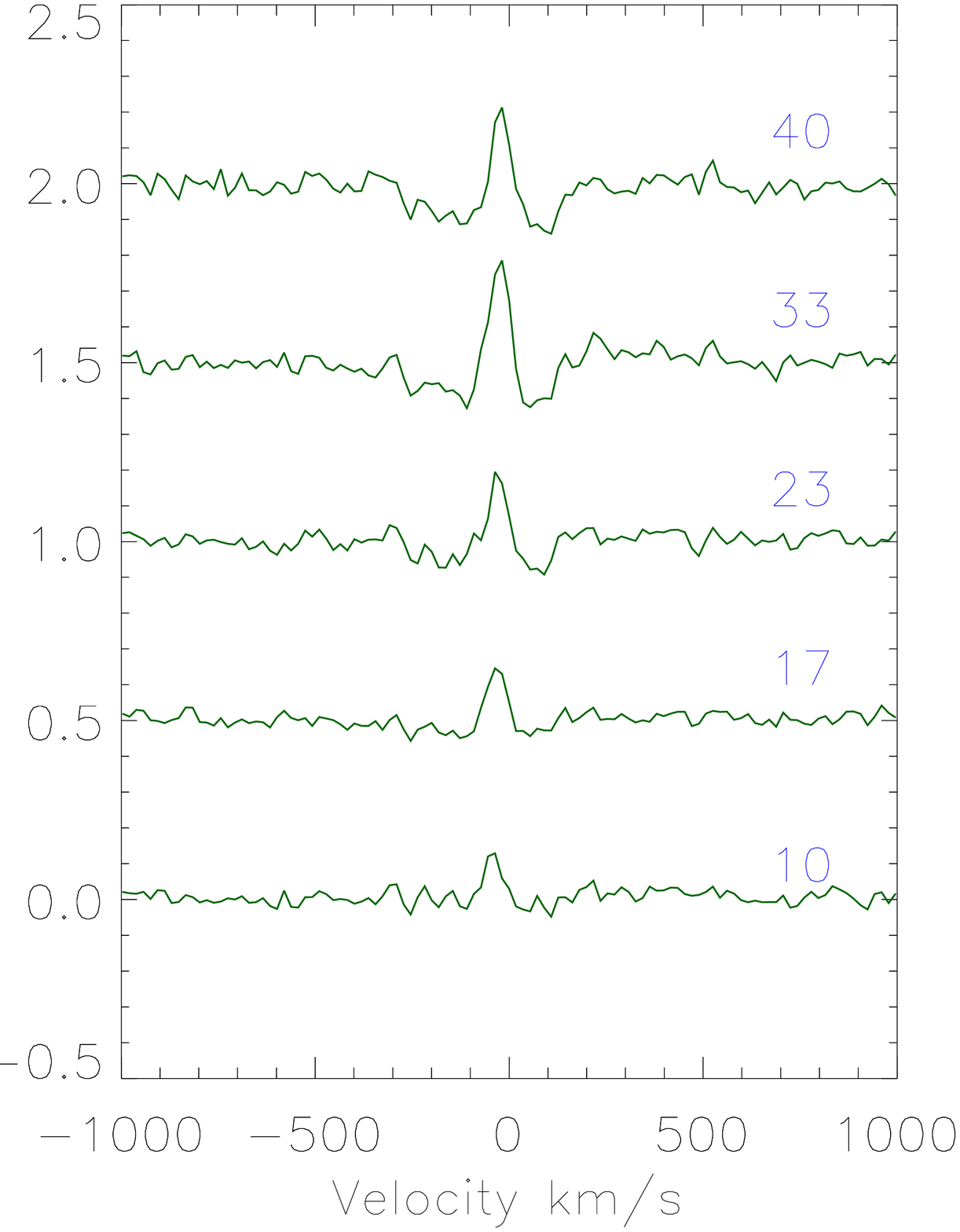}   \\ 
\end{tabular}
\caption{BF Ori 2001 observations.}
\label{fig:BFORI}
\end{figure}

\textbf{Stellar Properties:} BF Ori has a long history of large amplitude changes and has been classified as an UXors object \citep{1991A&AS...89..319B}. A period of 5.5 years has been found in these large variations \citep{1993Ap&SS.202..137S}. These variations are thought to be a result of viewing the star through the variable circumstellar disc \citep{1996ARep...40..171G}.  BF Ori has a stellar mass of 2.5 M$_{\odot}$ and a radius of 1.3 R$_{\odot}$ \citep{2006ApJ...653..657M,1992ApJ...397..613H}.

\noindent \textbf{Disc Properties:} An inner disc hole of 10 AU was found through SED modelling by \citet{1992ApJ...397..613H}.

\noindent \textbf{Accretion:} Using SED modelling, \citet{1992ApJ...397..613H} derived accretion rate of 1.1\,x\,10$^{-6}$M$_{\odot}$yr$^{-1}$.

\noindent \textbf{ISIS H$\alpha$ Observations:} BF Ori shows comparatively low levels of variations across the single observation block (Fig.\,\ref{fig:BFORI}). The H$\alpha$ profile has two peaks in emission with a central absorption feature. The largest changes take place within the central absorption feature, and take the form of a steady increase in emission in a narrow wavelength range. 

\noindent \textbf{Previous H$\alpha$ Observations:} A range of H$\alpha$ EW measurements have been reported,   10.0\AA~\citep{1988cels.book.....H}, 11.30\AA~\citep{1996A&AS..120..229R}, 3.70\AA~\citep{1998A&A...331..147C}, 6.70\AA~\citep{2004AJ....127.1682H} and 9.30\AA~\citep{2005A&A...436..209A}. The mean EW measured in the ISIS sample in 2001 is on the low end of this range at 3.88\AA, but coincides with the dip in strength in 1998 and 2004.  

\subsection{LkH$\alpha$ 215}
\textbf{Stellar Properties:} LkH$\alpha$ 215 is one of the larger stars in this sample, with a mass of 4.8 M$_{\odot}$ and radius of 5.4 R$_{\odot}$ \citep{2006ApJ...653..657M,1992ApJ...397..613H}.

\begin{figure}
\centering
\begin{tabular}{ccc}
\includegraphics[scale=0.22]{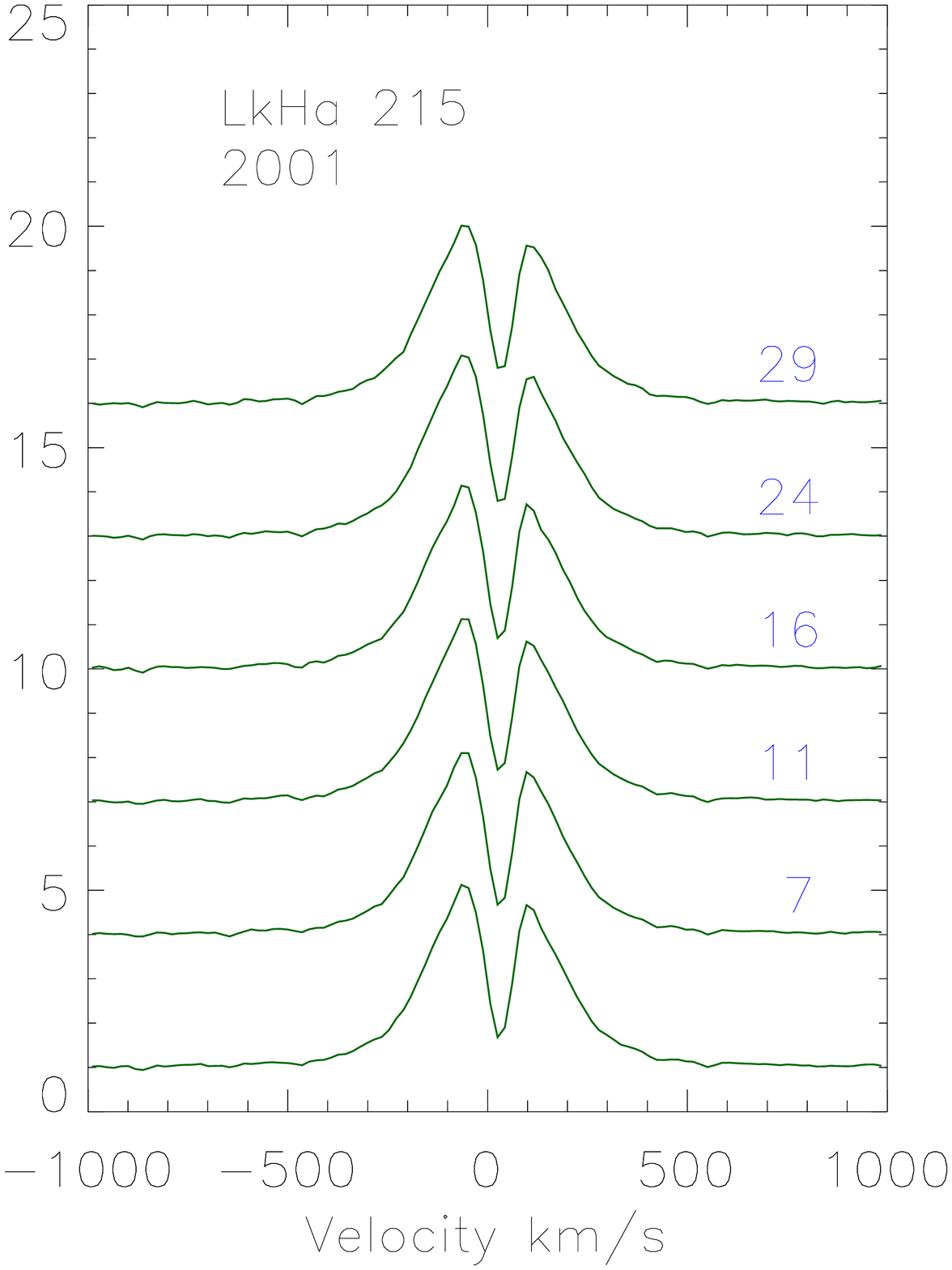}& 
\includegraphics[scale=0.22]{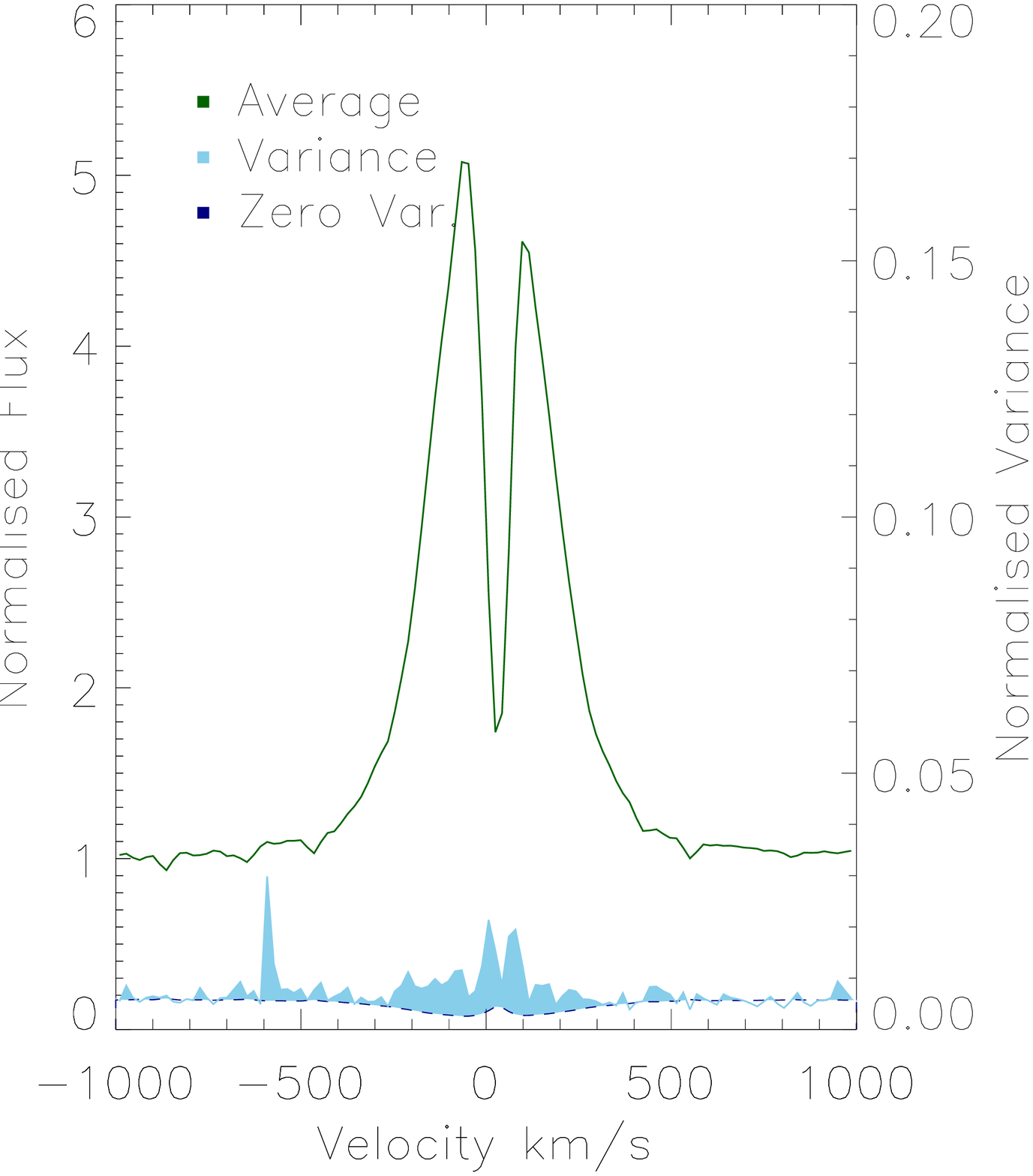} \\
\includegraphics[scale=0.22]{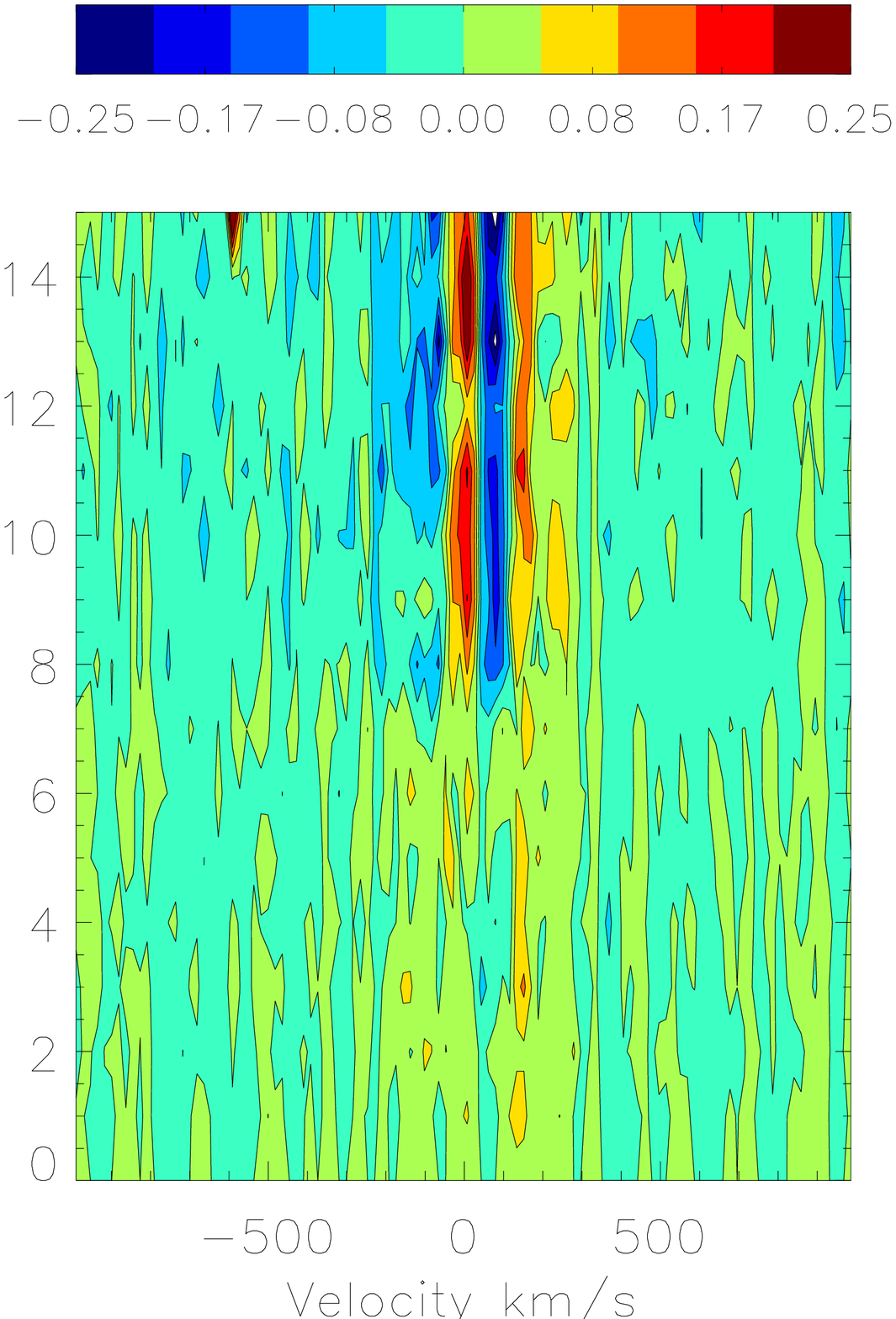} &
\includegraphics[scale=0.22]{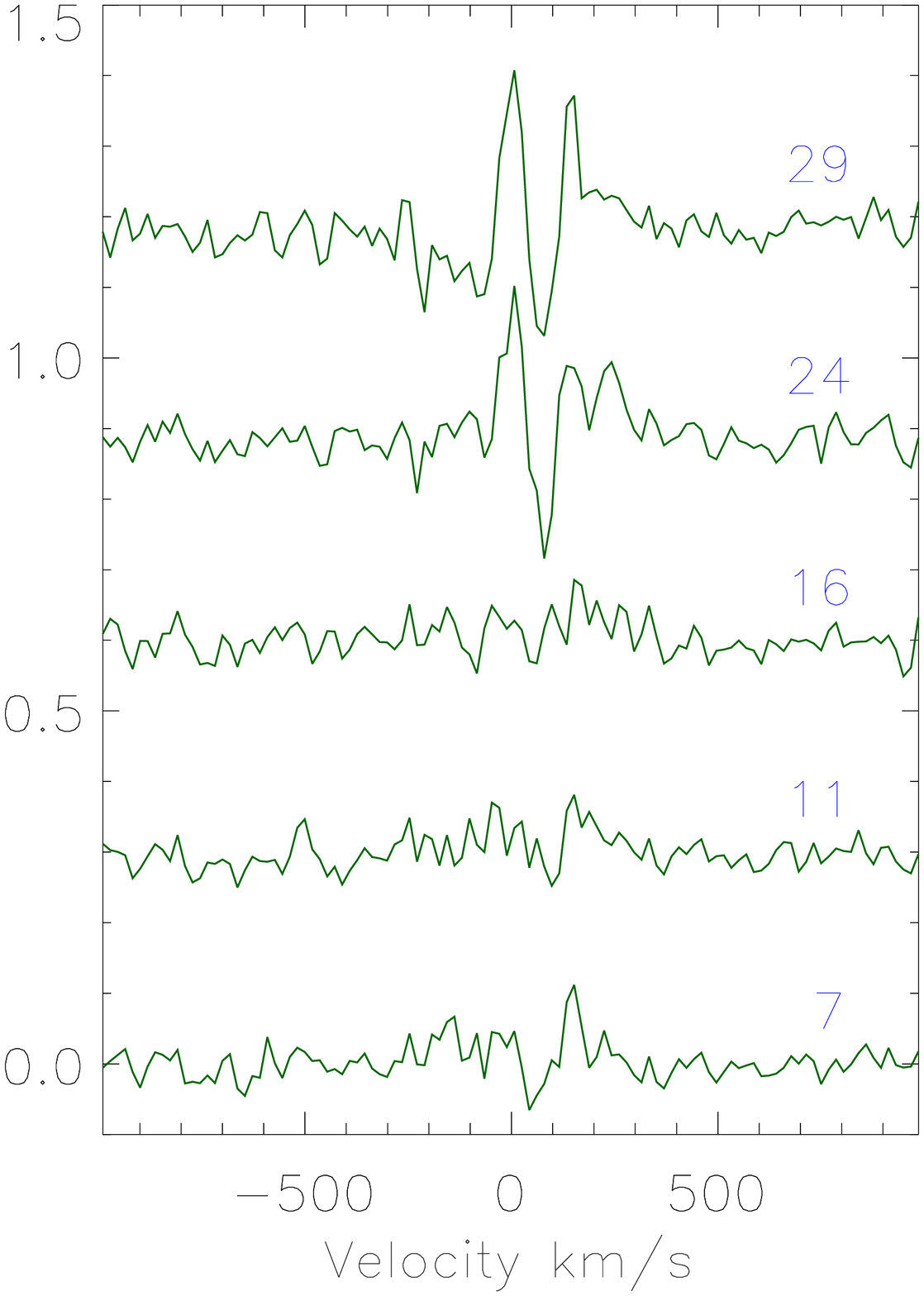}\\
\end{tabular}
\caption{LkH$\alpha$ 215 2001 observations.}
\label{fig:LKH}
\end{figure}

\noindent \textbf{Disc Properties:} \citet{2009A&A...497..117A} found an uncertain disc outer radius of $\sim$ 10\,AU, and a disc mass of 6\,x\,10$^{8}$ M$\odot$.

\noindent \textbf{ISIS H$\alpha$ Observations:} The H$\alpha$ emission in this object takes the form of a double peak profile (Fig.\,\ref{fig:LKH}). During the single observation block of this object, only very small changes in emission occur either side of the central absorption feature. 

\noindent \textbf{Previous H$\alpha$ Observations:} Previous observations of the H$\alpha$ emission have found a slightly weaker emission line than is reported here (mean of 3.45\,\AA), 25\,\AA~\citep{1988cels.book.....H}, 26.7\,\AA~\citep{1998A&A...331..147C} and 25.7\,\AA~\citep{2004AJ....127.1682H}.

\subsection{CO Ori}

\begin{figure}
\centering
\begin{tabular}{ccc}
\includegraphics[scale=0.22]{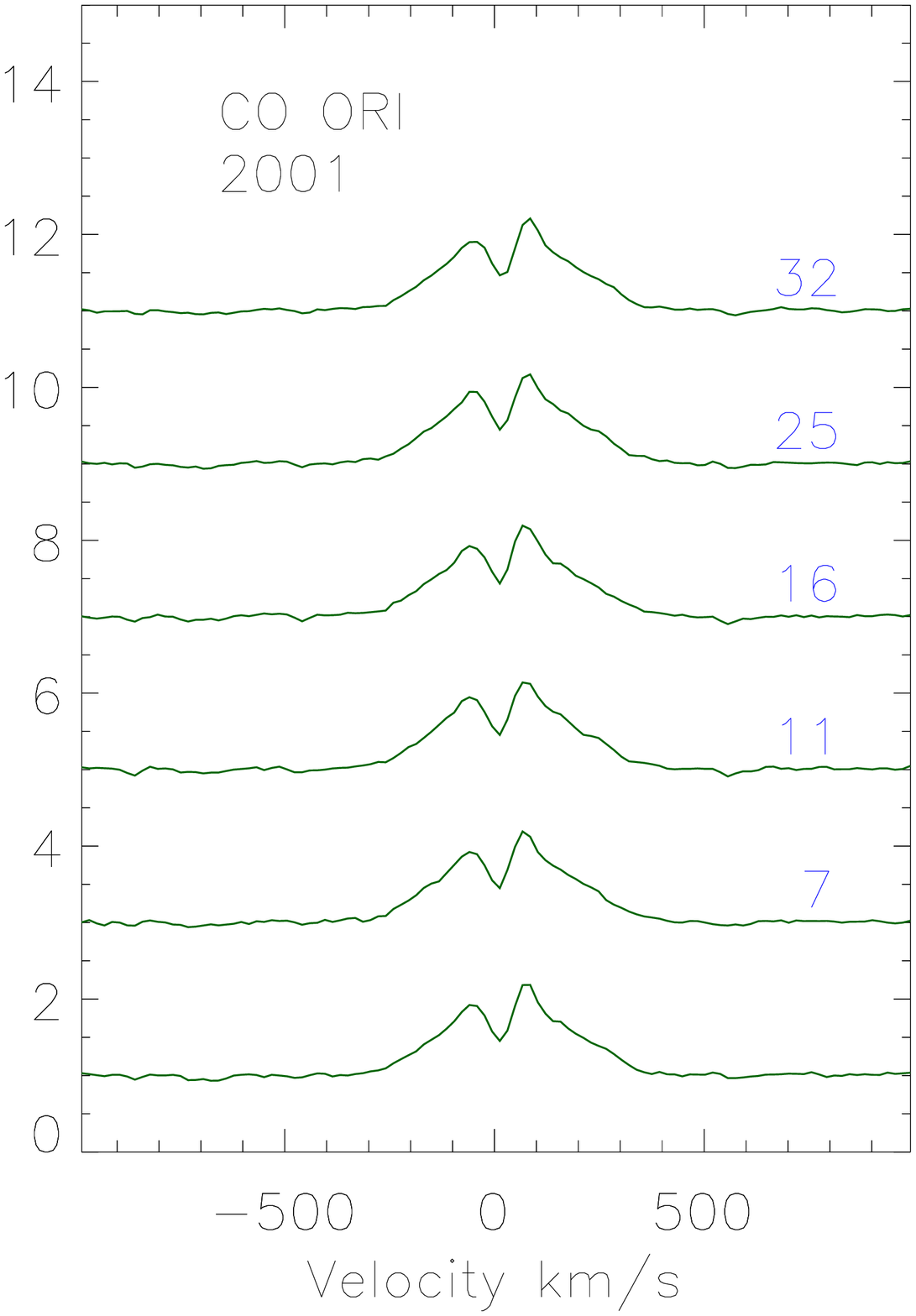} &
\includegraphics[scale=0.22]{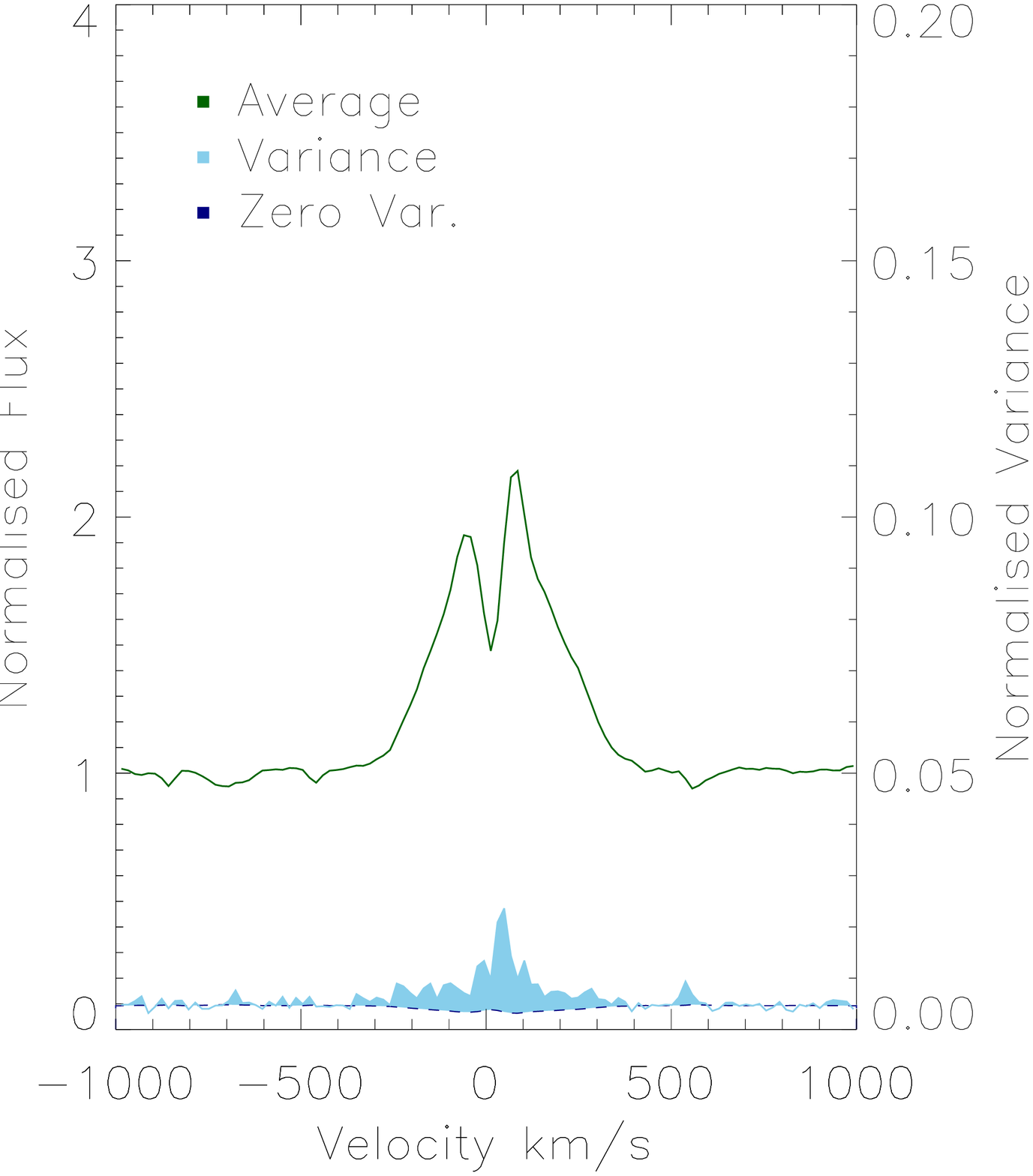} \\
\includegraphics[scale=0.22]{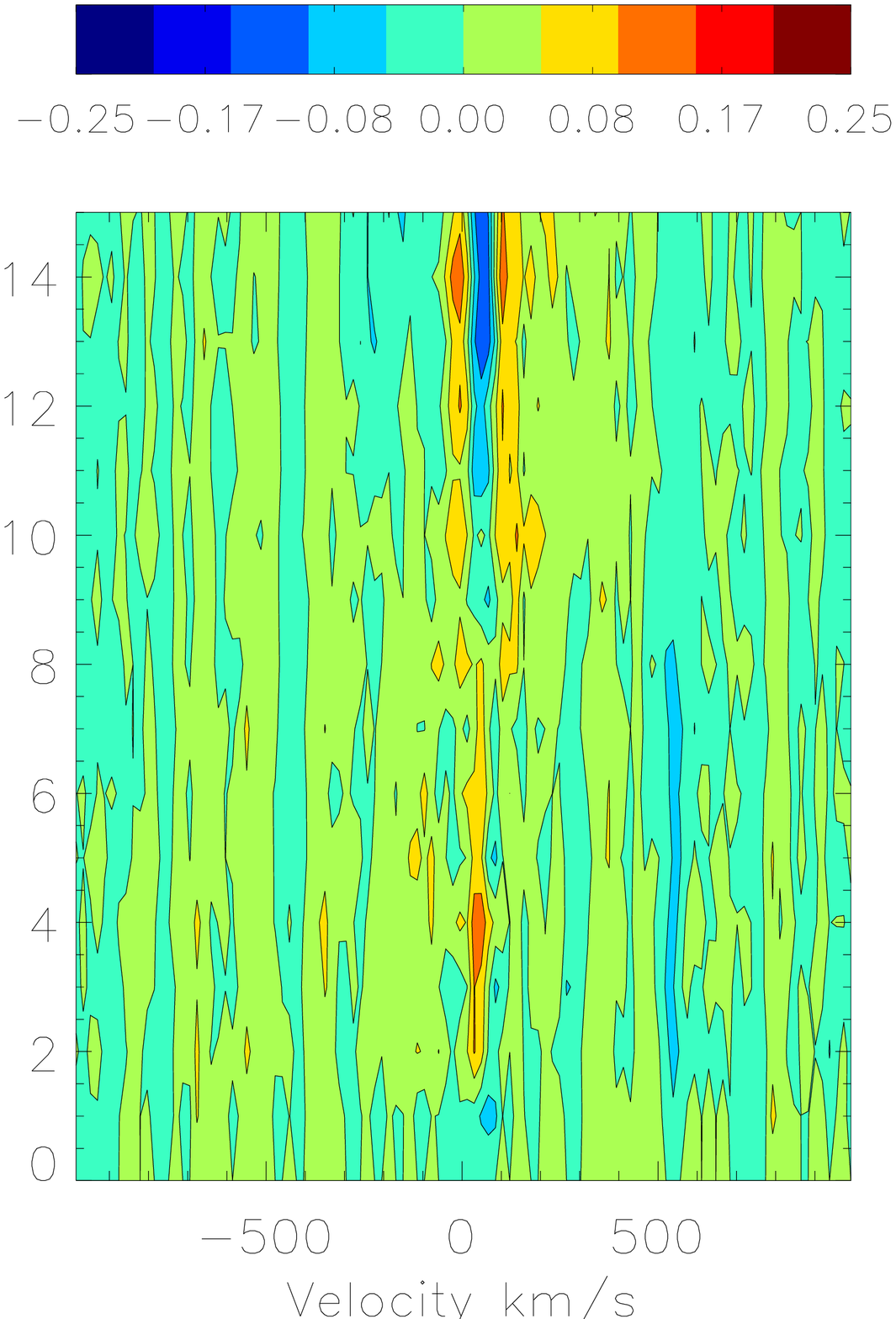}&
\includegraphics[scale=0.22]{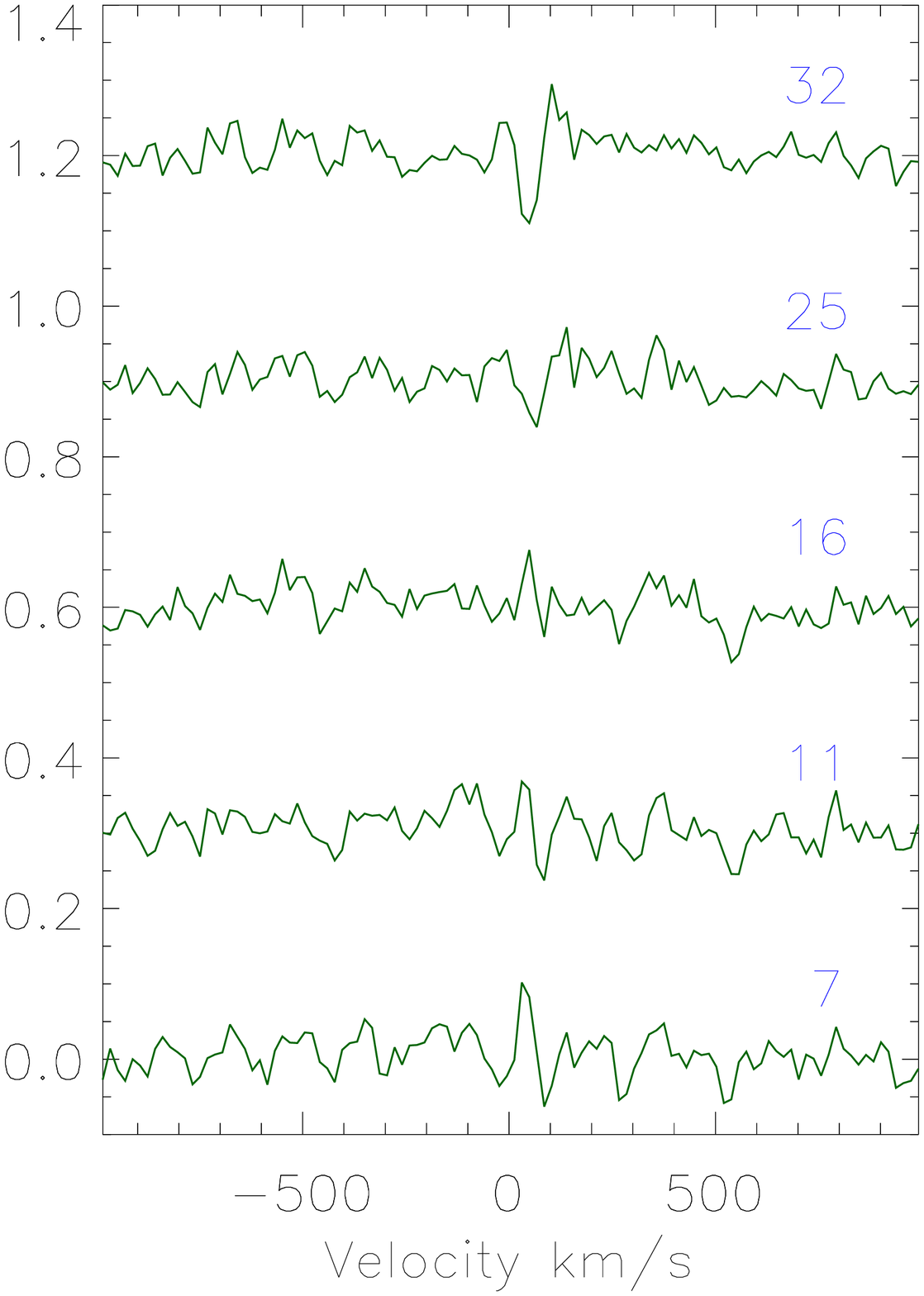}\\
\end{tabular}
\caption{CO Ori 2001 observations.}
\label{fig:COORI}
\end{figure}

\textbf{Stellar Properties:} CO Ori is one of the brightest T Tauri stars, however it has been poorly studied. It has previously shown a high and variable accretion rate \citep{2004AJ....128.1294C}, and also large photometric variations on time-scales of days \citep{1999AJ....118.1043H}, which led the authors to designate is as an UXors object.
\noindent \textbf{Disc Properties:} 

\noindent \textbf{ISIS H$\alpha$ Observations:}  The H$\alpha$ emission of CO Ori takes the form of a double peaked profile (Fig.\,\ref{fig:COORI}). The small changes in the profile across the hour of observations are concentrated in the two peaks of emission.

\noindent \textbf{Previous H$\alpha$ Observations:} CO Ori has previously been reported to have a weaker H$\alpha$ emission profile at 4.2\,\AA~and a pronounced P Cygni profile \citep{1996A&AS..120..229R}. About 10 years earlier it was observed to have stronger emission at 10\,\AA~\citep{1988cels.book.....H}.

The H$\alpha$ profile observed by \citet{2004AJ....128.1294C}, shows quite a different morphology than we observe. \citet{2004AJ....128.1294C} observed a extremely blue-shifted absorption feature and a single red-shifted emission peak. This suggested that the more central and weaker absorption feature that is observed in the ISIS data sometimes grows in strength and moves more towards the blue.

\subsection{MWC 480}
\textbf{Stellar Properties:} MWC 480 is a Herbig Ae star with a mass of 2.3 M$_{\odot}$ and a stellar radius of 2.1 R$_{\odot}$ \citep{1997ApJ...490..792M}. It is one of the few Herbig Ae stars that have had detections of strong kG magnetic fields on their surfaces \citep{2011A&A...536A..45H}.

\noindent \textbf{Accretion and Outflows:} Accretion in MWC 480 has been confirmed via both far-UV and X-ray detections. This target also drives a bipolar jet, but the rate of mass loss is lower than is considered normal for Herbig Ae stars \citep{2010ApJ...719.1565G}. An accretion rate of 3.8\,x\,10$^{-8}$ M$_{\odot}$yr$^{-1}$ was derived from the far-UV excess emission by \citet{2010ApJ...719.1565G}. A lower mean accretion rate was found in \citet{2013ApJ...776...44M}, 1.1\,x\,10$^{-7}$ M$_{\odot}$yr$^{-1}$, by fitting the Balmer excess. Across the few months of observations, they measure accretion rate changes between 5.24\,x\,10$^{-8}$ and 1.46\,x\,10$^{-7}$ M$_{\odot}$yr$^{-1}$. \textbf{The mean accretion measurement found in the ISIS sample, 
5.321\,x\,10$^{-7}$ M$_{\odot}$yr$^{-1}$, lies slightly above these accretion estimates. }

\noindent \textbf{Disc Properties:} A disc inclination angle of 30\degree~ has been determined \citep{1997ApJ...490..792M,2004ApJ...613.1049E}.

\begin{figure*}
\centering
\begin{tabular}{cccc}
\includegraphics[scale=0.22]{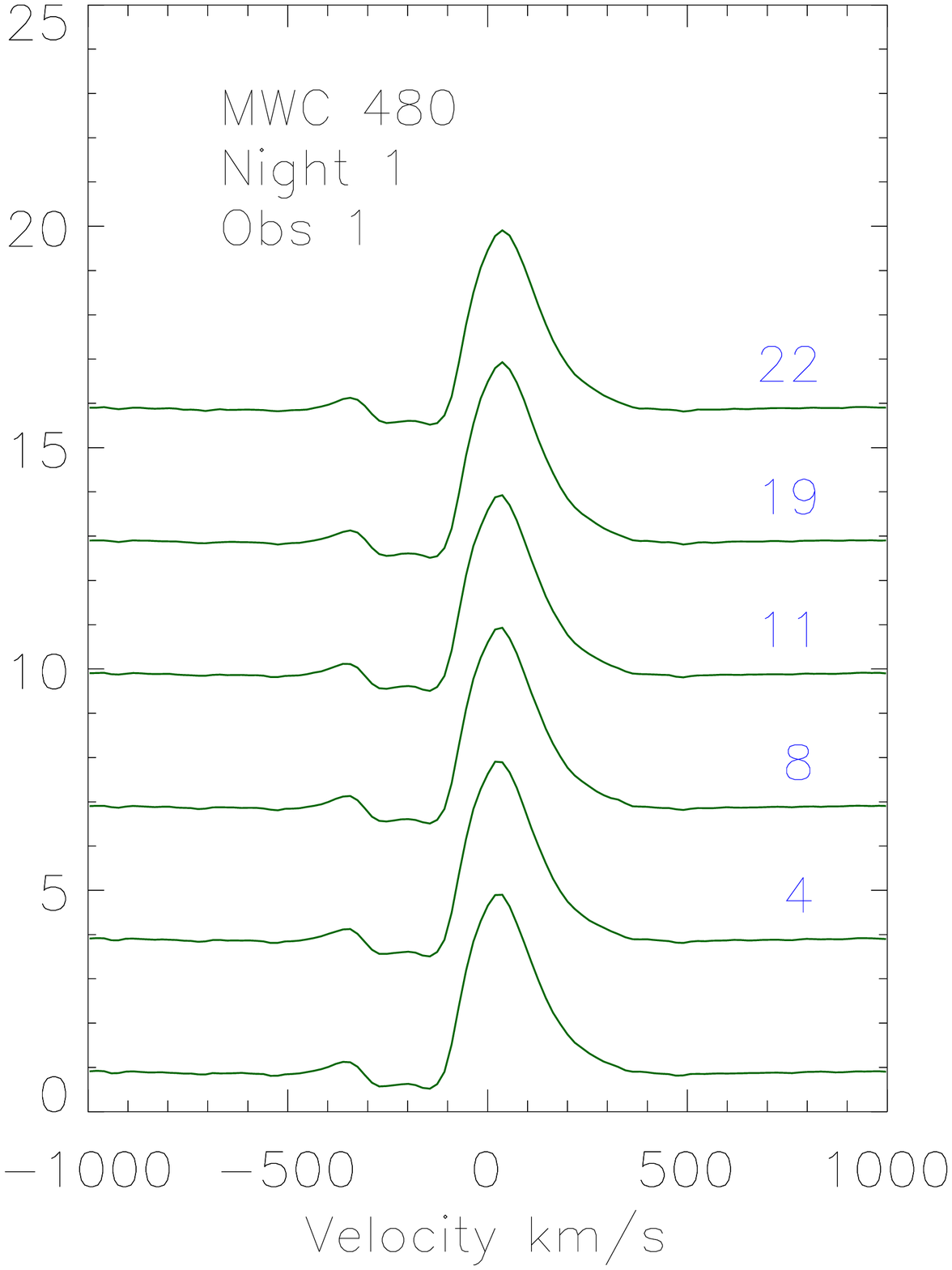} & \includegraphics[scale=0.22]{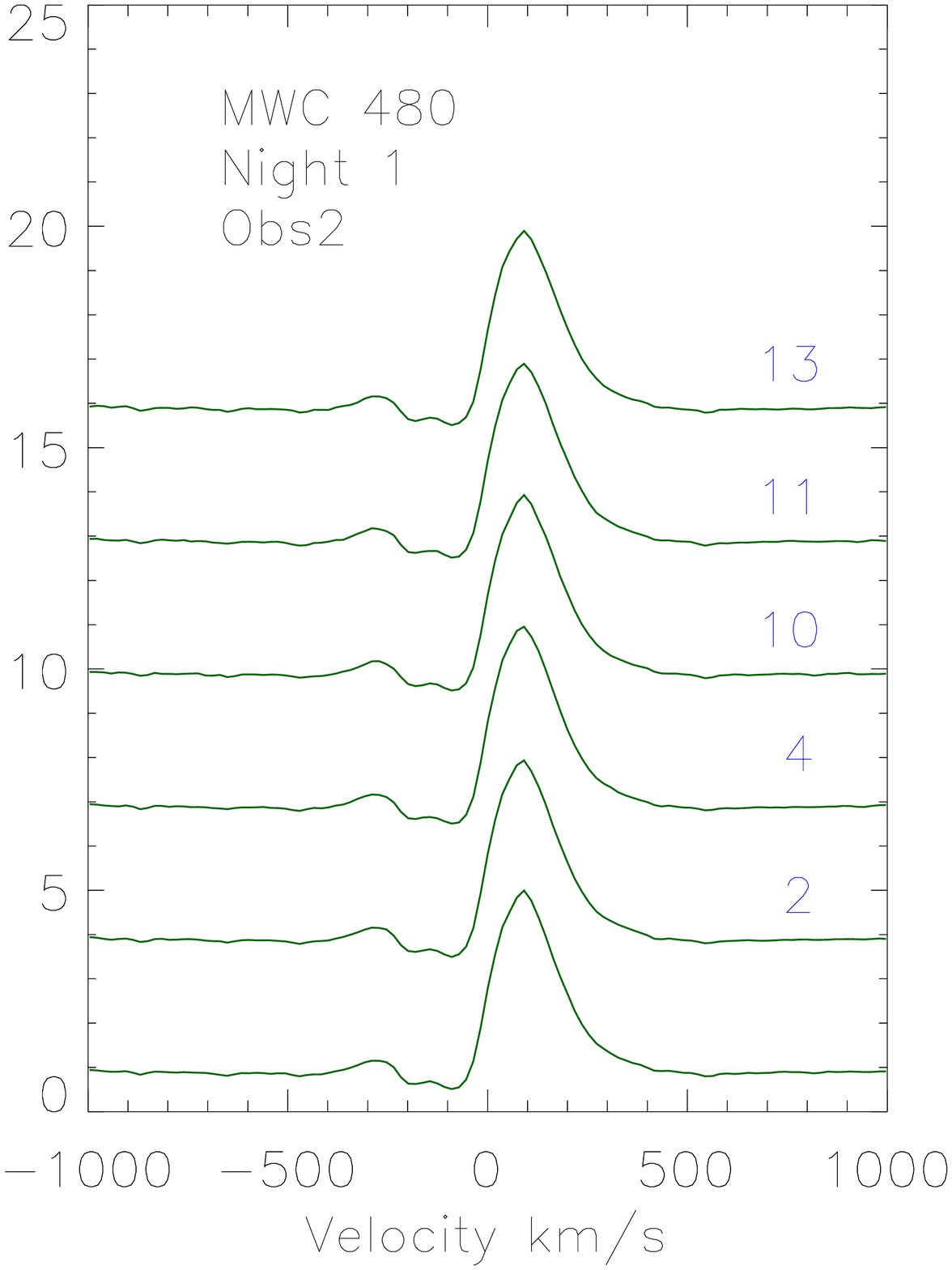} & \includegraphics[scale=0.22]{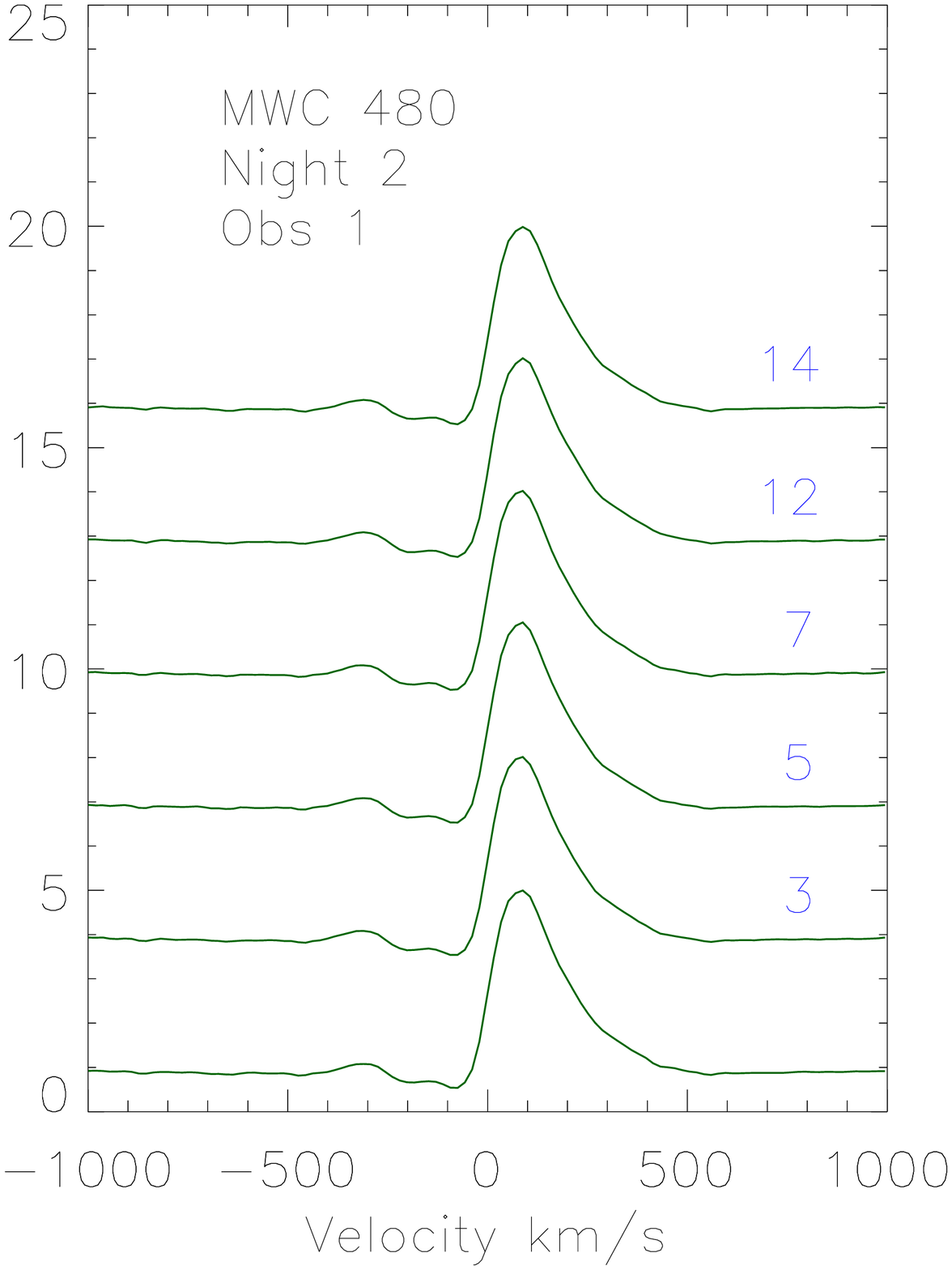} & \includegraphics[scale=0.22]{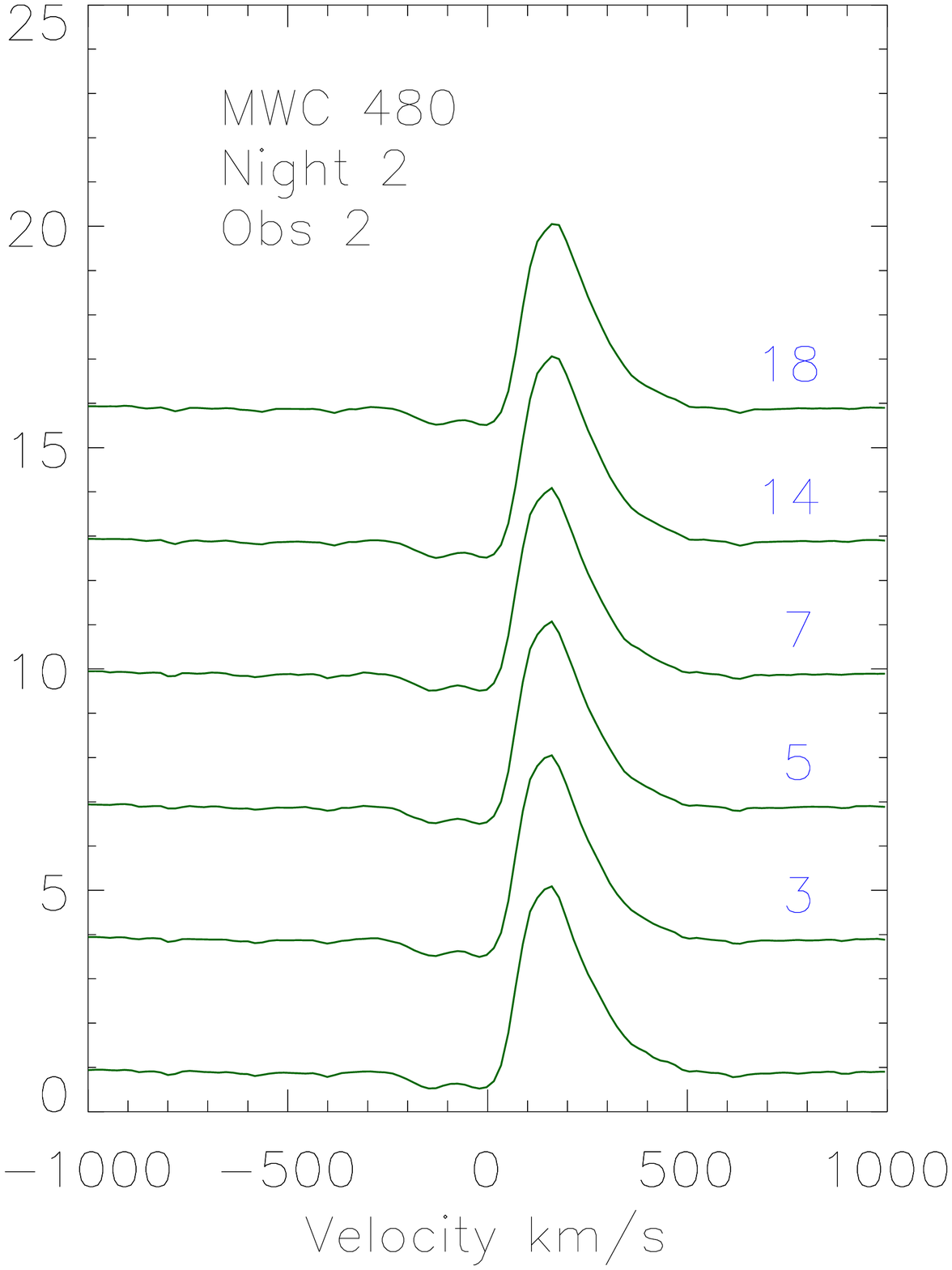} \\
\includegraphics[scale=0.22]{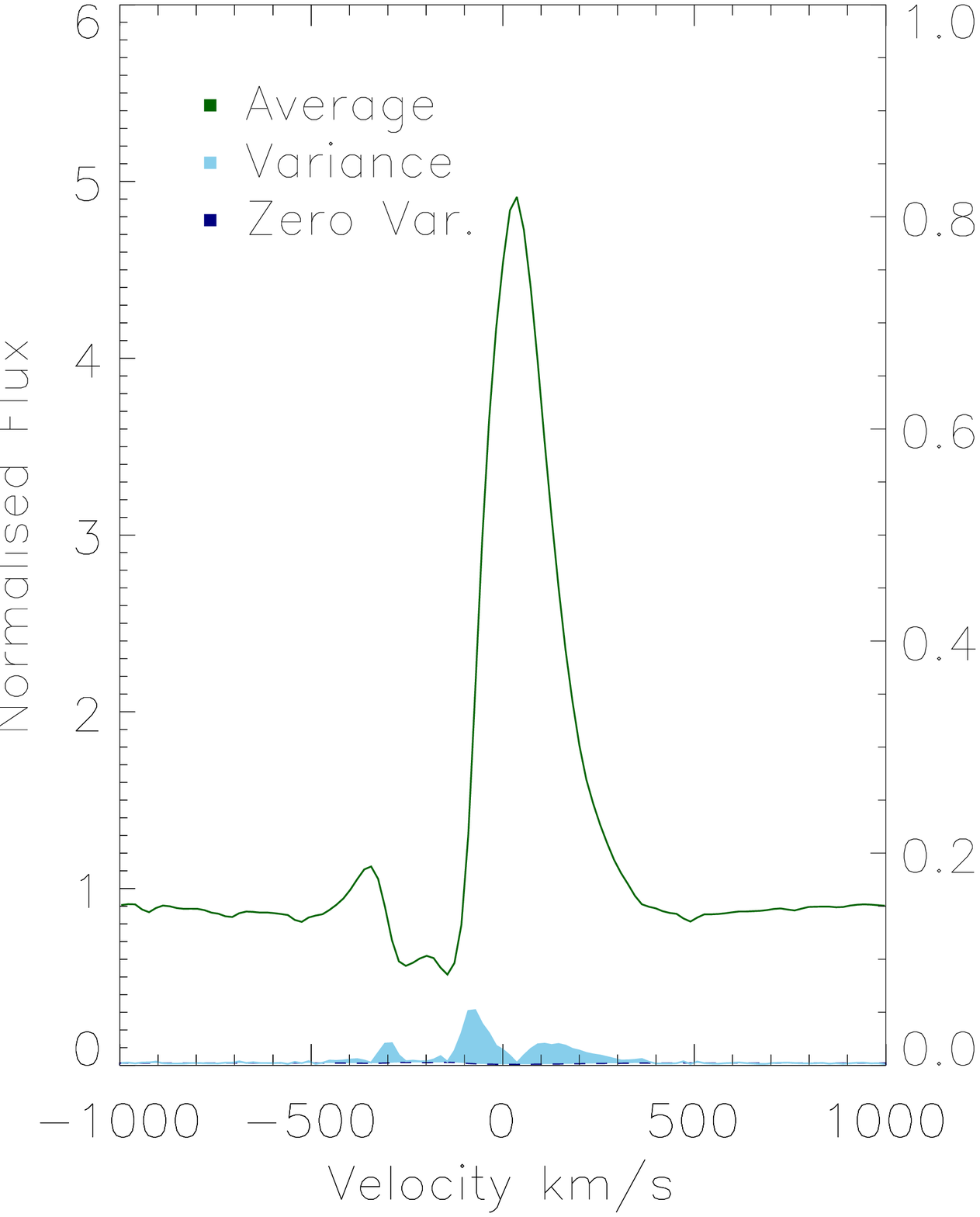} & \includegraphics[scale=0.22]{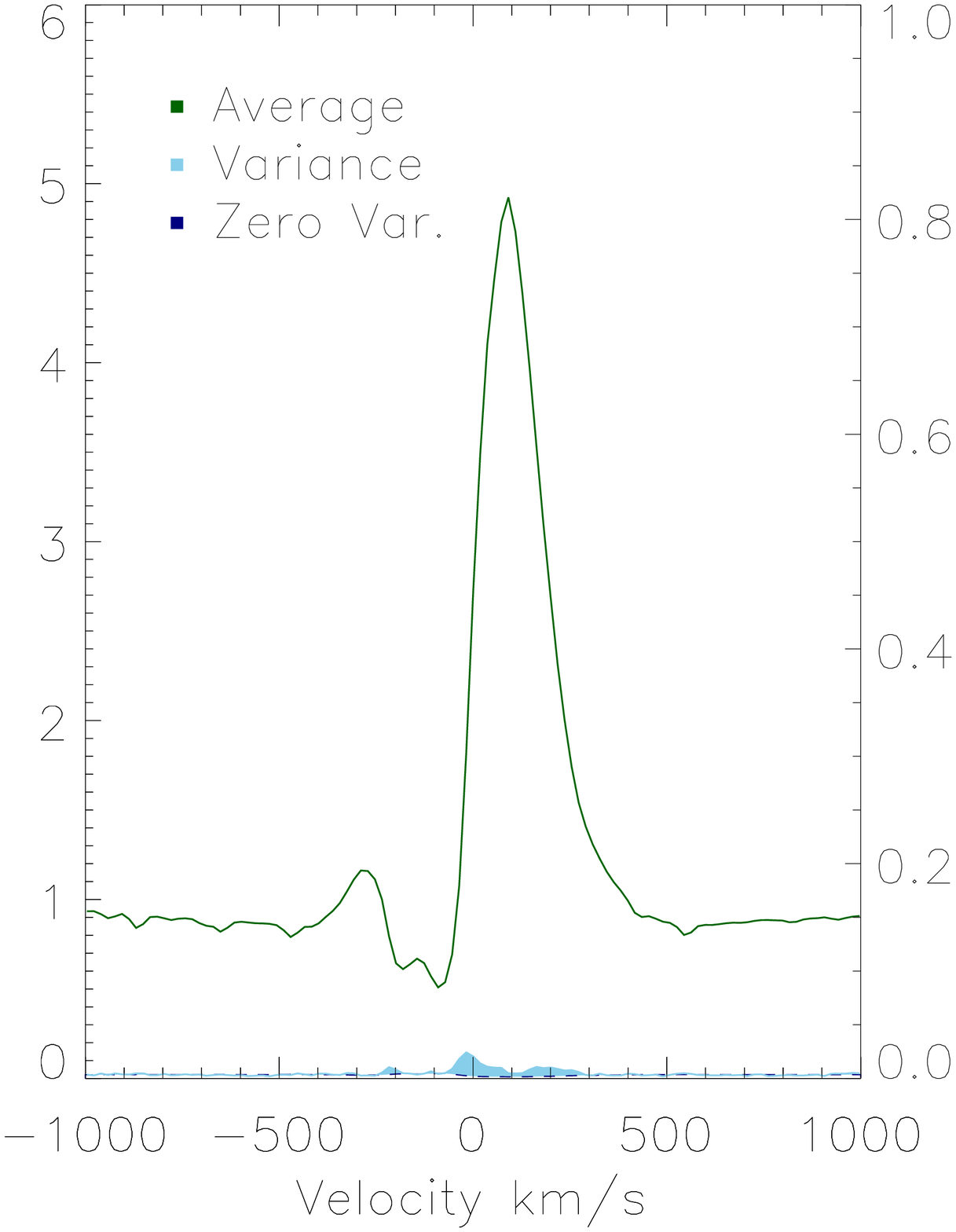}  & \includegraphics[scale=0.22]{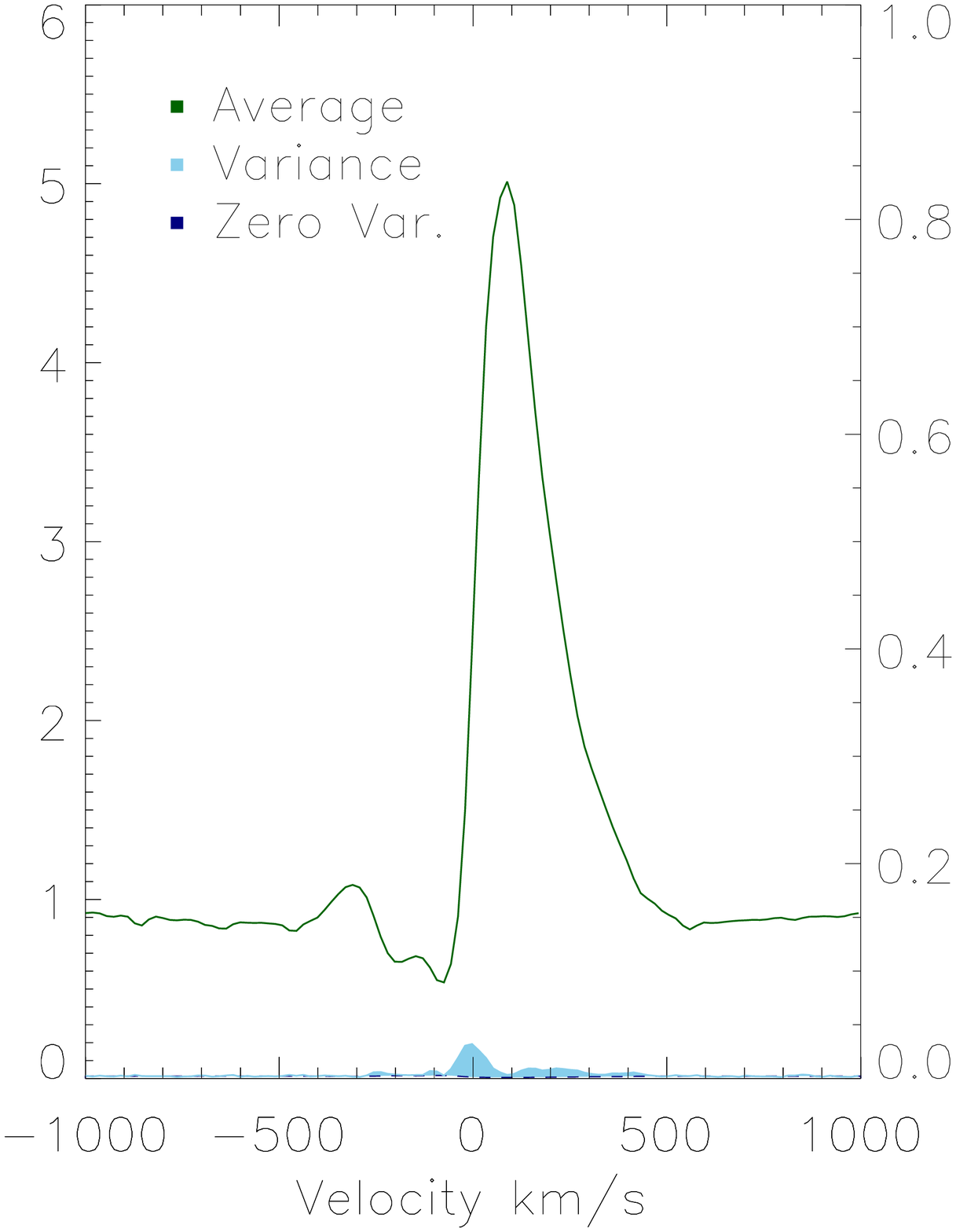} & \includegraphics[scale=0.22]{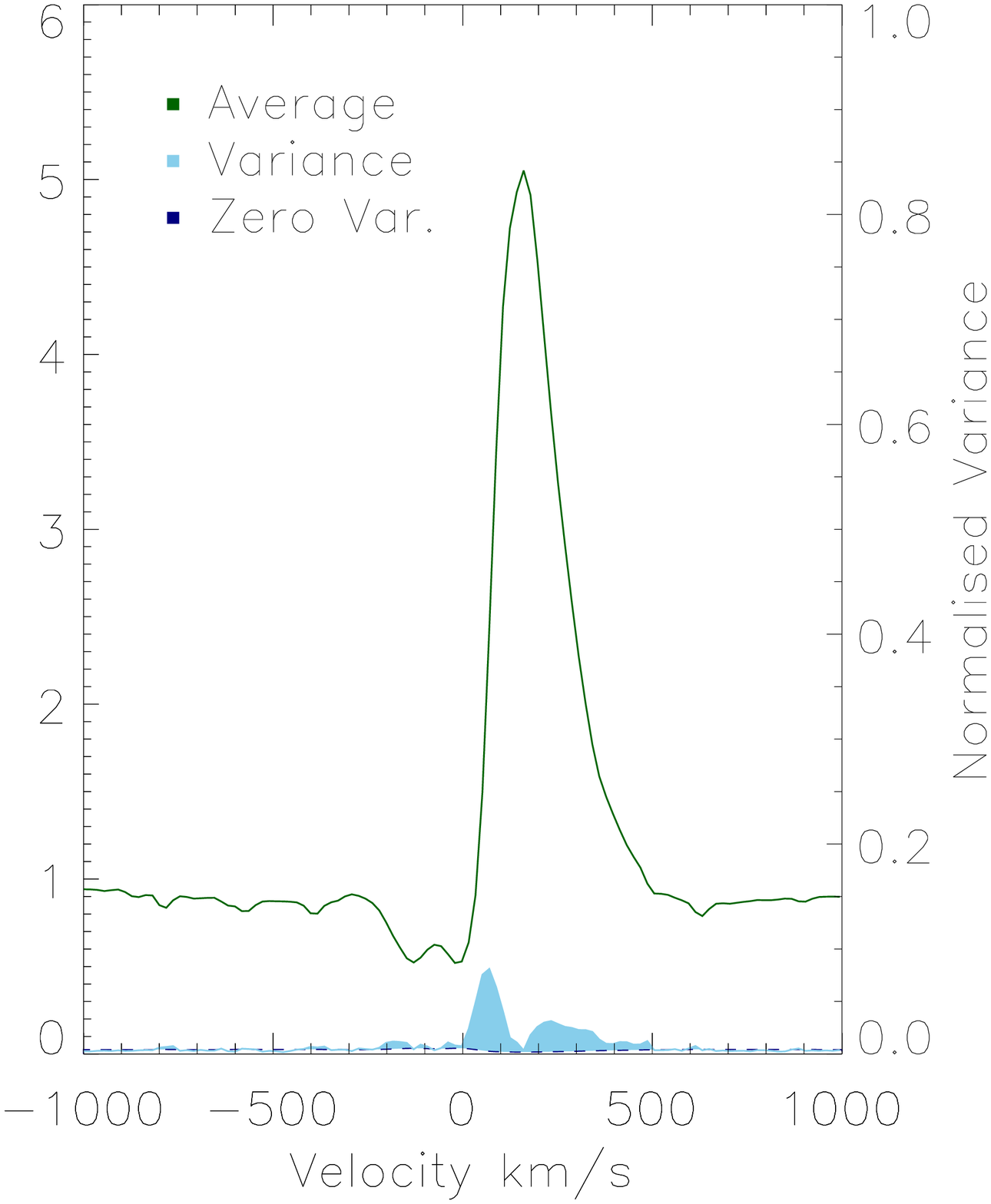} \\
\includegraphics[scale=0.22]{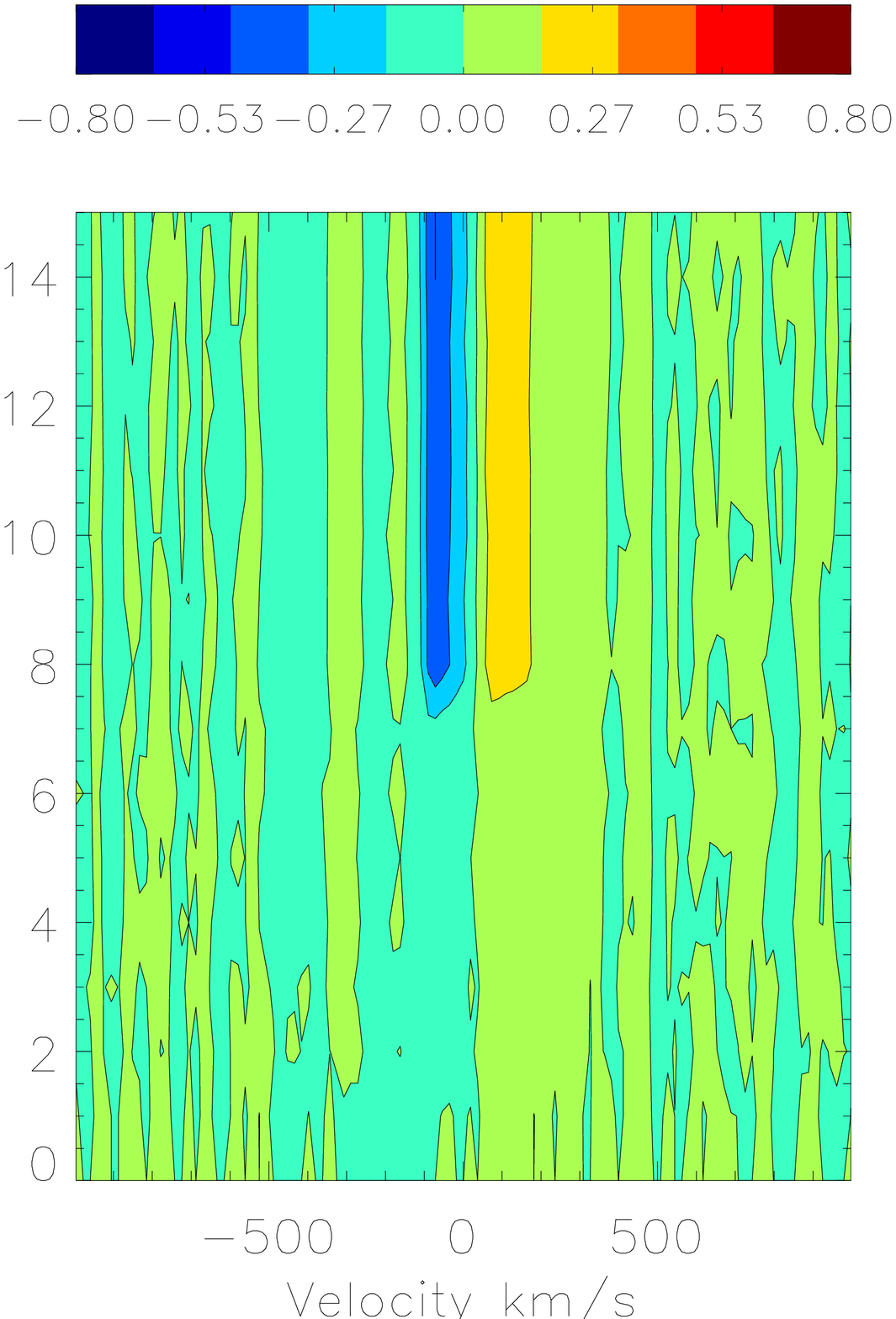} & \includegraphics[scale=0.22]{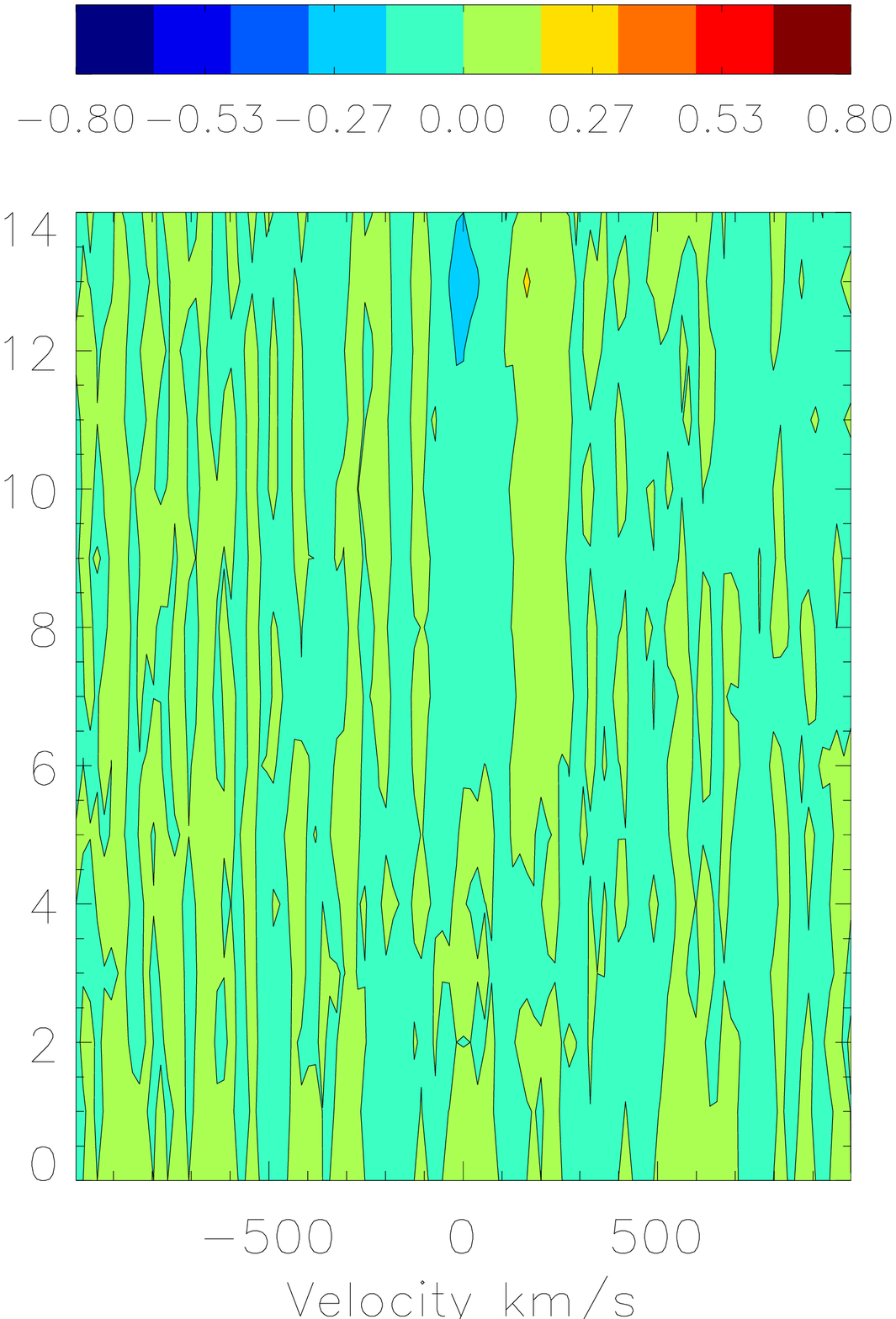} & \includegraphics[scale=0.22]{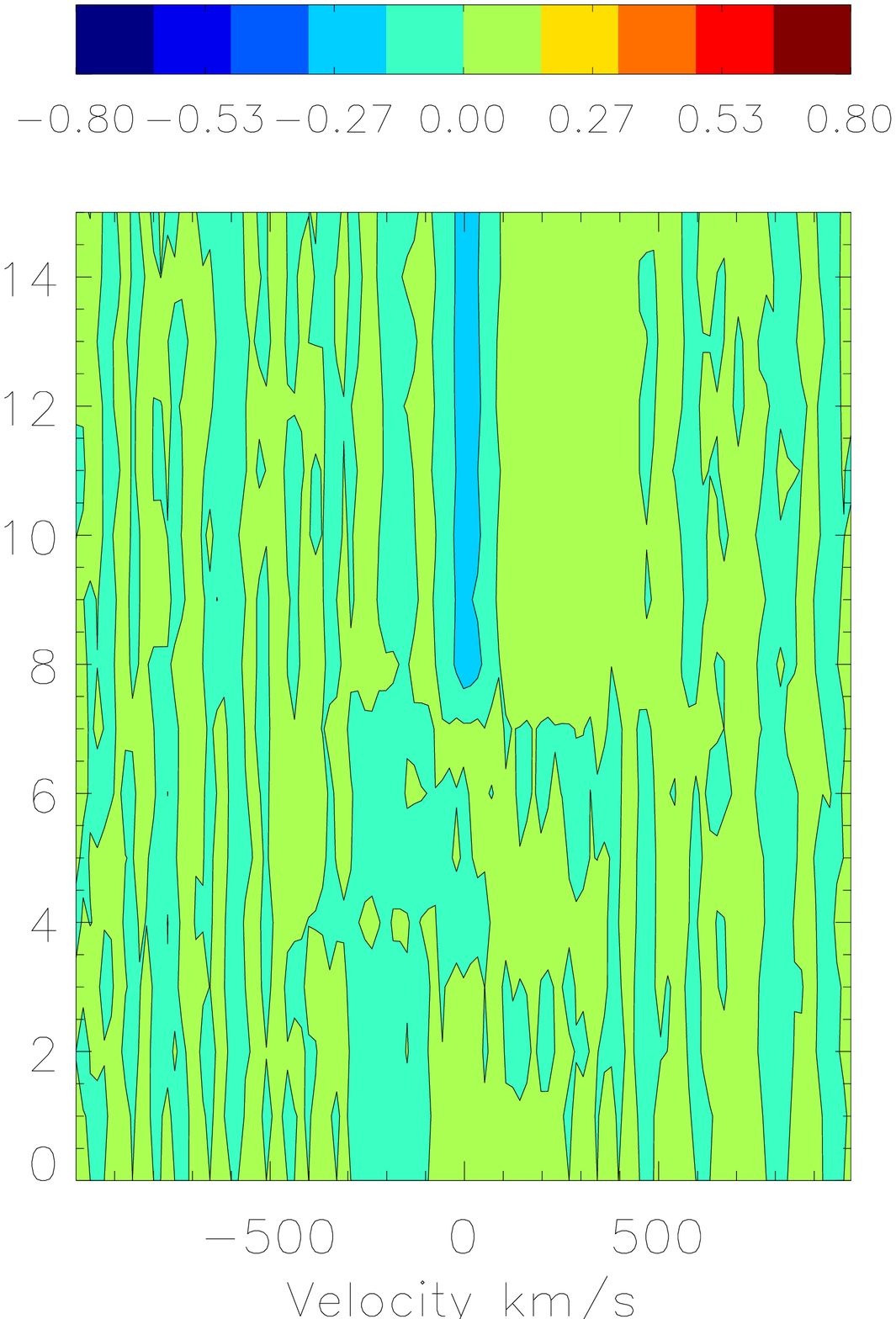} & \includegraphics[scale=0.22]{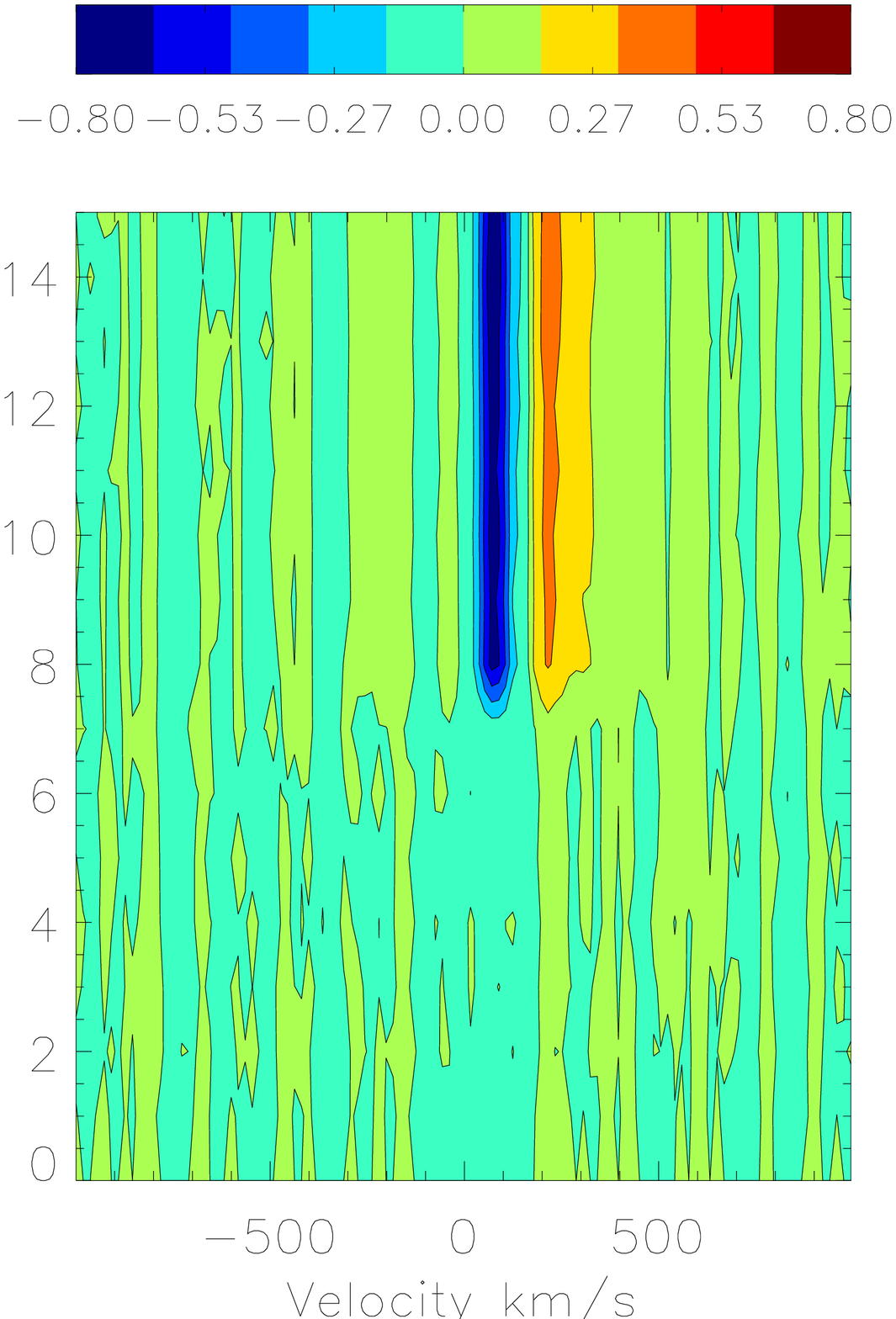} \\
\includegraphics[scale=0.22]{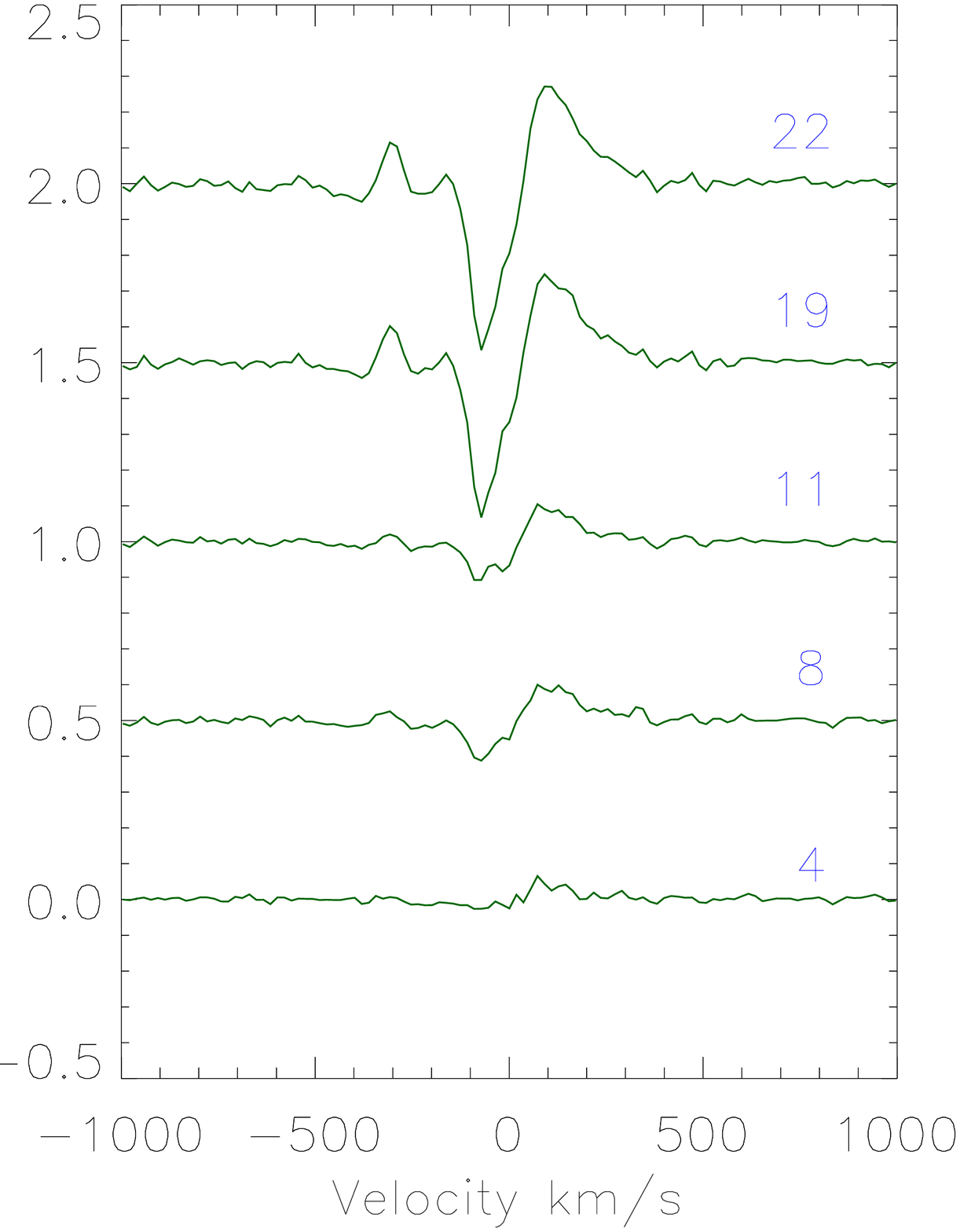} & \includegraphics[scale=0.22]{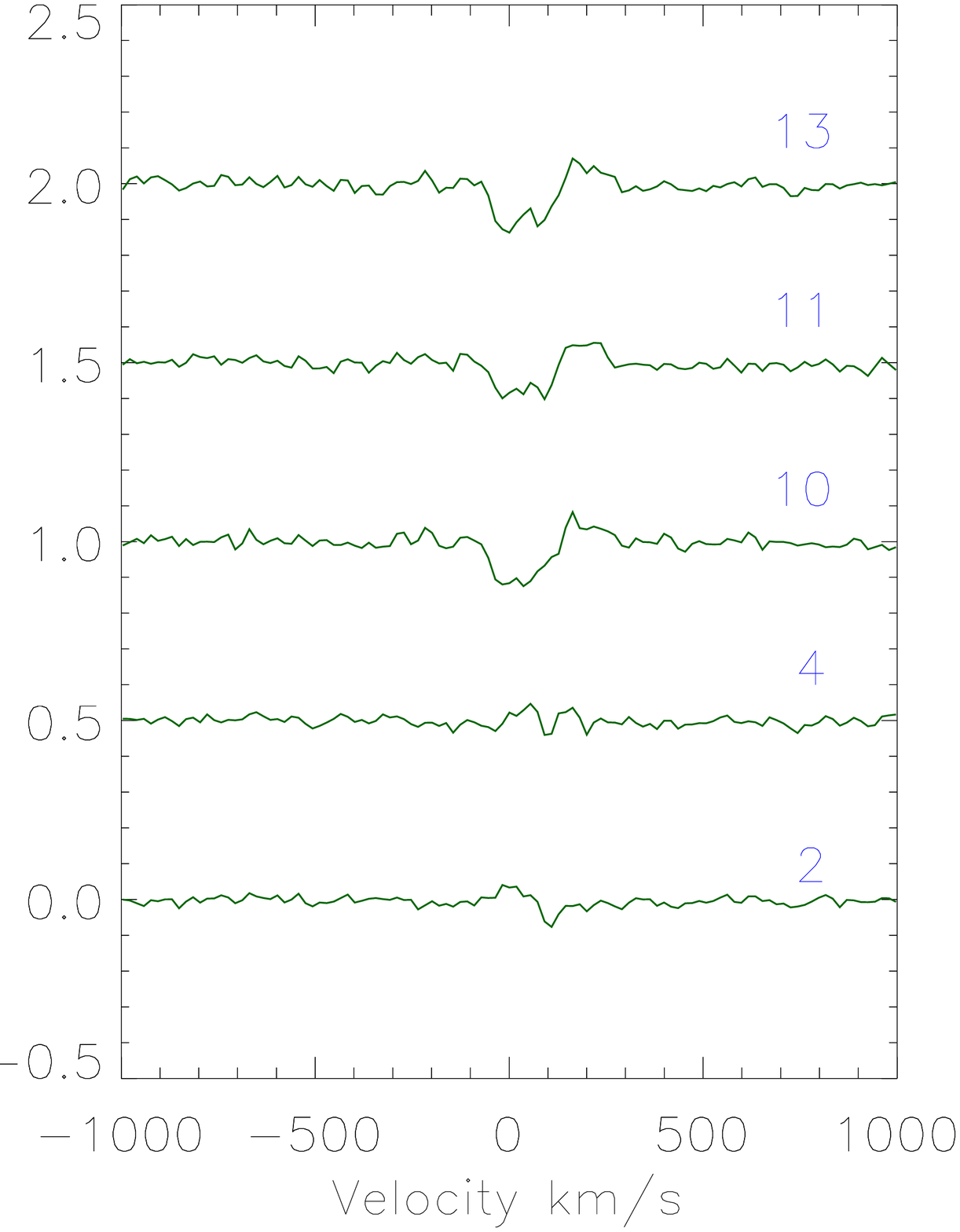} & \includegraphics[scale=0.22]{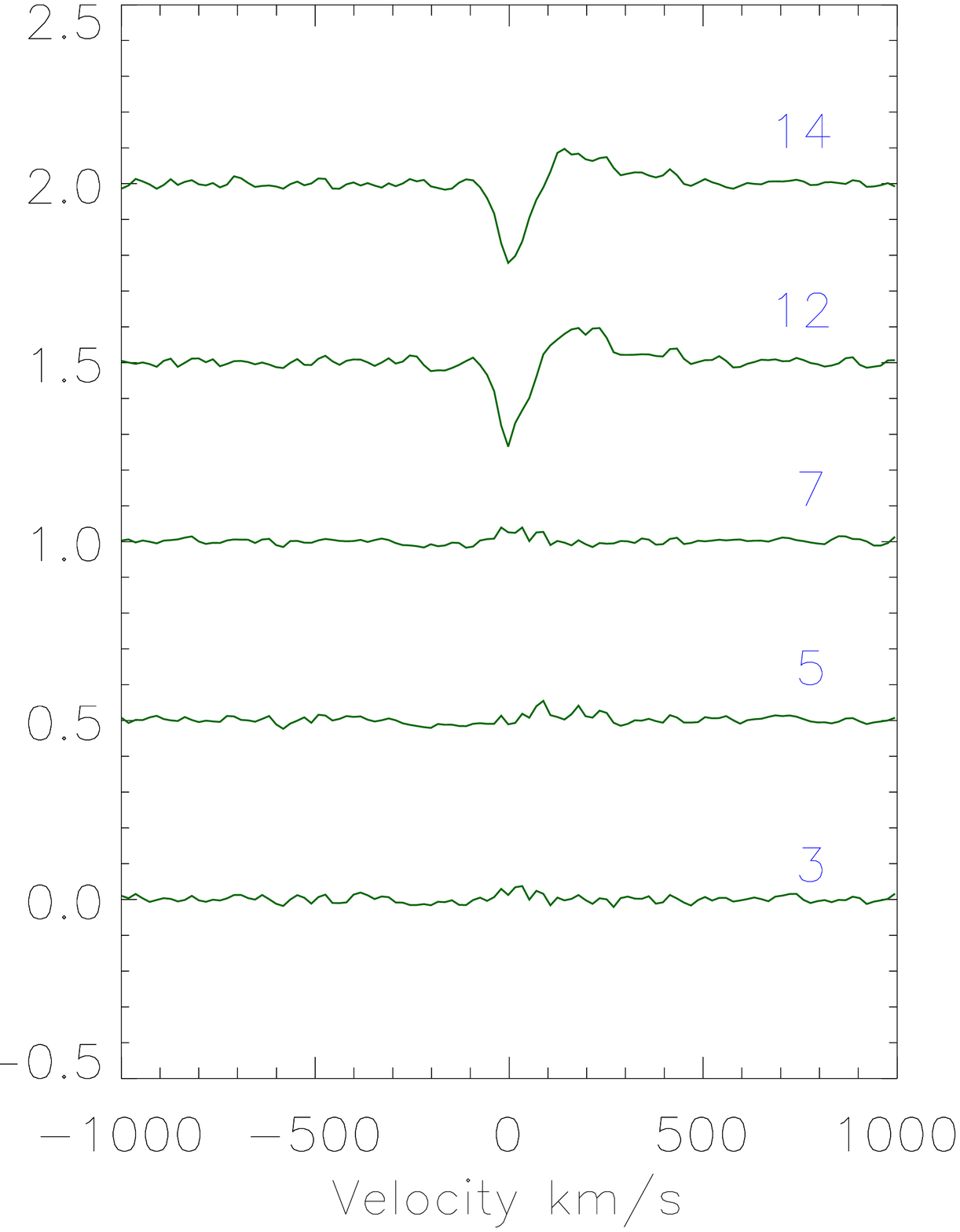} & \includegraphics[scale=0.22]{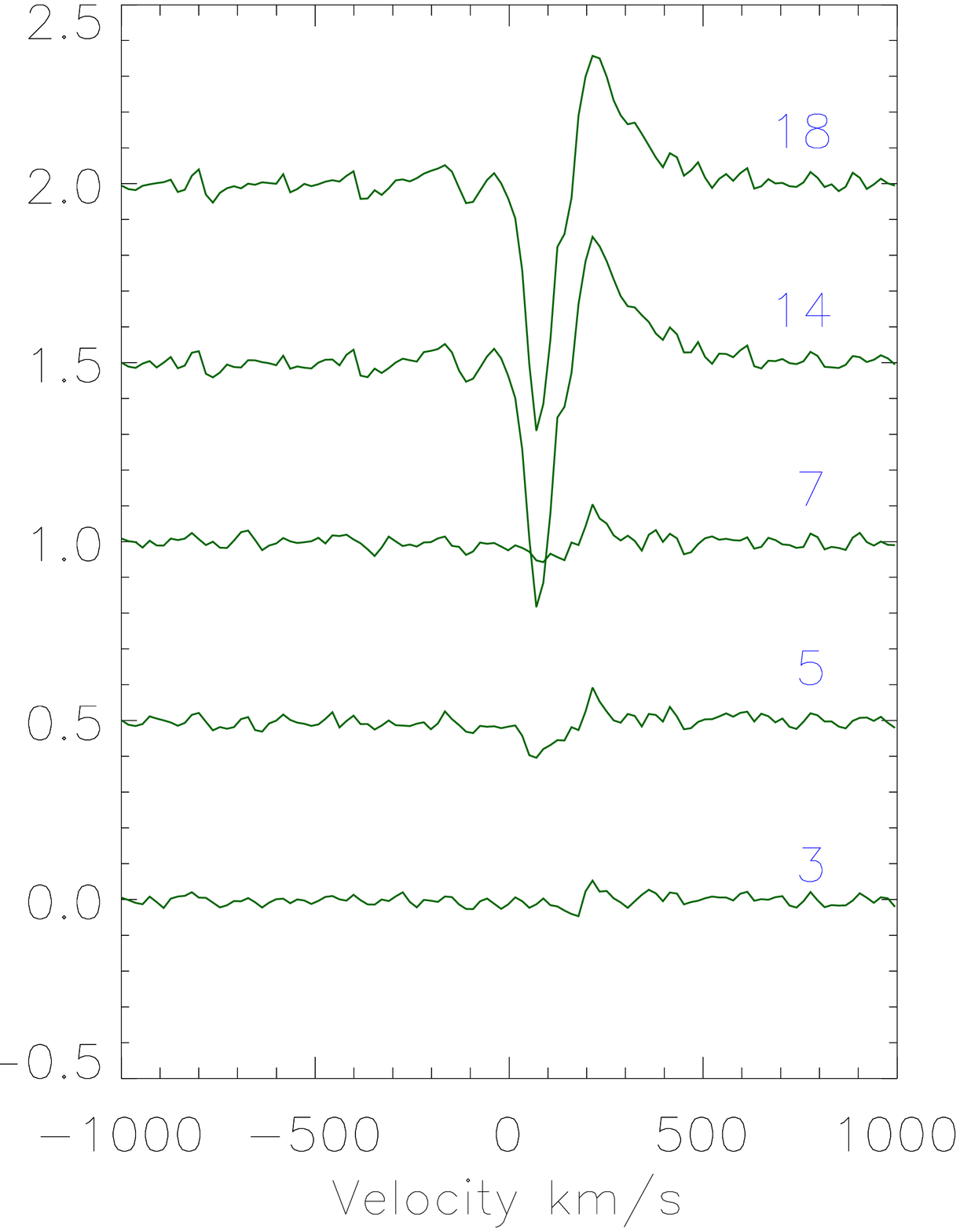}\\
\end{tabular}
\caption{MWC 480 2001 observations. }
\label{fig:MWC480_2001_1}
\end{figure*}

\noindent \textbf{ISIS H$\alpha$ Observations:}  The H$\alpha$ emission is very stable over the three observations blocks that span two nights (Fig.\,\ref{fig:MWC480_2001_1}). There is large blue-shifted absorption that extends below the level of the continuum. The small variations during the first and last observation block take place mainly in the the blue side of the emission peak. Measurements of the 10\%w are not taken due to the fact that the entire blue wing of the emission line is in absorption providing no way of telling (without interpolation) where the blue wing lies.

\begin{figure*}
\centering
\begin{tabular}{ccc}
\includegraphics[scale=0.37]{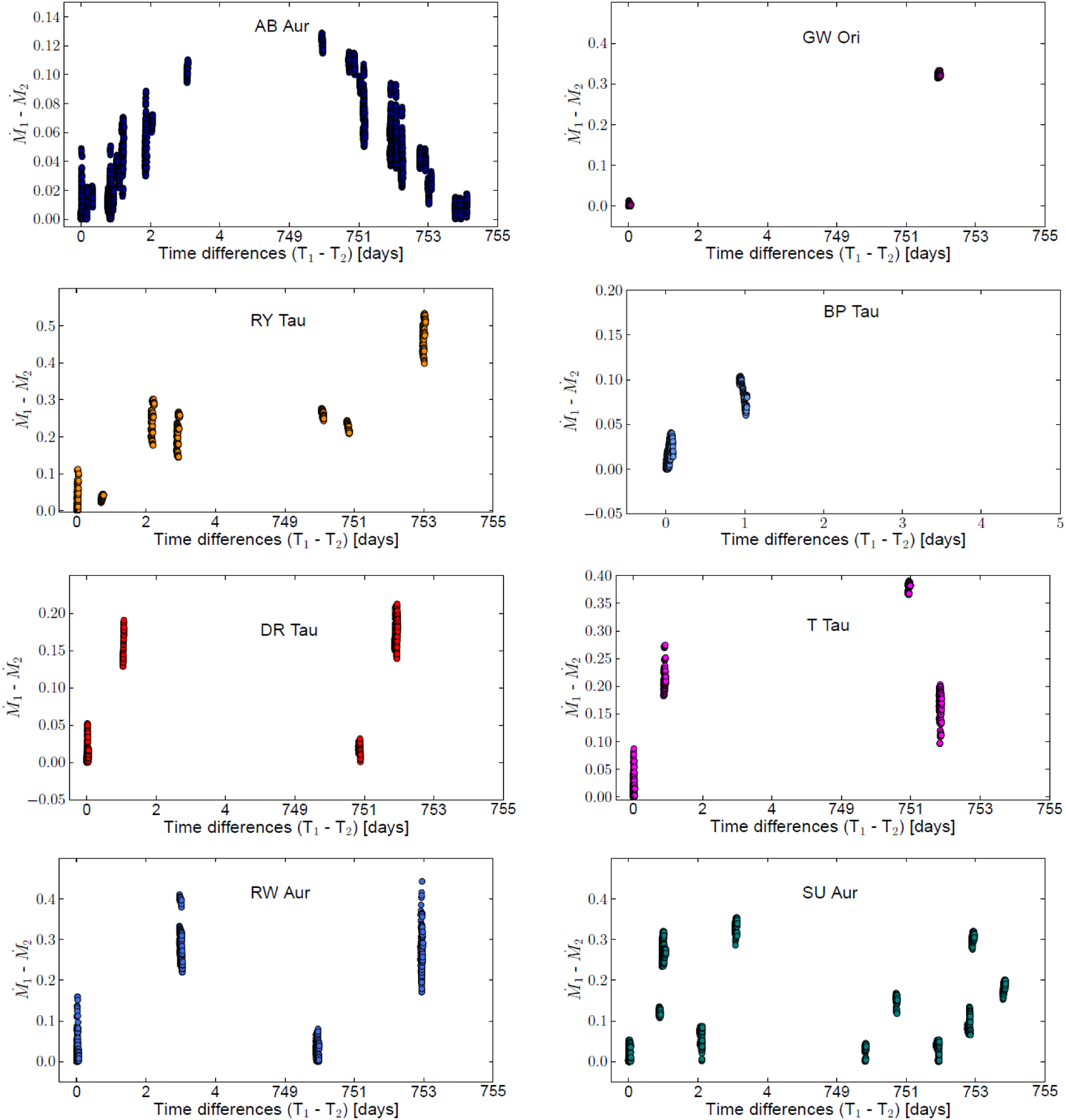} \\

\end{tabular}
\caption{Difference between two accretion rate measurements~\lbrack Log(M$_{\odot}$ yr$^{-1}$)\rbrack~versus their time difference \lbrack days\rbrack. This is done for every measurement for each object in order to cover all the possible time-scales in the sample. The same plot can be seen  in Fig.\,\ref{fig:accretion_timescales_mean} where the mean accretion rate difference in each time bin is plotted for half the sample. }
\label{fig:accretion_timescales_all_pointsA}
\end{figure*}

\begin{figure*}
\centering
\begin{tabular}{ccc}
 \includegraphics[scale=0.37]{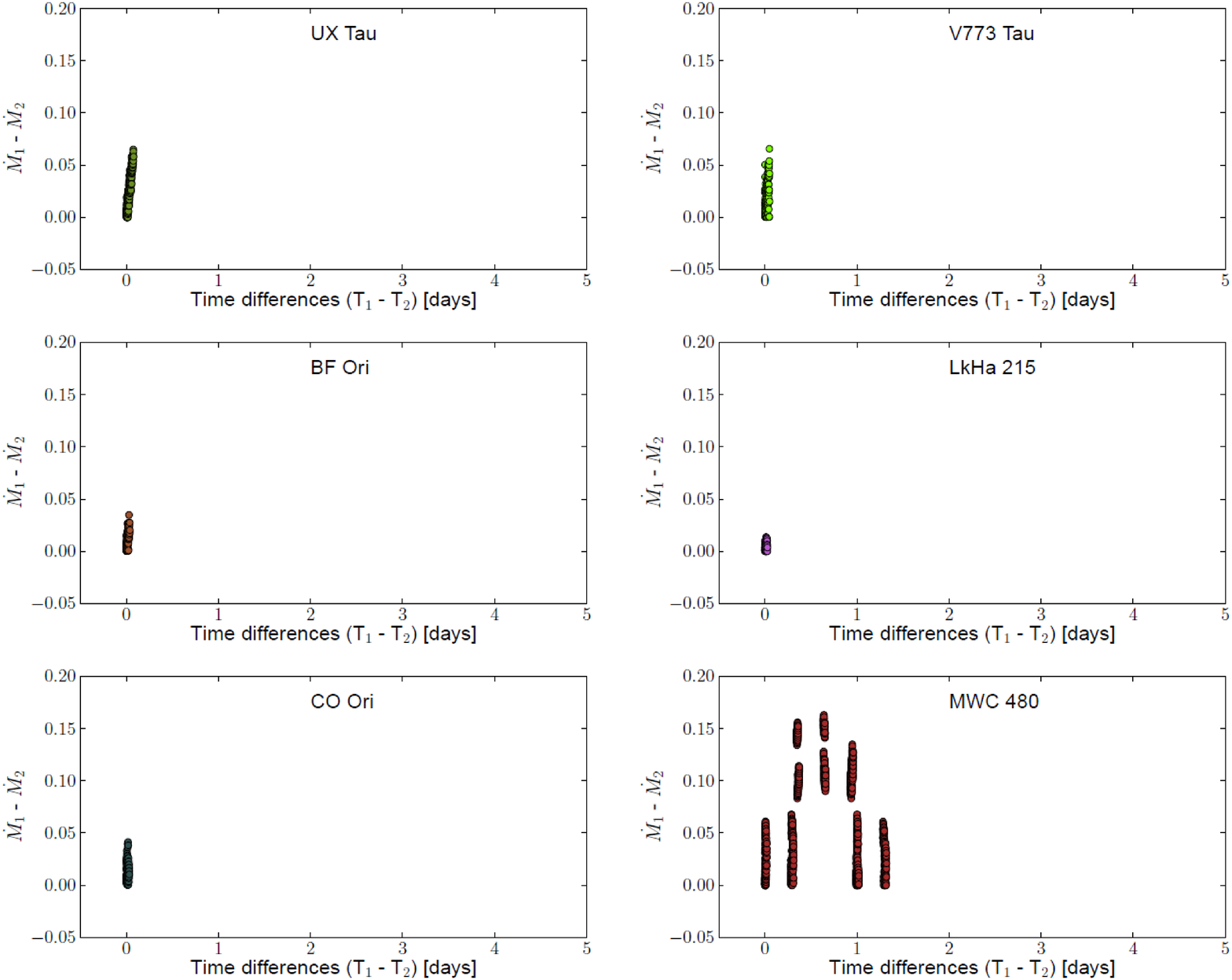}  \\
\end{tabular}
\caption{Continuation of Fig.\,\ref{fig:accretion_timescales_all_pointsA}. Difference between two accretion rate measurements~\lbrack Log(M$_{\odot}$ yr$^{-1}$)\rbrack~versus their time difference \lbrack days\rbrack. See Fig.\,\ref{fig:accretion_timescales_all_pointsA} for full caption. }
\label{fig:accretion_timescales_all_pointsB}
\end{figure*}

\newcommand\aj{AJ} 
\newcommand\actaa{AcA} 
\newcommand\araa{ARA\&A} 
\newcommand\apj{ApJ} 
\newcommand\apjl{ApJ} 
\newcommand\apjs{ApJS} 
\newcommand\aap{A\&A} 
\newcommand\aapr{A\&A~Rev.} 
\newcommand\aaps{A\&AS} 
\newcommand\mnras{MNRAS} 
\newcommand\pasa{PASA} 
\newcommand\pasp{PASP} 
\newcommand\pasj{PASJ} 
\newcommand\solphys{Sol.~Phys.} 
\newcommand\nat{Nature} 
\newcommand\bain{Bulletin of the Astronomical Institutes of the Netherlands}
\newcommand\memsai{Mem. Societa Astronomica Italiana}

\bibliographystyle{mn2e_new}
\bibliography{wht.bib}

\label{lastpage}

\end{document}